# *Plasmodium falciparum* Hsp70-x:

# A Heat Shock Protein at the Host - Parasite Interface

Rowan Hatherley[1], Gregory L. Blatch[2] and Özlem Tastan Bishop[1,3*]

[1]Rhodes University Bioinformatics (RUBi), Department of Biochemistry, Microbiology and Biotechnology, Rhodes University, Grahamstown, 6140, South Africa

[2]College of Health & Biomedicine, Victoria University, Melbourne, Victoria 8001, Australia

[3]Biological Sciences and Bioengineering, Faculty of Engineering and Natural Sciences, Sabanci University, Istanbul 34956, Turkey

Running title: *Plasmodium falciparum* Hsp70-x

*Address correspondence to this author at Rhodes University Bioinformatics (RUBi), Department of Biochemistry, Microbiology and Biotechnology, Rhodes University, Grahamstown, 6140, South Africa; Tel: +27-46-603-8956, Fax: +27-46-622-3984. E-mail: o.tastanbishop@ru.ac.za

# Abstract

*Plasmodium falciparum* 70 kDa heat shock proteins (PfHsp70s) are expressed at all stages of the pathogenic erythrocytic phase of the malaria parasite lifecycle. There are six PfHsp70s, all of which have orthologues in other plasmodial species, except for PfHsp70-x which is unique to *P. falciparum*. This paper highlights a number of original results obtained by a detailed bioinformatics analysis of the protein. Large scale sequence analysis indicated the presence of an extended transit peptide sequence of PfHsp70-x which potentially directs it to the endoplasmic reticulum (ER). Further analysis showed that PfHsp70-x does not have an ER-retention sequence, suggesting that the protein transits through the ER and is secreted into the parasitophorous vacuole (PV) or beyond into the erythrocyte cytosol. These results are consistent with experimental findings. Next, possible interactions between PfHsp70-x and exported *P. falciparum* Hsp40s or host erythrocyte DnaJs were interrogated by modeling and docking. Docking results indicated that interaction between PfHsp70-x and each of the Hsp40s, regardless of biological feasibility, seems equally likely. This suggests that J domain might not provide the specificity in the formation of unique Hsp70-Hsp40 complexes, but that the specificity might be provided by other domains of Hsp40s. By studying different structural conformations of PfHsp70-x, it was shown that Hsp40s can only bind when PfHsp70-x is in a certain conformation. Additionally, this work highlighted the possible dependence of the substrate binding domain residues on the orientation of the α-helical lid for formation of the substrate binding pocket.

# Introduction

Recent efforts at combating malaria have led to the number of annual deaths caused by this disease finally falling below the one million mark (Campbell & Steketee, 2011). Yet, the control of malaria is limited by resistance of the parasite to available drugs; thus there is continuous research to identify new drug targets and validate potential antimalarial compounds. Recently, heat shock proteins (Hsps) have received much attention as potential drug targets (Akide-Ndunge *et al.*, 2009; Banumathy, Singh, Pavithra, & Tatu, 2003; Chiang *et al.*, 2009; Kumar, Musiyenko, & Barik, 2003; Ramya, Surolia, & Surolia, 2006; Shonhai, 2010). There is evidence that despite the high levels of conservation, Hsps from the same family have sufficient structural and functional differences to be selectively targeted (Pesce, Cockburn, Goble, Stephens, & Blatch, 2010). Furthermore, highly conserved Hsps are often essential for cell survival, and therefore likely to evolve relatively slowly, making them less susceptible to variation under selection pressure. It is these features that make Hsps such as the 70 kDa family (Hsp70) attractive drug targets.

The most virulent malaria parasite, *Plasmodium falciparum (P. falciparum)*, has six Hsp70 (PfHsp70) proteins, three of which are believed to be localised in the cytosol (PfHsp70-1), in the endoplasmic reticulum (PfHsp70-2) and in the mitochondria (PfHsp70-3) of the parasite (Sargeant *et al.*, 2006). In 2012, PfHsp70-x, a second cytosolic protein, was shown to be exported to the host erythrocyte cytosol (Külzer *et al.*, 2012). Phylogenetic analysis of PfHsp70-y and PfHsp70-z, believed to localise in the ER and cytosol, respectively, revealed them to be closely related to Hsp110s of other organisms (Shonhai, Boshoff, & Blatch, 2007). Recently, Muralidharan, Oksman, Pal, Lindquist, and Goldberg (2012) characterised PfHsp70-z (named PfHsp110c), showing its role in suppressing protein aggregation during heat stress. Furthermore, knock-down of the PfHsp70-z gene left the parasite unable to survive at high temperatures, making this protein an attractive drug target. PfHsp70-1 has been biochemically characterised and is mainly localised in the cytosol and nucleus of the parasite. It has been shown to display chaperone functions, in both *in vitro* and *in vivo* assays (Matambo, Odunuga, Boshoff, & Blatch, 2004; Shonhai, Boshoff, & Blatch, 2005; Shonhai, Botha, de Beer, Boshoff, & Blatch, 2008). PfHsp70-1 is a ubiquitous, highly abundant protein, expressed at all stages of the erythrocytic phase of the parasite's lifecycle (Shonhai *et al*., 2007). PfHsp70-1 has been considered as a potential drug target, and small molecule inhibitors specific for PfHsp70-1 have recently been identified (Cockburn *et al*., 2011).

PfHsp70-x, with 74% sequence identity to PfHsp70-1, was first detected in 1989 (Kun & Muller-Hill, 1989), as a clone which reacted positively to sera from malaria immune patients during immunological screening. Yet, the expression of the protein was only confirmed in 2009 (Acharya *et al*., 2009). The findings suggested that the expression of PfHsp70-x is linked to the pathogenic stages of the parasite's life cycle, as it had never been detected in laboratory strains. Recently, PfHsp70-x was shown to be partly exported to the host erythrocyte, and partly present in the parasitophorous vacuole (PV) (Külzer *et al*., 2012), the parasite's membrane bound compartment, which is an important passageway to transport parasite encoded proteins into the host erythrocyte (Lingelbach & Joiner, 1998). It was reported that there are 49 PfHsp40s encoded on the *P. falciparum* genome, of which 18 are predicted to be exported into the erythrocyte (Njunge, Ludewig, Boshoff, Pesce, & Blatch, 2013). Two of the PfHsp40s, PfA and PfE, (PFA0660w and PFE0055c respectively) shown to be exported (Külzer *et al*., 2010), have also been found to occur in a common complex with PfHsp70-x (Külzer *et al*., 2012). Generally, the function of Hsp70 is largely linked to its ability to bind and release peptide substrates in an ATP-dependent cycle with help of Hsp40s and nucleotide exchange factors (NEFs) (Dragovic, Broadley, Shomura, Bracher, & Hartl, 2006; Harrison, Hayer-hartl, Liberto, Hartl, & Kuriyan, 1997; Hartl & Hayer-Hartl, 2002; Mayer & Bukau, 2005).

Hsp70s are structurally divided into two major domains: the 45 kDa N-terminal ATPase domain, which regulates the 25 kDa C-terminal substrate binding domain (SBD). The SBD is further sub-divided into a β-domain, containing the substrate binding pocket, and an α-helical lid, located closest to the C-terminus of the protein (Strub, Zufall, & Voos, 2003; Wang, Chang, & Wang, 1993). The SBD and the ATPase domain are joined by a flexible linker region, believed to facilitate inter-domain communication which has been found to be essential to Hsp70 function (Han & Christen, 2001).

At the interface of host - parasite interaction, PfHsp70-x has an important role, yet it is by far the less well-studied of the two cytosolic PfHsp70s, and therefore, it is important to analyse PfHsp70-x in detail. The interest of this paper is to investigate whether PfHsp70-x is a potential drug target by performing a detailed comparative bioinformatics analysis. The study was divided into two parts: 1) *Large scale sequence analysis:* PfHsp70-x was compared with other Hsp70s of *Plasmodium*, Apicomplexa and eukaryotes. To overcome the limitations of handling large data, we proceeded with an unconventional method of using consensus sequences, to simplify the alignments and make it easier to compare different groupings of sequences, as well as identify highly conserved sites. The investigation included sequence features of PfHsp70-x that are predicted to be involved in the interaction with associated proteins and peptide substrates during the ATPase cycle. 2) *Structural analysis:* PfHsp70-x was modeled with J domains of potential *Plasmodium* and human Hsp40 (PfHsp40 and HsHsp40 respectively) co-chaperons. Docking studies were also performed and compared to the homology models. Both aimed at an attempt to better understand the interactions that would aid with targeting the complexes towards anti-malarial drug design. To the authors' knowledge, docking and modeling studies for PfHsp70-x-J-domain complex calculations were not done previously. Further, PfHsp70-x was modeled in three different conformations representing different stages of ATPase cycle to understand the conformational dynamics of the protein. A number of novel results were obtained, including analysis of the Hsp70-Hsp40 complexes in the dynamics of ATPase cycle, for the first time.

# Methodology

*Data Retrieval*

The sequences of PfHsp70-1, PfHsp70-x, PfHsp70-2 and PfHsp70-3 (accession numbers: PF08_0054, MAL7P1.228, PFI0875w and PF11_0351 respectively) were retrieved from PlasmoDB (The Plasmodium Genome Database Collaborative, 2001). Using PfHsp70-1 and PfHsp70-x, PfHsp70-2 and PfHsp70-3, cytosolic (C-Hsp70), the endoplasmic reticulum (ER-Hsp70) and the mitochondrial (Mt-Hsp70) Hsp70s were searched respectively by NCBI-BLAST (Altschul *et al.*, 1997). An additional 89 sequences were retrieved, making up prokaryotic, eukaryotic, apicomplexan and plasmodial groups, for each type of Hsp70 (see S1). Sequence analyses included 13 eukaryote (non-apicomplexan), 9 apicomplexan (non-plasmodial) and 4 plasmodial (non-*falciparum*) C-Hsp70s; 12 eukaryote (non-apicomplexan), 6 apicomplexan (non-plasmodial), 4 plasmodial (non-*falciparum*) ER-Hsp70s; 15 eukaryote (non-apicomplexan), 8 apicomplexan (non-plasmodial), 3 *Plasmodium* (non-*falciparum*) Mt-Hsp70s as well as 15 prokaryote.

*Large Scale Sequence Analysis*

Sequence alignments were done using MAFFT (Katoh, Misawa, Kuma, & Miyata, 2002) and PROMALS3D (Pei, Kim, & Grishin, 2008). Alignments were performed using representative sequences from eukaryotes, Apicomplexa and *Plasmodium*, for each type of Hsp70. The results of the two alignment programs were compared. In both cases, residues were aligned identically, except highly un-conserved N-terminal and C-terminal regions (see S2). To be able to visualize and analyse the alignment results easily, consensus sequences were calculated. The sets of sequence alignments for Hsp70s from which the consensus sequences are calculated are given in S3A-S3J. The consensus sequences were then re-aligned for each type of Hsp70 with representatives of that Hsp70 type from *P. falciparum*. Alignments were visualised using BioEdit (Hall, 1999). Identity calculations were performed for all possible sequence pairs in each alignment. The lowest sequence identity pair for each alignment was then recorded (Table 1). All the large scale calculations were done by Python scripting (van Rossum & de Boer, 1991).

*Homology Modeling*

Homology modeling was performed for **a)** PfHsp70-x in three different conformations using three different templates (PDB IDs: 1YUW, 2KHO and 3D2F), identified using HHpred (Söding, 2005) **b)** The J domains of nine different Hsp40 co-chaperons, which included three Type II PfHsp40s (PF3D7_0113700/PFA0660w, PF3D7_0201800/PFB0090c and PF3D7_0501100.1/PFE0055c, denoted as PfA, PfB and PfE, respectively), two Type I human Hsp40s (HsDnaJA1 and HsDnaJA2), two Type II human Hsp40s (HsDnaJB1 and HsDnaJB4) and two Type III human Hsp40s (HsDnaJC5 and HsDnaJC13). Templates with the following PDB IDs were used for modeling: PfA - 1HDJ, PfB - 2CTR, PfE - 2CTP, HsDnaJA1 - 2O37, HsDnaJA2 - 1HDJ, HsDnaJB1 – 1HDJ, HsDnaJB4 – 1HDJ, HsDnaJC5 – 2CTW and HsDnaJC13 – 2CTW.

Alignments of template sequences to target sequences were done using PROMALS3D (Pei *et al.*, 2008). The alignment results for PfHsp70-x modeling were compared to the results of the large scale alignment, to make sure that they are accurate and if necessary, hand adjustments were made. Modeling was performed using MODELLER 9v7 (Sali & Blundell, 1993), with slow refinement. For each protein, 100 models were built. The best three models were selected by DOPE Z score calculations. These models were further evaluated using MetaMQAPII (Pawlowski, Gajda, Matlak, & Bujnicki, 2008). Final evaluations were performed using MetaMQAPII, Verify3D (Bowie, Luthy, & Eisenburg, 1991) and ProSA (Sippl, 1993; Wiederstein & Sippl, 2007). Model quality assessment results are presented in S4.

*Calculation of Protein Complexes*

Hsp70-x-Hsp40 protein complexes were calculated by two different approaches; homology modeling and docking. For homology modeling, a hybrid modeling approach was used (Tastan Bishop & Kroon, 2011). Due to the low sequence identity between auxilin and the different Hsp40s being studied, a set of hybrid templates were created for modeling the Hsp70-Hsp40 complexes. First, each Hsp40 J domain was modeled separately as monomers, as described above. Each J domain was then superimposed with auxilin of template structure (PDB ID 2QWO) to replace auxillin, using PyMOL (Schrödinger, LLC). These hybrid structures were used as templates for modeling PfHsp70-x and each J protein complexes, as described above. The resulting models underwent residue repacking and were then put through 100 cycles of energy minimisation, both done using PyRosetta (Chaudhury, Lyskov, & Gray, 2010).

For molecular docking studies, the HADDOCK web server was used (de Vries, van Dijk, & Bonvin, 2010). Each J domain model was submitted, along with PfHsp70-x, to the program. Residues actively involved in binding were set as those forming the HPD motif of each J protein and residues R201, N204, T207, I246 and V419 of PfHsp70-x. No passive residues were set for each docking run. The best docked structures were ranked by HADDOCK score, the weighted sum of the van der Waals energy, electrostatic energy, desolvation energy, the energy from restraint violations and the buried surface area.

*Identification of Important Residues*

Important residues were identified from large scale sequence analysis, as well as biochemical data from literature. The Protein Interactions Calculator (PIC) webserver (Tina, Bhadra, & Srinivasan, 2007) was used to identify residue interactions in each protein complex. Default settings were used when submitting models to the PIC server.

Further, SBD residues in contact with α-helical lid of PfHsp70-x, modeled after template 1YUW, were also identified as follows: The SBD and α-helical lid of this PfHsp70-x model were saved to a PDB file as two separate molecules, using PyMOL. Thereafter, a python script was run, which gave each molecule a different chain identifier. This PDB file was then submitted to the PIC webserver as mentioned above in order to identify interactions between the two sub-domains. These contact residues of the SBD were then compared to those in *E. coli* DnaK, that Zhu *et al*. (1996) identified as interacting with the substrate during peptide binding.

## Results and Discussion

*PfHsp70-x is Atypical of Cytosolic Hsp70s*

The initial part of the study was focused on large scale sequence analysis. The use of consensus sequences derived from large scale sequence alignments was found to be very useful for comparing the conservation of sequence features in different groups of related proteins. It also made it easy to identify sequence features of a protein that were distinct from features of other proteins that are supposedly related. PfHsp70-x and PfHsp70-1 were aligned with three consensus sequences of plasmodial, apicomplexan, and eukaryotic C-Hsp70s. Also included in this alignment was a consensus sequence of bacterial and archael DnaKs (Figure 1).

An initial notable feature in this alignment is that PfHsp70-x contains an extended N-terminal region that is 18 residues longer than any other C-Hsp70. An extended N-terminus is indicative of a transit peptide, which targets proteins to specific organelles (Patron & Waller, 2007). This is unusual, as C-Hsp70s usually do not have an N-terminal transit peptide sequence (Renner & Waters, 2007). It can also be seen that the *Plasmodium* consensus sequence contains a highly conserved, extended N-terminal region (Figure 1). This sequence differs to that of the PfHsp70-x. The signal peptide sequence responsible for targeting proteins to the ER has been described by von Heijne (1990) as having the following characteristics. Firstly, a positively-charged N-terminal region of up to five residues in length, followed by a hydrophobic patch of up to 15 residues in length, ending off with a more polar set of residues. PfHsp70-x contains two positively-charged residues (Lys2 and Lys4) within the first five residues of its sequence. This is followed by an unusual stretch of polar residues (Cys6, Ser7, Tyr8, His10 and Tyr11) before a stretch of eight hydrophobic residues. This region ends with 12 of the next 15 residues being polar. An alignment of PfHsp70-x with the transit peptides described by Renner & Waters (2007), as well as two known ER-Hsp70s (human Grp78 and PfHsp70-2) (see S5), indicates that this region of PfHsp70-x has a similar distribution of amino acids to the transit peptides of these ER-Hsp70s. This indicates that the extended N-terminus of PfHsp70-x potentially directs it to the ER. Recent experimental data has confirmed that the first 24 residues of PfHsp70-x play a role as an ER-type signal sequence allowing secretion into the PV (Külzer *et al.*, 2012). The subsequent eight residues are for secretion into host cell cytosol. Although it is indicated as a unique sequence in the paper, the large scale sequence analysis showed that 3 residues (Asn28, Glu31, Ser32) of this octameric export signal is conserved among *Plasmodium* species and further residue Ala25 conserved between *P. falciparum* and *P. knowlesi* (S2-S3). Furthermore, the fact that PfHsp70-x does not have an ER-retention sequence (e.g. the SDEL found at the C-terminus of PfHsp70-2/PfBiP), is consistent with findings that suggest it transits through the ER and is secreted into the PV or beyond into the erythrocyte cytosol.

In Figure 1, the residues highlighted in dark blue indicate those found only in PfHsp70-x at their given positions. PfHsp70-x contains 123 (18%) of these residues. PfHsp70-1, by comparison, only contains 14 (2%) (highlighted in red). PfHsp70-x also contains 32 (5%) residues that are conserved in eukaryotic, but not found in other plasmodial sequences (highlighted in yellow). PfHsp70-1 contains 2 (0.3%) of such residues. The C-terminal region of the C-Hsp70s has an unconserved, glycine-rich region, often containing at least one GGMP motif. A GGMP repeat is rare in non-parasitic Hsp70s (Matambo *et al.*, 2004; Wiser, Jennings, Lockyer, van Belkum, & van Doorn, 1995) and occurred at most once in the eukaryotic C-Hsp70s considered in this research. In the apicomplexan C-Hsp70s considered, there were several repeats of this motif, most prominent in plasmodial sequences. In PfHsp70-x, however, this region is not glycine-rich and does not contain any GGMP motifs. There is a stretch of 24 residues in the C-terminal region that share no resemblance to any of the other cytosolic proteins. Another interesting feature of PfHsp70-x is the carboxy-terminal EEVN motif, which differs to the EEVD motif, characteristic of cytosolic eukaryotic Hsp70s (Shonhai *et al.*, 2007).

Although PfHsp70-x appears to be cytosolic-like, these differences as described above indicate that it may have functions that differ to other C-Hsp70s. Thus, the analysis was extended to include all ER and Mt sequences (see S6). An indication of the level of conservation between the sequences aligned is given in Table 1, which reports the lowest sequence identity pair for each alignment. Across all alignments, each sequence was at least 50% identical to every other that was studied. The N-terminal extension of PfHsp70-x is approximately the same length as the transit peptides found in the ER-Hsp70 sequences analysed. PfHsp70-x was found to be more similar to C-Hsp70s than both ER-Hsp70s and Mt-Hsp70s. Based on the alignment, it seems that PfHsp70-x is likely to be cytosolic in origin, but is not under the same evolutionary constraints as other C-Hsp70s.

### *PfHsp70-x Displays Similar Interactions with Human and Parasitic Hsp40s*

Hsp70-Hsp40 protein complexes have been investigated as potential drug targets in various diseases including malaria (Shonhai, 2010). There has been recent evidence which indicates that PfHsp70-x localises with PfA and PfE in structures called J dots, found within the host cytosol (Külzer *et al.*, 2012). Additionally, PfB is predicted to be exported from the parasite (Botha, Pesce, & Blatch, 2007). In this paper, the interaction of PfHsp70-x with these three Type II *P. falciparum* Hsp40 co-chaperones, as well as two of each Type I, Type II and Type III host Hsp40s, was investigated at a structural level using protein complexes, calculated through homology modeling and structural optimisation. The two host Type II Hsp40s (HsDnaJB1 and HsDnaJB4) and Type III Hsp40s (HsDnaJC5 and HsDnaJC13) have both been found to be expressed within the erythrocyte (Pasini *et al.*, 2006; van Gestel *et al.*, 2010). HsDnaJA1 and HsDnaJA2 form a type of negative control, since there is no record of these proteins being expressed in the erythrocyte cytosol. Although, the absence of these proteins from the erythrocyte cytosol does not necessarily mean they are incapable of interacting with PfHsp70-x. Due to the low sequence identity between the auxilin J domain of the template structure, 2QWO, a set of hybrid templates were made to model each complex as explained in the methodology section. The structures were then optimised using PyRosetta, before interactions were analysed. Figure 2 shows the modeled interaction interface between PfHsp70-x and a J domain, and how the interactions of the nine sets of J domain proteins differed in the way they potentially interact with PfHsp70-x. A number of consistent interactions were observed (see S7). L411 of PfHsp70-x formed hydrophobic interactions with P12, part of the HPD motif (refer to Figure 2C, S7B and S8), in all the J domains studied. Residues L200 and R201 of PfHsp70-x formed hydrogen bonds, mainly with D13 in most proteins and residues A418 and V419 of PfHsp70-x commonly formed hydrogen bonds with K29 of the J domain. A more detailed account of these interactions are tabulated in the supplementary data S7A and S7B. These findings are mostly consistent with Jiang *et al.* (2007), however, they had reported that L200 should interact with H11 of the J domain, which was only seen in HsDnaJC5 and HsDnaJC13. This shift could be due to the variation in residues around the HPD motif of the J domain of auxilin, when compared to *P. falciparum* and human Hsp40s. An ionic interaction was predicted to occur between K420 of PfHsp70-x and acidic residues of the J domains. This was most consistently found to occur with the residue corresponding with E25 of the *P. falciparum* Hsp40s and the immediately adjacent acidic residue of the Type I human Hsp40s, HsDnaJA1 and HSDnaJA2. This interaction did not occur for either Type III host DnaJ. Ionic interactions were also predicted between K189 and R201 of PfHsp70-x and D13 of the J domain. These predicted interactions have not been previously reported. The data suggest that PfHsp70-x is capable of interacting with all three exported Hsp40s and the set of interactions with the human Hsp40s suggests that these may also interact with PfHsp70-x in a similar manner.

The structures solved by Jiang *et al.* (2007) were determined by cross-linking the residues equivalent to R201 of PfHsp70-x and the aspartic acid of the J domain HPD motif. This was done because work by Suh *et al.* (1998) revealed that these residues are likely to interact during J domain binding. It was mentioned, however, that it was possible that the J domain could freely rotate around this covalent bond when the structure was solved and that a non-functional orientation was calculated. As such, further analysis was conducted by docking each J domain to the ATPase domain of PfHsp70-x. The active residues given to guide the docking were those previously determined to be important to J domain binding. The results presented in Figure 3 indicate the top docking result obtained for each J protein. These reveal a number of orientations that fit the criteria of binding to R201, N204 and T207 (Suh *et al.*, 1998), as well as I246 (Jiang, Prasad, Lafer, & Sousa, 2005) of PfHsp70-x. Remarkably, in all orientations, the residue D13 of the HPD motif is positioned to interact with R201, N204 and T207 and each orientation rotates around this interaction (see S9C). All docked models had good hydrophobic and electrostatic interaction energies, determined by HADDOCK when compared to other complexes that were returned. In spite of these different orientations, the local contact areas on PfHsp70-x were fairly consistent. J domain residues which appeared to be most consistently involved in binding included those of the HPD motif, as well as R4, A7, M8, K14 and F28, indicated in Figures 3B and 3C, with all conserved interactions in the supplementary data S7C and S7D. Residues A7, M8 and F28 were common in hydrophobic interactions, mostly with I246 of PfHsp70-x. D13 formed hydrogen bonds with R201, N204 and T207 of PfHsp70-x and ionic interactions with K185 and R201. K14 was also involved in a large number of interactions, forming hydrogen bonds and ionic interactions with D216, E243, D244, E248, K20 and D421 of PfHsp70-x. H11 and P12 were found to form hydrogen bonds with F247 and R201, respectively and R4 was found to form both hydrogen bonds and ionic interactions with the adjacent residues, E243 and D244. In addition to the HPD motif, R4 has been found to be a highly conserved (R or K) and functionally important residue in many Hsp40 proteins of prokaryotic, mammalian and parasitic origin (Berjanskii *et al*., 2000; Genevaux, Schwager, Georgopoulos, & Kelley, 2002; Hennessy, Boshoff, & Blatch, 2005; Nicoll *et al*., 2007).

The six host Hsp40s had a 3-6 residue truncation within their J domains when compared to auxillin and the three Type II PfHsp40s considered in this study. Residues in this small loop region did display some interactions with PfHsp70-x (Figure 2B, Figure 3B). The docked models suggest this region to be more involved in the J domain-PfHsp70-x interaction than the hybrid models do. As this region does vary quite greatly in size and composition across the different J domains studied it is possible that it contributes to the specificity of this interaction. The correct interaction orientation needs to be determined, however, before this can be further investigated. This does also make it difficult to adequately compare the different groupings of Hsp40s used in this study. Although HsDnaJA1 and HsDnaJA2 were least likely to interact with PfHsp70-x (based on type and localisation), the results indicated that these were just as capable as any other Hsp40s tested. The docking results indicated that the two Type III Hsp40s tested (HsDnaJC5 and HsDnaJC13) may not bind as well as the other Hsp40s, since these displayed the fewest interactions after the HPD motif. However, without knowledge of which interactions are important to this binding process, it is difficult to comment on the significance of this difference.

The interactions observed by docking differ substantially from those present in the hybrid models. The latter of which mostly agreed with findings by Jiang *et al*. (2007). The docking results are similar only in the point of contact between the HPD motif and PfHsp70-x. Although, the residue D13 is just buried deeper within the ATPase domain interface, which allows interactions with N204 and T207. It is possible that this orientation was hindered by the mutation and cross-linking performed by Jiang *et al.* (2007), which may further explain the weaker J domain stimulation of BvHsc70 reported for their complex. The present work suggests an interaction interface which agrees more closely with work performed by Suh *et al.* (1998). Overall, an interaction between PfHsp70-x and each of these Hsp40s seems equally likely.

*Hsp40s can only bind when PfHsp70-x is in a Certain Conformation*

To get an idea of which J domain binding orientations were most accurate, the docked structures were superimposed with different conformations of PfHsp70-x, modeled based on templates 1YUW and 2KHO (Figure 4). J domain binding is supposed to regulate the interaction of the SBD and the ATPase domain of Hsp70s (Jiang *et al.*, 2007). All complex models were bound in orientations which clashed with this interaction (Figure 4B; S10A). If this is a competitive interaction, it is possible that the J domain displaces the SBD when interacting with the ATPase domain. Only three of the complex models did not interfere with the SBD of PfHsp70-x in its ADP-bound state (Figure 4C; S10B). These were models of PfHsp70-x complexed with HsDnaJA2, PfB and PfE. All three of these binding orientations involve interactions with the correct PfHsp70-x residues and do not clash with the ADP-bound form of PfHsp70-x.

*Substrate Binding is Dependent on the Orientation of the α-Helical Lid, Relative to the Substrate Binding Domain*

Residues involved in the formation of a substrate binding pocket were identified within PfHsp70-x, based on research done by Zhu *et al.* (1996) and Mayer *et al.* (2000). These residues are identical to those found in PfHsp70-1 and all *Plasmodium* and Apicomplexa C-Hsp70s used in this study. The only residue which differed between PfHsp70-x and the eukaryotic consensus sequence was L434, which was more commonly conserved as isoleucine. When these residues were mapped to the 3D structures of PfHsp70-x, it was revealed that the conformation of these residues changed with the structural conformation of the protein (Figure 5). The substrate binding pocket formed only when the α-helical lid of the SBD was in contact with the β sub-domain. When the lid was orientated away from the β sub-domain, the substrate binding pocket residues disassembled and the pocket closed. Another interesting effect was that the substrate-binding pocket was found to form independent of whether the SBD and ATPase domain were in contact with each other or if they were separated. As shown in (Figure 5C), even in the absence of a peptide substrate, the α-helical lid itself is bound by the SBD. This phenomenon was also found to occur in template 1YUW, used to model PfHsp70-x in this conformation. This further indicates that the orientation of the lid is important to the formation of the substrate binding pocket. The PIC webserver (Tina *et al*., 2007), was used once again, this time to identify which residues in the SBD made contact with the α-helical lid. When compared to residues which have been found to interact with the peptide substrate during substrate binding (Zhu *et al*. 1996), the substrate contact residues of PfHsp70-x include L434, T436, V440, T442, F549, T460, Q466, G468 and I471. The residues found to interact with the α-helical lid, did include some of these residues (V440, T460 and Q466), as well as adjacent residues A437, Q457, P467 and L470. While there is some overlap, it would seem that the lid is not bound as a peptide substrate would be, indicating that it may be merely present in the substrate binding pocket which forms, rather than bound by it. There has been much work done regarding the mechanism of substrate binding in Hsp70/DnaK and the α-helical lid has been found to be important in stabilising substrate binding (Moro, Fernández-Sáiz, & Muga, 2004). It has been shown by Pellecchia *et al*. (2000) that the removal of the α-helical lid from DnaK does not abolish substrate binding or ATP-induced release of substrate. However, research by Mayer *et al*. (2000) revealed that without the α-helical lid, DnaK is unable to refold denatured protein substrates. It has additionally been shown that the lid of DnaK will open to varying degrees to accommodate different types of substrate and will often not enclose the peptide (Schlecht, Erbse, Bukau, & Mayer, 2011). This indicates that the lid does not stabilise substrate binding by closing over the peptide substrate. The present findings for PfHsp70-x may further contribute to the understanding of the Hsp70 substrate-binding mechanism. These indicate that the residues of the SBD form a substrate binding pocket, based on the position of the alpha-helical lid, relative to the SBD. This would suggest that the role of the α-helical lid in substrate binding is an interaction with the β subdomain of the SBD, rather than the substrate itself.

# Conclusion

PfHsp70-x has been found to be a highly unusual protein, when compared at an amino acid sequence level to other cytosolic Hsp70s even from the *Plasmodium* genus. Although the fact that PfHsp70-x is exported into the host erythrocyte suggests a possible interaction with exported *P. falciparum* Hsp40s or host erythrocyte DnaJs, analysis conducted in this study was unable to provide any significant differences in the way these Hsp40s may interact with PfHsp70-x. This was largely due to inconclusive findings regarding the nature of the PfHsp70-x interaction with the J domains of these proteins. The loop region right after HPD motif varies quite greatly in size and composition across the different J domains studied and it is possible that it contributes to the specificity of the interactions. However, it is also possible that the domains of Hsp40s other than J domain play a role in the formation of specific Hsp70-x-Hsp40 complexes as indicated by Kampinga and Craig (2010). Homology models of the complexes were informative to a certain extent, as the template used for modeling was a crystal structure in which complex formation was formed by cross-linking. Docking results highlighted that J domain can rotate around HPD motif and present equally likely different orientations for docking. Further, by studying different structural conformations of PfHsp70-x, it was shown that Hsp40s can only bind when PfHsp70-x is in a certain conformation. Additionally, this work highlighted the possible dependence of the SBD residues on the orientation of the α-helical lid for formation of the substrate binding pocket. Since the α-helical lid is a very variable region across Hsp70s, it may present a preferable target for drug design. Overall, the unique substitutions found across the sequence of PfHsp70-x, often in regions highly conserved in other Hsp70s, may make this protein a potential target for drug design.

## Acknowledgements

RH thanks Rhodes University and National Research Foundation (NRF), South Africa, for financial support. Opinions expressed and conclusions arrived at, are those of the author and are not necessarily to be attributed to the NRF.

# Tables

**Table 1: Lowest sequence identity pairs from each alignment.** For each alignment performed, the sequence identity was calculated for each pair of sequences in the alignment. Listed below are the sequence pairs from each alignment with the lowest sequence identity, as well as the sequence identity for the pair.

| Hsp70 Sequences Aligned | Sequence 1 | Sequence 2 | Sequence Identity (%) |
| --- | --- | --- | --- |
| DnaKs (Prokaryotic) | *R. bacterium* DnaK | *M. tarda* DnaK | 51.62 |
| Cytosolic Eukaryotic | *A. thaliana* Hsp70B | *S. cerevisiae* Ssa1p | 69.98 |
| Cytosolic Apicomplexan | *T. annulata* Hsp70 | *C. parvum* Hsp70 | 70.86 |
| Cytosolic Plasmodial | *P. vivax* Hsp70 | *P. yoelii yoelii* Hsp70 | 95.80 |
| ER Eukaryotic | *C. elegans* Hsp-3 | *A. thaliana* BiP-2 | 64.55 |
| ER Apicomplexan | *E. tenella* BiP | *B. rodhaini* Grp78 | 59.46 |
| ER Plasmodial | *P. yoelii yoelii* ER-Hsp70 | *P. vivax* Grp78 | 85.21 |
| Mitochondrial Eukaryotic | *E. tenella* mtHsp70 | *A. echinatior* Hsc70-5 | 53.00 |
| Mitochondrial Apicomplexan | *T. annulata* mtHsp70 | *T. gondii* mtHsp70 | 64.06 |
| Mitochondrial Plasmodial | *P. knowlesi* mtHsp70 | *P. vivax* mtHsp70 | 97.44 |

**Figure Legends**

**Figure 1: Cytosolic Hsp70 sequence alignment.** PfHsp70-1 and PfHsp70-x were aligned with consensus sequences of cytosolic Hsp70 representatives of *Plasmodium*, Apicomplexa and eukaryotes, as well as prokaryotic DnaKs. Residues are highlighted as follows: Dark Blue: Unique to PfHsp70-x; Red: Unique to PfHsp70-1; Yellow: Residues present in either PfHsp70-x or PfHsp70-1 and eukaryotes, but not conserved in other *Plasmodium* Hsp70s; Green: Residues conserved in *Plasmodium*; Light Blue: Residues conserved in Apicomplexa; Purple: Residues which are highly conserved in eukaryotes, but not in PfHsp70-1 or PfHsp70-x. Consensus sequence letters in upper case were conserved in all sequences examined, whereas lower case letters indicate residues conserved in most sequences examined. A tilde (~) indicates a lack of consensus.

**Figure 2: Comparisons of PfHsp70-x binding to the J domains of different Hsp40s, based on Hybrid modeling.** A) Sequence regions of PfHsp70-x which were found to interact with the different J domains. The Hsp40 interacted with is indicated in parentheses. J proteins are denoted as follows: JA1 - HsDnaJA1; JA2 - HsDnaJA2; JB1 – HsDnaJB1; JB4 – HsDnaJB4; JC5 – HsDnaJC5; JC13 – HsDnaJC13. Residues highlighted in red and light blue, respectively, represent interactions found before and after residue repacking and energy minimisation using PyRosetta (refer to methodology). Residues highlighted in green are those which displayed interactions in both structures. B) Sequences are shown for portions of the J domains which displayed interactions with PfHsp70-x. Residues mapped to the structure in C are indicated over their position in the alignment. C) The interactions between PfHsp70-x and J proteins are shown, based on those which are most conserved across the 9 different J proteins. A section of the J domain is shown in cartoon form, with interacting residues displayed as text. Residues of PfHsp70-x are coloured blue, whereas J protein residues are black and numbered according to their position in alignment 2B. The different types of interactions are displayed alongside.

**Figure 3: Comparisons of PfHsp70-x binding to the J domains of different Hsp40s as determined by docking.** Each J domain was docked to PfHsp70-x and the interactions between the two proteins were determined using the PIC webserver. A) Sequence regions of PfHsp70-x which were found to interact with the different J domains. Sequences are named as in Fig. 2A. B) Portions of each J domain, highlighting residues predicted to interact with PfHsp70-x. C) The interactions between PfHsp70-x and J proteins are shown, as in Fig. 2C.

**Figure 4: Hsp40 binding within the ATPase cycle of Hsp70.** A) The ATPase cycle of Hsp70 is shown. The two circled representations of Hsp70 are those found in the PDB. The binding of a J protein occurs between these two states. B and C represent the docked complex of HsDnaJA2 and PfHsp70-x being superimposed with the full length model of PfHsp70-x in an ATP-bound and ADP-bound conformations, respectively. The PfHsp70-x domains are coloured as follows: Grey – ATPase domain; Yellow – β subdomain of the SBD; Pale blue – α helical lid of the SBD. The J domain of HsDnaJA2 is coloured teal.

**Figure 5: Formation of the substrate binding is dependent on the orientation of the α-helical lid, relative to the substrate binding domain.** A) PfHsp70-x is shown as it is modeled, based on a substrate-bound template, 2KHO (Left) and a template not involved in substrate binding, 3D2F (Right). The domains are coloured in a similar way to those in Figure 4. Residues involved in substrate binding were mapped to the protein and are coloured as follows: Blue – residues that bind substrate peptides; Red – hydrophobic arch residues; Green - residue that forms the hydrophobic pocket. B) Areas on the protein boxed in dashed lines are represented below each structure in surface view. C) The SBD of PfHsp70-x is also shown as modeled after template 1YUW, whose structure was solved in the absence of a peptide substrate. This is shown at two different angles in order to emphasise how the alpha-

helical lid is bound by the substrate binding pocket in the absence of substrate. Residues involved in substrate binding are shown as spheres.

**Figure 1: Cytosolic Hsp70 sequence alignment.**

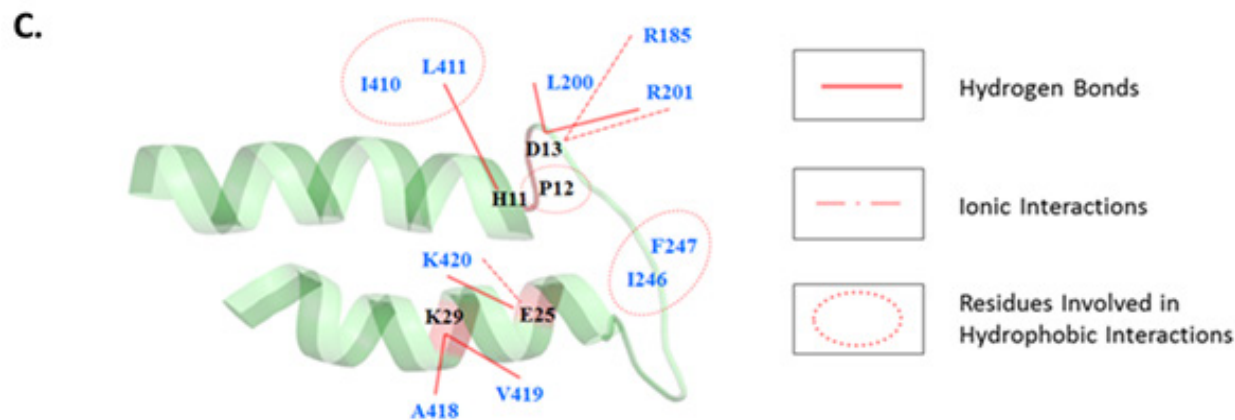

**Figure 2** Comparisons of PfHsp70-x binding to the J domains of different Hsp40s, based on Hybrid modeling.

**A.**

```
PfHsp70x(PfA)    179  YFNDSQRQATKDAGAIAGLNVLRIINEPTAAAIAYGLDKKGKGEQN  224
PfHsp70x(PfB)    179  YFNDSQRQATKDAGAIAGLNVLRIINEPTAAAIAYGLDKKGKGEQN  224
PfHsp70x(PfE)    179  YFNDSQRQATKDAGAIAGLNVLRIINEPTAAAIAYGLDKKGKGEQN  224
PfHsp70x(JA1)    179  YFNDSQRQATKDAGAIAGLNVLRIINEPTAAAIAYGLDKKGKGEQN  224
PfHsp70x(JA2)    179  YFNDSQRQATKDAGAIAGLNVLRIINEPTAAAIAYGLDKKGKGEQN  224
PfHsp70x(JB4)    179  YFNDSQRQATKDAGAIAGLNVLRIINEPTAAAIAYGLDKKGKGEQN  224
PfHsp70x(JB1)    179  YFNDSQRQATKDAGAIAGLNVLRIINEPTAAAIAYGLDKKGKGEQN  224
PfHsp70x(JC5)    179  YFNDSQRQATKDAGAIAGLNVLRIINEPTAAAIAYGLDKKGEQN   224
PfHsp70x(JC13)   179  YFNDSQRQATKDAGAIAGLNVLRIINEPTAAAIAYGLDKKGKGEQN  224

PfHsp70x(PfA)    240  LTLEDGIFEVKAT  252    404  AVQAAILSGDQSSAVKDLL  423
PfHsp70x(PfB)    240  LTLEDGIFEVKAT  252    404  AVQAAILSGDQSSAVKDLL  423
PfHsp70x(PfE)    240  LTLEDGIFEVKAT  252    404  AVQAAILSGDQSSAVKDLL  423
PfHsp70x(JA1)    240  LTLEDGIFEVKAT  252    404  AVQAAILSGDQSSAVKDLL  423
PfHsp70x(JA2)    240  LTLEDGIFEVKAT  252    404  AVQAAILSGDQSSAVKDLL  423
PfHsp70x(JB4)    240  LTLEDGIFEVKAT  252    404  AVQAAILSGDQSSAVKDLL  423
PfHsp70x(JB1)    240  LTLEDGIFEVKAT  252    404  AVQAAILSGDQSSAVKDLL  423
PfHsp70x(JC5)    240  LTLEDGIFEVKAT  252    404  AVQAAILSGDQSSAVKDLL  423
PfHsp70x(JC13)   240  LTLEDGIFEVKAT  252    404  AVQAAILSGDQSSAVKDLL  423
```

**B.**

```
              5         10        15        20        25        30        35        40        45        50
              R   AM    HPDK                           F
PfA    100  RAYLKLAMKWHPDKHVNKGSKVEAEEKFKNICEAYSVLSDNEKRVKYDLFGMDALKQS      158
PfB    106  KAYKKLAMKWHPDKHLNAASKKEADNMFKSISEAYEVLSDEEKRDIYDKYGEEGLDKY     164
PfE     98  KAYRKLAMKWHPDKHLNDEDKVEAERKFKLIGEAYEVLSDEEKRKNYDLFGQSGLGGT     156
JA1     24  KAYRKLALKYHPDKN------PNEGEKFKQISQAYEVLSDAKKRELYDKGGEQAIKEG     76
JA2     26  KAYRKLAKEYHPDKN------PNAGDKFKEISFAYEVLSNPEKRELYDRYGEQGLREG     78
JB4     22  KAYRKQALKFHPDKNKS----PQAEEKFKEVAEAYEVLSDPKKREIYDQFGEEGLKGG    75
JB1     22  RAYRRQALRYHPDKNKE----PGAEEKFKEIAEAYDVLSDPRKREIFDRYGEEGLKGS    75
JC5     33  KSYRKLALKYHPDKNPDN---PEAADKFKEINNAHAILTDATKRNIYDKY---------    79
JC13  1322  KAYFRLAQKYHPDKNPE-----GRDMFEKVNKAYEFLCTKSAKIVDGPDPENIILIL   1373
```

**C.**

Figure 3: Comparisons of PfHsp70-x binding to the J domains of different Hsp40s as determined by docking.

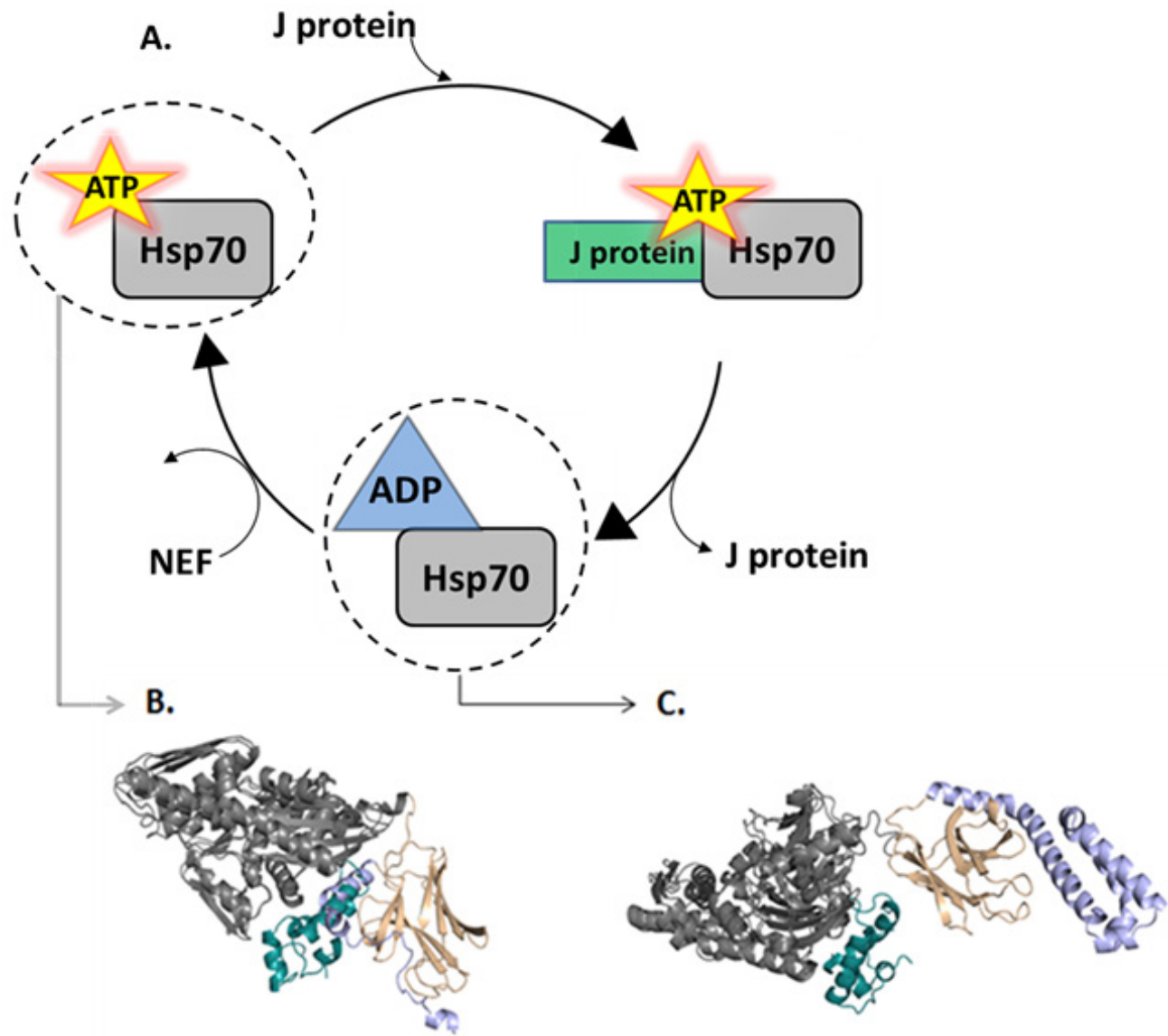

**Figure 4: Hsp40 binding within the ATPase cycle of Hsp70.**

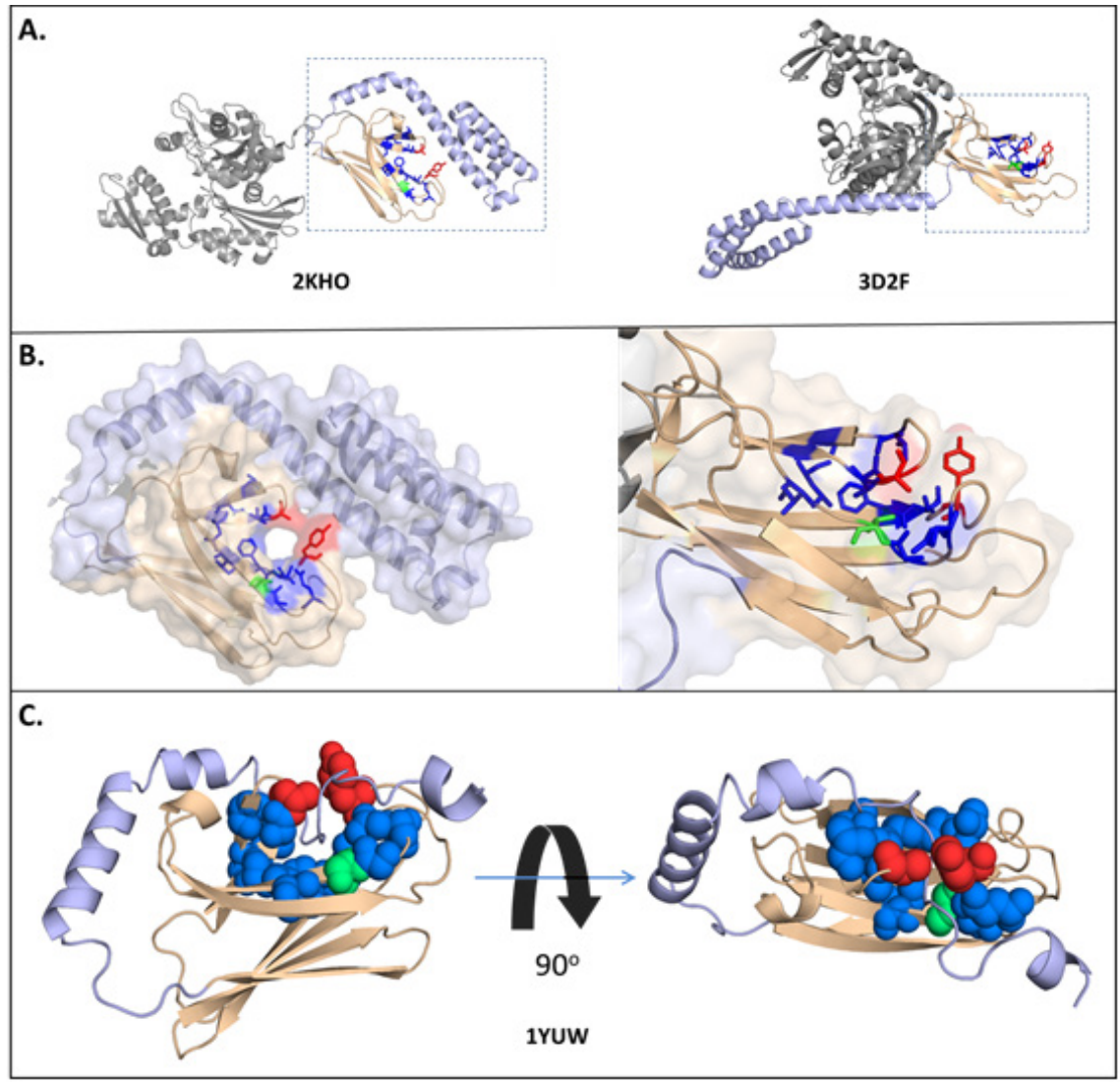

**Figure 5: Formation of the substrate binding is dependent on the orientation of the α-helical lid, relative to the substrate binding domain.**

## Supplementary Data S1:

**List of Hsp70 sequences used for alignments.** Listed is the name of the Hsp70s used in the alignments performed. Included is the organism the protein comes from, the Genbank, NCBI or Uniprot reference number of the protein, the predicted cellular localization of the protein and the alignment group it was included in when making consensus sequences.

# Supplementary Data S2:

**Supplementary Data S2A: Comparison of alignments of eukaryote C-Hsp70s as performed by MAFFT and PROMALS3D.** C-Hsp70s obtained from a variety of eukaryote sources were aligned using PROMALS3D and MAFFT. PfHsp70-1 and PfHsp70-x were included in this alignment. Each sequence aligned using PROMALS3D was placed directly above the same sequence aligned using MAFFT. Sequences aligned using PROMALS 3D have the suffix _Pm added to their names whereas those aligned using MAFFT have the suffix _Ma.

**Supplementary Data S2B: Comparison of alignments of eukaryote ER-Hsp70s as performed by MAFFT and PROMALS3D.** The eukaryotic ER-Hsp70s were aligned with PfHsp70-2 using PROMALS3D and MAFFT and annotated as in Supplementary Data S2A. Since the Alignments done by the two programs were of different lengths, the PROMALS3D sequences were moved forward until the GIDL motif of each sequence was aligned. A tilde (~) indicates that the sequence was moved one space forward.

**Supplementary Data S2C: Comparison of alignments of eukaryote Mt-Hsp70s as performed by MAFFT and PROMALS3D.** The eukaryotic Mt-Hsp70s were aligned with PfHsp70-3 using PROMALS3D and MAFFT and annotated as in Figure B.1. Sequences were hand-adjusted as in Supplementary Data S2A.

## Supplementary Data S3:

**Supplementary Data S3A: Alignment of eukaryote C-Hsp70 protein sequences.** Sequences aligned include those listed in Supplementary Data S1, whose predicted cellular localization was the cytosol and were in the eukaryotic alignment group. Alignment of these sequences was done using MAFFT. Residues highlighted in black represent those which are conserved at the level of identity, whereas those highlighted in grey represent those that were conserved at the level of similarity. A consensus sequence of these sequences is represented at the top of the alignment. Residues in upper case were conserved in all sequences whereas those in lower case were conserved in most sequences. A tilde (~) represents a variable region where there is no consensus residue.

**Supplementary Data S3B: Alignment of apicomplexan C-Hsp70 protein sequences.** Sequences aligned include those listed in Supplementary Data S1, whose predicted cellular localization was the cytosol and were in the Apicomplexa alignment group. This figure was otherwise constructed and highlighted as described in Supplementary Data S3A.

**Supplementary Data S3C: Alignment of *Plasmodium* C-Hsp70 protein sequences.** Sequences aligned include those listed in Supplementary Data S1 whose predicted cellular localization was the cytosol and were in the *Plasmodium* alignment group. This figure was otherwise constructed and highlighted as described in Supplementary Data S3A.

**Supplementary Data S3D: Alignment of eukaryote ER-Hsp70 protein sequences.** Sequences aligned include those listed in Supplementary Data S1 whose predicted cellular localization was the ER and were in the eukaryotic alignment group. This figure was otherwise constructed and highlighted as described in Supplementary Data S3A.

**Supplementary Data S3E: Alignment of apicomplexan ER-Hsp70 protein sequences.** Sequences aligned include those listed in Supplementary Data S1 whose predicted cellular localization was the ER and were in the Apicomplexa alignment group. This figure was otherwise constructed and highlighted as described in Supplementary Data S3A.

**Supplementary Data S3F: Alignment of *Plasmodium* ER-Hsp70 protein sequences.** Sequences aligned include those listed in Supplementary Data S1 whose predicted cellular localization was the ER and were in the *Plasmodium* alignment group. This figure was otherwise constructed and highlighted as described in Supplementary Data S3A.

**Supplementary Data S3G: Alignment of eukaryote Mt-Hsp70 protein sequences.** Sequences aligned include those listed in Supplementary Data S1 whose predicted cellular localization was the mitochondria and were in the eukaryotic alignment group. This figure was otherwise constructed and highlighted as described in Supplementary Data S3A.

**Supplementary Data S3H: Alignment of apicomplexan Mt-Hsp70 protein sequences.** Sequences aligned include those listed in Supplementary Data S1 whose predicted cellular localization was the mitochondria and were in the Apicomplexa alignment group. This figure was otherwise constructed and highlighted as described in Supplementary Data S3A.

**Supplementary Data S3I: Alignment of *Plasmodium* Mt-Hsp70 protein sequences.** Sequences aligned include those listed in Supplementary Data S1 whose predicted cellular localization was the mitochondria and were in the *Plasmodium* alignment group. This figure was otherwise constructed and highlighted as described in Supplementary Data S3A.

**Supplementary Data S3J: Alignment of prokaryote DnaK protein sequences.** Sequences aligned include those listed in Supplementary Data S1 that were in the prokaryote alignment group. This figure was otherwise constructed and highlighted as described in Supplementary Data S3A.

## Supplementary Data S4:

**Comparison of top models produced using original and modified templates.** Models were all produced using the same modeling procedure, with exception to the template used. Five different model quality assessment scores are reported for each set of models. DOPE Z score was calculated using MODELLER and the GDT_TS scores are values predicted by MetaMQAPII.

**Supplementary Data S5:**
**Alignment of N-terminal sequence of PfHsp70-x with other ER-transit peptides.** PfHsp70-x was aligned with the sequences of five different ER-Hsp70 sequences from *Chlamydomonas reinhardtii* (CrHsp70-4 and CrHsp70-5), *Ostreococcus lucimarinus* (OlHsp70-3), humans (HsGrp78) and *P. falciparum* (PfHsp70-2). Residues are coloured according to their physicochemical properties.

## Supplementary Data S6:

**Localization Alignment of Hsp70s.** The sequences of PfHsp70-1 and PfHsp70-x were aligned with the consensus sequences of cytosolic, ER- and Mt-Hsp70s of representative sequences of eukaryotic, apicomplexan and plasmodial species. The Residues are highlighted as follows: Dark Blue - Residues previously identified as being "unique" to PfHsp70-x, as indicated in Figure 1; Light Blue - Residues conserved specifically in C-Hsp70s. Green - Residues conserved specifically in ER-Hsp70s. Red - Residues conserved specifically in Mt-Hsp70s. Pink - PfHsp70-x residues not found in any of the cytosolic groups, but were highly conserved in ER-Hsp70s and/or Mt-Hsp70s. Yellow boxed line - gap regions common to more than one of the Hsp70 groups.

## Supplementary Data S7:

**Supplementary Data S7A: Conserved interactions between PfHsp70-x and each J protein, based on Hybrid modeling.** The different types of interactions are displayed for PfHsp70-x, modeled in complex with each of nine J proteins.

**Supplementary Data S7B: Conserved interactions between PfHsp70-x and each J protein, based on Hybrid modeling (Alignment renumbered).** The different types of interactions are displayed for PfHsp70-x, modeled in complex with each of nine J proteins. Residues of the J proteins have been renumbered, according to their position in the alignment of Figure 2B, in order to make these interactions comparable.

**Supplementary Data S7C: Conserved interactions between PfHsp70-x and each J protein, based on HADDOCK Docking.** The different types of interactions are displayed for PfHsp70-x, docked to each of nine J proteins, using the HADDOCK online webserver.

**Supplementary Data S7D: Conserved interactions between PfHsp70-x and each J protein, based on HADDOCK Docking (Alignment renumbered).** The different types of interactions are displayed for PfHsp70-x, docked to each of nine J proteins, using the HADDOCK online webserver. Residues of the J proteins have been renumbered, according to their position in the alignment of Figure 3B, in order to make these interactions comparable.

## Supplementary Data S8:

**Interaction interface between PfHsp70-x and a J domain as determined by Hybrid modelling.** A) The interaction interface between the ATPase domain of PfHsp70-x (grey) and a J domain (green). The area within the dashed box is shown in B. Some of the residues which showed conserved interactions are mapped to the structure in stick representation.

## Supplementary Data S9:

**Interaction interface between PfHsp70-x and a J domain as determined by HADDOCK docking.** A) Superimposition of the five different Type II J proteins docked in complex with PfHsp70-x, showing the different orientations adopted by the J domain proteins docked. These proteins are coloured as follows: Purple – PfA; Yellow – PfB; Orange – PfE; Red – HsDnaJB1; Green – HsDnaJB4. The ATPase domains of PfHsp70-x in each complex are coloured grey. B) The models are rotated 90° to indicate the interaction interface between the ATPase domain of PfHsp70-x and the J domains. The area within the dashed box is shown in C. Residues which showed conserved interactions are mapped to the structure in stick representation. The aspartic acids of each HPD motif are also shown as yellow sticks.

## Supplementary Data S10:

**Supplementary Data S10A:**

**Superimposition of J domain binding orientations to PfHsp70-x in an ATP-bound state.** The J domain of each J protein docked to the ATPase domain of PfHsp70-x was superimposed with the full length structure of PfHsp70-x, modeled after template 1YUW. The J domain docked is labelled next to each model. JA1 and JA2 represent HsDnaJA1 and HsDnaJA2, respectively.

**Supplementary Data S10B:**

**Superimposition of J domain binding orientations to PfHsp70-x in an ADP-bound state.** The J domain of each J protein docked to the ATPase domain of PfHsp70-x was superimposed with the full length structure of PfHsp70-x, modeled after template 2KHO. The J domain docked is labelled next to each model. JA1 and JA2 represent HsDnaJA1 and HsDnaJA2, respectively.

Supplementary Data S1

| Protein | Organism | Type of Organism | Reference/Accession no. | Predicted Cellular Localization | Alignment Group |
|---|---|---|---|---|---|
| Hsp70 | *Prorocentrum minimum* | Dinoflagellate | GB: ABI14407.1 | Cytosol | Eukaryote |
| Hsp70 | *Phytophthora infestans* | Oomycete | NCBI: XP_002902008.1 | Cytosol | Eukaryote |
| Hsc70-4 | *Acromyrmex echinatior* | Ant | GB: EGI62314.1 | Cytosol | Eukaryote |
| Hsp70 | *Xenopus laevis* | Frog | GB: AAH78115.1 | Cytosol | Eukaryote |
| Hsp70 | *Lates calcarifer* | Fish | GB: AEH27544.1 | Cytosol | Eukaryote |
| Hsc70 | *Haliotis diversicolor* | Sea Snail | GB: ACO36047.1 | Cytosol | Eukaryote |
| Hsp-1 | *Caenorhabditis elegans* | Roundworm | NCBI: NP_503068.1 | Cytosol | Eukaryote |
| Hsc70 | *Fenneropenaeus chinensis* | Prawn | GB: AAW71958.1 | Cytosol | Eukaryote |
| Hsp70-B | *Arabidopsis thaliana* | Plant | NCBI: NP_173055.1 | Cytosol | Eukaryote |
| Hsp70 | *Capsaspora owczarzaki* | Single-Celled Eukaryote | GB: EFW40320.1 | Cytosol | Eukaryote |
| Hsp70 | *Rattus norvegicus* | Rat | GB: CAA54424.1 | Cytosol | Eukaryote |
| Hsp70 | *Coturnix japonica* | Quail | GB: BAF38390.1 | Cytosol | Eukaryote |
| Ssa1p | *Saccharomyces cerevisiae* | Yeast | GB: EGA63465.1 | Cytosol | Eukaryote |
| Hsp70 | *Toxoplasma gondii* | Intracellular Parasite | GB: AAC26629.1 | Cytosol | Apicomplexa |
| HSP | *Cryptosporidium parvum* | Intracellular Parasite | GB: AAB16853.1 | Cytosol | Apicomplexa |
| HSP | *Cryptosporidium hominis* | Intracellular Parasite | NCBI: XP_666754.1 | Cytosol | Apicomplexa |
| Hsp70 | *Theileria parva* | Intracellular Parasite | GB: AAB53893.1 | Cytosol | Apicomplexa |
| Hsp70 | *Theileria annulata* | Intracellular Parasite | NCBI: XP_952563.1 | Cytosol | Apicomplexa |
| HSP | *Eimeria acervulina* | Intracellular Parasite | GB: CAA81135.1 | Cytosol | Apicomplexa |

Supplementary Data S1

| Protein | Organism | Type Of Organism | Reference/Accession no. | Predicted Cellular Localization | Alignment Group |
|---|---|---|---|---|---|
| Hsp70 | *Cyclospora cayetanensis* | Intracellular Parasite | GB: ADZ45037.1 | Cytosol | Apicomplexa |
| Hsp70 | *Neospora caninum* | Intracellular Parasite | GB: CBZ53608.1 | Cytosol | Apicomplexa |
| Hsp70 | *Babesia divergens* | Intracellular Parasite | GB: BAF02616.1 | Cytosol | Apicomplexa |
| Hsp70 | *Plasmodium vivax* | Intracellular Parasite | NCBI: XP_001614972.1 | Cytosol | Plasmodium |
| Hsp70 | *Plasmodium yoelii yoelii* | Intracellular Parasite | NCBI: XP_726754.1 | Cytosol | Plasmodium |
| Hsp70 | *Plasmodium berghei* | Intracellular Parasite | GB: AAL34314.1 | Cytosol | Plasmodium |
| Hsp70 | *Plasmodium knowlesi* | Intracellular Parasite | NCBI: XP_002258136.1 | Cytosol | Plasmodium |
| Grp78 precursor | *Homo sapiens* | Human | NCBI: NP_005338.1 | ER | Eukaryote |
| BiP-2 | *Arabidopsis thaliana* | Small Flowering Plant | NCBI: NP_851119.1 | ER | Eukaryote |
| BiP-3 precursor | *Phytophthora infestans* | Oomycete | NCBI: XP_002909118.1 | ER | Eukaryote |
| Hsc70-3 | *Acromyrmex echinatior* | Ant | GB: EGI70210.1 | ER | Eukaryote |
| Grp78 | *Fenneropenaeus chinensis* | Prawn | GB: ABM92447.1 | ER | Eukaryote |
| Hsp70 | *Capsaspora owczarzaki* | Single-Celled Eukaryote | GB: EFW43336.1 | ER | Eukaryote |
| Grp78 precursor | *Rattus norvegicus* | Rat | NCBI: NP_037215.1 | ER | Eukaryote |
| Hsp70 | *Coturnix japonica* | Quail | GB: BAF38391.1 | ER | Eukaryote |
| Hspa5 | *Xenopus laevis* | Frog | GB: AAH41200.1 | ER | Eukaryote |
| Kar2p | *Saccharomyces cerevisiae* | Yeast | GB: GAA24327.1 | ER | Eukaryote |
| BiP | *Oryzias latipes* | Fish | GB: BAL14281.1 | ER | Eukaryote |
| Hsp-3 | *Caenorhabditis elegans* | Roundworm | NCBI: NP_509019.1 | ER | Eukaryote |
| BiP | *Crypthecodinium cohnii* | Red Algae | GB: AAM02971.2 | ER | Eukaryote |
| Grp78 | *Aplysia californica* | Sea Slug | NCBI: NP_001191581.1 | ER | Eukaryote |

Supplementary Data S1

| Protein | Organism | Type Of Organism | Reference/Accession no. | Predicted Cellular Localization | Alignment Group |
|---|---|---|---|---|---|
| Hsp70 precursor | *Toxoplasma gondii* | Intracellular Parasite | GB: AAF23321.1 | ER | Apicomplexa |
| Hsp70 | *Cryptosporidium parvum* | Intracellular Parasite | NCBI: XP_628228.1 | ER | Apicomplexa |
| BiP | *Eimeria tenella* | Intracellular Parasite | GB: CAA91253.1 | ER | Apicomplexa |
| Hsp70 Precursor | *Neospora caninum* | Intracellular Parasite | GB: CBZ55124.1 | ER | Apicomplexa |
| Grp78 | *Babesia rodhaini* | Intracellular Parasite | GB: BAE16574.1 | ER | Apicomplexa |
| Grp78 | *Babesia microti* | Intracellular Parasite | GB: BAH28858.1 | ER | Apicomplexa |
| Grp78 | *Plasmodium vivax* | Intracellular Parasite | NCBI: XP_001614712.1 | ER | Plasmodium |
| HSP | *Plasmodium yoelii yoeli* | Iintracellular Parasite | NCBI: XP_725393.1 | ER | Plasmodium |
| HSP | *Plasmodium berghei* | Intracellular Parasite | NCBI: XP_677775.1 | ER | Plasmodium |
| HSP | *Plasmodium knowlesi* | Intracellular Parasite | NCBI: XP_002258451.1 | ER | Plasmodium |
| HspA9B | *Homo sapiens* | Human | Uniprot: P38646 | Mitochondria | Eukaryote |
| Mt-Hsp70 precursor | *Phytophthora infestans* | Oomycete | NCBI: XP_002898043.1 | Mitochondria | Eukaryote |
| Hsc70-5 | *Acromyrmex echinatior* | Ant | GB: EGI69033.1 | Mitochondria | Eukaryote |
| Hsp-6 | *Caenorhabditis elegans* | Roundworm | NCBI: NP_504291.1 | Mitochondria | Eukaryote |
| Mt-Hsp70-2 | *Arabidopsis thaliana* | Plant | NCBI: NP_196521.1 | Mitochondria | Eukaryote |
| Hsp70-2 | *Capsaspora owczarzaki* | Single-Celled Eukaryote | GB: EFW47073.1 | Mitochondria | Eukaryote |
| Mt-Hsp70 | *Rattus norvegicus* | Rat | NCBI: NP_001094128.1 | Mitochondria | Eukaryote |
| Ssc1p | *Saccharomyces cerevisiae* | Yeast | NCBI: NP_012579.1 | Mitochondria | Eukaryote |
| Hsp70-9 | *Xenopus laevis* | Frog | NCBI: NP_001080166.1 | Mitochondria | Eukaryote |
| Hsp70 | *Moina mongolica* | Crustacea | GB: ADA79523.1 | Mitochondria | Eukaryote |
| mortalin-2 | *Mya arenaria* | Clam | GB: AAP78491.2 | Mitochondria | Eukaryote |

Supplementary Data S1

| Protein | Organism | Type of Organism | Reference/Accession no. | Predicted Cellular Localization | Alignment Group |
|---|---|---|---|---|---|
| Hsp70-type | *Durinskia baltica* | Dinoflagellate | NCBI: YP_003735045.1 | Mitochondria | Eukaryote |
| Grp75 | *Sparus aurata* | Fish | GB: ABF70949.1 | Mitochondria | Eukaryote |
| Mt-Hsp70 precursor | *Gallus gallus* | Chicken | NCBI: NP_001006147.1 | Mitochondria | Eukaryote |
| Putative Hsp70 | *Toxoplasma gondii* | Intracellular Parasite | NCBI: XP_002367417.1 | Mitochondria | Apicomplexa |
| Mt-Hsp70 | *Cryptosporidium parvum* | Intracellular Parasite | NCBI: XP_626895.1 | Mitochondria | Apicomplexa |
| DnaK-like protein | *Cryptosporidium muris* | Intracellular Parasite | NCBI: XP_002142457.1 | Mitochondria | Apicomplexa |
| Hsp70 | *Theileria parva* | Intracellular Parasite | NCBI: XP_765320.1 | Mitochondria | Apicomplexa |
| Hsp70 | *Theileria annulata* | Intracellular Parasite | GB: AAQ63186.1 | Mitochondria | Apicomplexa |
| Organelle Hsp70 | *Eimeria tenella* | Intracellular Parasite | GB: CAA87086.1 | Mitochondria | Apicomplexa |
| hypothetical protein | *Neospora caninum* | Intracellular Parasite | GB: CBZ56259.1 | Mitochondria | Apicomplexa |
| Hsp70 | *Babesia bovis* | Intracellular Parasite | NCBI: XP_001609513.1 | Mitochondria | Apicomplexa |
| Hsp70 homologue | *Plasmodium vivax* | Intracellular Parasite | NCBI: XP_001615454.1 | Mitochondria | Plasmodium |
| Hsp70 | *Plasmodium yoelii yoelii* | Intracellular Parasite | NCBI: XP_727747.1 | Mitochondria | Plasmodium |
| Hsp70 homologue | *Plasmodium knowlesi* | Intracellular Parasite | NCBI: XP_002259347.1 | Mitochondria | Plasmodium |
| DnaK | *Rhodobacterales bacterium* | Bacteria | NCBI: ZP_01447398.1 | N/A | Prokaryote |
| DnaK | *Desulfurispirillum indicum* | Bacteria | NCBI: YP_004113662.1 | N/A | Prokaryote |
| DnaK | *Leptospirillum* | Bacteria | GB: EDZ39004.1 | N/A | Prokaryote |
| DnaK | *Methanolinea tarda* | Archae | NCBI: ZP_09043230.1 | N/A | Prokaryote |
| DnaK | *Methanohalophilus mahii* | Archae | NCBI: YP_003542757.1 | N/A | Prokaryote |
| DnaK | *Bartonella henselae* | Bacteria | NCBI: YP_032930.1 | N/A | Prokaryote |
| DnaK | *Ochrobactrum anthropi* | Bacteria | NCBI: YP_001369341.1 | N/A | Prokaryote |
| DnaK | *Thermodesulfovibrio yellowstonii* | Bacteria | NCBI: YP_002249549.1 | N/A | Prokaryote |

Supplementary Data S1

| Protein | Organism | Type of Organism | Reference/Accession no. | Predicted Cellular Localization | Alignment Group |
|---|---|---|---|---|---|
| DnaK | *Methanoculleus marisnigri* | Archae | NCBI: YP_001046976.1 | N/A | Prokaryote |
| DnaK | *Methanoplanus petrolearius* | Archae | NCBI: YP_003893884.1 | N/A | Prokaryote |
| DnaK | *Mesorhizobium alhagi* | Bacteria | NCBI: ZP_09291376.1 | N/A | Prokaryote |
| DnaK | *Chelativorans* | Bacteria | NCBI: YP_673244.1 | N/A | Prokaryote |
| DnaK | *Candidatus Nitrospira defluvii* | Bacteria | NCBI: YP_003797385.1 | N/A | Prokaryote |
| DnaK | *Methanosalsum zhilinae* | Archae | NCBI: YP_004616671.1 | N/A | Prokaryote |
| DnaK | *Methanosarcina mazei* | Archae | NCBI: NP_634529.1 | N/A | Prokaryote |

Supplementary Data S2A

```
                                          10        20        30        40        50        60        70        80        90       100       110       120
                                    ....|....|....|....|....|....|....|....|....|....|....|....|....|....|....|....|....|....|....|....|....|....|....|....|
       A._echinatior_Hsc70-4_Pm     M------------------------SKAPAVGIDLGTTYSCVGVFQHGKVEIIANDQGNRTTPSYVAFTDTERLIGDAAKNQVAMNPSNTIFDAKRLIGRRFDDTTVQSDMKHWP
       A._echinatior_Hsc70-4_Ma     M-----------------S------KAPAVGIDLGTTYSCVGVFQHGKVEIIANDQGNRTTPSYVAFTDTERLIGDAAKNQVAMNPSNTIFDAKRLIGRRFDDTTVQSDMKHWP
       S._cerevisiae_Ssa1p_Pm       M------------------------SKAVGIDLGTTYSCVAHFANDRVDIIANDQGNRTTPSFVAFTDTERLIGDAAKNQAAMNPSNTVFDAKRLIGRNFNDPEVQSDMKHFP
       S._cerevisiae_Ssa1p_Ma       M-----------------S------KAVGIDLGTTYSCVAHFANDRVDIIANDQGNRTTPSFVAFTDTERLIGDAAKNQAAMNPSNTVFDAKRLIGRNFNDPEVQSDMKHFP
       C._japonica_Hsp70_Pm         M------------------------SGKGPAIGIDLGTTYSCVGVFQHGKVEIIANDQGNRTTPSYVAFTDTERLIGDAAKNQVAMNPTNTIFDAKRLIGRKYDDPTVQSDMKHWP
       C._japonica_Hsp70_Ma         M-----------------S------GKGPAIGIDLGTTYSCVGVFQHGKVEIIANDQGNRTTPSYVAFTDTERLIGDAAKNQVAMNPTNTIFDAKRLIGRKYDDPTVQSDMKHWP
       F._chinensis_Hsc70_Pm        M------------------------AKAPAVGIDLGTTYSCVGVFQHGKVEIIANDQGNRTTPSYVAFTDTERLIGDAAKNQVAMNPNNTVFDAKRLIGRKFEDHTVQSDMKHWP
       F._chinensis_Hsc70_Ma        M------------------A-----KAPAVGIDLGTTYSCVGVFQHGKVEIIANDQGNRTTPSYVAFTDTERLIGDAAKNQVAMNPNNTVFDAKRLIGRKFEDHTVQSDMKHWP
       H._diversicolor_Hsc70_Pm     M------------------------AKAPAIGIDLGTTYSCVGVFQHGKVEIIANDQGNRTTPSYVAFTDTERLIGDAAKNQVAMNPENTIFDAKRLIGRRFDETNVQSDMKHWP
       H._diversicolor_Hsc70_Ma     M------------------A-----KAPAIGIDLGTTYSCVGVFQHGKVEIIANDQGNRTTPSYVAFTDTERLIGDAAKNQVAMNPENTIFDAKRLIGRRFDETNVQSDMKHWP
       R._norvegicus_Hsp70_Pm       M--------------------AANKGMAIGIDLGTTYSCVGVFQHGKVEIIANDQGNRTTPSYVAFTDTERLIGDAAKNQVAMNPQNTVFDAKRLIGRKFNDPVVQSDMKLWP
       R._norvegicus_Hsp70_Ma       M------------------A-ANGMAIGIDLGTTYSCVGVFQHGKVEIIANDQGNRTTPSYVAFTDTERLIGDAAKNQVAMNPQNTVFDAKRLIGRKFNDPVVQSDMKLWP
       Human_Hsp70-1_Pm             M------------------------AKAAAIGIDLGTTYSCVGVFQHGKVEIIANDQGNRTTPSYVAFTDTERLIGDAAKNQVALNPQNTVFDAKRLIGRKFGDPVVQSDMKHWP
       Human_Hsp70-1_Ma             M------------------A-----KAAAIGIDLGTTYSCVGVFQHGKVEIIANDQGNRTTPSYVAFTDTERLIGDAAKNQVALNPQNTVFDAKRLIGRKFGDPVVQSDMKHWP
       L._calcarifer_Hsp70_Pm       M----------------------SPAKGVAIGIDLGTTYSCVGVIFQHGKVEIIANDQGNRTTPSYVAFTDTERLIGDAAKNQVAMNPSNTIFDAKRLIGRKFDDSVVQSDMKHWP
       L._calcarifer_Hsp70_Ma       M------------------S---PAKGVAIGIDLGTTYSCVGVIFQHGKVEIIANDQGNRTTPSYVAFTDTERLIGDAAKNQVAMNPSNTIFDAKRLIGRKFDDSVVQSDMKHWP
       X._laevis_Hsp70_Pm           M---------------------ATKGVAVGIDLGTTYSCVGVFQHGKVEIIANDQGNRTTPSYVAFTDTERLIGDAAKNQVAMNPQNTVFDAKRLIGRKFNDPVVQCDLKHWP
       X._laevis_Hsp70_Ma           M------------------A-TKGVAVGIDLGTTYSCVGVFQHGKVEIIANDQGNRTTPSYVAFTDTERLIGDAAKNQVAMNPQNTVFDAKRLIGRKFNDPVVQCDLKHWP
       C._owczarzaki_Hsp70_Pm       M---------------------ATQLAVGIDLGTTYSCVGVFQHGKVEIIANDQGNRTTPSYVAFTDTERLIGDAAKNQVAMNPENTVFDAKRLIGRRFEDDPAVQSDMKHWP
       C._owczarzaki_Hsp70_Ma       M------------------A-TQLAVGIDLGTTYSCVGVFQHGKVEIIANDQGNRTTPSYVAFTDTERLIGDAAKNQVAMNPENTVFDAKRLIGRRFEDDPAVQSDMKHWP
       C._elegans_Hsp-1_Pm          M------------------------SKHNAVGIDLGTTYSCVGVFMHGKVEIIANDQGNRTTPSYVAFTDTERLIGDAAKNQVAMNPETNVFDAKRLIGRKFDDPAVQSDMKHWP
       C._elegans_Hsp-1_Ma          M-----------------S------KHNAVGIDLGTTYSCVGVFMHGKVEIIANDQGNRTTPSYVAFTDTERLIGDAAKNQVAMNPETNVFDAKRLIGRKFDDPAVQSDMKHWP
       P._infestans_Hsp70_Pm        M--------------------TQATSGYSVGIDLGTTYSCVGVWQNDRVEIIANDQGNRTTPSYVAFTDSERLIGDAAKNQVAMNAANTVFDAKRLIGRKFSDPVVQADIKHWP
       P._infestans_Hsp70_Ma        M-----------------T--QATSGYSVGIDLGTTYSCVGVWQNDRVEIIANDQGNRTTPSYVAFTDSERLIGDAAKNQVAMNAANTVFDAKRLIGRKFSDPVVQADIKHWP
       A._thaliana_Hsp70B_Pm        M--------------------ATKSEKAIGIDLGTTYSCVGVWMNDRVEIIENDQGNRTTPSYVAFTDTERLIGDAAKNQVALNPQNTVFDAKRLIGRKFSDPSVQSDILHWP
       A._thaliana_Hsp70B_Ma        M------------------A-TKSEKAIGIDLGTTYSCVGVWMNDRVEIIENDQGNRTTPSYVAFTDTERLIGDAAKNQVALNPQNTVFDAKRLIGRKFSDPSVQSDILHWP
       PfHsp70-x_Pm                 MKTKICSYIHYIVLFLIATTTVHTASNNAEESEVAIGIDLGTTYSCVCICRNGVVDIIANDQGNRTTPSYVAFTDTERLIGDAAKNQASRNPENTVFDAKRLIGRKFSETTVQSDMKHWP
       PfHsp70-x_Ma                 MKTKICSYIHYIVLFLIATTTVHTASNNAEESEVAIGIDLGTTYSCVCICRNGVVDIIANDQGNRTTPSYVAFTDTERLIGDAAKNQASRNPENTVFDAKRLIGRKFSETTVQSDMKHWP
       PfHsp70-1_Ma                 ------------------MASAKGSKPNLPESNIAIGIDLGTTYSCVGVWRNENVDIIANDQGNRTTPSYVAFTDTERLIGDAAKNQVARNPENTVFDAKRLIGRKFTESSVQSDMKHWP
       PfHsp70-1_Ma                 M----------------ASAKGSKPNLPESNIAIGIDLGTTYSCVGVWRNENVDIIANDQGNRTTPSYVAFTDTERLIGDAAKNQVARNPENTVFDAKRLIGRKFTESSVQSDMKHWP
```

```
                                    ....|....|....|....|....|....|....|....|....|....|....|....|....|....|....|....|....|....|....|....|....|....|....|....|
       A._echinatior_Hsc70-4_Pm     FTVVN-DGSKPKIKVSYKGEMKTFFPEEVSSMVLTKMKETAEAYLGKTITNAVITVPAYFNDSQRQATKDAGAIAGLNVLRIINEPTAAAIAYGLDKK--AAGEKNVLIFDLGGGTFDVS
       A._echinatior_Hsc70-4_Ma     FTVVN-DGSKPKIKVSYKGEMKTFFPEEVSSMVLTKMKETAEAYLGKTITNAVITVPAYFNDSQRQATKDAGAIAGLNVLRIINEPTAAAIAYGLDKK--AAGEKNVLIFDLGGGTFDVS
       S._cerevisiae_Ssa1p_Pm       FKLID-VDGKPQIQVEFKGETKNFTPEQISSMVLGKMKETAESYLGKAVNDAVVTVPAYFNDSQRQATKDAGTIAGLNVLRIINEPTAAAIAYGLDKK---GKEEVLIFDLGGGTFDVS
       S._cerevisiae_Ssa1p_Ma       FKLID-VDGKPQIQVEFKGETKNFTPEQISSMVLGKMKETAESYLGKAVNDAVVTVPAYFNDSQRQATKDAGTIAGLNVLRIINEPTAAAIAYGLDKK---GKEEVLIFDLGGGTFDVS
       C._japonica_Hsp70_Pm         FRVVN-EGGKPKVQVEYKGEMKTFFPEEISSMVLTKMKEIAEAYLGKKVQNAVITVPAYFNDSQRQATKDAGTITGLNMRIINEPTAAAIAYGLDKKGTRAGEKNVLIFDLGGGTFDVS
       C._japonica_Hsp70_Ma         FRVVN-EGGKPKVQVEYKGEMKTFFPEEISSMVLTKMKEIAEAYLGKKVQNAVITVPAYFNDSQRQATKDAGTITGLNMRIINEPTAAAIAYGLDKKGTRAGEKNVLIFDLGGGTFDVS
       F._chinensis_Hsc70_Pm        FTIIN-ESTKPKIQVEYKGDKKTYPEEISSMVLTKMKETAEAYLGSTVKDAVVTVPAYFNDSQRQATKDAGTISGLNVLRIINEPTAAAIAYGLDKK--VGGERNVLIFDLGGGTFDVS
       F._chinensis_Hsc70_Ma        FTIIN-ESTKPKIQVEYKGDKKTYPEEISSMVLTKMKETAEAYLGSTVKDAVVTVPAYFNDSQRQATKDAGTISGLNVLRIINEPTAAAIAYGLDKK--VGGERNVLIFDLGGGTFDVS
       H._diversicolor_Hsc70_Pm     FNVLS-DGGKPKIQVNYKDEPKTFYPEEISSMVLTKMKETAEQYLGKTITDAVVTVPAYFNDSQRQATKDAGTISGLNVLRIINEPTAAAIAYGLDKK--VGGERNVLIFDLGGGTFDVS
       H._diversicolor_Hsc70_Ma     FNVLS-DGGKPKIQVNYKDEPKTFYPEEISSMVLTKMKETAEQYLGKTITDAVVTVPAYFNDSQRQATKDAGTISGLNVLRIINEPTAAAIAYGLDKK--VGGERNVLIFDLGGGTFDVS
       R._norvegicus_Hsp70_Pm       FQVIN-EAGKPKVLVSYKGEKKAFYPEEISSMVLTKMKETAEALGHSVTNAVITVPAYFNDSQRQATKDAGVIAGLNVLRIINEPTAAAIAYGLDKG--SHGEREVLIFDLGGGTFDVS
       R._norvegicus_Hsp70_Ma       FQVIN-EAGKPKVLVSYKGEKKAFYPEEISSMVLTKMKETAEALGHSVTNAVITVPAYFNDSQRQATKDAGVIAGLNVLRIINEPTAAAIAYGLDKG--SHGEREVLIFDLGGGTFDVS
       Human_Hsp70-1_Pm             FQVIN-DGDKPKVQVSYKGDTKAFYPEEISSMVLTKMKEIAEAYLGYFVTNAVITVPAYFNDSQRQATKDAGVIAGLNVLRIINEPTAAAIAYGLDRT--GKGERNVLIFDLGGGTFDVS
       Human_Hsp70-1_Ma             FQVIN-DGDKPKVQVSYKGDTKAFYPEEISSMVLTKMKEIAEAYLGYFVTNAVITVPAYFNDSQRQATKDAGVIAGLNVLRIINEPTAAAIAYGLDRT--GKGERNVLIFDLGGGTFDVS
       L._calcarifer_Hsp70_Pm       FKVIS-DGGKPKIQVEYKGEDKAFYPEEISSMVLVKMKEIAEAYLGQKVSNAVITVPAYFNDSQRQATKDAGVIAGLNVLRIINEPTAAAIAYGLDKG--KSGERNVLIFDLGGGTFDVS
       L._calcarifer_Hsp70_Ma       FKVIS-DGGKPKIQVEYKGEDKAFYPEEISSMVLVKMKEIAEAYLGQKVSNAVITVPAYFNDSQRQATKDAGVIAGLNVLRIINEPTAAAIAYGLDKG--KSGERNVLIFDLGGGTFDVS
       X._laevis_Hsp70_Pm           FQVVS-DEGKPKVKVEYKGEEKSFFPEEISSMVLTKMKETAEAYLGHFVTNAVITVPAYFNDSQRQATKDAGVIAGLNILRIINEPTAAAIAYGLDKG--ARGEQNVLIFDLGGGTFDVS
       X._laevis_Hsp70_Ma           FQVVS-DEGKPKVKVEYKGEEKSFFPEEISSMVLTKMKETAEAYLGHFVTNAVITVPAYFNDSQRQATKDAGVIAGLNILRIINEPTAAAIAYGLDKG--ARGEQNVLIFDLGGGTFDVS
       C._owczarzaki_Hsp70_Pm       FKIVN-EATKPKIQVNYKGEEKVFSPEEISSMVLLKMKETAEAYLGKTINNAVVTVPAYFNDSQRQATKDAGTISGMNVLRIINEPTAAAIAYGLDKK--IGGERHVLIFDLGGGTFDVS
       C._owczarzaki_Hsp70_Ma       FKIVN-EATKPKIQVNYKGEEKVFSPEEISSMVLLKMKETAEAYLGKTINNAVVTVPAYFNDSQRQATKDAGTISGMNVLRIINEPTAAAIAYGLDKK--IGGERHVLIFDLGGGTFDVS
       C._elegans_Hsp-1_Pm          FKVISAECAKPKVQVEYKGENKIFTPEEISSMVLDKMKETAEALGTTVKDAVVTVPAYFNDSQRQATKDAGAIAGLNVLRIINEPTAAAIAYGLDKK--GHGERNVLIFDLGGGTFDVS
       C._elegans_Hsp-1_Ma          FKVISAECAKPKVQVEYKGENKIFTPEEISSMVLDKMKETAEALGTTVKDAVVTVPAYFNDSQRQATKDAGAIAGLNVLRIINEPTAAAIAYGLDKK--GHGERNVLIFDLGGGTFDVS
       P._infestans_Hsp70_Pm        FKITAGPCDKPQITVQFKGETKTFQPEEISSMVLIKMREVAEAFIGKEVKNAVITVPAYFNDSQRQATKDAGAIAGLNVLRIINEPTAAAIAYGLDKK---GGERNVLIFDLGGGTFDVS
       P._infestans_Hsp70_Ma        FKITAGPCDKPQITVQFKGETKTFQPEEISSMVLIKMREVAEAFIGKEVKNAVITVPAYFNDSQRQATKDAGAIAGLNVLRIINEPTAAAIAYGLDKK---GGERNVLIFDLGGGTFDVS
       A._thaliana_Hsp70B_Pm        FKVVSGPGKPKMIVVSYKNEEKQFSPEEISSMVLVKMKEVAEAFLGRTVKNAVVTVPAYFNDSQRQATKDAGAISGLNVLRIINEPTAAAIAYGLDKKGTKAGEKNVLIFDLGGGTFDVS
       A._thaliana_Hsp70B_Ma        FKVVSGPGKPKMIVVSYKNEEKQFSPEEISSMVLVKMKEVAEAFLGRTVKNAVVTVPAYFNDSQRQATKDAGAISGLNVLRIINEPTAAAIAYGLDKKGTKAGEKNVLIFDLGGGTFDVS
       PfHsp70-x_Pm                 FTVKGGSDGKPMIEVSYQGEKKTFFPEEISSMVLKMKEVAETYLGKEVKNAVITVPAYFNDSQRQATKDAGAIAGLNVLRIINEPTAAAIAYGLDKK--GKGEQNILIFDLGGGTFDVS
       PfHsp70-x_Ma                 FTVKGGSDGKPMIEVSYQGEKKTFFPEEISSMVLKMKEVAETYLGKEVKNAVITVPAYFNDSQRQATKDAGAIAGLNVLRIINEPTAAAIAYGLDKK--GKGEQNILIFDLGGGTFDVS
       PfHsp70-1_Ma                 FTVKSGVDEKPMIEVTYQGEKKLFFPEEISSMVLQKMKENAEAFLGKSIKNAVITVPAYFNDSQRQATKDAGTIAGLNVMRIINEPTAAAIAYGLHKK--GKGEKNILIFDLGGGTFDVS
       PfHsp70-1_Ma                 FTVKSGVDEKPMIEVTYQGEKKLFFPEEISSMVLQKMKENAEAFLGKSIKNAVITVPAYFNDSQRQATKDAGTIAGLNVMRIINEPTAAAIAYGLHKK--GKGEKNILIFDLGGGTFDVS
```

Supplementary Data S2A

```
                              ....|....|....|....|....|....|....|....|....|....|....|....|....|....|....|....|....|....|....|....|
A._echinatior_Hsc70-4_Pm      ILTIEDGIFEVKSTAGDTHLGGEDFDNRMVNHFVQEFKRKYK-KDLSSNKRALRRLRTACERAKRTLSSSTQASIEIDSLFEGIDFYTSVTRARFEELCADLFRSTLEPVEKALRDAKMD
A. echinatior_Hsc70-4_Ma      ILTIEDGIFEVKSTAGDTHLGGEDFDNRMVNHFVQEFKRKYK-KDLSSNKRALRRLRTACERAKRTLSSSTQASIEIDSLFEGIDFYTSVTRARFEELCADLFRSTLEPVEKALRDAKMD
S._cerevisiae_Ssa1p_Pm        LLSIEDGIFEVKATAGDTHLGGEDFDNRLVNHFIQEFKRKNK-KDLSTNQRALRRLRTACERAKRTLSSSAQTSVEIDSLFEGIDFYTSITRARFEELCADLFRSTLDPVEKVLRDAKLD
S. cerevisiae_Ssa1p_Ma        LLSIEDGIFEVKATAGDTHLGGEDFDNRLVNHFIQEFKRKNK-KDLSTNQRALRRLRTACERAKRTLSSSAQTSVEIDSLFEGIDFYTSITRARFEELCADLFRSTLDPVEKVLRDAKLD
C._japonica_Hsp70_Pm          ILTIEDGIFEVKSTAGDTHLGGEDFDNRMVNHFVEEFKRKHK-RDIAGNKRAVRRLRTACERAKRTLSSSTQASIEIDSLFEGIDFYTSITRARFEELNADLFRGTLEPVEKALRDAKLD
C. japonica_Hsp70_Ma          ILTIEDGIFEVKSTAGDTHLGGEDFDNRMVNHFVEEFKRKHK-RDIAGNKRAVRRLRTACERAKRTLSSSTQASIEIDSLFEGIDFYTSITRARFEELNADLFRGTLEPVEKALRDAKLD
F._chinensis_Hsc70_Pm         ILTIEDGIFEVKSTAGDTHLGGEDFDNRMVNHFIQEFKRKYK-KDFSENKRSLRRLRTACERAKRTLSSSTQASVEIDSLFEGIDFYTSITRARFEELCADLFRGTLEPVEKSLRDAKMD
F. chinensis_Hsc70_Ma         ILTIEDGIFEVKSTAGDTHLGGEDFDNRMVNHFIQEFKRKYK-KDFSENKRSLRRLRTACERAKRTLSSSTQASVEIDSLFEGIDFYTSITRARFEELCADLFRGTLEPVEKSLRDAKMD
H._diversicolor_Hsc70_Pm      ILTIEDGIFEVKSTAGDTHLGGEDFDNRMVNHFVEEFKRKHK-KDISDNKRAVRRLRTACERAKRTLSSSTQASIEIDSLFEGVDYYTSITRARFEELNADLFRGTLEPVEKALRDAKAD
H. diversicolor_Hsc70_Ma      ILTIEDGIFEVKSTAGDTHLGGEDFDNRMVNHFIQEFKRKHK-KDISDNKRAVRRLRTACERAKRTLSSSTQASIEIDSLFEGVDYYTSITRARFEELNADLFRGTLEPVEKALRDAKAD
R._norvegicus_Hsp70_Pm        ILTIDDGIFEVKATAGDTHLGGEDFDNRLVSHFVEEFKRKHK-KDISQNKRAVRRLATACERAKRTLSSSTQANLEIDSLYEGIDFYTSITRARFEELCADLFRGTLEPVEKSLRDAKMD
R. norvegicus_Hsp70_Ma        ILTIDDGIFEVKATAGDTHLGGEDFDNRLVSHFVEEFKRKHK-KDISQNKRAVRRLATACERAKRTLSSSTQANLEIDSLYEGIDFYTSITRARFEELCADLFRGTLEPVEKSLRDAKMD
Human_Hsp70-1_Pm              ILTIDDGIFEVKATAGDTHLGGEDFDNRLVNHFVEEFKRKHK-KDISQNKRAVRRLRTACERAKRTLSSSTQASLEIDSLFEGIDFYTSITRARFEELCSDLFRSTLEPVEKALRDAKLD
Human_Hsp70-1_Ma              ILTIDDGIFEVKATAGDTHLGGEDFDNRLVNHFVEEFKRKHK-KDISQNKRAVRRLRTACERAKRTLSSSTQASLEIDSLFEGIDFYTSITRARFEELCSDLFRSTLEPVEKALRDAKLD
L._calcarifer_Hsp70_Pm        ILTIEDGIFEVKATAGDTHLGGEDFDNRMVNHFVEEFKRKHK-KDISQNKRALRRLRTACERAKRTLSSSQASIEIDSLFEGIDFYTSITRARFEELCSDLFRGTLEPVEKALRDAKMD
L. calcarifer_Hsp70_Ma        ILTIEDGIFEVKATAGDTHLGGEDFDNRMVNHFVEEFKRKHK-KDISDNKRALRRLRTACERAKRTLSSSQASIEIDSLFEGIDFYTSITRARFEELCSDLFRGTLEPVEKALRDAKMD
X._laevis_Hsp70_Pm            ILTIDDGIFEVKATAGDTHLGGEDFDNRMVNHFIEEFKRKHK-KDISQNKRALRRLRTACERAKRTLSSSQASIEIDSLFEGIDFYTAITRARFEELCSDLFRGTLEPVEKALRDAKLD
X. laevis_Hsp70_Ma            ILTIDDGIFEVKATAGDTHLGGEDFDNRMVNHFVEEFKRKHK-KDISQNKRALRRLRTACERAKRTLSSSQASIEIDSLFEGIDFYTAITRARFEELCSDLFRGTLEPVEKALRDAKLD
C._owczarzaki_Hsp70_Pm        VLTIEDGIFEVKSTAGDTHLGGEDFDNRMVNHFVQEFKRKEK-KDLSTSARALRRLRTACERAKRTLSSSTEASIEIDSLFEGVDFYTKITRARFEELCADLFRGTLDPVEKSLRDAKMD
C. owczarzaki_Hsp70_Ma        VLTIEDGIFEVKSTAGDTHLGGEDFDNRMVNHFVQEFKRKEK-KDLSTSARALRRLRTACERAKRTLSSSTEASIEIDSLFEGVDFYTKITRARFEELCADLFRGTLDPVEKSLRDAKMD
C._elegans_Hsp-1_Pm           ILTIEDGIFEVKSTAGDTHLGGEDFDNRMVNHFCAEFKRKHK-KDLASNPRALRRLRTACERAKRTLSSSTQASIEIDSLFEGIDFYTNITRARFEELCADLFRSTMDPVEKSLRDAKMD
C. elegans_Hsp-1_Ma           ILTIEDGIFEVKSTAGDTHLGGEDFDNRMVNHFCAEFKRKHK-KDLASNPRALRRLRTACERAKRTLSSSTQASIEIDSLFEGIDFYTNITRARFEELCADLFRSTMDPVEKSLRDAKMD
P._infestans_Hsp70_Pm         LLSIEEGIFEVKATAGDTHLGGEDFDNRLVEHFVQEFKRKHR-KDLTQNQRALRRLRTACERAKRTLSSSAQAYIEIDSLFDCVDNSTITRARFEDMCGDYFRKIMEPVEKVLRDAKLS
P. infestans_Hsp70_Ma         LLSIEEGIFEVKATAGDTHLGGEDFDNRLVEHFVQEFKRKHR-KDLTQNQRALRRLRTACERAKRTLSSSAQAYIEIDSLFDCVDNSTITRARFEDMCGDYFRKIMEPVEKVLRDAKLS
A._thaliana_Hsp70B_Pm         LLTIEEGVFEVKATAGDTHLGGEDFDNRLVNHFVAEFRRKHK-KDIAGNARALRRLRTACERAKRTLSSTAQTTIEIDSLHEGIDFYATISRARFEEMNMDLFRKCMDPVEKVLKDAKLD
A. thaliana_Hsp70B_Ma         LLTIEEGVFEVKATAGDTHLGGEDFDNRLVNHFVAEFRRKHK-KDIAGNARALRRLRTACERAKRTLSSTAQTTIEIDSLHEGIDFYATISRARFEEMNMDLFRKCMDPVEKVLKDAKLD
PfHsp70-x_Pm                  LLTLEDGIFEVKATSGDTHLGGEDFDNKLVNFCVQDFKKNGG-KDVSKNSKSLRRLRTCCEKAKRVLSSSAQATIEVDSLFGIDYNVNITRAKFEELCMDQFRNTLIPVEKVLKDAKMD
PfHsp70-x_Ma                  LLTLEDGIFEVKATSGDTHLGGEDFDNKLVNFCVQDFKKNGG-KDVSKNSKSLRRLRTCCEKAKRVLSSSAQATIEVDSLFGIDYNVNITRAKFEELCMDQFRNTLIPVEKVLKDAKMD
PfHsp70-1_Ma                  LLTIEDGIFEVKATAGDTHLGGEDFDNRLVNFCVEDFKRKNRGKDLSKNSRALRRLRTCCERAKRTLSSSTQATIEIDSLFEGIDYSVTVSRARFEELCIDYFRDTLIPVEKVLKDAMMD
PfHsp70-1_Ma                  LLTIEDGIFEVKATAGDTHLGGEDFDNRLVNFCVEDFKRKNRGKDLSKNSRALRRLRTCCERAKRTLSSSTQATIEIDSLFEGIDYSVTVSRARFEELCIDYFRDTLIPVEKVLKDAMMD

                              ....|....|....|....|....|....|....|....|....|....|....|....|....|....|....|....|....|....|....|....|
A._echinatior_Hsc70-4_Pm      KAQVHSIVLVGGSTRIPKIQKLLQDFFNGKELNKSINPDEAVAYGAAVQAAILHG-DKSEEVQDLLLLDVTPLSLGIETAGGVMTTLIKRNTTIPTKQTQTFTTYSDNQPGVLIQVYEGE
A. echinatior_Hsc70-4_Ma      KAQVHSIVLVGGSTRIPKIQKLLQDFFNGKELNKSINPDEAVAYGAAVQAAILHG-DKSEEVQDLLLLDVTPLSLGIETAGGVMTTLIKRNTTIPTKQTQTFTTYSDNQPGVLIQVYEGE
S._cerevisiae_Ssa1p_Pm        KSQVDEIVLVGGSTRIPKVQKLVTDYFNGKEFNRSINPDEAVAYGAAVQAAILTG-DESSKTQDLLLLDVAPLSLGIETAGGVMTKLIERNSTIPTKKSEIFSTYADNQPGVLIQVFEGE
S. cerevisiae_Ssa1p_Ma        KSQVDEIVLVGGSTRIPKVQKLVTDYFNGKEFNRSINPDEAVAYGAAVQAAILTG-DESSKTQDLLLLDVAPLSLGIETAGGVMTKLIERNSTIPTKKSEIFSTYADNQPGVLIQVFEGE
C._japonica_Hsp70_Pm          KGQIEIVLVGGSTRIPKIQKLLQDFFNGKELNKSINPDEAVAYGAAVQAAILMG-DNSENVQDLLLLDVTPLSLGIETAGGVMTALIKRNTTIPTKQTQTFTTYSDNQNSVLVQVYEGE
C. japonica_Hsp70_Ma          KGQIEIVLVGGSTRIPKIQKLLQDFFNGKELNKSINPDEAVAYGAAVQAAILMG-DNSENVQDLLLLDVTPLSLGIETAGGVMTALIKRNTTIPTKQTQTFTTYSDNQNSVLVQVYEGE
F._chinensis_Hsc70_Pm         KAQIHDIVLVGGSTRIPKIQKLLQDFFNGKELNKSINPDEAVAYGAAVQAAILCG-DKSEAVQDLLLLDVTPLSLGIETAGGVMTALIKRNTTIPTKQTQTFTTYSDNQPGVLIQVYEGE
F. chinensis_Hsc70_Ma         KAQIHDIVLVGGSTRIPKIQKLLQDFFNGKELNKSINPDEAVAYGAAVQAAILCG-DKSEAVQDLLLLDVTPLSLGIETAGGVMTALIKRNTTIPTKQTQTFTTYSDNQPGVLIQVYEGE
H._diversicolor_Hsc70_Pm      KVSIHDIVLVGGSTRIPKIQKLLQDFFNGKELCKSINPDEAVAYGAAVQAAILHG-DKSEEVQDLLLLDVTPLSLGIETAGGVMTVLIKRNTTIPTKQTQTFTTYSDNQPGVLIQVEGE
H. diversicolor_Hsc70_Ma      KVSIHDIVLVGGSTRIPKIQKLLQDFFNGKELCKSINPDEAVAYGAAVQAAILHG-DKSEEVQDLLLLDVTPLSLGIETAGGVMTVLIKRNTTIPTKQTQTFTTYSDNQPGVLIQVEGE
R._norvegicus_Hsp70_Pm        KAKIHDIVLVGGSTRIPKVQKLLQDYFNGRDLNKSINPDEAVAYGAAVQAAILMG-DKSEKVQDLLLLDVAPLSLGLETAGGVMTVLIKRNSTIPTKQIFTTYSDNQPGVLIQVYEGE
R. norvegicus_Hsp70_Ma        KAKIHDIVLVGGSTRIPKVQKLLQDYFNGRDLNKSINPDEAVAYGAAVQAAILMG-DKSEKVQDLLLLDVAPLSLGLETAGGVMTVLIKRNSTIPTKQIFTTYSDNQPGVLIQVYEGE
Human_Hsp70-1_Pm              KAQIHDLVLVGGSTRIPKVQKLLQDFFNGRDLNKSINPDEAVAYGAAVQAAILMG-DKSENVQDLLLLDVAPLSLGLETAGGVMTALIKRNSTIPTKQTIFTTYSDNQPGVLIQVYEGE
Human_Hsp70-1_Ma              KAQIHDLVLVGGSTRIPKVQKLLQDFFNGRDLNKSINPDEAVAYGAAVQAAILMG-DKSENVQDLLLLDVAPLSLGLETAGGVMTALIKRNSTIPTKQTIFTTYSDNQPGVLIQVYEGE
L._calcarifer_Hsp70_Pm        KAQIHDIVLVGGSTRIPKIQKLLQDFFNGRELNKSINPDEAVAYGAAVQAAILTG-DTSENVQDLLLLDVAPLSLGIETAGGVMTSLIKRNTTIPTKQAQVFTTYSDNQPGVLIQVYEGE
L. calcarifer_Hsp70_Ma        KAQIHDIVLVGGSTRIPKIQKLLQDFFNGRELNKSINPDEAVAYGAAVQAAILTG-DTSENVQDLLLLDVAPLSLGIETAGGVMTSLIKRNTTIPTKQAQVFTTYSDNQPGVLIQVYEGE
X._laevis_Hsp70_Pm            KSQIHEIVLVGGSTRIPKVQKLLQDFFNGRELNKSINPDEAVAYGAAVQAAILMG-DKSENVQDLLLLDVAPLSLGLETAGGVMTVLIKRNTTIPTKQTQTFTTYSDNQPGVLIQVEGE
X. laevis_Hsp70_Ma            KSQIHEIVLVGGSTRIPKVQKLLQDFFNGRELNKSINPDEAVAYGAAVQAAILMG-DKSENVQDLLLLDVAPLSLGLETAGGVMTVLIKRNTTIPTKQTQTFTTYSDNQPGVLIQVEGE
C._owczarzaki_Hsp70_Pm        KGTIDDIVLVGGSTRIPKVQKLLQDFFNGKELNKSINPDEAVAYGAAVQAAILSG-DKSEAVQDLLLLDVAPLSLGLETAGGVMTTLIKRNTTIPTKQTQTFTTYADNQPGVLIQVYEGE
C. owczarzaki_Hsp70_Ma        KGTIDDIVLVGGSTRIPKVQKLLQDFFNGKELNKSINPDEAVAYGAAVQAAILSG-DKSEAVQDLLLLDVAPLSLGLETAGGVMTTLIKRNTTIPTKQTQTFTTYADNQPGVLIQVYEGE
C._elegans_Hsp-1_Pm           KSQVHDIVLVGGSTRIPKVQKLLSDLFSGKELNKSINPDEAVAYGAAVQAAILSG-DKSEAVQDLLLLDVAPLSLGLETAGGVMTALIKRNTTIPTKTAQTFTTYSDNQPGVLIQVYEGE
C. elegans_Hsp-1_Ma           KSQVHDIVLVGGSTRIPKVQKLLSDLFSGKELNKSINPDEAVAYGAAVQAAILSG-DKSEAVQDLLLLDVAPLSLGLETAGGVMTALIKRNTTIPTKTAQTFTTYSDNQPGVLIQVYEGE
P._infestans_Hsp70_Pm         KSQVHEVVLVGGSTRIPKVQQLLSDFFNGKEENKSINPDEAVAYGATVQAAILSGNDSSEKLQDLLLLDVTPLSLGLETAGGVMTTLIARNTTVPTKKSQTESTYADNQPGVLIQVYEGE
P. infestans_Hsp70_Ma         KSQVHEVVLVGGSTRIPKVQQLLSDFFNGKEENKSINPDEAVAYGATVQAAILSGNDSSEKLQDLLLLDVTPLSLGLETAGGVMTTLIARNTTVPTKKSQTESTYADNQPGVLIQVYEGE
A._thaliana_Hsp70B_Pm         KSSVHDVVLVGGSTRIPKIQQLLQDFFNGKELCKSINPDEAVAYGAAVQAAILTG-EGSEKVQDLLLLDVAPLSLGLETAGGVMTVLIPRNTTVPCKEQVFSTYADNQPGVLIQVYEGE
A. thaliana_Hsp70B_Ma         KSSVHDVVLVGGSTRIPKIQQLLQDFFNGKELCKSINPDEAVAYGAAVQAAILTG-EGSEKVQDLLLLDVAPLSLGLETAGGVMTVLIPRNTTVPCKEQVFSTYADNQPGVLIQVYEGE
PfHsp70-x_Pm                  KSQVHEIVLVGGSTRIPKIQQLIKDFFNGKEPCKAINPDEAVAYGAAVQAAILSG-DQSSAVKDLLLLDVCPLSLGLETAGGVMTKLIERNTTIPTKKNQIFTTYADNQPGVLIQVYEGE
PfHsp70-x_Ma                  KSQVHEIVLVGGSTRIPKIQQLIKDFFNGKEPCKAINPDEAVAYGAAVQAAILSG-DQSSAVKDLLLLDVCPLSLGLETAGGVMTKLIERNTTIPTKKNQIFTTYADNQPGVLIQVYEGE
PfHsp70-1_Ma                  KKSVHEVVLVGGSTRIPKIQTLIKEFFNGKEACRSINPDEAVAYGAAVQAAILSG-DQSNAVQDLLLLDVCSLSLGLETAGGVMTKLIERNTTIPAKKSQIFTTYADNQPGVLIQVYEGE
PfHsp70-1_Ma                  KKSVHEVVLVGGSTRIPKIQTLIKEFFNGKEACRSINPDEAVAYGAAVQAAILSG-DQSNAVQDLLLLDVCSLSLGLETAGGVMTKLIERNTTIPAKKSQIFTTYADNQPGVLIQVYEGE
```

Supplementary Data S2A

```
                                  ....|....|....|....|....|....|....|....|....|....|....|....|....|....|....|....|....|....|....|....|
      A._echinatior_Hsc70-4_Pm    RAMTKDNNILGKFELTGIPPAPRGVPQIEVTFDIDANGILNVSAIEKSTGKENKITITNDKGRLSKEDIERMVNEAEKYRSEDEQRERISAKNALESYCFNMKSTMEDDRVKDKIEASD
      A. echinatior_Hsc70-4_Ma    RAMTKDNNILGKFELTGIPPAPRGVPQIEVTFDIDANGILNVSAIEKSTGKENKITITNDKGRLSKEDIERMVNEAEKYRSEDEQRERISAKNALESYCFNMKSTMEDDRVKDKIEASD
      S._cerevisiae_Ssa1p_Pm      RAKTKDNNLLGKFELSGIPPAPRGVPQIEVTFDVDSNGILNVSAVEKSTGKSNKITITNDKGRLSKEDIEKMVAEAEKYKEEDEKESQRIASKNQLESIAYSLKNTISEA--GDKLEQAD
      S. cerevisiae_Ssa1p_Ma      RAKTKDNNLLGKFELSGIPPAPRGVPQIEVTFDVDSNGILNVSAVEKSTGKSNKITITNDKGRLSKEDIEKMVAEAEKYKEEDEKESQRIASKNQLESIAYSLKNTISEA--GDKLEQAD
      C._japonica_Hsp70_Pm        RAMTKDNNLLGKFDLTGIPPAPRGVPQIEVTFDIDANGILNVSAVDKSTGKENKITITNDKGRLSKDDIDRMVQEAEKYKAEDEANRDRVGAKNSLESYTYNMKQTVEDDKLKGKISDQQ
      C. japonica_Hsp70_Ma        RAMTKDNNLLGKFDLTGIPPAPRGVPQIEVTFDIDANGILNVSAVDKSTGKENKITITNDKGRLSKDDIDRMVQEAEKYKAEDEANRDRVGAKNSLESYTYNMKQTVEDDKLKGKISDQQ
      F._chinensis_Hsc70_Pm       RAMTKDNNLLGKFELSGIPPAPRGVPQIEVTFDIDASGILNVSAVDKSTGKENKITITNDKGRLSKEEIERMVQDAEKYKADDEKQRDRISAKNSLESYCFNMKSTVEDEKEKEKISEED
      F. chinensis_Hsc70_Ma       RAMTKDNNLLGKFELSGIPPAPRGVPQIEVTFDIDASGILNVSAVDKSTGKENKITITNDKGRLSKEEIERMVQDAEKYKADDEKQRDRISAKNSLESYCFNMKSTVEDEKGKEKISEED
      H._diversicolor_Hsc70_Pm    RAMTKDNNILGKFELTGIPPAPRGVPQIEVTFDIDANGILNVSAVDKSTMKENKITITNDKGRLSKEEIERMVNEAENYKAEDEKQKDRIQAKNGLESYAFNMKSTVEDEKLKDKISEDD
      H. diversicolor_Hsc70_Ma    RAMTKDNNILGKFELTGIPPAPRGVPQIEVTFDIDANGILNVSAVDKSTMKENKITITNDKGRLSKEEIERMVNEAENYKAEDEKQKDRIQAKNGLESYAFNMKSTVEDEKLKDKISEDD
      R._norvegicus_Hsp70_Pm      RAMTRDNNLLGRFDLTGIPPAPRGVPQIEVTFDIDANGILNVTAMDKSTGKANKITITNDKGRLSKEEIERMVQEAERYKAEDEGQREKIAAKNALESYAFNMKSAVGDEGLKDKISESD
      R. norvegicus_Hsp70_Ma      RAMTRDNNLLGRFDLTGIPPAPRGVPQIEVTFDIDANGILNVTAMDKSTGKANKITITNDKGRLSKEEIERMVQEAERYKAEDEGQREKIAAKNALESYAFNMKSAVGDEGLKDKISESD
      Human_Hsp70-1_Pm            RAMTKDNNLLGRFELSGIPPAPRGVPQIEVTFDIDANGILNVTATDKSTGKANKITITNDKGRLSKEEIERMVQEAEKYKAEDEVQRERVSAKNALESYAFNMKSAVEDEGLKGKISEAD
      Human_Hsp70-1_Ma            RAMTKDNNLLGRFELSGIPPAPRGVPQIEVTFDIDANGILNVTATDKSTGKANKITITNDKGRLSKEEIERMVQEAEKYKAEDEVQRERVSAKNALESYAFNMKSAVEDEGLKGKISEAD
      L._calcarifer_Hsp70_Pm      RAMTKDNNLLGKFELTGIPPAPRGVPQIEVTFDIDAHGILNVSAVDKSTGKENKITITNDKGRLSKEEIERMVQDADKYKAEDDLQRDKISAKNSLESYAFNMKSSVQDENMKGKISEED
      L. calcarifer_Hsp70_Ma      RAMTKDNNLLGKFELTGIPPAPRGVPQIEVTFDIDAHGILNVSAVDKSTGKENKITITNDKGRLSKEEIERMVQDADKYKAEDDLQRDKISAKNSLESYAFNMKSSVQDENMKGKISEED
      X._laevis_Hsp70_Pm          RAMTKDNNLLGKFELSGIPPAPRGVPQIEVTFDIDANGILNVEKSSGKQNKITITNDKGRLSKDDIEKMVQEAEKYKADDAQRERVDAKNALESYAFNLRKCMVEDENVKGKISDED
      X. laevis_Hsp70_Ma          RAMTKDNNLLGKFELSGIPPAPRGVPQIEVTFDIDANGILNVEKSSGKQNKITITNDKGRLSKDDIEKMVQEAEKYKADDAQRERVDAKNALESYAFNLRKCMVEDENVKGKISDED
      C._owczarzaki_Hsp70_Pm      RAMTRDNNLLGKFELSGIPPAPRGVPQIEVTFDIDANGILNVSAVDKSTGKVNKITITNDKGRLSKEEIERMVAEADKYKAQDEAQRERVAGKNALESYCFNMKQAVDDDNLSKKLSDED
      C. owczarzaki_Hsp70_Ma      RAMTRDNNLLGKFELSGIPPAPRGVPQIEVTFDIDANGILNVSAVDKSTGKVNKITITNDKGRLSKEEIERMVAEADKYKAQDEAQRERVAGKNALESYCFNMKQAVDDDNLSKKLSDED
      C._elegans_Hsp-1_Pm         RAMTKDNNLLGKFELSGIPPAPRGVPQIEVTFDIDANGILNVSATDKSTGKQNKITITNDKGRLSKDDIERMVNEAEKYKADDEAQKDRIGAKNGLESYAFNLKQTIEDEKLKDKISPED
      C. elegans_Hsp-1_Ma         RAMTKDNNLLGKFELSGIPPAPRGVPQIEVTFDIDANGILNVSATDKSTGKQNKITITNDKGRLSKDDIERMVNEAEKYKADDEAQKDRIGAKNGLESYAFNLKQTIEDEKLKDKISPED
      P._infestans_Hsp70_Pm       RTMTFDNNLLGKFNLDGIPPMPRGVPQIDVTFDIDANGILNVSAVEKSTGKENKITITNDKGRLSQAEIERMVAEAEKYKSEDEANKVRIEAKNALENYAYSLRNSLNDEKMKEKIPEAD
      P. infestans_Hsp70_Ma       RTMTFDNNLLGKFNLDGIPPMPRGVPQIDVTFDIDANGILNVSAVEKSTGKENKITITNDKGRLSQAEIERMVAEAEKYKSEDEANKVRIEAKNALENYAYSLRNSLNDEKMKEKIPEAD
      A._thaliana_Hsp70B_Pm       RARTRDNNLLGTFELKGIPPAPRGVPQINVCFDIDANGILNVSAEDKTAGVKNQITITNDKGRLSKEEIEKMVQDAEKYKAEDEQVKKKVEAKNSLENYAYNMRNTIKDEKLAQKLTQED
      A. thaliana_Hsp70B_Ma       RARTRDNNLLGTFELKGIPPAPRGVPQINVCFDIDANGILNVSAEDKTAGVKNQITITNDKGRLSKEEIEKMVQDAEKYKAEDEQVKKKVEAKNSLENYAYNMRNTIKDEKLAQKLTQED
      PfHsp70-x_Pm                RAMTKDNNLLGKFQLEGIPPAPRSVPQIEVTFDIDANGILNVTALDKGTGKQNQITITNDKGRLSKDDIDRMVNDAEKYKEEDEQNKNRIEARNLENYCYNVKNTLQDENLKTKIPKDD
      PfHsp70-x_Ma                RAMTKDNNLLGKFQLEGIPPAPRSVPQIEVTFDIDANGILNVTALDKGTGKQNQITITNDKGRLSKDDIDRMVNDAEKYKEEDEQNKNRIEARNLENYCYNVKNTLQDENLKTKIPKDD
      PfHsp70-1_Ma                RALTKDNNLLGKFHLDGIPPAPRKVPQIEVTFDIDANGILNVTAVEKSTGKQNHITITNDKGRLSQDEIDRMVNDAEKYKAEDEENRKRIEARNSLENYCYGVKSSLEDQKIKEKLQPAE
      PfHsp70-1_Ma                RALTKDNNLLGKFHLDGIPPAPRKVPQIEVTFDIDANGILNVTAVEKSTGKQNHITITNDKGRLSQDEIDRMVNDAEKYKAEDEENRKRIEARNSLENYCYGVKSSLEDQKIKEKLQPAE

                                  ....|....|....|....|....|....|....|....|....|....|....|....|....|....|....|....|....|....|....|....|
      A._echinatior_Hsc70-4_Pm    KEKVLSKCNEVISWLDANQLAEKEEFADKQKELEALCNPIVTKLYQSGGAP-----GGFPGAGG----------------AGANPGAGGAGPTIEEVD
      A. echinatior_Hsc70-4_Ma    KEKVLSKCNEVISWLDANQLAEKEEFADKQKELEALCNPIVTKLYQSGGAP-----GGFPGA-----------------GGAGANPGAGGAGPTIEEVD
      S._cerevisiae_Ssa1p_Pm      KDTVTKKAEETISWLDSNTTASKEEFDDKLKELQDIANPIMSKLYQAXGAP-GGAAGGAPGGFP--------------GGAPPAPEAEGPTVEEVD
      S. cerevisiae_Ssa1p_Ma      KDTVTKKAEETISWLDSNTTASKEEFDDKLKELQDIANPIMSKLYQAXGAP-GGAAGGAPGGFP--------------GGAPPAPEAEGPTVEEVD
      C._japonica_Hsp70_Pm        KQKVLDKCREVISWLDRNQMAEKEEYEHKQKELEKICNPIVTKLYQGAGGA--------------------------------GACGSGGPTIEEVD
      C. japonica_Hsp70_Ma        KQKVLDKCREVISWLDRNQMAEKEEYEHKQKELEKICNPIVTKLYQGAGGA--------------------------------GACGSGGPTIEEVD
      F._chinensis_Hsc70_Pm       RNKILETCNETEKWLDVNQLGEKEEYEHKQKEIEQVCNPIITKMYAAAGGA---PPGGMPGGFPG---------GAPGAGGAAPGACGSSGPTIEEVD
      F. chinensis_Hsc70_Ma       RNKILETCNETIKWLDVNQLGEKEEYEHKQKEIEQVCNPIITKMYAAAGGA---PPGGMPGGFPGGAPG----------AGGAAPGACGSSGPTIEEVD
      H._diversicolor_Hsc70_Pm    KKTITDKCNDVISWLDSNQLAEKDEFEHKQKELEGVCNPIITKLYQAAGGA-----GGMPNFNPG-------AAGAGGAGGAQTGCSSGPTIEEVD
      H. diversicolor_Hsc70_Ma    KKTITDKCNDVISWLDSNQLAEKDEFEHKQKELEGVCNPIITKLYQAAGGA-----GGMPNFNPGA------AGAGGAGGAQTGCSSGPTIEEVD
      R._norvegicus_Hsp70_Pm      KKKIILDKCSEVLSWLEANQLAEKEEFDHKRKELENMCNPIITKLYQSGCTG---PTCAPGYT--------------------PGRARTGPTIEEVD
      R. norvegicus_Hsp70_Ma      KKKIILDKCSEVLSWLEANQLAEKEEFDHKRKELENMCNPIITKLYQSGCTG---PTCAPGY--------------------TPGRARTGPTIEEVD
      Human_Hsp70-1_Pm            KKKVLDKCQEVISWLDANTLAEKDEFEHKRKELEQVCNPIISGLYQGAGGP---GPGGFG---------------------AQGPKGGSGSGPTIEEVD
      Human_Hsp70-1_Ma            KKKVLDKCQEVISWLDANTLAEKDEFEHKRKELEQVCNPIISGLYQGAGGP---GPGGF-----------------------GAQGPKGGSGSGPTIEEVD
      L._calcarifer_Hsp70_Pm      QKKVIEKCDETITWLENNQLADKEEYQHQKELEKVCNPIISKLYQ------GGMPASS---------------CREEARAGSCGPTIEEVD
      L. calcarifer_Hsp70_Ma      QKKVIEKCDETITWLENNQLAEKEEYQHQKELEKVCNPIISKLYQGGMPA-------------------------SSCREEARAGSCGPTIEEVD
      X._laevis_Hsp70_Pm          KRTISEKCTQVISWLENNQLAEKEEYAFQKDLEKVCPIITKLYQGSVPG--GVPGGMPGSSC---------------GAQARQCGSSGPTIEEVD
      X. laevis_Hsp70_Ma          KRTISEKCTQVISWLENNQLAEKEEYAFQKDLEKVQPIITKLYQGSVPG--GVPGGMPGS-----------------SCGAQARQCGSSGPTIEEVD
      C._owczarzaki_Hsp70_Pm      KKTVIEKVEEAMKWLEANQLAEKEEFEHRLKELEKACGSPIIAKAYQGGAAP-----GGMPGAEGG-----------WQGPGAGATADPAAGPTIEEVD
      C. owczarzaki_Hsp70_Ma      KKTVIEKVEAMKWLEANQLAEKEEFEHRLKELEKACSPIIAKAYQGGAAP-----GGMPGAE-----------GGWQGPGAGATADPAAGPTIEEVD
      C._elegans_Hsp-1_Pm         KKKIEDKCDEILKWLDSNQLAEKEEFEHQQKDLEGLANPIISKLYQSAGGA--PPGAAPGG--------------------AAGGAGPTIEEVD
      C. elegans_Hsp-1_Ma         KKKIEDKCDEILKWLDSNQLAEKEEFEHQQKDLEGLANPIISKLYQSAGGA---PPGAAPGG--------------------AAGGAGPTIEEVD
      P._infestans_Hsp70_Pm       KKVVDDKVTEVIQWMDAHQSSEKEEYESKQKELESVANPVLQKVYASAGGAGDGMPGSMPNDMPG--------------TDSRSSGAEQGPKIEEVD
      P. infestans_Hsp70_Ma       KKVVDDKVTEVIQWMDAHQSSEKEEYESKQKELESVANEVLQKVYASAGGAGDGMPGSMPNDMPG--------------TDSRSSGAEQGPKIEEVD
      A._thaliana_Hsp70B_Pm       KQKIEKAIDETTEWIEGNQLAEVDEFEYKLKELEGICNPIISKMYQGGAAA----GGMPTD-----------------GDFSSCAAGPKIEEVD
      A. thaliana_Hsp70B_Ma       KQKIEKAIDETEWIGQNQLADVDEFEYKLKELEGICNPIISKLYQGGAAA-----GGMPTD-----------------GDFSSCAAGPKIEEVD
      PfHsp70-x_Pm                SEKCMKTVKSVLDWLEKNQTAEETEEYNEKEKDISSVYNPIMTKIYQGASAQ---EQKAEAT--------------NLRGRNSENKEAQNNGPTVEEVN
      PfHsp70-x_Ma                SEKCMKTVKSVLDWLEKNQTAEETEEYNEKEKDISSVYNPIMTKIYQGASAQ---EPQKAEA--------------TNLRGRNSENKEAQNNGPTVEEVN
      PfHsp70-1_Ma                IETCMKTITTTILEWLEKNQLAGKDEYEAKQKEAESVCAPIMSKIYQDAAGAAGGMPGGMPGGMPGGMPGGMNFPGGMPGAGMPGNAPAGSGPTVEEVD
      PfHsp70-1_Ma                IETCMKTITTTILEWLEKNQLAGKDEYEAKQKEAESVCAPIMSKIYQDAAGAAGGMPGGMPGGMPGGMPGGMNFPGGMPGAGMPGNAPAGSGPTVEEVD
```

Supplementary Data S2B

```
                                      10        20        30        40        50        60        70        80        90       100       110       120
                                ....|....|....|....|....|....|....|....|....|....|....|....|....|....|....|....|....|....|....|....|....|....|....|....|
     Human_Grp78_Pm             ------------------------MK-------LSLVAAMLLLLSAA--RAEEED---KKEDV-GTVVGIDLGTTYSCVGVFKNGRVEIIANDQGNRITPSYVAFTPE
     Human_Grp78_Ma             ----------------MKLSLVAA------MLLLLSAA----RAEE--------ED--KKEDV-GTVVGIDLGTTYSCVGVFKNGRVEIIANDQGNRITPSYVAFTPE
     R._norvegicus_Grp78_Pm     ------------------------MK---------FTVVAAALLLLCAV--RAEEED---KKEDV-GTVVGIDLGTTYSCVGVFKNGRVEIIANDQGNRITPSYVAFTPE
     R._norvegicus_Grp78_Ma     ----------------MKFTVVAA------ALLLLCAV----RAEE--------ED--KKEDV-GTVVGIDLGTTYSCVGVFKNGRVEIIANDQGNRITPSYVAFTPE
     O._latipes_Bip_Pm          ------------------------MK----------LLWAVMLVVGAV--FAEEED---KKDSV-GTVIGIDLGTTYSCVGVFKNGRVEIIANDQGNRITPSYVAFTSE
     O._latipes_Bip_Ma          ----------------MK--LLWA------VMLVVGAV----FAEE--------ED--KKDSV-GTVIGIDLGTTYSCVGVFKNGRVEIIANDQGNRITPSYVAFTSE
     F._chinensis_Grp78_Pm      ------------------------MR------CWTALAAIAAVMAV--AATAKD---KKEEV-GTVIGIDLGTTYSCVGVFKNGRVEIIANDQGNRITPSYVAFTAD
     F._chinensis_Grp78_Ma      ----------------MRCWTALA------AIAAVMAV----AATA--------KD--KKEEV-GTVIGIDLGTTYSCVGVFKNGRVEIIANDQGNRITPSYVAFTAD
     C._elegans_Hsp-3_Pm        ------------------------MK--------TLFLLGLIALSAVSVYCEEEKTEKKETKY-GTIIGIDLGTTYSCVGVYKNGRVEIIANDQGNRITPSYVAFSGD
     C._elegans_Hsp-3_Ma        ----------------MKTLFL---------LGLIALSA-----VSVYCEEEEKTEKKETKY-GTIIGIDLGTTYSCVGVYKNGRVEIIANDQGNRITPSYVAFSGD
     A._thaliana_BiP-2_Pm       ---------------------MARSFGANSTVVLAIIFFGCLFAFS--TAKEEA---TKL-GSVIGIDLGTTYSCVGVYKNGRVEIIANDQGNRITPSWVSFTD-
     A._thaliana_BiP-2_Ma       --------------MARSFGANSTVVLAIIFFGCLFA--------FSTA---KE--EATKL-GSVIGIDLGTTYSCVGVYKNGHVEIIANDQGNRITPSWVSFTD-
     S._cerevisiae_Kar2p_Pm     --------------MFFNRLSAGKLLVPLSVVLYALFVVILPLQNSFHSSNVL--VRGADD----VENY-GTVIGIDLGTTYSCVAVMKNGKTEILANEQGNRITPSYVAFTDD
     S._cerevisiae_Kar2p_Ma     -MFFNRLSAGKLLVPLSVVLYA------------LFVVILPLQNSFHSSNVLVRG--------AD--DVENY-GTVIGIDLGTTYSCVAVMKNGKTEILANEQGNRITPSYVAFTDD
     A._californica_Grp78_Pm    ---------------------MD-RFTPFFLLVILFSSNLLVRAD--GDEEEDEGDKKKSEV-GTVIGIDLGTTYSCVGVFKNGRVDIIANDQGNRITPSYVAFTAD
     A._californica_Grp78_Ma    --------------MDRFTPFF-------LLVILFSS------NLLVRADGDEEEDEGDKKKSEV-GTVIGIDLGTTYSCVGVFKNGRVDIIANDQGNRITPSYVAFTAD
     X._laevis_Hspa5_Pm         ---------------------MV--------TMKLFALVLLVSASVFAADDDD---KKEDV-GTVVGIDLGTTYSCVGVFKNGRVEIIANDQGNRITPSYVAFTPE
     X._laevis_Hspa5_Ma         --------------MVTMKLFA--------LVLLVSASV----FAADD------DD--KKEDV-GTVVGIDLGTTYSCVGVFKNGRVEIIANDQGNRITPSYVAFTPE
     A._echinatior_Hsc70-3_Pm   ----------------------MK---------GVISLLIVGLLAVAT--IAKEDK-EKEDI-GTVIGIDLGTTYSCVGVYKNGRVEIIANDQGNRITPSYVAFTND
     A._echinatior_Hsc70-3_Ma   ---------------MKGVISL---------LIVGLLAV-----ATIA---KED-KEKEDI-GTVIGIDLGTTYSCVGVYKNGRVEIIANDQGNRITPSYVAFTND
     C._owczarzaki_Hsp70-Like_Pm ---------------------MR------VAYLASLALVALTLAL--VNADET---KQREY-GTVIGIDLGTTYSCVGVFKNGRVEIIPNDQGNRITPSYVAFTDD
     C._owczarzaki_Hsp70-Like_Ma --------------MRVAYLAS-------LALVALTL-----ALVNAD-------ET--KQREY-GTVIGIDLGTTYSCVGVFKNGRVEIIPNDQGNRITPSYVAFTDD
     C._cohnii_BiP_Pm           --------GTSVDVLSFGKLSAWDTAPLFAMARLHWAGIFVLALCVA--VAKDDE-----KKIDGPVIGIDLGTTYSCVGVIYKNGRNRITPSYVAFTED
     C._cohnii_BiP_Ma           GTSVDVLSFGKLSAWDTAPLFAMARLHWAG-------IFVLALAC------VAVA-----KD--DEKKIDGPVIGIDLGTTYSCVGVIYKNGRVEIIPNDQGNRITPSYVAFTED
     PfHsp70-2_Pm               ---------------------MK-----QIRPYILLLIVSLLKFI--SAVD-------SNIEGPVIGIDLGTTYSCVGVFKNGRVEILNNELGNRITPSYVSFVD-
     PfHsp70-2_Ma               --------------MKQIRPYI-------LLLIVSLLK-------FISAV--------DSNIEGPVIGIDLGTTYSCVGVFKNGRVEILNNELGNRITPSYVSFVD-

     Human_Grp78_Pm             -GERLIGDAAKNQLTSNPENTVFDAKRLIGRTWNDPSVQQDIKFLPFKVVEKKTKPYIQVDIGGGQTKTFAPEEISAMVLTKMKETAEAYLGKKVTHAVVTVPAYFNDAQRQATKDAGTI
     Human_Grp78_Ma             -GERLIGDAAKNQLTSNPENTVFDAKRLIGRTWNDPSVQQDIKFLPFKVVEKKTKPYIQVDIGGGQTKTFAPEEISAMVLTKMKETAEAYLGKKVTHAVVTVPAYFNDAQRQATKDAGTI
     R._norvegicus_Grp78_Pm     -GERLIGDAAKNQLTSNPENTVFDAKRLIGRTWNDPSVQQDIKFLPFKVVEKKTKPYIQVDIGGGQTKTFAPEEISAMVLTKMKETAEAYLGKKVTHAVVTVPAYFNDAQRQATKDAGTI
     R._norvegicus_Grp78_Ma     -GERLIGDAAKNQLTSNPENTVFDAKRLIGRTWNDPSVQQDIKFLPFKVVEKKTKPYIQVDIGGGQTKTFAPEEISAMVLTKMKETAEAYLGKKVTHAVVTVPAYFNDAQRQATKDAGTI
     O._latipes_Bip_Pm          -GERLIGDAAKNQLTSNPENTVFDAKRLIGRSWGDPSVQQDIKYFPFKVIEKKSKPHIQVDIGGGQMKTFAPEEISAMVLTKMKETAEAYLGKKVTHAVVTVPAYFNDAQRQATKDAGTI
     O._latipes_Bip_Ma          -GERLIGDAAKNQLTSNPENTVFDAKRLIGRSWGDPSVQQDIKYFPFKVIEKKSKPHIQVDIGGGQMKTFAPEEISAMVLTKMKETAEAYLGKKVTHAVVTVPAYFNDAQRQATKDAGTI
     F._chinensis_Grp78_Pm      -GERLIGDSAKNQLTTNPENTVFDAKRLIGREWTDKSVQHDIQFFPFKVINKNDKPHIKVATVQGE-KVFAAEEISAMVLGKMKETAEAYLGKPVTHAVVTVPAYFNDAQRQATKDAGTI
     F._chinensis_Grp78_Ma      -GERLIGDSAKNQLTTNPENTVFDAKRLIGREWTDKSVQHDIQFFPFKVINKNDKPHIKVATVQGE-KVFAAEEISAMVLGKMKETAEAYLGKPVTHAVVTVPAYFNDAQRQATKDAGTI
     C._elegans_Hsp-3_Pm        QGDRLIGDAAKNQLTTNPENTIFDAKRLIGRDYNDKTVQADIKHWPFKVIDKSNKPSVEVKVGSDN-KQFTPEEVSAMVLVKMKETAESYLGKEVKNAVVTVPAYFNDAQRQATKDAGTI
     C._elegans_Hsp-3_Ma        QGDRLIGDAAKNQLTTNPENTIFDAKRLIGRDYNDKTVQADIKHWPFKVIDKSNKPSVEVKVGSDN-KQFTPEEVSAMVLVKMKETAESYLGKEVKNAVVTVPAYFNDAQRQATKDAGTI
     A._thaliana_BiP-2_Pm       -SERLIGEAAKNQAAVNPERTVFDVKRLIGRKFEDKEVQKDRKLVPYQIVNKDGKPYIQVKIKDGETKVFSPEEISAMILTKMKETAEAYLGKKIKDAVVTVPAYFNDAQRQATKDAGVI
     A._thaliana_BiP-2_Ma       -SERLIGEAAKNQAAVNPERTVFDVKRLIGRKFEDKEVQKDRKLVPYQIVNKDGKPYIQVKIKDGETKVFSPEEISAMILTKMKETAEAYLGKKIKDAVVTVPAYFNDAQRQATKDAGVI
     S._cerevisiae_Kar2p_Pm     --ERLIGDAAKNQVAANPQNTIFDIKRLIGLKYNDRSVQKDIKHLPFNVVNKDGKPAVEVSVKGEK-KVFTPEEISGMILGKMKQIAEDYLGTKVTHAVVTVPAYFNDAQRQATKDAGTI
     S._cerevisiae_Kar2p_Ma     --ERLIGDAAKNQVAANPQNTIFDIKRLIGLKYNDRSVQKDIKHLPFNVVNKDGKPAVEVSVKGEK-KVFTPEEISGMILGKMKQIAEDYLGTKVTHAVVTVPAYFNDAQRQATKDAGTI
     A._californica_Grp78_Pm    -GERLIGDAAKNQLTSNPENTIFDVKRLIGRTFDDKSVQHDIKFYPFKVTNANNKPHIQAATGEGD-RSFAPEEISAMVLSKMRDIAEEYLGKKITNAVVTVPAYFNDAQRQATKDAGTI
     A._californica_Grp78_Ma    -GERLIGDAAKNQLTSNPENTIFDVKRLIGRTFDDKSVQHDIKFYPFKVTNANNKPHIQAATGEGD-RSFAPEEISAMVLSKMRDIAEEYLGKKITNAVVTVPAYFNDAQRQATKDAGTI
     X._laevis_Hspa5_Pm         -GERLIGDAAKNQLTSNPENTVFDAKRLIGRTWNDPSVQQDIKYLPFKVIEKKTKPYIQVNIG-DQMKTFAPEEISAMVLVKMKETAEAYLGKKVTHAVVTVPAYFNDAQRQATKDAGVI
     X._laevis_Hspa5_Ma         -GERLIGDAAKNQLTSNPENTVFDAKRLIGRTWNDPSVQQDIKYLPFKVIEKKTKPYIQVNIG-DQMKTFAPEEISAMVLVKMKETAEAYLGKKVTHAVVTVPAYFNDAQRQATKDAGVI
     A._echinatior_Hsc70-3_Pm   -GERLIGDAAKNQLTTNPENTVFDAKRLIGREWSDPTVQHDVKFFPFPVIEKNNKPHIKVETSQGE-KVFAPEEISAMVLGKMKETAEAYLGKKVTHAVVTVPAYFNDAQRQATKDAGTI
     A._echinatior_Hsc70-3_Ma   -GERLIGDAAKNQLTTNPENTVFDAKRLIGREWSDPTVQHDVKFFPFPVIEKNNKPHIKVETSQGE-KVFAPEEISAMVLGKMKETAEAYLGKKVTHAVVTVPAYFNDAQRQATKDAGTI
     C._owczarzaki_Hsp70-Like_Pm-GERLIGDAAKNQLTANPFNTIFDAKRLIGREFNEKSIQADIKLWPFKVVNKNAKPEYIKVATSAGD-KVLSPEEVSAMILTKMKETAEAYLGHKVTHAVVTVPAYFNDAQRQATKDAGTI
     C._owczarzaki_Hsp70-Like_Ma-GERLIGDAAKNQLTANPFNTIFDAKRLIGREFNEKSIQADIKLWPFKVVNKNAKPEYIKVATSAGD-KVLSPEEVSAMILTKMKETAEAYLGHKVTHAVVTVPAYFNDAQRQATKDAGTI
     C._cohnii_BiP_Pm           --ERLIGEAAKNQATINPAQTLFDVKRLIGRRFKDSTVQKDIKLLPFDIVDKNGKPQISVKVKGET-KQMAPEEVSSMVLTKMKETAENYLGKEVKHAVITVPAYFNDAQRQSTKDAGVI
     C._cohnii_BiP_Ma           --ERLIGEAAKNQATINPAQTLFDVKRLIGRRFKDSTVQKDIKLLPFDIVDKNGKPQISVKVKGET-KQMAPEEVSSMVLTKMKETAENYLGKEVKHAVITVPAYFNDAQRQSTKDAGVI
     PfHsp70-2_Pm               -GERKVGEAAKLEATLHPTQTVFDVKRLIGRKFDDQEVVKDRSLLEYEIVNNQGKPNIKVQIKDKD-TTFAPEQISAMVLEKMKETAQSFLGKPVKNAVVTVPAYFNDAQRQATKDAGTI
     PfHsp70-2_Ma               -GERKVGEAAKLEATLHPTQTVFDVKRLIGRKFDDQEVVKDRSLLEYEIVNNQGKPNIKVQIKDKD-TTFAPEQISAMVLEKMKETAQSFLGKPVKNAVVTVPAYFNDAQRQATKDAGTI
```

Supplementary Data S2B

```
                                    ....|....|....|....|....|....|....|....|....|....|....|....|....|....|....|....|....|....|....|....|
Human_Grp78_Pm                      AGLNVMRIINEPTAAAIAYGLDKREGEKNILVFDLGGGTFDVSLLTIDNGVFEVVATNGDTHLGGEDFDQRVMEHFIKLYKKKTGKDVRKDNRAVQKLRREVEKAKRALSSQHQARIEIE
Human_Grp78_Ma                      AGLNVMRIINEPTAAAIAYGLDKREGEKNILVFDLGGGTFDVSLLTIDNGVFEVVATNGDTHLGGEDFDQRVMEHFIKLYKKKTGKDVRKDNRAVQKLRREVEKAKRALSSQHQARIEIE
R._norvegicus_Grp78_Pm              AGLNVMRIINEPTAAAIAYGLDKREGEKNILVFDLGGGTFDVSLLTIDNGVFEVVATNGDTHLGGEDFDQRVMEHFIKLYKKKTGKDVRKDNRAVQKLRREVEKAKRALSSQHQARIEIE
R._norvegicus_Grp78_Ma              AGLNVMRIINEPTAAAIAYGLDKREGEKNILVFDLGGGTFDVSLLTIDNGVFEVVATNGDTHLGGEDFDQRVMEHFIKLYKKKTGKDVRKDNRAVQKLRREVEKAKRALSSQHQARIEIE
O._latipes_Bip_Pm                   AGLNVMRIINEPTAAAIAYGLDKKEGEKNILVFDLGGGTFDVSLLTIDNGVFEVVATNGDTHLGGEDFDQRVMEHFIKLYKKKTGKDVRKDNRAVQKLRREVEKAKRALSAQHQARIEIE
O._latipes_Bip_Ma                   AGLNVMRIINEPTAAAIAYGLDKDGEKNILVFDLGGGTFDVSLLTIDNGVFEVVATNGDTHLGGEDFDQRVMEHFIKLYKKKTGKDVRKDNRAVQKLRREVEKAKRALSAQHQARIEIE
F._chinensis_Grp78_Pm               AGLTVMRIINEPTAAAIAYGIDKKDGEKNILVFDLGGGTFDVSLLTIDSGVFEVVATNGDTHLGGEDFDQRVMDHFIKLYKKKKGKDIRRDNRAVQKLRREVEKAKRSLSSSHQVRIEIE
F._chinensis_Grp78_Ma               AGLTVMRIINEPTAAAIAYGIDKKDGEKNILVFDLGGGTFDVSLLTIDSGVFEVVATNGDTHLGGEDFDQRVMDHFIKLYKKKKGKDIRRDNRAVQKLRREVEKAKRSLSSSHQVRIEIE
C._elegans_Hsp-3_Pm                 AGLNVVRIINEPTAAAIAYGLDKKDGERNILVFDLGGGTFDVSMLTIDNGVFEVLATNGDTHLGGEDFDQRVMEYFIKLYKKKSGKDLRKDKRAVQKLRREVEKAKRALSTQHQTKVEIE
C._elegans_Hsp-3_Ma                 AGLNVVRIINEPTAAAIAYGLDKKDGERNILVFDLGGGTFDVSMLTIDNGVFEVLATNGDTHLGGEDFDQRVMEYFIKLYKKKSGKDLRKDKRAVQKLRREVEKAKRALSTQHQTKVEIE
A._thaliana_BiP-2_Pm                AGLNVARIINEPTAAAIAYGLDKKGEKNILVFDLGGGTFDVSVLTIDNGVFEVLSTNGDTHLGGEDFDHRIMEYFIKLIKKKHQKDISDNKALGKLRRECERAKRALSSQHQVRVEIE
A._thaliana_BiP-2_Ma                AGLNVARIINEPTAAAIAYGLDKKGEKNILVFDLGGGTFDVSVLTIDNGVFEVLSTNGDTHLGGEDFDHRIMEYFIKLIKKKHCKDISKDNKALGKLRRECERAKRALSSQHQVRVEIE
S._cerevisiae_Kar2p_Pm              AGLNVLRIVNEPTAAAIAYGLDKSDKEHQIIVYDLGGGTFDVSLLSIENGVFEVQATSGDTHLGGEDFDYKIVRQIIKAFKKHGIDVSDNNKALAKLRREAEKAKRALSSQMSTRIEID
S._cerevisiae_Kar2p_Ma              AGLNVLRIVNEPTAAAIAYGLDKSDKEHQIIVYDLGGGTFDVSLLSIENGVFEVQATSGDTHLGGEDFDYKIVRQIIKAFKKHGIDVSDNNKALAKLRREAEKAKRALSSQMSTRIEID
A._californica_Grp78_Pm             AGLNVMRIINEPTAAAIAYGLDKKEGEKNILVFDLGGGTFDVSLLTIDNGVFEVVSTNGDTHLGGEDFDQRVMEHFIKLYKKKKGKDIRKDNRAVQKLRREVEKAKRALSSAHQVRLEIE
A._californica_Grp78_Ma             AGLNVMRIINEPTAAAIAYGLDKKEGEKNILVFDLGGGTFDVSLLTIDNGVFEVVSTNGDTHLGGEDFDQRVMEHFIKLYKKKKGKDIRKDNRAVQKLRREVEKAKRALSSAHQVRLEIE
X._laevis_Hspa5_Pm                  AGLNVMRIINEPTAAAIAYGLDKREGEKNILVFDLGGGTFDVSLLTIDNGVFEVVATNGDTHLGGEDFDQRVMEHFIKLYKKKTGKDVRKDNRAVQKLRREVEKAKRALSAQHQSRIEIE
X._laevis_Hspa5_Ma                  AGLNVMRIINEPTAAAIAYGLDKREGEKNILVFDLGGGTFDVSLLTIDNGVFEVVATNGDTHLGGEDFDQRVMEHFIKLYKKKTGKDVRKDNRAVQKLRREVEKAKRALSAQHQSRIEIE
A._echinatior_Hsc70-3_Pm            SGLVMRIINEPTAAAIAYGLDKKDGEKNVLVFDLGGGTFDVSLLTIDNGVFEVVATNGDTHLGGEDFDQRVMDHFIKLYKKKKGKDIRKDNRAVQKLRREVEKAKRALSASHQVRIEIE
A._echinatior_Hsc70-3_Ma            SGLVMRIINEPTAAAIAYGLDKKDGEKNVLVFDLGGGTFDVSLLTIDNGVFEVVATNGDTHLGGEDFDQRVMDHFIKLYKKKKGKDIRKDNRAVQKLRREVEKAKRALSASHQVRIEIE
C._owczarzaki_Hsp70-Like_Pm         AGLTVMRIINEPTAAAIAYGLDKNDGEKNILVFDLGGGTFDVSLLTIDNGVFEVVATNGDTHLGGQDFDQRIMQHRMRQYEKKGKDISSDVRAVQKLRREAEKAKRALSSQHSARLEIE
C._owczarzaki_Hsp70-Like_Ma         AGLTVLRIINEPTAAAIAYGLDKNDGEKNILVFDLGGGTFDVSLLTIDNGVFEVVATNGDTHLGGQDFDQRIMQHRMRQYKEKKGKDISSDVRAVQKLRREAEKAKRALSSQHSARLEIE
C._cohnii_BiP_Pm                    AGLNVIRIINEPTAAAIAYGLDKK-AEKNILVYDLGGGTFDVSLLTIDNGVFEVVATNGDTHLGGEDFDQRVMQHPMKVFQKKHSKDMSKDKRAIQKLRREVEKAKRALSSTHQARLEIE
C._cohnii_BiP_Ma                    AGLNVIRIINEPTAAAIAYGLDKK-AEKNILVYDLGGGTFDVSLLTIDNGVFEVVATNGDTHLGGEDFDQRVMQHPMKVFQKKHSKDMSKDKRAIQKLRREVEKAKRALSSTHQARLEIE
PfHsp70-2_Pm                        AGLNIVRIINEPTAAALAYGLDKKE-ETSILVYDLGGGTFDVSILVIDNGVFEVYATAGNTHLGGEDFDQRVMDYFIKMFKKKNNIDLRTDKRAIQKLRKEVEIAKRNLSVVHSTQIEIE
PfHsp70-2_Ma                        AGLNIVRIINEPTAAALAYGLDKKE-ETSILVYDLGGGTFDVSILVIDNGVFEVYATAGNTHLGGEDFDQRVMDYFIKMFKKKNNIDLRTDKRAIQKLRKEVEIAKRNLSVVHSTQIEIE

                                    ....|....|....|....|....|....|....|....|....|....|....|....|....|....|....|....|....|....|....|....|
Human_Grp78_Pm                      SFYEGEDFSETLTRAKFEELNMDLFRSTMKPVQKVLEDSDLKKSDIDEIVLVGGSTRIPKIQQLVKEFFNGKEPSRGINPDEAVAYGAAVQAGVLSGD--QDTGDLVLLDVCPLTLGIET
Human_Grp78_Ma                      SFYEGEDFSETLTRAKFEELNMDLFRSTMKPVQKVLEDSDLKKSDIDEIVLVGGSTRIPKIQQLVKEFFNGKEPSRGINPDEAVAYGAAVQAGVLSGD--QDTGDLVLLDVCPLTLGIET
R._norvegicus_Grp78_Pm              SFFEGEDFSETLTRAKFEELNMDLFRSTMKPVQKVLEDSDLKKSDIDEIVLVGGSTRIPKIQQLVKEFFNGKEPSRGINPDEAVAYGAAVQAGVLSGD--QDTGDLVLLDVCPLTLGIET
R._norvegicus_Grp78_Ma              SFFEGEDFSETLTRAKFEELNMDLFRSTMKPVQKVLEDSDLKKSDIDEIVLVGGSTRIPKIQQLVKEFFNGKEPSRGINPDEAVAYGAAVQAGVLSGD--QDTGDLVLLDVCPLTLGIET
O._latipes_Bip_Pm                   SFFEGEDFSETLTRAKFEELNMDLFRSTMKPVQKVLEDSDLKKSEIDEIVLVGGSTRIPKIQQLVKEFFNGKEPSRGINPDEAVAYGAAVQAGVLSGE--EDTGDVVLLDVCPLTLGIET
O._latipes_Bip_Ma                   SFFEGEDFSETLTRAKFEELNMDLFRSTMKPVQKVLEDSDLKKSEIDEIVLVGGSTRIPKIQQLVKEFFNGKEPSRGINPDEAVAYGAAVQAGVLSGE--EDTGDVVLLDVCPLTLGIET
F._chinensis_Grp78_Pm               SFFEGDDFSETLTRAKFEELNMDLFRSTMKPVQKVLEDSDLQKKEIDEIVLVGGSTRIPKIQQLVKEFFGKEPSRGINPDEAVAYGAAVQAGVLSGE--DDTNDLVLLDVNPLTLGIET
F._chinensis_Grp78_Ma               SFFEGDDFSETLTRAKFEELNMDLFRSTMKPVQKVLEDSDLQKKEIDEIVLVGGSTRIPKIQQLVKEFFGKEPSRGINPDEAVAYGAAVQAGVLSGE--DDTNDLVLLDVNPLTLGIET
C._elegans_Hsp-3_Pm                 SLFDGEDFSETLTRAKFEELNMDLFRATLKPVQKVLEDSDLKKDVHEIVLVGGSTRIPKVQQLIKEFFNGKEPSRGINPDEAVAYGAAVQAGVLSGE--EDTGEIVLLDVNPLTMGIET
C._elegans_Hsp-3_Ma                 SLFDGEDFSETLTRAKFEELNMDLFRATLKPVQKVLEDSDLKKDVHEIVLVGGSTRIPKVQQLIKEFFNGKEPSRGINPDEAVAYGAAVQAGVLSGE--EDTGEIVLLDVNPLTMGIET
A._thaliana_BiP-2_Pm                SLFDGVDLSEFLTRARFEELNNDLFRKTMCPVKKAMDDAGLQKSQIDEIVLVGGSTRIPKVQQLLKDFFEGKEPNKGVNPDEAVAVQEGILSGEGGDETKDILLLDVAPLTLGIET
A._thaliana_BiP-2_Ma                SLFDGVDLSEFLTRARFEELNNDLFRKTMCPVKKAMDDAGLQKSQIDEIVLVGGSTRIPKVQQLLKDFFEGKEPNKGVNPDEAVAVQEGILSGEGGDETKDILLLDVAPLTLGIET
S._cerevisiae_Kar2p_Pm              SFVDGIDLSETLTRAKFEELNLDLFKKTLKPVEKVLQDSGLEKKDVDDIVLVGGSTRIPKVQQLLESYFDGKKASKGVNPDEAVAYGAAVQAGVLSGE--EGVEDIVLLDVNALTIGIET
S._cerevisiae_Kar2p_Ma              SFVDGIDLSETLTRAKFEELNLDLFKKTLKPVEKVLQDSGLEKKDVDDIVLVGGSTRIPKVQQLLESYFDGKKASKGVNPDEAVAYGAAVQAGVLSGE--EGVEDIVLLDVNALTIGIET
A._californica_Grp78_Pm             SFFDGEDFSESLTRAKFEELNMDLFRSTMKPVKQVLEDADLKTDDIDEIVLVGGSTRIPKVQQLVKEYFNGKEPSRGINPDEAVAYGAAVQAGVLSGE--EDTGDLVLLDVNPLTMGIET
A._californica_Grp78_Ma             SFFDGEDFSESLTRAKFEELNMDLFRSTMKPVKQVLEDADLKTDDIDEIVLVGGSTRIPKVQQLVKEYFNGKEPSRGINPDEAVAYGAAVQAGVLSGE--EDTGDLVLLDVNPLTMGIET
X._laevis_Hspa5_Pm                  SFFEGEDFSETLTRAKFEELNMDLFRSTMKPVQKVLDDADLKKSDIDEIVLVGGSTRIPKIQQLVKEFFNGKEPSRGINPDEAVAYGAAVQAGVLSGD--QDTGDLVLLDVCPLTLGIET
X._laevis_Hspa5_Ma                  SFFEGEDFSETLTRAKFEELNMDLFRSTMKPVQKVLDDADLKKSDIDEIVLVGGSTRIPKIQQLVKEFFNGKEPSRGINPDEAVAYGAAVQAGVLSGD--QDTGDLVLLDVCPLTLGIET
A._echinatior_Hsc70-3_Pm            SFFEGEDFSETLTRAKFEELNMDLFRSTLKPVQKVLEDSDMSKKDVDEIVLVGGSTRIPKVQQLVKEFFGKEPSRGINPDEAVAYGAAVQAGVLSGE--QDTDAIVLLDVNPLTMGIET
A._echinatior_Hsc70-3_Ma            SFFEGEDFSETLTRAKFEELNMDLFRSTLKPVQKVLEDSDMSKKDVDEIVLVGGSTRIPKVQQLVKEFFGKEPSRGINPDEAVAYGAAVQAGVLSGE--QDTDAIVLLDVNPLTMGIET
C._owczarzaki_Hsp70-Like_Pm         AFFDGEDFSETLTRAKFEELNADLFRSTLAPVKKVLEDSGLRKEEIDEIVLVGGSTRIPKVQKDFFNGKEPNRGINPDEAVAYGAAVQAGVLGGE--ESTGDLVLLDVNPLTLGIET
C._owczarzaki_Hsp70-Like_Ma         AFFDGEDFSETLTRAKFEELNADLFRSTLAPVKKVLEDSGLRKEEIDEIVLVGGSTRIPKVQKDFFNGKEPNRGINPDEAVAYGAAVQAGVLGGE--ESTGDLVLLDVNPLTLGIET
C._cohnii_BiP_Pm                    ALFDGVDFSETLTRARFEEINNDLFKNTLGPVKQVIEDSGLKKTQIDEIVLVGGSTRIPKVQQLIKDFFNGKEPNRGINPDEAVAYGAAVQAGILSGE--GGQDILLLDVTPLTLGIET
C._cohnii_BiP_Ma                    ALFDGVDFSETLTRARFEEINNDLFKNTLGPVKQVIEDSGLKKTQIDEIVLVGGSTRIPKVQQLIKDFFNGKEPNRGINPDEAVAYGAAVQAGILSGE--GG-QDILLLDVTPLTLGIET
PfHsp70-2_Pm                        DIVEGHNFSETLTRAKFEELNDDLFRETLEPVKKVIDDAKYEKSKIDEIVLVGGSTRIPKIQQIIKEFFNGKEPNRGINPDEAVAYGAAIQAGIIIGE---ELQDVVLLDVTPLTLGIET
PfHsp70-2_Ma                        DIVEGHNFSETLTRAKFEELNDDLFRETLEPVKKVIDDAKYEKSKIDEIVLVGGSTRIPKIQQIIKEFFNGKEPNRGINPDEAVAYGAAIQAGIIIGE---ELQDVVLLDVTPLTLGIET
```

Supplementary Data S2B

```
Human_Grp78_Pm             VGGVMTKLIPRNTVVPTKKSQIFSTASDNQPTVTIKVYEGERPLTKDNHLLGTFDLTGIPPAPRGVPQIEVTFEIDVNGILRVTAEDKGTGNKNKITITNDQNRLTPEEIERMVNDAEKF
Human_Grp78_Ma             VGGVMTKLIPRNTVVPTKKSQIFSTASDNQPTVTIKVYEGERPLTKDNHLLGTFDLTGIPPAPRGVPQIEVTFEIDVNGILRVTAEDKGTGNKNKITITNDQNRLTPEEIERMVNDAEKF
R._norvegicus_Grp78_Pm     VGGVMTKLIPRNTVVPTKKSQIFSTASDNQPTVTIKVYEGERPLTKDNHLLGTFDLTGIPPAPRGVPQIEVTFEIDVNGILRVTAEDKGTGNKNKITITNDQNRLTPEEIERMVNDAEKF
R._norvegicus_Grp78_Ma     VGGVMTKLIPRNTVVPTKKSQIFSTASDNQPTVTIKVYEGERPLTKDNHLLGTFDLTGIPPAPRGVPQIEVTFEIDVNGILRVTAEDKGTGNKNKITITNDQNRLTPEEIERMVNDAEKF
O._latipes_Bip_Pm          VGGVMTKLIPRNTVVPTKKSQIFSTASDNQPTVTIKVYEGERPLTKDNHLLGTFDLTGIPPAPRGVPQIEVTFEIDVNGILRVTAEDKGTGNKNKITITNDQNRLTPEDIERMVNDAERF
O._latipes_Bip_Ma          VGGVMTKLIPRNTVVPTKKSQIFSTASDNQPTVTIKVYEGERPLTKDNHLLGTFDLTGIPPAPRGVPQIEVTFEIDVNGILRVTAEDKGTGNKNKITITNDQNRLTPEDIERMVNDAERF
F._chinensis_Grp78_Pm      VGGVMTKLIPRNTVIPTKKSQIFSTASDNQHVTIQVFEGERPMTKDNHILGKFDLTGIPPAPRGVPQIEVTFEIDANGILQVSAEGKGTGNKEKITITNDQNRLTPEDIERMIKDAEVF
F._chinensis_Grp78_Ma      VGGVMTKLIPRNTVIPTKKSQIFSTASDNQHVTIQVFEGERPMTKDNHILGKFDLTGIPPAPRGVPQIEVTFEIDANGILQVSAEGKGTGNKEKITITNDQNRLTPEDIERMIKDAEVF
C._elegans_Hsp-3_Pm        VGGVMTKLIGRNTVIPTKKSQVFSTAADNQPTVTIQVFEGERPMTKDNHQLGKFDLTGLPPAPRGVPQIEVTFEIDVNGIHVTAEDKGTGNKNKITITNDQNRLSPEDIERMINDAEKF
C._elegans_Hsp-3_Ma        VGGVMTKLIGRNTVIPTKKSQVFSTAADNQPTVTIQVFEGERPMTKDNHQLGKFDLTGLPPAPRGVPQIEVTFEIDVNGIHVTAEDKGTGNKNKITITNDQNRLSPEDIERMINDAEKF
A._thaliana_BiP-2_Pm       VGGVMTKLIPRNTVIPTKKSQVFTTYQDQQLTVSIQVFEGERSLTKDCRLLGKFDLTGVPPAPRGILPQIEVTFEVDANGILNVKAEDKASGKSEKITITNEKGRLSQEEIDRMVKFAEF
A._thaliana_BiP-2_Ma       VGGVMTKLIPRNTVIPTKKSQVFTTYQDQQLTVSIQVFEGERSLTKDCRLLGKFDLTGVPPAPRGILPQIEVTFEVDANGILNVKAEDKASGKSEKITITNEKGRLSQEEIDRMVKPAEF
S._cerevisiae_Kar2p_Pm     TGGVMTPLIKRNTAIPTKKSQIFSTAVDNQPTVMIKVYEGERAMSKDNNLLGKFELTGIPPAPRGVPQIEVTFALDANGIIKVSADKGTGCKSESITITNDKGRLTQEEIDRMVEFAKF
S._cerevisiae_Kar2p_Ma     TGGVMTPLIKRNTAIPTKKSQIFSTAVDNQPTVMIKVYEGERAMSKDNNLLGKFELTGIPPAPRGVPQIEVTFALDANGIIKVSADKGTGCKSESITITNDKGRLTQEEIDRMVEFAKF
A._californica_Grp78_Pm    VGGVMTKLIPRNTVIPTKKSQIFSTAADNQPTVTIQVYEGERSMTKDNHLLGKFDLTGIPPAPRGVPQIEVTFEIDVNGILRVTAEDKGTGSKNQIVIQNDQNRLSPEDIERMINDAEKY
A._californica_Grp78_Ma    VGGVMTKLIPRNTVIPTKKSQIFSTAADNQPTVTIQVYEGERSMTKDNHLLGKFDLTGIPPAPRGVPQIEVTFEIDVNGILRVTAEDKGTGSKNQIVIQNDQNRLSPEDIERMINDAEKY
X._laevis_Hspa5_Pm         VGGVMTKLIPRNTVVPTKKSQIFSTASDNQPTVTIKVYEGERPLTKDNHLLGTFDLTGIPPAPRGVPQIEVTFEIDVNGILRVTAEDKGTGNKNKITITNDQNRLTPEEIERMVTDAEKF
X._laevis_Hspa5_Ma         VGGVMTKLIPRNTVVPTKKSQIFSTASDNQPTVTIKVYEGERPLTKDNHLLGTFDLTGIPPAPRGVPQIEVTFEIDVNGILRVTAEDKGTGNKNKITITNDQNRLTPEEIERMVTDAEKF
A._echinatior_Hsc70-3_Pm   VGGVMTKLIPRNTVIPTKKSQIFSTASDSQHVTIQVFEGERPMTKDNHLLGKFDLTGIPPAPRGIPQIEVTFEIDANGILQVSAEDKGTGNREKIVITNDRNLTPDDIERMIKDAEKF
A._echinatior_Hsc70-3_Ma   VGGVMTKLIPRNTVIPTKKSQIFSTASDSQHVTIQVFEGERPMTKDNHLLGKFDLTGIPPAPRGIPQIEVTFEIDANGILQVSAEDKGTGNREKIVITNDRNLTPDDIERMIKDAEKF
C._owczarzaki_Hsp70-Like_Pm VGGVMTKLIPRNSVIPTKKSQIFSTAADNQPTVTIQVYEGERSLTKDCHTLGKFDLTGIPPAARGVPQIEVTFEIDANGILRVSAEDKGTCKSEKITITNDQNRLTPEEIERMVNDAERF
C._owczarzaki_Hsp70-Like_Ma VGGVMTKLIPRNSVIPTKKSQIFSTAADNQPTVTIQVYEGERSLTKDCHTLGKFDLTGIPPAARGVPQIEVTFEIDANGILRVSAEDKGTCKSEKITITNDQNRLTPEEIERMVNDAERF
C._cohnii_BiP_Pm           VGGVMTKLINRNTVIPTKKSQTFSTYQDNQPAVNIQVFEGERPMTKDNHLLGKFELCGIPPAPRGCPQIEVTFEIDSNGILNVGAEDKATGKEKITITNDKGRLTEEQIEKMIKEAEQF
C._cohnii_BiP_Ma           VGGVMTKLINRNTVIPTKKSQTFSTYQDNQPAVNIQVFEGERPMTKDNHLLGKFELGGIPPAPRGCPQIEVTFEIDSNGILNVGAEDKATGKEKITITNDKGRLTEEQIEKMIKEAEQF
PfHsp70-2_Pm               VGGIMTQLIKRNTVIPTKKSQTFSTYQDNQPAVLIQVFEGERALTKDNHLLGKFEISGIPPAORGVPKIEVTFTVDKNGILHVEAEDKGTCKSRGITITNDKGRLSKEQIEKMINDAEKF
PfHsp70-2_Ma               VGGIMTQLIKRNTVIPTKKSQTFSTYQDNQPAVLIQVFEGERALTKDNHLLGKFEISGIPPAORGVPKIEVTFTVDKNGILHVEAEDKGTCKSRGITITNDKGRLSKEQIEKMINDAEKF

Human_Grp78_Pm             AEEDKKLKERIDTRNELESYAYSLKNQIG---DKEKLGGKLSSEDKETMEKAVEEKIEWLE-SHQD-ADIEDFKAKKKELEEIVQPIISKLYGSAC--PPP----TGEEDTA-------E
Human_Grp78_Ma             AEEDKKLKERIDTRNELESYAYSLKNQIG---DKEKLGGKLSSEDKETMEKAVEEKIEWLE-SHQD-ADIEDFKAKKKELEEIVQPIISKLYGSAC--P----PPTGEEDTA------EK
R._norvegicus_Grp78_Pm     AEEDKKLKERIDTRNELESYAYSLKNQIG---DKEKLGGKLSPEDKETMEKAVEEKIEWLE-SHQD-ADIEDFKAKKKELEEIVQPIISKLYGSGC--PPP----TGEEDTS-------E
R._norvegicus_Grp78_Ma     AEEDKKLKERIDTRNELESYAYSLKNQIG---DKEKLGGKLSPEDKETMEKAVEEKIEWLE-SHQD-ADIEDFKAKKKELEEIVQPIISKLYGSGC--P----PPTGEEDTS------EK
O._latipes_Bip_Pm          ADEDKRLKERIDARNELESYAYSLKNQIG---DKEKLGGKLSDDDKETIEKAVEEKIEWME-SHQD-AETEDFQAKKKELEEVVQPIITKLYGSAGG-PPP----EGAESEA-------E
O._latipes_Bip_Ma          ADEDKRLKERIDARNELESYAYSLKNQIG---DKEKLGGKLSDDDKETIEKAVEEKIEWME-SHQD-AETEDFQAKKKELEEVVQPIITKLYGSAGG-GP----PPEGAESEA------EK
F._chinensis_Grp78_Pm      ADEDKKLKERVESRNELESYAYSLKNQIN---DKEKLGSKLSDEDKEKMDEVIEEKIKWLE-DNPE-ADAEDYKTQKKELEDVVQPIITKLYQOSGEAPPP----TEDEENY-------E
F._chinensis_Grp78_Ma      ADEDKKLKERVESRNELESYAYSLKNQIN---DKEKLGSKLSDEDKEKMDEVIEEKIKWLE-DNPE-ADAEDYKTQKKELEDVVQPIITKLYQOSGEAP----PPTEDEENY------EK
C._elegans_Hsp-3_Pm        AEDDKKVKDKAEARNELESYAYNLKNQIE---DKEKLGGKIDEDDKKTIEEAVEEAISWLG-SNAE-ASAEELKEQKKDLESKVQPIVSKLYKDACAGGEE---APEEGSD-------D
C._elegans_Hsp-3_Ma        AEDDKKVKDKAEARNELESYAYNLKNQIE---DKEKLGGKIDEDDKKTIEEAVEEAISWLG-SNAE-ASAPELKEQKKDLESKVQPIVSKLYKDACAGG----EEAPEEGSD------DK
A._thaliana_BiP-2_Pm       AEEDKKVKEKIDARNALETYVYNMKNQVS---DKDKLADKIEGDEKEKIEAATKEALEWLD-ENQN-SEKEEYDEKLKEVEAVCNPIITAVYQRSGGAPGAGGESSTEEDE-------S
A._thaliana_BiP-2_Ma       AEEDKKVKEKIDARNALETYVYNMKNQVS---DKDKLADKIEGDEKEKIEAATKEALEWLD-ENQN-SEKEEYDEKLKEVEAVCNPIITAVYQRSGGAPGAGGESSTEEDE------SH
S._cerevisiae_Kar2p_Pm     ASEDASIKAKVEARNKLENYAHSLKNQVN-----GDLGEKIEEDKETLLDAANDVLEWLD-DNFETAIAEDFDEKFESLSKVAYPITSKLYGADGSGAA----DYDDEDEDDDGDYFE
S._cerevisiae_Kar2p_Ma     ASEDASIKAKVEARNKLENYAHSLKNQVN-----GDLGEKIEEDKETLLDAANDVLEWLD-DNFETAIAEDFDEKFESLSKVAYPITSKLYGADGSG----AADYDDEDEDDDGDYFEH
A._californica_Grp78_Pm    ADEDKKVKEKVDAKNELESYAYSLKNQIG---DKEKLGAKLSDEDKEKITEAVDEAIKWLE-SNAE-AESEAFNEKKTELEGIVQPIMTKLYEQSGGAPPP----SGEEESEE-----AE
A._californica_Grp78_Ma    ADEDKKVKEKVDAKNELESYAYSLKNQIG---DKEKLGAKLSDEDKEKITEAVDEAIKWLE-SNAE-AESEAFNEKKTELEGIVQPIMTKLYEQSGGAP----PPSGEEESEE----EAEK
X._laevis_Hspa5_Pm         AEEDKKLKERIDTRNELESYAYSLKNQIG---DKEKLGGKLSSEDKETIEKAVEEKIEWLE-SHQD-ADIEDFKAKKKELEEIVQPIVGKLYGAGA-PPP----EGAEET-------E
X._laevis_Hspa5_Ma         AEEDKKLKERIDTRNELESYAYSLKNQIG---DKEKLGGKLSSEDKETIEKAVEEKIEWLE-SHQD-ADIEDFKAKKKELEEIVQPIVGKLYGAGA-AP----PPEGAE-ET------EK
A._echinatior_Hsc70-3_Pm   ADDDKKLKERVEARNDLESYAYSLKNQLA---DKEKLGSKVSEADKATMEEAIDEKIKWLE-DNQD-TEPEEYKKQKKELTDIVQPIIAKLYQGAGGVPP----TGGEDDD------M
A._echinatior_Hsc70-3_Ma   ADDDKKLKERVEARNDLESYAYSLKNQLA---DKEKLGSKVSEADKATMEEAIDEKIKWLE-DNQD-TEPEEYKKQKKELTDIVQPIIAKLYQGAGGV----PPTGGEDDD------MR
C._owczarzaki_Hsp70-Like_Pm ADQDKKVKERVDAKNELESYAYNLRNQIK---DEDKLGGKIDEEDKETITAVDEIAWLDQSGAE-ATTEELKEKAAFEKIVQPIVSKLYQGAGGAP----PSG---DDE---------G
C._owczarzaki_Hsp70-Like_Ma ADQDKKVKERVDAKNELESYAYNLRNQIK---DEDKLGGKIDEEDKETITAVDEIAWLDQSGAE-ATTEELKEKAAFEKIVQPIVSKLYQGAGGA----PSG---DDE---------GH
C._cohnii_BiP_Pm           ADEDKKVKERVDAKNSFDGYIHSMRSATEGSGDNKGLSEKMDEDEKILEALKDGQSWLD-SNPE-GDADIKEFHKEVEGICAPIVSKLYGVGGGAGA----ADEDE---------A
C._cohnii_BiP_Ma           ADEDKKVKERVDAKNSFDGYIHSMRSATEGSGDNKGLSEKMDEDEKILEALKDGQSWLD-SNPE-GDADIKEFHKEVEGICAPIVSKLYGVGGGA----GAADEDEDE--------AH
PfHsp70-2_Pm               ADEDKNLREKVEAKNNLDNYIQSMKATVE---DKDKLADKIEKEDKNTILSAVKDAEDWLN-NNSN-ADSEALKQKLKDLEAVCQPIIVKLYGQPGGPSPQ---PSGDEDV-------D
PfHsp70-2_Ma               ADEDKNLREKVEAKNNLDNYIQSMKATVE---DKDKLADKIEKEDKNTILSAVKDAEDWLN-NNSN-ADSEALKQKLKDLEAVCQPIIVKLYGQPGGPS----PQPSGDEDV------DS
```

Supplementary Data S2B

```
Human_Grp78_Pm             KDEL
Human_Grp78_Ma             DEL
R._norvegicus_Grp78_Pm     KDEL
R._norvegicus_Grp78_Ma     DEL
O._latipes_Bip_Pm          KDEL
O._latipes_Bip_Ma          DEL
F._chinensis_Grp78_Pm      KDEL
F._chinensis_Grp78_Ma      DEL
C._elegans_Hsp-3_Pm        KDEL
C._elegans_Hsp-3_Ma        DEL
A._thaliana_BiP-2_Pm       HDEL
A._thaliana_BiP-2_Ma       DEL
S._cerevisiae_Kar2p_Pm     HDEL
S._cerevisiae_Kar2p_Ma     DEL
A._californica_Grp78_Pm    KDEL
A._californica_Grp78_Ma    DEL
X._laevis_Hspa5_Pm         KDEL
X._laevis_Hspa5_Ma         DEL
A._echinatior_Hsc70-3_Pm   RDEL
A._echinatior_Hsc70-3_Ma   DEL
C._owczarzaki_Hsp70-Like_Pm  HDEF
C._owczarzaki_Hsp70-Like_Ma  DEF
C._cohnii_BiP_Pm           HDEL
C._cohnii_BiP_Ma           DEL
PfHsp70-2_Pm               SDEL
PfHsp70-2_Ma               DEL
```

Supplementary Data S2C

```
                                        10        20        30        40        50        60        70        80        90       100       110       120
                                 ....|....|....|....|....|....|....|....|....|....|....|....|....|....|....|....|....|....|....|....|....|....|....|....|
         PfHsp70-3_Pm            MASLN--------------------KKNIVKI-LERCVKNTLLSEKSRSLCTSKINRNRASGDIIGIDLGTTNSCVAIMEGKQGKVIENSEGRTTPSVVAFTNDNQRLVGIVAKRQA
         PfHsp70-3_Ma            ~~~~MAS------------------LNKKNIVKILERCVKNTLLSEKSRSLCTSKINRNRASGDIIGIDLGTTNSCVAIMEGKQGKVIENSEGRTTPSVVAFTNDNQRLVGIVAKRQA
         C._owczarzaki_Hsp70-2_Pm MLKAM--------------------SVARL-ASSGMTMRQSAAKPIAAALSRSYSSAVKGQVIGIDLGTTNSCVAVMEGKTPKVIENAEGARTTPSVVAFTEEGERLVGIPARRQA
         C._owczarzaki_Hsp70-2_Ma ~~~~MLKAMSV--------------ARLASSGMTMRQSAAKPIAAALSRSYSSAVKGQVIGIDLGTTNSCVAVMEGKTPKVIENAEGARTTPSVVAFTEEGERLVGIPARRQA
         G._gallus_Mt-Hsp70   _Pm MISAS---------RAAAARLPLLLPRGGPVPAVPGLAQTFWNGLSQNVLRAASSRKYASEAIKGAVIGIDLGTTNSCVAVMEGKTAKVLENSEGARTTPSVVAFTADGERLVGMPAKRQA
         G._gallus_Mt-Hsp70   _Ma ~~~~MISASRAAA-RLPLLLPRGGPVPAVPGLAQTFWNGL----SQNVLRAASSRKYASEAIKGAVIGIDLGTTNSCVAVMEGKTAKVLENSEGARTTPSVVAFTADGERLVGMPAKRQA
         Human_Mt-Hsp70       _Pm MISAS---------RAAAARLVGAAASRGPTAA---RHQDSWNGLSHEAFRLVSRRDYASEAIKGAVVGIDLGTTNSCVAVMEGKQAKVLENAEGARTTPSVVAFTADGERLVGMPAKRQA
         Human_Mt-Hsp70       _Ma ~~~~MISASRAAAARLVGAAASRGPTAA---RHQDSWNGL----SHEAFRLVSRRDYASEAIKGAVVGIDLGTTNSCVAVMEGKQAKVLENAEGARTTPSVVAFTADGERLVGMPAKRQA
         R._norvegicus_Mt-Hsp70  _Pm MISAS--------RAAAARLVGTTASRSPAAA---RHQDGWNGLSHEVFRFVSRRDYASEAIKGAVIGIDLGTTNSCVAVMEGKQAKVLENSEGARTTPSVVAFTPDGERLVGMPAKRQA
         R._norvegicus_Mt-Hsp70  _Ma ~~~~MISASRAAAARLVGTTASRSPAAA---RHQDGWNGL----SHEVFRFVSRRDYASEAIKGAVIGIDLGTTNSCVAVMEGKQAKVLENSEGARTTPSVVAFTKDGERLVGMPAKRQA
         S._aurata_Grp75_Pm      ---------------------------------------------------ASEAIKGAVIGIDLGTTNSCVAVMEGKQAKVLENAEGARTTPSVIAFTAEGERLVGMPAKRQS
         S._aurata_Grp75_Ma      ---------------------------------------------------ASEAIKGAVIGIDLGTTNSCVAVMEGKQAKVLENAEGARTTPSVIAFTAEGERLVGMPAKRQS
         X._laevis_Hsp70-9_Pm    MLCAG------------------GSLASPLSHVSRI-LRKGCFNGLSQDVLQSVFKRDYASESVKGAVIGIDLGTTNSCVAVMEGKQAKVLENSEGARTTPSVVAFSSEGERLVGMPAKRQA
         X._laevis_Hsp70-9_Ma    ~~~~MLCAGGSLASPLSHVSRI---------LRKGCFNGL----SQDVLQSVFKRDYASESVKGAVIGIDLGTTNSCVAVMEGKQAKVLENSEGARTTPSVVAFSSEGERLVGMPAKRQA
         M._mongolica_Hsp70_Pm   MLQAA---RLCTRQVRECHGLISDPNRVRNWSTI-------GKNLIASSTWGVQCRFKSDGVKGEVIGVDLGTTNSCVAIMEGKTPKVIENAEGSRTTPSVIAFTKGGERLAGMPAKRQA
         M._mongolica_Hsp70_Ma   ~~~~MLQAARLCTRQVRECHGLISDPNRVRNWSTIGKNLI------ASSTWGVQCRFKSDGVKGEVIGVDLGTTNSCVAIMEGKTPKVIENAEGSRTTPSVIAFTKGGERLAGMPAKRQA
         A._echinatior_Hsc70-5_Pm --------------------------------------------------------MEGSRTTPSYVAFSKDDERLVGIAAKRQA
         A._echinatior_Hsc70-5_Ma --------------------------------------------------------MEGSRTTPSYVAFSKDDERLVGIAAKRQA
         M._arenaria_mortalin-2_Pm MFTSA---RHTSRSLKKTFGT----HQARMISNI-CKKILAGEEIKSKSGLYLLPSRLRSDKVKGYVIGIDLGTTNSCVAIMEGKTGKVLENAEGARTTPSVVAFTKDGERLVGMPAKRQA
         M._arenaria_mortalin-2_Ma ~~~~MFTSARHTSRSLKKTFGT---HQARMISNICKKILAGEEIKSKSGLYLLPSRLRSDKVKGYVIGIDLGTTNSCVAIMEGKQAKVLENAEGARTTPSVVAFTKDGERLVGMPAKRQA
         C._elegans_Hsp-6_Pm     MLSAR--------------------SFLSSARTIARSSLMSARSLSDKFKGTVIGIDLGTTNSCVSIMEGKPKVIENAEGVRTTPSTVAFTADGERLVGEAPAKRQA
         C._elegans_Hsp-6_Ma     ~~~~MLSARSFLSSA--------------RTIARSSLMSARSLSDKFKGTVIGIDLGTTNSCVSIMEGKPKVIENAEGVRTTPSTVAFTADGERLVGEAPAKRQA
         S._cerevisiae_Ssc1p_Pm  MLAAK-------------------NILNR-SSLSSSFRIATRLQSTKVQGSVIGIDLGTTNSAVAIMEGKVPKTIENAEGSRTTPSVVAFTKEGERLVGIPAKRQA
         S._cerevisiae_Ssc1p_Ma  ~~~~MLAAKNILNRS---------------SLSSSFRIATRLQSTKVQGSVIGIDLGTTNSAVAIMEGKVPKVIENAEGSRTTPSVVAFTKEGERLVGIPAKRQA
         A._thaliana_HSO70-2_Pm  MATAALLRSIRRREVVSSPFSAYRCLSSS-------GKASLNSSYLGQNFRSFSRAFSSKPAGNDVIGIDLGTTNSCVAVMEGKNPKVIENAEGARTTPSVVAFNTKGELVCTPAKRQA
         A._thaliana_HSO70-2_Ma  ~~~~MATAALLRSIRRREVVSSPF--SAYRCLSSSGKASLNSSYLGQNFRS-FSRAFSSKPAGNDVIGIDLGTTNSCVAVMEGKNPKVIENAEGARTTPSVVAFNTKGELVCTPAKRQA
         P._infestans_Mt-Hsp70   _Pm M---------------------------------------------------------FSAAAGSEVIGIDLGTTNSCVAVMEGKTARVLENSEGARTTPSVVAILDNDERLVGMPAKRQA
         P._infestans_Mt-Hsp70   _Ma --------------------------------------------------------MFSAAAGSEVIGIDLGTTNSCVAVMEGKTARVLENSEGARTTPSVVAILDNDERLVGMPAKRQA
         E._tenella_Mt-Hsp70  _Pm MRGAVALSAARALWAAAVP------PQPRGPPKE-QRVFSAVRTAAVGTLSSLAGRRGFSGVRGDVVIGIDLGTTNSCVAVMEGSQPKVLENSEGMRTTPSVVAFTKDGORLVGVVAKRQA
         E._tenella_Mt-Hsp70  _Ma ~~~~~~~~~~~~~~~MRGAVALSAARALWAAAVP---PQPRGPPKEQRVFSAVRTAAVGTLSSLAGRRGFSGVRGDVVIGIDLGTTNSCVAVMEGSQPKVLENSEGMRTTPSVVAFTKDGORLVGVVAKRQA
         D._baltica_Hsp70-type_Pm ------------------------------------------------------MAKVVGIDLGTTNSVVAAIEGQQPSVITNAEGLRTTPSIVAYTKKQELVCQIAKRQA
         D._baltica_Hsp70-type_Ma ~~~~---------------------------------------------------MAKVVGIDLGTTNSVVAAIEGQQPSVITNAEGLRTTPSIVAYTKKQELVCQIAKRQA
```

```
         PfHsp70-3_Pm            ITNPENTVYATKRFIGRKYEDATKKEQNLPYKIVRASNG-DAWIE--AQGKKYSPSQIGACVLEKMKETAENYLGRKVHQAVITVPAYFNDSQRQATKDAGKIAGLDVLRIINEPTAA
         PfHsp70-3_Ma            ITNPENTVYATKRFIGRKYEDATKKEQNLPYKIVRASNG-DAWIE--AQGKKYSPSQIGACVLEKMKETAENYLGRKVHQAVITVPAYFNDSQRQATKDAGKIAGLDVLRIINEPTAA
         C._owczarzaki_Hsp70-2_Pm VTNPENTFTATKRLIGRQFDDSEVQRERKLVAYEIVKHTNG-DAWVK--SRDKTYSPSQIGAFVLIKMKETAESYLNTKVENAVTVPAYFNDSQRQATKDAGQISGLNVLRVINEPTAA
         C._owczarzaki_Hsp70-2_Ma VTNPENTFTATKRLIGRQFDDSEVQRERKLVAYEIVKHTNG-DAWVK--SRDKTYSPSQIGAFVLIKMKETAESYLNTKVENAVTVPAYFNDSQRQATKDAGQISGLNVLRVINEPTAA
         G._gallus_Mt-Hsp70   _Pm VTNPENTFYATKRLIGRRFDDSEVKKDIKNVPFKIVRASNG-DAWVE--AHGKLYSPSQIGAFVLMKMKETAENYLGHPAKNAVITVPAYFNDSQRQATKDAGQISGLNVLRVINEPTAA
         G._gallus_Mt-Hsp70   _Ma VTNPENTFYATKRLIGRRFDDSEVKKDIKNVPFKIVRASNG-DAWVE--AHGKLYSPSQIGAFVLMKMKETAENYLGHPAKNAVITVPAYFNDSQRQATKDAGQISGLNVLRVINEPTAA
         Human_Mt-Hsp70       _Pm VTNPENTFYATKRLIGRRYMDDPEVQKDIKNVPFKIVRASNG-DAWVE--AHGKLYSPSQIGAFVLMKMKETAENYLGHTAKNAVITVPAYFNDSQRQATKDAGQISGLNVLRVINEPTAA
         Human_Mt-Hsp70       _Ma VTNPENTFYATKRLIGRRYMDDPEVQKDIKNVPFKIVRASNG-DAWVE--AHGKLYSPSQIGAFVLMKMKETAENYLGHTAKNAVITVPAYFNDSQRQATKDAGQISGLNVLRVINEPTAA
         R._norvegicus_Mt-Hsp70  _Pm VTNPENTFYATKRLIGRRYMDDPEVQKDIKNVPFKIVRASNG-DAWVE--AHGKLYSPSQIGAFVLMKMKETAENYLGHTAKNAVITVPAYFNDSQRQATKDAGQISGLNVLRVINEPTAA
         R._norvegicus_Mt-Hsp70  _Ma VTNPENTFYATKRLIGRRYMDDPEVQKDIKNVPFKIVRASNG-DAWVE--AHGKLYSPSQIGAFVLMKMKETAENYLGHTAKNAVITVPAYFNDSQRQATKDAGQISGLNVLRVINEPTAA
         S._aurata_Grp75_Pm      VTNPENTLYATKRLIGRRFDDPEVQKDMKNVPYKIVRASNG-DAWVE--ARGKMYSPSQGAFVLMKMKETAENYLGTKVKNAVTVPAYFNDSQRQATKDAGQISGLNVLRVINEPTAA
         S._aurata_Grp75_Ma      VTNPENTLYATKRLIGRRFDDPEVQKDMKNVPYKIVRASNG-DAWVE--ARGKMYSPSQAGAFVLMKMKETAENYLGTKVKNAVTVPAYFNDSQRQATKDAGQISGLNVLRVINEPTAA
         X._laevis_Hsp70-9_Pm    VTNPENTFYATKRLIGRRFDAEVQKDIKNVPFKIVRASNG-DAWVE--SHGKLYSPSQIGAFVLIKMKETAENYLGHSAKNAVITVPAYFNDSQRQATKDAGQISGLNVLRVINEPTAA
         X._laevis_Hsp70-9_Ma    VTNPENTFYATKRLIGRRFDAEVQKDIKNVPFKIVRASNG-DAWVE--SHGKLYSPSQIGAFVLIKMKETAENYLGHSAKNAVITVPAYFNDSQRQATKDAGQISGLNVLRVINEPTAA
         M._mongolica_Hsp70_Pm   VTNAQNTLYATKRLIGRRFDDPEVKKDMKNVSYKIVRASNG-DAWVE--AQGKMYEPSQIGAFVLVKMKETAEAYLGQPVKNAVITVPAYFNDSQRQATKDAGQISGLNVLRTINEPTAA
         M._mongolica_Hsp70_Ma   VTNAQNTLYATKRLIGRRFDDPEVKKDMKTVSYKIVRASNG-DAWVE--AQGKMYEPSQIGAFVLVKMKETAEAYLGQPVKNAVITVPAYFNDSQRQATKDAGQISGLNVLRTINEPTAA
         A._echinatior_Hsc70-5_Pm VTNSVNTFYATKRLIGRRFEDPEVKKRMKSVYKIVRASNG-DAWVQG-ADGKMYSPSQIGAFVLMKMKETAAAYLNTSVKNAVITVPAYFNDSQRQATKDAGQIAGLNVLRVINEPTAA
         A._echinatior_Hsc70-5_Ma VTNSVNTFYATKRLIGRRFEDPEVKKRMKSVYKIVRASNG-DAWVQG-ADGKMYSPSQIGAFVLMKMKETAAAYLNTSVKNAVITVPAYFNDSQRQATKDAGQIAGLNVLRVINEPTAA
         M._arenaria_mortalin-2_Pm VTNAANTLHATKRLIGRRFEDKEVKKDMETVPYKIVRANNG-DAWVE--AHGKTYSPSQIGAFVLMKMKETADNYLGQPVKNAVTVPAYFNDSQRQATKDAGQISGLNVLRVINEPTAA
         M._arenaria_mortalin-2_Ma VTNAANTLHATKRLIGRRFEDKEVKKDMETVPYKIVRANNG-DAWVE--AHGKTYSPSQIGAFVLMKMKETADNYLGQPVKNAVTVPAYFNDSQRQATKDAGQISGLNVLRVINEPTAA
         C._elegans_Hsp-6_Pm     VTNSANTLFATKRLIGRRYEDPEVQKDLKVVPYKIVKASNG-DAWVE--AQGKVYSPSQVGAFVLMKMKETAESYLGTTVNNAVTVPAYFNDSQRQATKDAGQISGLNVLRVINEPTAA
         C._elegans_Hsp-6_Ma     VTNSANTLFATKRLIGRRYEDPEVQKDLKVVPYKIVRASNG-DAWVE--AQGKVYSPSQVGAFVLMKMKETAESYLGTTVNNAVTVPAYFNDSQRQATKDAGQISGLNVLRVINEPTAA
         S._cerevisiae_Ssc1p_Pm  VVNPENTLFATKRLIGRRFEDAEVQRDIKOVPYKIVKHSNG-DAWVE--ARGQTYSPAQIGGFVLNKMKETAEALGKPVKNAVTVPAYFNDSQRQATKDAGQIVGLNVLRVVNEPTAA
         S._cerevisiae_Ssc1p_Ma  VVNPENTLFATKRLIGRRFEDAEVQRDIKOVPYKIVKHSNG-DAWVE--ARGQTYSPAQIGGFVLNKMKETAEALGKPVKNAVTVPAYFNDSQRQATKDAGQIVGLNVLRVVNEPTAA
         A._thaliana_HSO70-2_Pm  VTNPANTVSGTKRLIGRKFDDPQTQKEMKMVPYKIVRAENG-DAWVE--ANQQYSPSQIGAFILLKMKETAEAYLGKSVTKAVVTVPAYFNDAQRQATKDAGRIAGLDVERIINEPTAA
         A._thaliana_HSO70-2_Ma  VTNPANTVSGTKRLIGRKFDDPQTQKEMKMVPYKIVRAENG-DAWVE--ANQQYSPSQIGAFILLKMKETAEAYLGKSVTKAVVTVPAYFNDAQRQATKDAGRIAGLDVERIINEPTAA
         P._infestans_Mt-Hsp70   _Pm VTNPENTFYAVKRLIGRKFEDKEPQEVSKVVSYKIVKGNGKDAWVE--AKKQKYSPSQIGSMVLMKMKETADGELGKPITQAVVTVPAYFNDSQRQATKDAGKIAGLDVLRIINEPTAA
         P._infestans_Mt-Hsp70   _Ma VTNPENTFYAVKRLIGRKFEDKEPQEVSKVVSYKIVKGNGKDAWVE--AKKQKYSPSQIGSMVLMKMKETADGELGKPITQAVVTVPAYFNDSQRQATKDAGKIAGLDVLRIINEPTAA
         E._tenella_Mt-Hsp70  _Pm ITNPENTFSTKRLIGRSFDEEATAKERKILPYKVIRADNG-DAWVE--GWGKKYSPSQIGAFVLMKMKETAESYLGRDVNQAVITVPAYFNDSQRQATKDAGKIAGLDVLRIINEPTAA
         E._tenella_Mt-Hsp70  _Ma ITNPENTFSTKRLIGRSFDEEATAKERKILPYKVIRADNG-DAWVE--GWGKKYSPSQIGAFVLMKMKETAESYLGRDVNQAVITVPAYFNDSQRQATKDAGKIAGLDVLRIINEPTAA
         D._baltica_Hsp70-type_Pm VTNPENTBFSVKRFIGSKE--EISAEAKQLPYKVTKDSND-NIKIKCPALAKEFSPEEISAQVLRKLINDAKTYLSQDVTQAVITVPAYFNDSQRQATMDAGKIACIEVLRIINEPTAA
         D._baltica_Hsp70-type_Ma VTNPENTBFSVKRFIGSK--EESISAEAKQLPYKVTKDSND-NIKIKCPALAKEFSPEEISAQVLRKLINDAKTYLSQDVTQAVITVPAYFNDSQRQATMDAGKIACIEVLRIINEPTAA
```

Supplementary Data S2C

```
PfHsp70-3_Pm                    ALAFGLEK-SDGKVIAVYDLGGGTFDISILEILSGVFEVKATNGNTSLGGEDFDQRILEYFISEFKKKENIDLKNDKLALQRLREAAETAKIELSSKTQTEINLPFITANQTGPKHLQIK
PfHsp70-3_Ma                    ALAFGLEK-SDGKVIAVYDLGGGTFDISILEILSGVFEVKATNGNTSLGGEDFDQRILEYFISEFKKKENIDLKNDKLALQRLREAAETAKIELSSKTQTEINLPFITANQTGPKHLQIK
C._owczarzaki_Hsp70-2_Pm        ALAYGMDR-SDDKIIAVYDLGGGTFDVSILEIQKGVFEVKATNGDTFLGGEDFDNHLVQFLLEEFKKQHGMDLSKDTVALQRLREAAEKAKIELSSTNQTEVNLPYITADAKGPKHFVHK
C._owczarzaki_Hsp70-2_Ma        ALAYGMDR-SDDKIIAVYDLGGGTFDVSILEIQKGVFEVKATNGDTFLGGEDFDNHLVQFLLEEFKKQHGMDLSKDTVALQRLREAAEKAKIELSSTNQTEVNLPYITADAKGPKHFVHK
G._gallus_Mt-Hsp70   _Pm        ALAYGLDK-SDKVIAVYDLGGGTFDISILEIQKGVFEVKSTNGDTFLGGEDFDQALLQYIVKEFKRETSVDLTKDNMALQRVREASEKAKCELSSSVQTDINLPYLTMDASGPKHLNMK
G._gallus_Mt-Hsp70   _Ma        ALAYGLDK-SDKVIAVYDLGGGTFDISILEIQKGVFEVKSTNGDTFLGGEDFDQALLQYIVKEFKRETSVDLTKDNMALQRVREASEKAKCELSSSVQTDINLPYLTMDASGPKHLNMK
Human_Mt-Hsp70       _Pm        ALAYGLDK-SEDKVIAVYDLGGGTFDISILEIQKGVFEVKSTNGDTFLGGEDFDQALLRHIVKEFKRETGVDLTKDNMALQRVREAAEKAKCELSSSVQTDINLPYLTMDSSGPKHLNMK
Human_Mt-Hsp70       _Ma        ALAYGLDK-SEDKVIAVYDLGGGTFDISILEIQKGVFEVKSTNGDTFLGGEDFDQALLRHIVKEFKRETGVDLTKDNMALQRVREAAEKAKCELSSSVQTDINLPYLTMDSSGPKHLNMK
R._norvegicus_Mt-Hsp70   _Pm    ALAYGLDK-SEDKVIAVYDLGGGTFDISILEIQKGVFEVKSTNGDTFLGGEDFDQALLRHIVKEFKRETGVDLTKDNMALQRVREAAEKAKCELSSSVQTDINLPYLTMDASGPKHLNMK
R._norvegicus_Mt-Hsp70   _Ma    ALAYGLDK-SEDKVIAVYDLGGGTFDISILEIQKGVFEVKSTNGDTFLGGEDFDQALLRHIVKEFKRETGVDLTKDNMALQRVREAAEKAKCELSSSVQTDINLPYLTMDASGPKHLNMK
S._aurata_Grp75_Pm              ALAYGLDK-TQDKIIAVYDLGGGTFDISVLEIQKGVFEVKSTNGDTFLGGEDFDQBLIRHIVKEFKRESGVDLTKDSMALQRVREAAEKAKCELSSSIQTDINLPYCTMDASGPKHLNMK
S._aurata_Grp75_Ma              ALAYGLDK-TQDKIIAVYDLGGGTFDISVLEIQKGVFEVKSTNGDTFLGGEDFDQBLIRHIVKEFKRESGVDLTKDSMALQRVREAAEKAKCELSSSIQTDINLPYCTMDASGPKHLNMK
X._laevis_Hsp70-9_Pm            ALAYGLDK-SDKVIAVYDLGGGTFDISILEIQKGVFEVKSTNGDTFLGGEDFDQALLQHIVKQFKRESGVDLTKDNMALQRVREAAEKAKCELSSSIQTDINLPYLTMDASGPKHLNMK
X._laevis_Hsp70-9_Ma            ALAYGLDK-SDKVIAVYDLGGGTFDISILEIQKGVFEVKSTNGDTFLGGEDFDQALLQHIVKQFKRESGVDLTKDNMALQRVREAAEKAKCELSSSIQTDINLPYLTMDASGPKHLNMK
M._mongolica_Hsp70_Pm           ALAYGMDK-SEDKIIAVYDLGGGTFDISILEIQKGVFEVKSTNGDTFLGGEDFDNALVNFLVAERRDQGLDVTKDPMAMQRVKEAAEKAKIELSSSMQTDINLPYLTMDASGPKHMNLK
M._mongolica_Hsp70_Ma           ALAYGMDK-SEDKIIAVYDLGGGTFDISILEIQKGVFEVKSTNGDTFLGGEDFDNALVNFLVAERRDQGLDVTKDPMAMQRVKEAAEKAKIELSSSMQTDINLPYLTMDASGPKHMNLK
A._echinatior_Hsc70-5_Pm        ALAYGMDK-QEDKIIAVYDLGGGTFDISILEIQKGVFEVKSTNGDTFLGGEDFDNALVNYLVSEFKKKEQGIDVTKDAMAMQRLKEASEKAKIELSSSIQTDINLPYLTMDSSGPKHLNLK
A._echinatior_Hsc70-5_Ma        ALAYGMDK-QEDKIIAVYDLGGGTFDISILEIQKGVFEVKSTNGDTFLGGEDFDNALVNYLVSEFKKEQGIDVTKDAMAMQRLKEASEKAKIELSSSIQTDINLPYLTMDSSGPKHLNLK
M._arenaria_mortalin-2_Pm       ALAYGMDK-TGDKIIAVYDLGGGTFDISILEIQKGVFEVKSTNGDTFLGGEDFDNVLVSYLAKEFQKDQGIDVTKDNMAMQRLREAAEKAKIELSSSIQTDINLPYLTMDACGPKHMNMK
M._arenaria_mortalin-2_Ma       ALAYGMDK-TGDKIIAVYDLGGGTFDISILEIQKGVFEVKSTNGDTFLGGEDFDNVLVSYLAKEFQKDQGIDVTKDNMAMQRLREAAEKAKIELSSSIQTDINLPYLTMDACGPKHMNMK
C._elegans_Hsp-6_Pm             ALAYGLDKDAGDKIIAVYDLGGGTFDVSILEIQKGVFEVKSTNGDTFLGGEDFDHALVHHLVGEFKKEQGVDLTKDPQAMQRLREAAEKAKCELSSTIQTDINLPYITMDQSGPKHLNLK
C._elegans_Hsp-6_Ma             ALAYGLDKDAGDKIIAVYDLGGGTFDVSILEIQKGVFEVKSTNGDTFLGGEDFDHALVHHLVGEFKKEQGVDLTKDPQAMQRLREAAEKAKCELSSTIQTDINLPYITMDQSGPKHLNLK
S._cerevisiae_Ssc1p_Pm          ALAYGLEK-SDSKVVAVFDLGGGTFDISILDIDNGVFEVKSTNGDTFLGGEDFDIYLLRELVSRFKTETGIDLENDRMAIQRLREAAEKAKIELSSTVSTEINLPFITADASGPKHINMK
S._cerevisiae_Ssc1p_Ma          ALAYGLEK-SDSKVVAVFDLGGGTFDISILDIDNGVFEVKSTNGDTFLGGEDFDIYLLRELVSRFKTETGIDLENDRMAIQRLREAAEKAKIELSSTVSTEINLPFITADASGPKHINMK
A._thaliana_HSO70-2_Pm          AISYCMTN-KEG-LIAVFDLGGGTFDVSVLEISNGVFEVKATNGDTFLGGEDFDNALLDFLVNEFKTTEGIDLAKDRLALQRLREAAEKAKIELSSTSQTEINLPFITADASGAKEFNIT
A._thaliana_HSO70-2_Ma          AISYCMTN-KEG-LIAVFDLGGGTFDVSVLEISNGVFEVKATNGDTFLGGEDFDNALLDFLVNEFKTTEGIDLAKDRLALQRLREAAEKAKIELSSTSQTEINLPFITADASGAKEFNIT
P._infestans_Mt-Hsp70    _Pm    ALAYGMDK-ADGKVIAVFDLGGGTFDVSILEIGGGVFEVKSTNGDTFLGGEDFDEELLRYLVNEFKKETSIDLSGDNIAMQRLREAAEKAKREIDGLAQTDISLPFITADATGPKHLNMK
P._infestans_Mt-Hsp70    _Ma    ALAYGMDK-ADGKVIAVFDLGGGTFDVSILEIGGGVFEVKSTNGDTFLGGEDFDEELLRYLVNEFKKETSIDLSGDNIAMQRLREAAEKAKREIDGLAQTDISLPFITADATGPKHLNMK
E._tenella_Mt-Hsp70      _Pm    ALAYCMEK-EDGRTIAVYDLGGGTFDVSILEILGGVFEVKATNGNTSLGGEDFDQKVLQFLVNEFKKKEGIDLSKDRIALQRLREAAETAKIELSSKLSTEINLPFITADQSGPKHLQVS
E._tenella_Mt-Hsp70      _Ma    ALAYCMEK-EDGRTIAVYDLGGGTFDVSILEILGGVFEVKATNGNTSLGGEDFDQKVLQFLVNEFKKKEGIDLSKDRIALQRLREAAETAKIELSSKLSTEINLPFITADQSGPKHLQVS
D._baltica_Hsp70-type_Pm        SLAYGLDK-KQNETILVFDLGGGTFDVSILEVGDGIFEVLATAGDTNLGGDDFDKVLVTWLMNDFKQKEGIDLSTDIQALQRLTEAAEKAMELSTVDKTNISLPFITADQTGPKHIDKE
D._baltica_Hsp70_Ma             SLAYGLDK-KQNETILVFDLGGGTFDVSILEVGDGIFEVLATAGDTNLGGDDFDKVIVTWLMNDFKQKEGIDLSTDIQALQRLTEAAEKAMELSTVDKTNISLPFITADQTGPKHIDKE
```

```
PfHsp70-3_Pm                    LTRAKLEFICRDLLKGTIEPCEKCIKDADVKKEINEIILVGGMTRMPKVTDTVKQIFQNNPSKGVNPDEAVALGAAIQGGVLKCEIKDLLLLDVIPLSLGIETLGGVFTKLINRNTTIP
PfHsp70-3_Ma                    LTRAKLEFICRDLLKGTIEPCEKCIKDADVKKEINEIILVGGMTRMPKVTDTVKQIFQNNPSKGVNPDEAVALGAAIQGGVLKCEIKDLLLLDVIPLSLGIETLGGVFTKLINRNTTIP
C._owczarzaki_Hsp70-2_Pm        LTRAKFESIVGSLVQKTIDPCRKCIKDAGLEKSQIGEVLLVGGMTRMPKVVDTVRELFGREPSKGVNPDEAVAVGAAIQGGVLAGDVTDVLLLDVTPLSLGIETSGGVFSRLINRNTTIP
C._owczarzaki_Hsp70-2_Ma        LTRAKFESIVGSLVQKTIDPCRKCIKDAGLEKSQIGEVLLVGGMTRMPKVVDTVRELFGREPSKGVNPDEAVAVGAAIQGGVLAGDVTDVLLLDVTPLSLGIETSGGVFSRLINRNTTIP
G._gallus_Mt-Hsp70   _Pm        LSRSQFEGIVADLIKRTVAPCQKAMQDAEVSKSDIGEVILVGGMTRMPKVQQTVQDLFGRAPSKAVNPDEAVAIGAAIQGGVLAGDVTDVLLLDVTPLSLGIETLGGVFTKLINRNTTIP
G._gallus_Mt-Hsp70   _Ma        LSRSQFEGIVADLIKRTVAPCQKAMQDAEVSKSDIGEVILVGGMTRMPKVQQTVQDLFGRAPSKAVNPDEAVAIGAAIQGGVLAGDVTDVLLLDVTPLSLGIETLGGVFTKLINRNTTIP
Human_Mt-Hsp70       _Pm        LTRAQFEGIVTDLIRRTIAPCQKAMQDAEVSKSDIGEVILVGGMTRMPKVQQTVQDLFGRAPSKAVNPDEAVAIGAAIQGGVLAGDVTDVLLLDVTPLSLGIETLGGVFTKLINRNTTIP
Human_Mt-Hsp70       _Ma        LTRAQFEGIVTDLIRRTIAPCQKAMQDAEVSKSDIGEVILVGGMTRMPKVQQTVQDLFGRAPSKAVNPDEAVAIGAAIQGGVLAGDVTDVLLLDVTPLSLGIETLGGVFTKLINRNTTIP
R._norvegicus_Mt-Hsp70   _Pm    LTRAQFEGIVTDLIKRTIAPCQKAMQDAEVSKSDIGEVILVGGMTRMPKVQQTVQDLFGRAPSKAVNPDEAVAIGAAIQGGVLAGDVTDVLLLDVTPLSLGIETLGGVFTKLINRNTTIP
R._norvegicus_Mt-Hsp70   _Ma    LTRAQFEGIVTDLIKRTIAPCQKAMQDREVSKSDIGEVILVGGMTRMPKVQQTVQDLFGRAPSKAVNPDEAVAIGAAIQGGVLAGDVTDVLLLDVTPLSLGIETLGGVFTKLINRNTTIP
S._aurata_Grp75_Pm              LSRAQFEGIVADLIRRTVAPCQKAMQDAEVSKGDIGEVLLVGGMSRMPKVQQTVQDLFGRAPSKSVNPDEAVAIGAAIQGGVLAGDVTDVLLLDVTPLSLGIETLGGVFTKLINRNTTIP
S._aurata_Grp75_Ma              LSRAQFEGIVADLIRRTVAPCQKAMQDAEVSKGDIGEVLLVGGMSRMPKVQQTVQDLFGRAPSKSVNPDEAVAIGAAIQGGVLAGDVTDVLLLDVTPLSLGIETLGGVFTKLINRNTTIP
X._laevis_Hsp70-9_Pm            LTRSQFEGIVTDLIKRTVAPFQKAMQDAEVGKSDIGEVLLVGGMTRMPKVQQTVQDLFGRAPSKAVNPDEAVAIGAAIQGGVLAGDVTDVLLLDVTPLSLGIETLGGVFTKLIGRNTTIP
X._laevis_Hsp70-9_Ma            LTRSQFEGIVTDLIKRTVAPFQKAMQDAEVGKSDIGEVLLVGGMTRMPKVQQTVQDLFGRAPSKAVNPDEAVAIGAAIQGGVLAGDVTDVLLLDVTPLSLGIETLGGVFTKLIGRNTTIP
M._mongolica_Hsp70_Pm           LSRAKFELIVGDLIKRTVAPCQKALKDAEVGKNEIGDVLLVGGMTRMPKVQDTVKEIFGRVPSKAVNPDEAVAVGAAIQGGVLAGVTDILLLDVTPLSLGIETLGGVFTKLIQRNTTIP
M._mongolica_Hsp70_Ma           LSRAKFELIVGDLIKRTVAPCQKALKDAEVGKNEIGDVLLVGGMTRMPKVQDTVKEIFGRVPSKAVNPDEAVAVGAAIQGGVLAGVTDILLLDVTPLSLGIETLGGVFTKLIQRNTTIP
A._echinatior_Hsc70-5_Pm        LSRSKFESLVNDLIKRTVQPCQKALSDAEVTKSDIGEVLLVGGMTRVPKVQQTVQEIFGROPSKAVNPDEAVAGAAVQGGVLAGDVTDVLLLDVTPLSLGIETLGGVFTRLISRNTTIP
A._echinatior_Hsc70-5_Ma        LSRSKFESLVNDLIKRTVQPCQKALSDAEVTKSDIGEVLLVGGMTRVPKVQQTVQEIFGROPSKAVNPDEAVAGAAVQGGVLAGDVTDVLLLDVTPLSLGIETLGGVFTRLISRNTTIP
M._arenaria_mortalin-2_Pm       LSRAKFESLVDDLIKRTVGPCNKALDAEIKKSDIGDVLLVGGMTRMPKVQQVVQEVFGRAPCKSVNPDEAVAIGAAIQGGVLAGDVTDVLLLDVTPLSLGIETLGGVFSRLITRNTTIP
M._arenaria_mortalin-2_Ma       LSRAKFESLVDDLIKRTVGPCNKALDAEIKKSDIGDVLLVGGMTRMPKVQQVVQEVFGRAPCKSVNPDEAVAIGAAIQGGVLAGDVTDVLLLDVTPLSLGIETLGGVFSRLITRNTTIP
C._elegans_Hsp-6_Pm             LTRAKFEQIVGDLIKRTIEPCRKALRDAEVKSSQIADVLLVGGMSRMPKVQATVQEIFGKVPSKAVNPDEAVAMGAAIQGAVLAGDVTDVLLLDVTPLSLGIETLGGIMTKLITRNTTIP
C._elegans_Hsp-6_Ma             LTRAKFEQIVGDLIKRTIEPCRKALRDAEVKSSQIADVLLVGGMSRMPKVQATVQEIFGKVPSKAVNPDEAVAMGAAIQGAVLAGDVTDVLLLDVTPLSLGIETLGGIMTKLITRNTTIP
S._cerevisiae_Ssc1p_Pm          FSRAQFEDITAPLVKRVDPVKKALKDAGLSTSDISEVLLVGGMSRMPKVVETVKTVKSLFGKDPSKAVNPDEAVAIGAAVQGAVLSGEVTDVLLLDVTPLSLGIETLGGVFTRLIPRNTTIP
S._cerevisiae_Ssc1p_Ma          FSRAQFEDITAPLVKRVDPVKKALKDAGLSTSDISEVLLVGGMSRMPKVVETVKSLFGKDPSKAVNPDEAVAIGAAVQGAVLSGEVTDVLLLDVTPLSLGIETLGGVFTRLIPRNTTIP
A._thaliana_HSO70-2_Pm          LTRSRFEEITAVNHLIERTRDPCKNCIKDAGISAKEVDEVLLVGGMTRVPKVQSIVAEIFGKSPSKGVNPDEAAGAAIQGGILRGDVKELLLLDVTPLSLGIETLGGVFTRLIPRNTTIP
A._thaliana_HSO70-2_Ma          LTRSRFEEITAVNHLIERTRDPCKNCIKDAGISAKEVDEVLLVGGMTRVPKVQSIVAEIFGKSPSKGVNPDEAAGAAIQGGILRGDVKELLLLDVTPLSLGIETLGGVFTRLIPRNTTIP
P._infestans_Mt-Hsp70    _Pm    ITRAIFEKLVGKLIERTMGPCKKCVKDAGLDKSEINEVILVGGMSRMPKVQTTVEEFFGKKPSKGVNPDEVVAMGAAIQGGVLRGDVKDILLLDVTPLSLGIETLGGVFTKLIPRNTTIP
P._infestans_Mt-Hsp70    _Ma    ITRAIFEKLVGKLIERTMGPCKKCVKDAGLDKSEINEVILVGGMSRMPKVQTTVEEFFGKKPSKGVNPDEVVAMGAAIQGGVLRGDVKDILLLDVTPLSLGIETLGGVFTKLIPRNTTIP
E._tenella_Mt-Hsp70      _Pm    LSRAHLERIVGALLQQSIEPCEKCIRDAGVQKADLSDVILVGGMTRMPKVAEVVKNIFHKEPSKGVNPDEAVAAGAAIQAGVLKCEIKDLLLLDVCPLSLGIETLGGVFTRLINRNTTIP
E._tenella_Mt-Hsp70      _Ma    LSRAHLERIVGALLQQSIEPCEKCIRDAGVQKADLSDVILVGGMTRMPKVAEVVKNIFHKEPSKGVNPDEAVAAGAAIQAGVLKCEIKDLLLLDVCPLSLGIETLGGVFTRLINRNTTIP
D._baltica_Hsp70-type_Pm        LTRETFEKICEKLIDRCRIPVEKALNDARLDKSDINEVVLVGGSTRIPATQQLVESLTGKKPNQSVNPDEVVAIGAAIQAGILAGEIKDILLLDVTPLSLGVETLGGVMTKIIARNTTIP
D._baltica_Hsp70_Ma             LTRETFEKICEKLIDRCRIPVEKALNDARLDKSDINEVVLVGGSTRIPATQQLVESLTCKKPNQSVNPDEVVAIGAAIQAGILAGEIKDILLLDVTPLSLGVETLGGVMTKIIARNTTIP
```

Supplementary Data S2C

[Multiple sequence alignment of Hsp70 proteins from various organisms, showing two blocks of aligned sequences. Sequences include PfHsp70-3, C. owczarzaki Hsp70-2, G. gallus Mt-Hsp70, Human Mt-Hsp70, R. norvegicus Mt-Hsp70, S. aurata Grp75, X. laevis Hsp70-9, M. mongolica Hsp70, A. echinatior Hsc70-5, M. arenaria mortalin-2, C. elegans Hsp-6, S. cerevisiae Ssc1p, A. thaliana HSO70-2, P. infestans Mt-Hsp70, E. tenella Mt-Hsp70, and D. baltica Hsp70-type, each with _Pm and _Ma variants.]

Supplementary Data S3A

```
                                10        20        30        40        50        60        70        80        90       100       110       120
                           ....|....|....|....|....|....|....|....|....|....|....|....|....|....|....|....|....|....|....|....|....|....|....|....|
Consensus              1   Ma---kg~avGIDLGTTYSCVgvfqhgkVeIIaNDQGNRTTPSyVAFTDtERLIGDAAKNQvAmNpcNTvFDAKRLIGRkfddpvvQsDmkhwPFkvin-~g~KPkiqVeyKge~KtFyP   112
Human Hsp70-1          1   MA---KAAAIGIDLGTTYSCVGVFQHGKVEIIANDQGNRTTPSYVAFTDTERLIGDAAKNQVALNPQNTVFDAKRLIGRKFGDPVVQSDMKHWPFQVIN-DGDKPKVQVSYKGDTKAFYP   116
R. norvegicus Hsp70    1   MA-ANKGMAIGIDLGTTYSCVGVFQHGKVEIIANDQGNRTTPSYVAFTDTERLIGDAAKNQVAMNPQNTVFDAKRLIGRKFENDPVVQSDMKLWPFQVIN-EAGKPKVLVSYKGEKKAFYP   118
X. laevis Hsp70        1   MA--TKGVAVGIDLGTTYSCVGVFQHGKVEIIANDQGNRTTPSYVAFTDTERLIGDAAKNQVAMNPNNTVFDAKRLIGRKFNDPVVQCDLKHWPFQVVS-DEGKPKVKVEYKGEEKSFFP   117
L. calcarifer Hsp70    1   MS-PAKGVAIGIDLGTTYSCVGIFQHGKVEIIANDQGNRTTPSYVAFTDTERLIGDAAKNQVALNPSNTVFDAKRLIGRKFDDSVVQSDMKHWPFKVIS-DGGKPKIQVEYKGEDKAFYP   118
A. echinatior Hsc70-4  1   MS---KAPAVGIDLGTTYSCVGVFQHGKVEIIANDQGNRTTPSYVAFTDTERLIGDAAKNQVAMNPSNTIFDAKRLIGRRFDDTTVQSDMKHWPFTVVN-DGSKPKIKVSYKGEMKTFFP   116
F. chinensis Hsc70     1   MA---KAPAIGIDLGTTYSCVGVFQHGKVEIIANDQGNRTTPSYVAFTDTERLIGDAAKNQVAMNPNNTVFDAKRLIGRKFEDHTVQSDMKHWPFTIIN-ESTKPKIQVEYKGDKKTFYP   116
H. diversicolor Hsc70  1   MA---KAPAIGIDLGTTYSCVGVFQHGKVEIIANDQGNRTTPSYVAFTDTERLIGDAAKNQVAMNPENTIFDAKRLIGRRFDETNVQSDMKHWPFNVLS-DGGKPKIQVNYKDEPKTFYP   116
C. japonica Hsp70      1   MS--GKGPAIGIDLGTTYSCVGVFQHGKVEIIANDQGNRTTPSYVAFTDTERLIGDAAKNQVAMNPLNTIFDAKRLIGRKYDDPTVQSDMKHWPFRVVN-EGGKPKVQVEYKGEMKTFFP   117
C. elegans Hsp-1       1   MS---KHNAVGIDLGTTYSCVGVFMHGKVEIIANDQGNRTTPSYVAFTDTERLIGDAAKNQVAMNPHNTVFDAKRLIGRKFDDPAVQSDMKHWPFKVISAEGAKPKVQVEYKGENKIFTP   117
C. owczarzaki Hsp70    1   MA---TQLAVGIDLGTTYSCVGVFQHGKVEIIANDQGNRTTPSYVAFTDTERLIGDAAKNQVAMNPENTVFDAKRLIGRFRDDPAVQSDMKHWPFIVN-EATKPKIQVNKGEEKVFSP   116
A. thaliana Hsp70B     1   MA-TKSEKAIGIDLGTTYSCVGVWMNDRVEIIPNDQGNRTTPSYVAFTDTERLIGDAAKNQVALNPQNTVFDAKRLIGRKFSDPSVQSDILHWPFKVVSGPGEKPMIVVSYKNEEKQFSP   119
S. cerevisiae Ssa1p    1   MS-----KAVGIDLGTTYSCVAHFANDRVDIIANDQGNRTTPSFVAFTDTERLIGDAAKNQAAMNPSNTVFDAKRLIGRNFNDPEVQGDMKHFPFLID-VDGKPQIQVEFKGETKNFTP   114
P. infestans T30-4 Hsp70 1 MTQATSGYSVGIDLGTTYSCVGVWQNDRVEIIANDQGNRTTPSYVAFTDSERLIGDAAKNQVAMNAANTVFDAKRLIGRKFSDPVVQADIKHWPFKITAGPGDKPQITVQFKGETKTFQP   120
```

```
                                130       140       150       160       170       180       190       200       210       220
                           ....|....|....|....|....|....|....|....|....|....|....|....|....|....|....|....|....|....|....|....|....|....|
Consensus              113 EeiSSMVLtKMKEtAEaylGktv~nAViTVPAYFNDSQRQATKDAGtIaGlNvlRIINEPTAAAIAYGLDkk-~~ggErnVLIFDLGGGTFDVSiLtIedGiFEVKaTAGDTHLGGEDFD   228
Human Hsp70-1          117 EEISSMVLTKMKETAEAYLGYPVTNAVITVPAYFNDSQRQATKDAGVIAGLNVLRIINEPTAAAIAYGLDRT--GKGERNVLIFDLGGGTFDVSILTIDDGIFEVKATAGDTHLGGEDFD   234
R. norvegicus Hsp70    119 EEISSMVLTKMKETAEAFLGHSVTNAVITVPAYFNDSQRQATKDAGVIAGLNVLRIINEPTAAAIAYGLDKG--SHGERHVLIFDLGGGTFDVSILTIDDGIFEVKATAGDTHLGGEDFD   236
X. laevis Hsp70        118 EEISSMVLTKMKETAEAYLGHPVTNAVITVPAYFNDSQRQATKDAGVIAGLNILRIINEPTAAAIAYGLDKG--ARGEQNVLIFDLGGGTFDVSILTIDDGIFEVKATAGDTHLGGEDFD   235
L. calcarifer Hsp70    119 EEISSMVLVKMKIAEAYLGQKVSNAVITVPAYFNDSQRQATKDAGVIAGLNVLRIINEPTAAAIAYGLDKK--KSGERNVLIFDLGGGTFDVSILTIEDGIFEVKATAGDTHLGGEDFD   236
A. echinatior Hsc70-4  117 EEVSSMVLTKMKETAEAYLGKTITNAVITVPAYFNDSQRQATKDAGAIAGLNVLRIINEPTAAAIAYGLDKK--AAGEKNVLIFDLGGGTFDVSILTIEDGIFEVKSTAGDTHLGGEDFD   234
F. chinensis Hsc70     117 EEISSMVLIKMKETAEAYLGSTVKDAVVTVPAYFNDSQRQATKDAGTISGLNVLRIINEPTAAAIAYGLDKK--VGGERNVLIFDLGGGTFDVSILTIEDGIFEVKSTAGDTHLGGEDFD   234
H. diversicolor Hsc70  117 EEISSMVLTKMKETAEQYLGKTITDAVVTVPAYFNDSQRQATKDAGTISGLNVLRIINEPTAAAIAYGLDKK--VGGERNVLIFDLGGGTFDVSILTIEDGIFEVKSTAGDTHLGGEDFD   234
C. japonica Hsp70      118 EEISSMVLTKMEIAEAYLGKKVQNAVITVPAYFNDSQRQATKDAGTITGLNVMRIINEPTAAAIAYGLDKKGTRAEKNVLIFDLGGGTFDVSILTIEDGIFEVKSTAGDTHLGGEDFD   237
C. elegans Hsp-1       118 EEISSMVLLKMKETAEAFLGTVKDAVVTVPAYFNDSQRQATKDAGAIAGLNVLRIINEPTAAAIAYGLDKK--GHGERNVLIFDLGGGTFDVSILTIEDGIFEVKSTAGDTHLGGEDFD   235
C. owczarzaki Hsp70    117 EEISSMVLLKMKETAEAYLGKTINNAVTVPAYFNDSQRQATKDAGTISGMNVLRIINEPTAAAIAYGLDKK--IGGERHVLIFDLGGGTFDVSVLTIEDGIFEVKSTAGDTHLGGEDFD   234
A. thaliana Hsp70B     120 EEISSMVLVKMKEVAEAFLGRTVNAVVTVPAYFNDSQRQATKDAGAISGLNVLRIINEPTAAAIAYGLDKKGTKAGEKNVLIFDLGGGTFDVSLLTIEEGVFEVKATAGDTHLGGEDFD   239
S. cerevisiae Ssa1p    115 EQISSMVLGKMKETAESYLGAKVNDAVVTVPAYFNDSQRQATKDAGTIAGLNVLRIINEPTAAAIAYGLDKK---GKEEHVLIFDLGGGTFDVSLLSIEDGIFEVKATAGDTHLGGEDFD   231
P. infestans T30-4 Hsp70 121 EEISSMVLIKMREVAEAFIGKEVKNAVITVPAYFNDSQRQATKDAGAIAGLNVLRIINEPTAAAIAYGLDKK---GGERNVLIFDLGGGTFDVSLLSIEEGIFEVKATAGDTHLGGEDFD   237
```

```
                                240       250       260       270       280       290       300       310       320       330       340
                           ....|....|....|....|....|....|....|....|....|....|....|....|....|....|....|....|....|....|....|....|....|....|....|....|
Consensus              229 NRmVnHFvqEFkRKhkkDisonkRalRRLrTACERAKRTLSSstqasiEIDSLfeGiDfytsitRARFEelcaDlFRgtlePVEKaLrDAK-dKaqihdiVLVGGSTRIPKvQkLlqDfF   347
Human Hsp70-1          235 NRLVNHFVEEFKRKHKKDISQNKRAVRRLRTACERAKRTLSSSTQASLEIDSLFEGIDFYTSITRARFEELCSDLFRSTLEPVEKALRDAKLDKAQIHDLVLVGGSTRIPKVQKLLQDFF   354
R. norvegicus Hsp70    237 NRLVSHFVEEFKRKHKKDISQNKRAVRRLATACERAKRTLSSSTQANLEIDSLYEGIDFYTSITRARFEELCADLFRGTLEPVEKSLRDAKIKHDIVLVGGSTRIPKVQKLLQDYF   356
X. laevis Hsp70        236 NRMVNHFVEEFKRKHKKDISQNKRALRRLRTACERAKRTLSSSSQASIEIDSLFEGIDFYTAITRARFEELCSDLFRGTLEPVEKSLRDAKLDKSQIHEIVLVGGSTRIPKVQKLLQDFF   355
L. calcarifer Hsp70    237 NRMVNHFVEEFKRKHKKDISQNKRALRRLRTACERAKRTLSSSSQASIEIDSLFEGIDFYTSITRARFEELCSDLFRGTLEPVEKALRDAKLDKAQIQKLLQDFF   356
A. echinatior Hsc70-4  235 NRMVNHFVQEFKRKYKKDLSSNKRAVRRLRTACERAKRTLSSSTQASIEIDSLFEGIDFYTSVTRARFEELCADLFRSTLEPVEKALRDAKMDKAQVHSIVLVGGSTRIPKIQKLLQDFF   354
F. chinensis Hsc70     235 NRMVNHFIQEFKRKYKKDPSENKRSLRRLRTACERAKRTLSSSTQASVEIDSLFEGIDFYTSITRARFEELCADLFRGTLEPVEKSLRDAKMDKAQIHDLVLVGGSTRIPKIQKLLQDFF   354
H. diversicolor Hsc70  235 NRMVNHFIQEFKRKHKKDISQNKRALRRLRTACERAKRTLSSSTQASIEIDSLFEGVDYYTSITRARFEELNADLFRGTLEPVEKALRDAKVSIHDIVLVGGSTRIPKIQKLLQDFF   354
C. japonica Hsp70      238 NRMVNHFVEEFKRHKRDIAGNKRAVRRLRTACERAKRTLSSSTQASVEIDSLFEGIDFYTSITRARFEEINADLFRGTLEPVEKALRDAKLDKGQIQEIVLVGGSTRIPKIQKLLQDFF   357
C. elegans Hsp-1       236 NRMVNHFCAEFKRKHKKDLASNPRALRRLRTACERAKRTLSSSSQASIEIDSLFEGIDFYNITRARFEELCADLFRSTMDPVEKSLRDAKMDKSQVHDIVLVGGSTRIPKVQKLLSDLF   355
C. owczarzaki Hsp70    235 NRMVNHFVQEFKRKEKKDLSTSARALRRLRTACERAKRTLSSSTEASIEIDSLFEGVDFYTKITRARFEELCADLFRGTLDPVEKSLRDAKMDKGTIDDIVLVGGSTRIPKVQKLQDFF   354
A. thaliana Hsp70B     240 NRLMVHFVEAEFRKRKHKDIAGNARALRRLRTACERAKRTLSSTAQTTIEIDSLHEGIDFYATISRARFEEMNMLFRKCMDVKLKDAKLDKSSVHDVVLVGGSTRIPKIQQLLDFF   359
S. cerevisiae Ssa1p    232 NRLVNHFIQEFKRKNKKDLSTNQRALRRLRTACERAKRTLSSSAQTSVEIDSLFEGIDFYTSITRARFEELCADLFRSTLDPVEKVLRDAKLDKSQVDEIVLVGGSTRIPKVQKLVTDYF   351
P. infestans T30-4 Hsp70 238 NRLVEHFVQEFKRKHRKDLTQNQRALRRLRTACERAKRTLSSSAQAYIEIDSLFDGVDFNSTITRARFEDMCGDYFRKTMEPVEKVLRDAKLSKGQVHEVVLVGGSTRIPKVQQLLSDFF   357
```

Supplementary Data S3A

```
                              ....|....|....|....|....|....|....|....|....|....|....|....|
Consensus               348   nGkelnkSINPDEAVAYGAaVQAAILnG-dkSe~vQDLLLLLDVaPLSLGiETAGGVMT~LIkRNtTiPtKqtqtFtTYsDNQpgVLiQVyEGERamTkDNNlLGkFeLsGIPPaPRGVPQ   464
Human Hsp70-1           355   NGRDLNKSINPDEAVAYGAAVQAAILMG-DKSENVQDLLLLLDVAPLSLGLETAGGVMTALIKRNSTIPTKQTQIFTTYSDNQPGVLIQVYEGERAMTKDNNLLGRFELSGIPPAPRGVPQ   473
R. norvegicus Hsp70     357   NGRDLNKSINPDEAVAYGAAVQAAILMG-DKSEKVQDLLLLLDVAPLSLGLETAGGVMTVLIKRNSTIPTKQTQIFTTYSDNQPGVLIQVYEGERAMTRDNNLLGRFDLTGIPPAPRGVPQ   475
X. laevis Hsp70         356   NGRELNKSINPDEAVAYGAAVQAAILMG-DKSENVQDLLLLLDVAPLSLGLETAGGVMTVLIKRNTTIPTKQTQTFTTYSDNQPGVLIQVFEGERAMTKDNNLLGKFELTGIPPAPRGVPQ   474
L. calcarifer Hsp70     357   NGRELNKSINPDEAVAYGAAVQAAILTG-DTSGNVQDLLLLLDVAPLSLGIETAGGVMTSLIKRNTTIPTKQAQVFTTYSDNQPGVLIQVYEGERAMTKDNNLLGKFELTGIPPAPRGVPQ   475
A. echinatior Hsc70-4   355   NGKELNKSINPDEAVAYGAAVQAAILHG-DKSEEVQDLLLLLDVTPLSLGIETAGGVMTTLIKRNTTIPTKQTQTFTTYSDNQPGVLIQVYEGERAMTKDNNILGKFELTGIPPAPRGVPQ   473
F. chinensis Hsc70      355   NGKELNKSINPDEAVAYGAAVQAAILCG-DKSEAVQDLLLLLDVTPLSLGIETAGGVMTALIKRNTTIPTKQTQTFTTYSDNQPGVLIQVYEGERAMTKDNNLLGKFELSGIPPAPRGVPQ   473
H. diversicolor Hsc70   355   NGKELCKSINPDEAVAYGAAVQAAILHG-DKSEEVQDLLLLLDVTPLSLGLETAGGVMTVLIKRNTTIPTKQTQTFTTYSDNQPGVLIQVFEGERAMTKDNNLLGKILGKFDLTGIPPAPRGVPQ   473
C. japonica Hsp70       358   NGKELNKSINPDEAVAYGAAVQAAILMG-DNSENVQDLLLLLDVTPLSLGIETAGGVMTTLIKRNTTIPTKQTQTFTTYSDNQNSVLVQVYEGERAMTKDNNLLGKFDLTGIPPAPRGVPQ   476
C. elegans Hsp-1        356   SGKELNKSINPDEAVAYGAAVQAAILSG-DKSEAVQDLLLLLDVAPLSLGIETAGGVMTALIKRNTTIPTKTAQTFTTYSDNQPGVLIQVYEGERAMTKDNNLLGKFELSGIPPAPRGVPQ   474
C. owczarzaki Hsp70     355   NGKELNKSINPDEAVAYGAAVQAAILSG-DKSEAVQDLLLLLDVAPLSLGIETAGGVMTTLIKRNTTIPTKQTQTFTTYADNQPGVLIQVYEGERAMTRDNNLLGKFELSGIPPAPRGVPQ   473
A. thaliana Hsp70B      360   NGKELCKSINPDEAVAYGAAVQAAILHG-EGSEKVQDLLLLLDVTPLSLGLETAGGVMTVLIPRNTTVPCKKEQVFSTYADNQPGVLIQVCERARTRDNNLLGTFELKGIPPAPRGVPQ    478
S. cerevisiae Ssa1p     352   NGKEPNRSINPDEAVAYGAAVQAAILTG-DESSKTQDLLLLLDVAPLSLGLETAGGVMTKLIPRNSTIPTKKSEIFSTYADNQPGVLIQVFEGERAKTKDNNLLGKFELSGIPPAPRGVPQ   470
P. infestans T30-4 Hsp70 358  NGKEPNKSINPDEAVAYGATVQAAILSGNDSSEKLQDLLLLLDVTPLSLGLETAGGVMTTLIARNTTVPTKKSQTFSTYADNQPGVLIQVFEGERTMTRDNNLLGKFNLDGIPPMPRGVPQ   477

                              ....|....|....|....|....|....|....|....|....|....|....|....|
Consensus               465   IeVtFDiDanGILNVsAvdKstgk~NkITITNDKGRLSkeeIerMVqeAekykaeDeaqr~risaKNaLEsyafnmkstvedeklkdKiseeDkkkvldkc~evisWldanqlaekeEfe   581
Human Hsp70-1           474   IEVTFDIDANGILNVTATDKSTGKANKITITNDKGRLSKEEIERMVQEAEKYKAEDEVQREVSAKNALESYAFNMKSAVEDEGLKDKISEADKKVLDKCQEVISWLDANTLAEKDEFE     593
R. norvegicus Hsp70     476   IEVTFDIDANGILNVTAMDKSTGKANKITITNDKGRLSKEEIERMVQEAERYKAEDEGQREKIAAKNALESYAFNMKSAVGDEGLKDKISESDKKILDKCSEVLSWLEANQLAEKEEFD   595
X. laevis Hsp70         475   IEVTFDIDANGILNVSAVESSGKQNKITITNDKGRLSKEDIEKMVQEAEKYKADDERVDAKNALESYAFNMKSVQDENMKGKISEDDKRTISEKCTQVISWLENNQLAEEKEYQ      594
L. calcarifer Hsp70     476   IEVTFDIDAHGILNVSAVDKSTGKANKITITNDKGRLSKEEIERMVQDADKYKAEDDLQRDRISAKNSLESYAFNMKSSVQDENMKGKISEEDQKKVIEKCDETITWLENNQLADKEEYQ  595
A. echinatior Hsc70-4   474   IEVTFDIDANGILNVSAIEKSTGKENKITITNDKGRLSKEDIERMVNEAEKYRSEDEQQRERISAKNALESYCFNMKSTMEDDKVKDKIEASDKEKVLSKCNEVISWLDANQLAEKEEFA  593
F. chinensis Hsc70      474   IEVTFDIDASGILNVSDVKSTGKENKITITNDKGRLSKEEIERMVQDAEKYKADDEKQRDRISAKNSLESYCFNMKSTVEDEKFEKISEEDRNKILETCNETIKWLDMNQLGEKEEYE   593
H. diversicolor Hsc70   474   IEVTFDIDANGILNVSAVDKSTMKENKITITNDKGRLSKEEIERMVNEAENYKAEDEKQKDRIQAKNGLESYAFNMKSKISEDTIKCNDVISWLDSNQLAEKDEFE               593
C. japonica Hsp70       477   IEVTFDIDANGILNVSAVDKSTGKENKITITNDKGRLSKDDIDRMVQEAEKYKAEDEANRDRVGAKNSLESYTYNMKQTVEDDKLKGKISDQDKQKVLDKCREVISWLDMNQMAEKEEYQ  596
C. elegans Hsp-1        475   IEVTFDIDANGILNVSATDKSTGKQNKITITNDKGRLSKDDIERMVNEAEKYKADDEAQKDRIGAKNGLESYAFNLKQTIEDEKLKDKISPEDKKKIEDKCDEILKWLDSNQTAEKEEFE  594
C. owczarzaki Hsp70     474   IEVTFDIDANGILNVSAVDKSTGKVNKITITNDKGRLSKEEIERMVAEADKYKAQDEAQRERVAGKNALESYCFNMKQAVDDNISKKLSDEDKKTVTEKVEEAMKWLEANQLAEKEEFE   593
A. thaliana Hsp70B      479   INVCFDIDANGILNVSAEDKTAGVKNQITITNDKGRLSKEEIEKMVKDAEKYKAEDEQVKKKVEAKNSLENYAYNMRNTIKDEKIAQKLTQEDKQIEKAIDETIEWIEGNQLAEVDEFE   598
S. cerevisiae Ssa1p     471   IEVTFDVDSNGILNVSAVEKGTGKSNKITITNDKGRLSKEDIEKMVAEAEKFKEEDEKESQRIASKNQLESIAYSLKNTISEA--GDKLEQADKDTVTKKAEETISWLDSNTTASKEEFD   588
P. infestans T30-4 Hsp70 478  IDVTFDIDANGILNVSAVEKSTGKENKITITNDKGRLSQAEIERMVAEAEKYKSEDEANKVRIEAKNALENYAYSLRNSLNDEKMKPKIPEADKKVVDDKVTEVIQWMDAHQSSEKEEYE  597

                              ....|....|....|....|....|....|....|....|....|....|....|
Consensus               582   hkqKele~~cnPiitklYqg~gga~~~~~ggmp~~~~~~~~~~~~g~ag~~ag~~~sgGPtiEEVD   621
Human Hsp70-1           594   HKRKELEQVCNPIISGLYQGAGGP---GPGGF------------GAQGPKGGS----GSGPTIEEVD   641
R. norvegicus Hsp70     596   HKRKELENMCNPIITKLYQSGCTG----PTCAP-----------GYTP-GRA----RTGPTIEEVD   641
X. laevis Hsp70         595   FQQKDLEKVCQPIITKLYQSGVPG---GVPGGMP------GSSC-GAQARQGG----SSGPTIEEVD  647
L. calcarifer Hsp70     596   HQQKELEKVCNPIISKLYQ---------GGMP--------ASSC--REEARAG----SGPTIEEVD   639
A. echinatior Hsc70-4   594   DKQKELEALCNPIVTKLYQSGCAP-----GGFP-------GAGG-AGANPGAG----GAGPTIEEVD  643
F. chinensis Hsc70      594   HKQKELEQVCNPIITKMYAAGGA---PPGGMPGGFPG--GAPGAGGAAPGAGG----SSGPTIEEVD  652
H. diversicolor Hsc70   594   HKQKELEGVCNPIITKLYQAAGGA------GGMPNFNPGAAGAGGAGAQT-GGS---SGGPTIEEVD  651
C. japonica Hsp70       597   HKQKELEKLCNPIVTKLYQGAGGA-------------------------GAGG----AGGPTIEEVD  634
C. elegans Hsp-1        595   HQQKDLEGLANPIISKLYQSAGGA---PPGAAP----------GGAAGG----AGGPTIEEVD      640
C. owczarzaki Hsp70     594   HRLKELEKACSPIIAKAYQGGAAP-----GGMP-------GAEG-GWQGPGAGATADPAAGPTIEEVD  648
A. thaliana Hsp70B      599   YKLKELEGICNPIISKMYQGGAAA------GGMPTD----------GDFSSSGA----AGGPKIEEVD  646
S. cerevisiae Ssa1p     589   DKLKELQDIANPIMSKLYQAXCAP-GGAAGGAPGGFP--------GAPPAPE-----AEGPTVEEVD   642
P. infestans T30-4 Hsp70 598  SKQKELESVANPVLQKVYASAGGAGDGMPGSMPNDMP--------GTDSRSSGA----EQGPKIEEVD  653
```

Supplementary Data S3B

```
                                10        20        30        40        50        60        70        80        90       100       110       120
                       ....|....|....|....|....|....|....|....|....|....|....|....|....|....|....|....|....|....|....|....|....|....|....|....|
Consensus            1 M---~egpAiGIDLGTTYSCVgVwknd~VeIipNDQGNRTTPSYVAFTdTERLiGDAAKNQvARNPENTvFDAKRLIGRkFDDpaVQsDMkHWPFkV~aGpggKP~IeVtyqGekKtFHpE 114
T. gondii Hsp70      1 M--ADSPAVGIDLGTTYSCVGVWKNDAVEIIANDQGNRTTPSYVAFTDTERLVGDAAKNQVARNPENTIFDAKRLIGRKFDDPSVQSDMKHWPFKVIAGPGDKPLIEVTYQGEKKTFHPE 118
N. caninum Hsp70     1 M--ADSPAVGIDLGTTYSCVGVWKNDAVEIIANDQGNRTTPSYVAFTDTERLVGDAAKNQVARNPENTIFDAKRLIGRKFDDPSVQSDMKHWPFKVIAGPGDKPLIEVTYQGEKKTFHPE 118
E. acervulina Hsp70  1 M--SEAPAVGIDLGTTYSCVGVWKNDGVEIIANDQGNRTTPSYVAFTDTERLVGDAAKNQVARNPENTVFDAKRLIGRKFDDPAVQADMKHWPFETVKAGPGDKPLIEVNYQGSKKTFHPE 118
C. cayetanensis Hsp70 1 M--AEAPAVGIDLGTTYSCVGVWKNEGVEIIANDQGNRTTPSYVAFTDTERLVGDAAKNQVARNPENTVFDAKRLIGRKFDDPAVQSDMKHWPFTVKAGSGGKPLIEVNYQGATKTFHPE 118
C. parvum Hsp70      1 MTSSEGPAIGIDLGTTYSCVGVWRNDTVDIVPNDQGNRTTPSYVAFTETERLIGDAAKNQVARNPENTVFDAKRLIGRKFDDQAVQSDMTHWPFKVVRGPKDKPIISVNYLGEKKEFHAE 120
C. hominis Hsp70     1 MTSSEGPAIGIDLGTTYSCVGVWRNDTVDIVPNDQGNRTTPSYVAFTETERLIGDAAKNQVARNPENTVFDAKRLIGRKFDDQAVQSDMTHWPFKVVRGPKDKPIISVNYLGEKKEFHAE 120
T. parva Hsp70       1 M---TGPAIGIDLGTTYSCVAVYKDNNVEIIPNDQGNRTTPSYVAFTDTERLIGDAAKNQEARNPENTIFDAKRLIGRKFDDRTVQEDMKHWPFKVTNGPNGKPNIEVTFQGEKKTFHAE 117
T. annulata Hsp70    1 M---TGPAIGIDLGTTYSCVAVYKDNNVEIIPNDQGNRTTPSYVAFTDTERLIGDAAKNQEARNPENTIFDAKRLIGRKFDDRTVQEDMKHWPFKVTNGPNGKPNIEVTFQGEKKTFHAE 117
B. divergens Hsp70   1 M---DGIAIGIDLGTTYSCVGVYKDNNVEIIPNDQGNRTTPSYVAFTDTERLIGDAAKNQEARNPENTVFDAKRLIGRRFDDPTVQDDMKHWPFKVINGVGGKPTIEVTFQGQKKTFHPE 117

                       ....|....|....|....|....|....|....|....|....|....|....|....|....|....|....|....|....|....|....|....|....|....|....|....|
Consensus          115 EiSaMVL~KMKEIaEaylGk~vKeaViTVPAYFNDSQRQATKDAGtIAGLnVmRIINEPTAAAIAYGLDKKG~gEmNVLIFDlGGGTFDVSlLTIEDGIFEVKATAGDTHLGGEDFDNrL 231
T. gondii Hsp70    119 EVSAMVLGKMKEIAEAYLGKEVKEAVITVPAYFNDSQRQATKDAGTIAGLSVLRIINEPTAAAIAYGLDKKCGEMNVLIFDMGGGTFDVSLLTIEDGIFEVKATAGDTHLGGEDFDNRL 238
N. caninum Hsp70   119 EVSAMVLGKMKEIAEAYLGKDVKEAVITVPAYFNDSQRQATKDAGTIAGISVLRIINEPTAAAIAYGLDKKCGEMNVLIFDMGGGTFDVSLLTIEDGIFEVKATAGDTHLGGEDFDNRL 238
E. acervulina Hsp70 119 EISAMVLMKMKEIAEAFIGKEVKEAVITVPAYFNDSQRQATKDAGTIAGLNVLRIINEPTAAAIAYGLDKKHGEMNVLIFDMGGGTFDVSLLTIEDGIFEVKATAGDTHLGGEDFDNRL 238
C. cayetanensis Hsp70 119 EISAMVLVKMKEIAESFVGKEVKEAVITVPAYFNDSQRQATKDAGTIAGLNVLRIINEPTAAAIAYGLDKKQGEMNVLIFDMGGGTFDVSLLTIEDGIFEVKATAGDTHLGGEDFDNRL 238
C. parvum Hsp70    121 EISAMVLQKMKEISEAYLGRQIKNAVVTVPAYFNDSQRQATKDAGAIAGLNVMRIINEPTAAAIAYGLDKKGTGERNVLIFDLGGGTFDVSLLTIEDGIFEVKATAGDTHLGGEDFDNRL 240
C. hominis Hsp70   121 EISAMVLQKMKEISEAYLGRQIKNAVVTVPAYFNDSQRQATKDAGAIAGLNVMRIINEPTAAAIAYGLDKKGTGERNVLIFDLGGGTFDVSLLTIEDGIFEVKATAGDTHLGGEDFDNRL 240
T. parva Hsp70     118 EISSMVLTKMKEIAEAFLGKSVKDAVITVPAYFNDSQRQATKDAGTIAGLNVMRIINEPTAAAIAYGLDKKGGEKNVLIFDLGGGTFDVSILTIEDGIFEVKATAGDTHLGGEDFDNLL 237
T. annulata Hsp70  118 EISSMVLTKMKEIAESFLGKSVKDAVITVPAYFNDSQRQATKDAGTIAGLNVMRIINEPTAAAIAYGLDKKGGEKNVLIFDLGGGTFDVSILTIEDGIFEVKATAGDTHLGGEDFDNLL 237
B. divergens Hsp70 118 EISSMVLIKMKEIAELYLGKTVKDAVITVPAYFNDSQRQATKDAGTIAGLNVMRIINEPTAAAIAYGLDKKGSTEKNVLIFDLGGGTFDVSILTIEDGIFEVKATAGDTHLGGEDFDNLL 237

                       ....|....|....|....|....|....|....|....|....|....|....|....|....|....|....|....|....|....|....|....|....|....|....|....|
Consensus          232 VefCvqDFkRkNrgkdistNsRALRRLRTqCERaKRtLSSSTQATIElDSLfEGIDYsvsiSRARFEElCmdyFR~tL~PVEKvLkdsgiDKRsvheVVLVGGSTRIPKiQqlIqeFFNg 349
T. gondii Hsp70    239 VDFCVQDFKRKNRGKDISTNSRALRRLRTQCERTKRTLSSSTQATIEIDSLFEGIDYSVSISRARFEELCMDYFRNSLLPVEKVLKDSGIDKRSVSEVVLVGGSTRIPKIQQLITDFFNG 358
N. caninum Hsp70   239 VDFCVQDFKRKNRGKDISTNSRALRRLRTQCERTKRTLSSSTQATIEIDSLFEGIDYSVSISRARFEELCMDYFRNSLLPVEKVLKDSGIDKRSVSEVVLVGGSTRIPKIQQLITDFFNG 358
E. acervulina Hsp70 239 VDFCVQDFKRKNRSKDPSTNSRALRRLRTQCERAKRTLSSSTQATIEIDSLFEGIDYSVSISRARFEELCMDYFRNSLVPVEKVLKDSGIDKRSVHEVVLVGGSTRIPKIQQLIQEFFNG 358
C. cayetanensis Hsp70 239 VDFCMQDFKRKNRSKDLSNSRALRRLRTQCERAKRTLSSSTQATIEIDSLFEGIDYSVSLSRARFEELCMDYFRSSLVPVEKVLKDAAIDKRSVHEVVLVGGSTRIPKIQQIQEFFND 358
C. parvum Hsp70    241 VEFCVQDFKRKNRGMDLTTNARALRRLRTQCERAKRTLSSSTQATIELDSLYEGIDYSVAISRARFEELCADYFRATIAPVEKVLKDAGMDKRSVHDVVLVGGSTRIPKVQALIQEFFNG 360
C. hominis Hsp70   241 VEFCVQDFKRKNRGMDLTSNARALRRLRTQCERAKRTLSSSTQATIELDSLYEGIDYSVAISRARFEELCADYFRATIAPVEKVLKDAGMDKRSVHDVVLVGGSTRIPKVQALIQEFFNG 360
T. parva Hsp70     238 VEHCVRDFMRLNNGKNISSNKRALRRLRTHCERAKRVLSSSTQATIELDSLYEGIDYNTTISRARFEELCNEKFRSTLVPVEKALESSGLDKRSIHEVVLVGGSTRIPKIQTLIKNFFNG 357
T. annulata Hsp70  238 VEHCVRDFMRLNNGKNISSNKRALRRLRTHCERAKRVLSSSTQATIELDSLYEGIDYNTTISRARFEELCNEKFRSTLVPVEKALESSGLDKRSIHEVVLVGGSTRIPKIQTLIKNFFNG 357
B. divergens Hsp70 238 VEHCVRDFMRMNNGKNIATNKRALRRLRTHCERAKRVLSSSTQATIELDSLFEGIDYNTTISRARFEEMCGEKFRGTLIPVEKALESSGLDKRIHEVVLVGGSTRIPKIQQLIKDFFNG 357
```

Supplementary Data S3B

```
                                    ....|....|....|....|....|....|....|....|....|....|....|....|....|....|....|....|....|....|....|....|....|....|....|....|
Consensus              350 KEPcrsINPDEAVAYGAAVQAAILkG~qssqvQdLLLLDVAPLSLGLETAGGVMTkLIeRNTTIPtKKsQvFTTyaDnQpGVlIQVfeGERAMTKDNnLLGKFHLdGIpPAPRGVPQIEV 468
T. gondii Hsp70        359 KEPCRSINPDEAVAYGAAVQAAILKGVTSSQVQDLLLLDVAPLSLGLETAGGVMTKLIERNTTIPTKKSQTFTTYADNQPGVLIQVYEGERAMTKDNNLLGKFHLDGIPPAPRGVPQIEV 478
N. caninum Hsp70       359 KEPCRSINPDEAVAYGAAVQAAILKGVTSSQVQDLLLLDVAPLSLGLETAGGVMTKLIERNTTIPTKKSQTFTTYADNQPGVLIQVYEGERAMTKDNNLLGKFHLDGIPPAPRGVPQIEV 478
E. acervulina Hsp70    359 KEPCRSINPDEAVAYGAAVQAAILKGVNSQVQDLLLLDVAPLSLGLETAGGVMTKLIERNTTIPTKKSQIFTTYADNQPGVLIQVFEGERAMTKDNNLLGKFHLDGIPPAPRGVPQIEV 478
C. cayetanensis Hsp70  359 KEPCRSINPDEAVAYGAAVQAAILKGVNNSQVQDLLLLDVAPLSLGLETAGGVMTKLIERNTTIPTKKSQVFTTYADNQPGVLIQVFEGERAMTKDNNLLGKFHLDGIPPAPRGVPQIEV 478
C. parvum Hsp70        361 KEPCKAINPDEAVAYGAAVQAAIINGEQSSAVQDLLLLDVAPLSLGLETAGGVMTKLIERNTTIPAKKTQVFTTYADNQSGVLIQVYEGERAMTKDNHLLGKFHLDGIPPAPRGVPQIEV 480
C. hominis Hsp70       361 KEPCKAINPDEAVAYGAAVQAAIINGEQSSAVQDLLLLDVAPLSLGLETAGGVMTKLIERNTTIPAKKTQVFTTYADNQSGVLIQVYEGERAMTKDNHLLGKFHLDGIPPAPRGVPQIEV 480
T. parva Hsp70         358 KEPCRSINPDEAVAYGAAVQAAILSGNQSEKIQELLLLDVAPLSLGLETAGGVMTVLIKRNTTIPTKKNQIFTTNEDRQEGVLIQVKEGERAMTKDNNLLGKFHLTGIAPAPRGVPQIEV 477
T. annulata Hsp70      358 KEPCRSINPDEAVAYGAAVQAAIISGNQSEKIQELLLLDVAPLSLGLETAGGVMTVLIKRNTTIPTKKNQIFTTNEDRQEGVLIQVFEGERAMTKDNNLLGKFHITGIAPAPRGVPQIEV 477
B. divergens Hsp70     358 KEPSRSINPDEAVAYGAAVQAAILSGDQSGKIQELLLLDVAPLSLGLETAGGVMTVLIKRNTTIPTKKTQVFTTNEDRQEGVFIQVFEGERAMTKDNNLLGKFHITGIAPAPRGVPQIEV 477

                                    ....|....|....|....|....|....|....|....|....|....|....|....|....|....|....|....|....|....|....|....|....|....|....|....|
Consensus              469 TFDIDANGIlNVtA~dKsTGKSnqiTITNDKGRLSqseIdRMVaeAEKyKaEDeqnr~rvEakn~LENYcYsMr~Tl~dekvKdKlsa~erd~A~kaIqdaldWlekNqlAekeEfEakq 580
T. gondii Hsp70        479 TFDIDANGIMNVTAQDKSTGKSNQITITNDKGRLSASEIDRMXQEAEKYKAEDEQNKHRVEAKNGLENYCYHMRQTLDDEKLKDKISSSEDRDTANKAIQEFALDWIDKNQLAEKEEFEAKQ 598
N. caninum Hsp70       479 TFDIDANGIMNVTAQDKSTGKSNQITITNDKGRLSASEIDRMVEAEKYKAEDEQNRHRVEAKNGLENYCYHMRQTLDDEKLKDKISSEDRETANKAIQDALDWIDKNQLAEKEEFEAKQ 598
E. acervulina Hsp70    479 TFDIDANGIMNVTATEKNTGKSNQITITNDKGRLSQEEIDRMVAEAEKYKAEDEANKQRVEAKNALENYCYSMRGTMEDEKIKDKISAEDRETASAIQKALDWIDKNQLAEKEEFEAKQ 598
C. cayetanensis Hsp70  479 TFDIDANGIMNVTATEKNTGKSNQITITNDKGRLSQSEIDRMVAEAEKYKAEDDANKQRVESKNALENYCYSMRSTMEDEKIKDKVSDNDREAATSAIQKTLDWIPDKNHLAEKEEFEAKR 598
C. parvum Hsp70        481 TFDIDANGILNVSAVDKSTGKSSKITITNDKGRLSKDDIERMVNDAEKYKGEDEQNRLKIEAKNSLENYLYNMRNTIQEPKVKEKLSQSEIDEAEKKIKDALDWLEHNQTAEKDEFEHQQ 600
C. hominis Hsp70       481 TFDIDANGILNVSAVDKSTGKSSKITITNDKGRLSKDDIERMVNDAEKYKGEDEQNRLKIEAKNSLENYLYNMRNTIQEPKVKEKLSQSEIDEAEKKIKDALDWLEHNQTAEKDEFEHQQ 600
T. parva Hsp70         478 TFDIDANGILNVTAMDKSTGKSEHVTITNDKGRLSQEEIDRMVEEAEKYKEEDEKRRKCVESKHLENYCYSMKNTLSEDQVKQKIGADEVDNALNTITPALKNVETNQLAEHDEFEDKL 597
T. annulata Hsp70      478 TFDIDANGILNVTAMDKSTGKSEHVTITNDKGRLSQEEIDRMVEEAEKYKEEDEKRRCVESRHLENYCYSMKNTLSEDQVKQKIGADEVDSALSTITDALKNVEANQLAEHDEYEDKL 597
B. divergens Hsp70     478 TFDIDANGILNVTAMDKSTGKSEHVTITNDKGRLSTADIERMVAEAEKFKEEDENRRSCVEAKHQLENYCYSMRSTLGDDNVKTKLFAGEIEEALKAIEEAIKWLEGNQTATKEEFEYKL 597

                                    ....|....|....|....|....|....|....|....|....|....|....|....|....|....|....|....|....|....|....|
Consensus              581 KevE~vc~Pl~tKlYqagaaa------------------------ggmpggmpgGmpggmp~~gmggsgGPTVEEVD 628
T. gondii Hsp70        599 KEVESVCTPIITKLYQAGAAA-----GGMPGGM-GGMPGGM-GGMPGGM-GGMPGGM-GGMPGAMGGSGGPTVEEVD 667
N. caninum Hsp70       599 KEVESVCTPIITKLYQAGAAA-GGMPGGMPGGM-GGMPGGM-GGMPGGM-GGMPGGMPGGMG-GMGGSGGPTVEEVD 671
E. acervulina Hsp70    599 KEVEAVCTPIVTKMYQSAAGA---------------------QGGMPGGMP-DMSA-AAGAAGGPTVEEVD 646
C. cayetanensis Hsp70  599 KEVEAVCAPIVTKMYQAAAGS-----------------GGMPGGMPGGMP-DMS--GAGGAGGPTVEEVD 648
C. parvum Hsp70        601 KEIETHMNPIMMKIYSAE----GGMPGGMPGGMPGMPGGMPGGMPGGMPGGMPGGMPGGM-GMPGSNGPTVEEVD 673
C. hominis Hsp70       601 KEIETHMNPIMMKIYSAEGGMPGGMPGGMPGGMPGGMPGGMPGGMPGGMPGGMPGGMPGGMG-GMPGSNGPTVEEVD 677
T. parva Hsp70         598 KHVEGVCNPIVTKLYQSGGAP---------------GAGPDMGACFPGGAP--PPSSSSGPTVEEVD 647
T. annulata Hsp70      598 KHVEGVCNPIVTKLYQSGAAP---------------G-GPDMSGGFPGGAA--PPPQSSGPTVEEVD 646
B. divergens Hsp70     598 KEVEKVCQPLATKMYQAAGAS---------------GMPTGGPGGFGGAAP--GAGTSSGPTVEEVD 647
```

Supplementary Data S3C

```
                                    10        20        30        40        50        60        70        80        90        100       110       120
                                    ....|....|....|....|....|....|....|....|....|....|....|....|....|....|....|....|....|....|....|....|....|....|....|....|
Consensus                  1   MA~~KAsKPNLPESNIAIGIDLGTTYSCVGVWRNENVDIIANDQGNRTTPSYVAFTDTERLIGDAAKNQVARNPENTVFDAKRLIGRKFTESSVQSDMKHWPFTVKSG~~EKPMIEV~YQ  115
P. vivax Hsp70             1   MASGKASKPNLPESNIAIGIDLGTTYSCVGVWRNENVDIIANDQGNRTTPSYVAFTDTERLIGDAAKNQVARNPENTVFDAKRLIGRKFTESSVQSDMKHWPFTVKSGVDEKPMIEVSYQ  120
P. knowlesi Hsp70          1   MASGKAAKPNLPESNIAIGIDLGTTYSCVGVWRNENVDIIANDQGNRTTPSYVAFTDTERLIGDAAKNQVARNPENTVFDAKRLIGRKFTESSVQSDMKHWPFTVKSGVDEKPMIEVSYQ  120
P. yoelii yoelii Hsp70     1   MANAKASKPNLPESNIAIGIDLGTTYSCVGVWRNENVDIIANDQGNRTTPSYVAFTDTERLIGDAAKNQVARNPENTVFDAKRLIGRKFTESSVQSDMKHWPFTVKSGIEEKPMIEVVYQ  120
P. berghei Hsp70           1   MANAKA-KPNLPESNIAIGIDLGTTYSCVGVWRNENVDIIANDQGNRTTPSYVAFTDTERLIGDAAKNQVARNPENTVFDAKRLIGRKFTESSVQSDMKHWPFTVKSGIEEKPMIEVVYQ  119

                                    ....|....|....|....|....|....|....|....|....|....|....|....|....|....|....|....|....|....|....|....|....|....|....|....|
Consensus                116   GEKKLFHPEEISSMVLQKMKENAEAFLGKSIKNAVITVPAYFNDSQRQATKDAGTIAGLNVMRIINEPTAAAIAYGLHKKGKGEKNILIFDLGGGTFDVSLLTIEDGIFEVKATAGDTHL  235
P. vivax Hsp70           121   GEKKLFHPEEISSMVLQKMKENAEAFLGKSIKNAVITVPAYFNDSQRQATKDAGTIAGLNVMRIINEPTAAAIAYGLHKKGKGEKNILIFDLGGGTFDVSLLTIEDGIFEVKATAGDTHL  240
P. knowlesi Hsp70        121   GEKKLFHPEEISSMVLQKMKENAEAFLGKSIKNAVITVPAYFNDSQRQATKDAGTIAGLNVMRIINEPTAAAIAYGLHKKGKGEKNILIFDLGGGTFDVSLLTIEDGIFEVKATAGDTHL  240
P. yoelii yoelii Hsp70   121   GEKKLFHPEEISSMVLQKMKENAEAFLGKSIKNAVITVPAYFNDSQRQATKDAGTIAGLNVMRIINEPTAAAIAYGLHKKGKGEKNILIFDLGGGTFDVSLLTIEDGIFEVKATAGDTHL  240
P. berghei Hsp70         120   GEKKLFHPEEISSMVLQKMKENAEAFLGKSIKNAVITVPAYFNDSQRQATKDAGTIAGLNVMRIINEPTAAAIAYGLHKKGKGEKNILIFDLGGGTFDVSLLTIEDGIFEVKATAGDTHL  239

                                    ....|....|....|....|....|....|....|....|....|....|....|....|....|....|....|....|....|....|....|....|....|....|....|....|
Consensus                236   GGEDFDNRLVNFCVEDFKRKNRGKDLSKNSRALRRLRTQCERAKRTLSSSTQATIEIDSLFEGIDYSVTVSRARFEELCIDYFRDTLIPVEKVLKDAMMDKKSVHEVVLVGGSTRIPKIQ  355
P. vivax Hsp70           241   GGEDFDNRLVNFCVEDFKRKNRGKDLSKNSRALRRLRTQCERAKRTLSSSTQATIEIDSLFEGIDYSVTVSRARFEELCIDYFRDTLIPVEKVLKDAMMDKKSVHEVVLVGGSTRIPKIQ  360
P. knowlesi Hsp70        241   GGEDFDNRLVNFCVEDFKRKNRGKDLSKNSRALRRLRTQCERAKRTLSSSTQATIEIDSLFEGIDYSVTVSRARFEELCIDYFRDTLIPVEKVLKDAMMDKKSVHEVVLVGGSTRIPKIQ  360
P. yoelii yoelii Hsp70   241   GGEDFDNRLVNFCVEDFKRKNRGKDLSKNSRALRRLRTQCERAKRTLSSSTQATIEIDSLFEGIDYSVTVSRARFEELCIDYFRDTLIPVEKVLKDAMMDKKSVHEVVLVGGSTRIPKIQ  360
P. berghei Hsp70         240   GGEDFDNRLVNFCVEDFKRKNRGKDLSKNSRALRRLRTQCERAKRTLSSSTQATIEIDSLFEGIDYSVTVSRARFEELCIDYFRDTLIPVEKVLKDAMMDKKSVHEVVLVGGSTRIPKIQ  359

                                    ....|....|....|....|....|....|....|....|....|....|....|....|....|....|....|....|....|....|....|....|....|....|....|....|
Consensus                356   TLIKEFFNGKEACRSINPDEAVAYGAAVQAAILSGDQSNAVQDLLLLDVCSLSLGLETAGGVMTKLIERNTTIPAKKSQIFTTYADNQPGVLIQVYEGERALTKDNNLLGKFHLDGIPPA  475
P. vivax Hsp70           361   TLIKEFFNGKEACRSINPDEAVAYGAAVQAAILSGDQSNAVQDLLLLDVCSLSLGLETAGGVMTKLIERNTTIPAKKSQIFTTYADNQPGVLIQVYEGERALTKDNNLLGKFHLDGIPPA  480
P. knowlesi Hsp70        361   TLIKEFFNGKEACRSINPDEAVAYGAAVQAAILSGDQSNAVQDLLLLDVCSLSLGLETAGGVMTKLIERNTTIPAKKSQIFTTYADNQPGVLIQVYEGERALTKDNNLLGKFHLDGIPPA  480
P. yoelii yoelii Hsp70   361   TLIKEFFNGKEACRSINPDEAVAYGAAVQAAILSGDQSNAVQDLLLLDVCSLSLGLETAGGVMTKLIERNTTIPAKKSQIFTTYADNQPGVLIQVYEGERALTKDNNLLGKFHLDGIPPA  480
P. berghei Hsp70         360   TLIKEFFNGKEACRSINPDEAVAYGAAVQAAILSGDQSNAVQDLLLLDVCSLSLGLETAGGVMTKLIERNTTIPAKKSQIFTTYADNQPGVLIQVYEGERALTKDNNLLGKFHLDGIPPA  479

                                    ....|....|....|....|....|....|....|....|....|....|....|....|....|....|....|....|....|....|....|....|....|....|....|....|
Consensus                476   PRKVPQIEVTFDIDANGILNVTAVEKSTGKQNHITITNDKGRLSPEEIDRMVNDAEKYKAEDEENKkRIEARNSLENYCYGVKSSLEDQKIKEKLQP~E~ETCMKs~tsILEWLEKNQLA  592
P. vivax Hsp70           481   PRKVPQIEVTFDIDANGILNVTAVEKSTGKQNHITITNDKGRLSPEEIDRMVNDAEKYKAEDEENKKRIEARNSLENYCYGVKSSLEDQKIKEKLQPAEIETCMKCITTILEWLEKNQLA  600
P. knowlesi Hsp70        481   PRKVPQIEVTFDIDANGILNVTAVEKSTGKQNHITITNDKGRLSPEEIDRMVNDAEKYKAEDEENKNRIEARNSLENYCYGVKSSLEDQKIKEKLQPSEIETCMKSISSILEWLEKNQLA  600
P. yoelii yoelii Hsp70   481   PRKVPQIEVTFDIDANGILNVTAVEKSTGKQNHITITNDKGRLSPEEIDRMVNDAEKYKAEDEENKKRIEARNSLENYCYGVKSSLEDQKIKEKLQPNEVETCMKSVTSILEWLEKNQLA  600
P. berghei Hsp70         480   PRKVPQIEVTFDIDANGILNVTAVEKSTGKQNHITITNDKGRLSPEEIDRMVNDAEKYKAEDEENKKRIEARNSLENYCYGVKSSLEDQKIKEKLQPNEVETCMKSVTSILEWLEKNQLA  599

                                    ....|....|....|....|....|....|....|....|....|....|....|....|....|....|....|....|....|....|....|....|....|....|....|....|
Consensus                592   ~K~EYE~KQKEAE~VC~PIMSKIYQDag~Aa~~~~~~~~GGMPGGMPGGMPGGMP~GGMPGGMNF----PGGMPGG~G~P~GAPAGSGPTVEEVD  665
P. vivax Hsp70           601   SKEEYESKQKEAESVCAPIMSKIYQDVGAAGGMPGGMPGGMPGGMPGGMPGGMPGGGMPGGMNF----PGGMPGG-GMPGGAPAGSGPTVEEVD  690
P. knowlesi Hsp70        601   SKEEYESKQKEAESVCAPIMSKIYQDAAGAA--------GGMPGGMPGGMPGGMPGGMPGGMNF----PGGMPGG-GMPGGAPAGSGPTVEEVD  682
P. yoelii yoelii Hsp70   601   GKDEYEAKQKEAEAVCSPIMSKIYQDAGAAA--------GGMPGGMPGGMPGGMP-GGMPGGMNF----PGGMPGGMGAPAGAPAGSGPTVEEVD  682
P. berghei Hsp70         600   GKDEYEAKQKEAEAVCSPIMSKIYQDAGAAG-GMPGGMPGGMPGGMPGGMPGGMP-GGMPGGMNFPGGMPGGMPGGMGAPAGAPAGSGPTVEEVD  692
```

# Supplementary Data S3D

```
                                                    10        20        30        40        50        60        70        80        90       100       110       120
                                           ....|....|....|....|....|....|....|....|....|....|....|....|....|....|....|....|....|....|....|....|....|....|....|....|....
Consensus                      1           M------k~~~~aa---------~~~~ll---------av-----aae~~~~d--kked~GtviGIDLGTTYSCVgvfKNGrveIiaNdQGNRITPSyVaFt~d-geRLIGdaAKNQltsN    75
Human Grp78                    1           M------KLSLVAA-------MLLLLS----------AA-----RAEE---ED--KKEDVGTVVGIDLGTTYSCVGVFKNGRVEIIANDQGNRITPSYVAFTPE-GERLIGDAAKNQLTSN    87
R. norvegicus Grp78            1           M------KFTVVAA-------ALLLLC----------AV-----RAEE---ED--KKEDVGTVVGIDLGTTYSCVGVFKNGRVEIIANDQGNRITPSYVAFTPE-GERLIGDAAKNQLTSN    87
X. laevis Hspa5                1           M------VTMKLFA-------LVLLVS----------AS---VFAADD---DD--KKEDVGTVVGIDLGTTYSCVGVFKNGRVEIIANDQGNRITPSYVAFTPE-GERLIGDAAKNQLTSN    89
O. latipes Bip                 1           M------K--LLWA-------VMLVVG----------AV-----FAEE---ED--KKDSVGTVIGIDLGTTYSCVGVFKNGRVEIIANDQGNRITPSYVAFTSE-GERLIGDAAKNQLTSN    85
A. echinatior Hsc70-3          1           M------KGVISL--------LIVGLL----------AV-----ATIA---KEDKEKEDIGTVIGIDLGTTYSCVGVFKNGRVEIIANDQGNRITPSYVAFTND-GERLIGDAAKNQLTTN    88
F. chinensis Grp78             1           M------RCWTALA-------AIAAVM----------AV-----AATA---KD--KEEVGTVIGIDLGTTYSCVGVFKNGRVEIIANDQGNRITPSYVAFTAG-GERLIGDSAKNQLTSN    88
A. californica Grp78           1           M------DRFTPFF-------LLVILF----------SSNLLVRADGDEEDEGDKKSEVGTVIGIDLGTTYSCVGVFKNGRVDIIANDQGNRITPSYVAFTAD-GERLIGDAAKNQLTSN    97
C. elegans Hsp-3               1           M------KTLFL---------LGLIAL----------SA---VSVYCEEEEKTEKKTYGTIIGIDLGTTYSCVGVYKNGRVEIIANDQGNRITPSYVAFSGDQGDRLIGDAAKNQLTIN    93
C. owczarzaki Hsp70-Like       1           M------RVAYLAS-------LALVAL----------TL---ALVNAD-----ETKQREYGTVIGIDLGTTYSCVGVFKNGRVEIIPDQGNRITPSYVAFTDD-GERLIGDAAKNQLTAN    89
A. thaliana BiP-2              1           M------ARSFGANSTVVLAIIFFGCL----------FA-----FSTA---KE--EATKLGSVIGIDLGTTYSCVGVYKNGHVEIIANDQGNRITPSWVGFTDS--ERLIGEAAKNQAAVN    93
C. cohnii BiP                  1           M------ARLHWAG-------IFVLAL----------AC-----VAVA--KDDEKKID-GPVIGIDLGTTYSCVGIYKNGRVEIIPNDQGNRITPSYVAFTED--ERLIGEAAKNQATIN    87
S. cerevisiae Kar2p            1           MFFNRLSAGKLLVPLSVVLYALFVVILPLQNSFHSSNV-----LVRG---AD--DVENYGTVIGIDLGTTYSCVAVMKNGKTEILANEQGNRITPSYVAFTDD--ERLIGDAAKNQVAAN   108

                                                   ....|....|....|....|....|....|....|....|....|....|....|....|....|....|....|....|....|....|....|....|....|....|....|....|
Consensus                     76           PenTvFDaKRLIGrtw~dksvQqDikfl Pfkvv~kn~KPyiqv~ig~ge~kvfapEEiSaMvLtKMketAEaYLGkkvthAVvTVPAYFNDaQRQaTKDaGtIaGLnVmRIiNEPTAAAI   189
Human Grp78                   88           PENTVFDAKRLIGRTWNDPSVQQDIKFLPFKVVEKKTKPYIQVDIGGGQTKTFAPEEISAMVLTKMKETAEAYLGKKVTHAVVTVPAYFNDAQRQATKDAGTIAGLNVMRIINEPTAAAI   207
R. norvegicus Grp78           88           PENTVFDAKRLIGRTWNDPSVQQDIKFLPFKVVEKKTKPYIQVDIGGGQTKTFAPEEISAMVLTKMKETAEAYLGKKVTHAVVTVPAYFNDAQRQATKDAGTIAGLNVMRIINEPTAAAI   207
X. laevis Hspa5               90           PENTVFDAKRLIGRTWNDPSVQQDIKYLPFKVIEKKTKPYIQVNIG-DQMKTFAPEEISAMVLTKMKETAEAYLGKKVTHAVVTVPAYFNDAQRQATKDAGTIAGLNVMRIINEPTAAAI   208
O. latipes Bip                86           PENTVFDAKRLIGRSWGDSSVQQDIKFYFPFKVIEKKSKPHICIDIGGGQMKTFAPEEISAMVLTKMKETAEAYLGKKVTHAVVTVPAYFNDAQRQATKDAGTIAGLNVMRIINEPTAAAI   205
A. echinatior Hsc70-3         89           PENTVFDAKRLIGREWSDPTVQHDVKFFPEPVIEKNNKPHIKVETSQGE-KVFAPEEISAMVLGKMKETAEAYLGKKVTHAVVTVPAYFNDAQRQATKDAGTISGLVVMRIINEPTAAAI   207
F. chinensis Grp78            88           PENTVFDAKRLIGREWTDKSVQHDIQFFPKVINKNDKPHIKVATVQGE-KVFAAEEISAMVLGKMKETAEAYLGKPVTHAVVTVPAYFNDAQRQATKDAGIIAGITVMRIINEPTAAAI   206
A. californica Grp78          98           PENTIFDVKRLIGRTFDDSVQHDIKFYPPFKVTNANNKPHIQAATGEGD-RSFAPEEISAMVLSKMRDIAEEYLGKKITNAVVTVPAYFNDAQRQATKDAGTIAGLNVMRIINEPTAAAI   216
C. elegans Hsp-3              94           PENTIFDAKRLIGRDYNDKTVQADIKHWPFKDSKNKPSVEVKVGSDN-KQFTPEEVSAMVLKMKEIAESYLGKEVKNAVVTVPAYFNDAQRQATKDAGTIAGLNVVRIINEPTAAAI   212
C. owczarzaki Hsp70-Like      90           PFNTIFDAKRLIGREFNEKSIQADIKLWPFKVVNKNAKPYIKVATSAGD-KVLSPEEVSAMILLKMKETAEAYLGHKVTHAVVTVPAYFNDAQRQATKDAGTIAGLTVLRIINEPTAAAI   208
A. thaliana BiP-2             94           PERTVFDVKRLIGRKFEDKEVQKDRKLVPYQIVNKDGKPYIQVKIKDGETKVISPEEISAMILTKMKETAEAYLGKKIKDAVVTVPAYFNDAQRQATKDAGVIAGLNVARIINEPTAAAI   213
C. cohnii BiP                 88           PAQTLFDVKRLIGRRFKDSTVQKDIKLLPFDIVDKNGKPQIISVKVK--GETKQMAPEEVSSMVLTKMKETAENYLGKEVHAVITVPAYFNDQQRQSTKDAGVIAGLRIINEPTAAAI   206
S. cerevisiae Kar2p          109           PQNTIFDIKRLIGLKYNDRSVQKDIKHLPENVVNKDGKPKAVEVSVK-GEKKVFTPEEISGMILGKMKQIAEDYLGTKVHAVVTVPAYFNDAQRQATKDAGTIAGLNVLRIVNEPTAAAI   227

                                                   ....|....|....|....|....|....|....|....|....|....|....|....|....|....|....|....|....|....|....|....|....|....|....|....|
Consensus                    190           AYGlDKkdgEknilVfDLGGGTFDVSlLtIdnGVFEVvaTnGDTHLGGeDFDqrvmehfiklykkK~gkD~rkdnrAvqKLrREvEkAKRaLSsqhqariEIesff~GeDfSEtLTRakF   306
Human Grp78                  208           AYGLDKREGEKNILVFDLGGGTFDVSLLTIDNGVFEVVATNGDTHLGGEDFDQRVMEHFIKLYKKKTGKDVRKDNRAVQKLRREVEKAKRALSSQHQARIEIESFYEGEDFSETLTRAKF   327
R. norvegicus Grp78          208           AYGLDKREGEKNILVFDLGGGTFDVSLLTIDNGVFEVVATNGDTHLGGEDFDQRVMEHFIKLYKKKTGKDVRKDNRAVQKLRREVEKAKRALSSQHQARIEIESFFEGEDFSETLTRAKF   327
X. laevis Hspa5              209           AYGLDKREGEKNILVFDLGGGTFDVSLLTIDNGVFEVVATNGDTHLGGEDFDQRVMEHFIKLYKKKSGKDVRKDNRAVQKLRREVEKAKRALSAQHQSRIEIESFSEGEDFSETLTRAKF   328
O. latipes Bip               206           AYGLDKKDGEKNILVFDLGGGTFDVSLLTIDNGVFEVVATNGDTHLGGEDFDQRVMEHFIKLYKKKTGKDVRKDNRAVQKLRREVEKAKRALSAQHQARIEIESFFEGEDFSETLTRAKF   325
A. echinatior Hsc70-3        208           AYGLDKKDGEKNVLVFDLGGGTFDVSLLTIDNGVFEVVATNGDTHLGGEDFDQRVMDHFIKLYKKKGKDIRKDNRAVQKLRREVEKAKRALSASHQVRIEIESFFEGEDFSETLTRAKF   327
F. chinensis Grp78           207           AYGIDKKDGEKNILVFDLGGGTFDVSLLTIDSGVFEVVATNGDTHLGGEDFDQRVMDHFIKLYKKKGKDIRRDNRAVQKLRREVEKAKRSLSSSHQVRIEIESFFEGDDFSETLTRAKF   326
A. californica Grp78         217           AYGLDKKEGEKNILVFDLGGGTFDVSLLTIDNGVFEVVSTNGDTHLGGEDFDQRVMEHFIKLYKKKGKDIRKDNRAVQKLRREVEKAKRALSSAHQVRLEIESFFEGEDFSESLTRAKF   336
C. elegans Hsp-3             213           AYGLDKKDGERNILVFDLGGGTFDVSMLTIDNGVFEVLATNGDTHLGGEDFDQRVMEYFIKLYKKKSGKDLRKDRAVQKLRREVEKAKRALSTQHQTKVEIESLFDGEDFSETLTRAKF   332
C. owczarzaki Hsp70-Like     209           AYGLDKKEGEKNILVFDLGGGTFDVSLLTIDNGVFEVVATNGDTHLGGQDFDQRIVMEHFMRQYKEKGKDISSSDVRKARAVQKLRREAEAKRALSSQHSARLEIEAFFDGEDFSETLTRAKF   328
A. thaliana BiP-2            214           AYGLDKKGGEKNILVFDLGGGTFDVSVLTIDNGVFEVLSTNGDTHLGGEDFDHRIMEYFIKLIKKHGKDISKDNKALGKLRREECERAKRALSSQHQVRVEIESLSDGVDLSEPLTRAFE   333
C. cohnii BiP                207           AYGLDKK-AEKNILVYDLGGGTFDVSLLTIDNGVFEVVATNGDTHLGGEDFDQVHFMKVFQKKHSKDMSKDRAIQKLRERAKRALSSQHQARLEIEALFDGVDFSETLTRARF   325
S. cerevisiae Kar2p          228           AYGLDKSDKEHQIIVYDLGGGTFDVSLLSIENGVFEVQATSGDTHLGGEDFDYKIVRQLIKAFKKKHGIDVSDNNKALAKIKREAEKAKRALSSQMSTRIEIDSFVDGCIDLSETLTRAKF   347
```

Supplementary Data S3D

```
                               ....|....|....|....|....|....|....|....|....|....|....|....|....|....|....|....|....|....|....|....|....|....|....|....|
Consensus                  307 EElNmDLFrsTmkPVqkvleDsdlkksdideIVLVGGSTRIPKvQQLvkefFnGKepsrGiNPDEAVAYGAAVQaGvlsGe--edtgdlvLLDV~pLTlGIETvGGVMTkLIpRNtviPT 423
Human Grp78                328 EELNMDLFRSTMKPVQKVLEDSDLKKSDIDEIVLVGGSTRIPKIQQLVKEFFNGKEPSRGINPDEAVAYGAAVQAGVLSGD--QDTGDLVLLDVCPLTLGIETVGGVMTKLIPRNTVVPT 445
R. norvegicus Grp78        328 EELNMDLFRSTMKPVQKVLEDSDLKKSDIDEIVLVGGSTRIPKIQQLVKEFFNGKEPSRGINPDEAVAYGAAVQAGVLSGD--QDTGDLVLLDVCPLTLGIETVGGVMTKLIPRNTVVPT 445
X. laevis Hspa5            329 EELNMDLFRSTMKPVQKVLDDADLKKSDIDEIVLVGGSTRIPKIQQLVKEFFNGKEPSRGINPDEAVAYGAAVQAGVLSGD--QDTGDLVLLDVCPLTLGIETVGGVMTKLIPRNTVVPT 446
O. latipes Bip             326 EELNMDLFRSTMKPVQKVLEDSDLKKSEIDEIVLVGGSTRIPKIQQLVKEFFNGKEPSRGINPDEAVAYGAAVQAGVLSGE--EDTGDVVLLDVCPLTLGIETVGGVMTKLIPRNTVVPT 443
A. echinatior Hsc70-3      328 EELNMDLFRSTLKPVQKVLEDSDMSKKDVDEIVLVGGSTRIPKVQQLVKEFFGKEPSRGINPDEAVAYGAAVQAGVLSGE--QDTDAIVLLDVNPLTMGIETVGGVMTKLIPRNTVIPT 445
F. chinensis Grp78         327 EELNMDLFRSTMKPVQKVLEDSDLQKSDIDEIVLVGGSTRIPKIQQLVKEFFGKEPSRGINPDEAVAYGAAVQAGVLSGE--DDTNDLVLLDVNPLTLGIETVGGVMTKLIPRNTVIPT 444
A. californica Grp78       337 EELNMDLFRSTMKPVKQVLEDADLKTDDIDEIVLVGGSTRIPKVQQLVKEYFNGKEPSRGINPDEAVAYGAAVQAGVLSGE--EDTGDLVLLDVNPLTMGIETVGGVMTKLIPRNTVIPT 454
C. elegans Hsp-3           333 EELNMDLFRATLKPVQKVLEDSDLKKDDVHEIVLVGGSTRVPQIIKEFFNGKEPSRGINPDEAVAYGAAVQGVISGE--EDTGEIVLLDVNPLTMGIETVGGVMTKLIGRNTVIPT 450
C. owczarzaki Hsp70-Like   329 EELNADLFRSTLAPVKKVLEDSGIRKEEDEIVLVGGSTRVPQQLVKDFFNGKEPNRGINPDEAVAYGAAVQAGVLGGE--ESTGDLVLLDVNPLTGIETVGGVMTKLIPRNSVIPT 446
A. thaliana BiP-2          334 EELNMDLFRKTMGPVKAMDDAGLQKSQIDEIVLVGGSTRIPKVQQILKDFFGKEPNKGVNPDEAVAYGAAVQGILSGEGGDTKDILLLDVAPLTLGIETVGGVMTKLIPRNTVIPT 453
C. cohnii BiP              326 EEINNDLFKNTLGPVKQVIEDSGLKKTQIDEIVLVGGSTRVPQQLIKDFFNGKEPNRGINPDEAVAYGAAVQAGILSGE--GGQ-DILLLDVTPLTLGIETVGGVMTKLINRNTVIPT 442
S. cerevisiae Kar2p        348 EELNLDLFKKTLKPVEKVLQDSGLEKKDVDDIVLVGGSTRIPKVQQLLESYFDGKKASKGINPDEAVAYGAAVQAGVLSGE--EGVEDIVLLDVNALTLGIETTGGVMTPLIKRNTAIPT 465

                               ....|....|....|....|....|....|....|....|....|....|....|....|....|....|....|....|....|....|....|....|....|....|
Consensus                  424 KKSQiFsTasDnQptVtIqVyEGERp~tKDnhlLGkFdLtGiPPApRGvPQIEVTFeiDvNGIL~VtAedKgtGnk~kItItNdqnRLtpeeIerMvndAEkfAdeDkklKervdarNel 540
Human Grp78                446 KKSQIFSTASDNQPTVTIKVYEGERPLTKDNHLLGTFDLTGIPPAPRGVPQIEVTFEIDVNGILRVTAEDKGTGNKNKITITNDQNRLTPEEIERMVNDAEKFAEEDKKLKERIDTRNEL 565
R. norvegicus Grp78        446 KKSQIFSTASDNQPTVTIKVYEGERPLTKDNHLLGTFDLTGIPPAPRGVPQIEVTFEIDVNGILRVTAEDKGTGNKNKITITNDQNRLTPEEIERMVNDAEKFAEEDKKLKERIDTRNEL 565
X. laevis Hspa5            447 KKSQIFSTASDNQPTVTIKVYEGERPLTKDNHLLGTFDLTGIPPAPRGVPQIEVTFEIDVNGILRVTAEDKGTGNKNKITITNDQNRLTPEEIERMVTDAEKFAEEDKKLKERIDTRNEL 566
O. latipes Bip             444 KKSQIFSTASDNQPTVTIKVYEGERPLTKDNHLLGTFDLTGIPPAPRGVPQIEVTFEIDVNGILRVTAEDKGTGNKNKITITNDQNRLTPEDIERMVNDAERFADEDKRLKERIDARNEL 563
A. echinatior Hsc70-3      446 KKSQIFSTASDSQHTVTIQVEGERPMTKDNHLLGKFDLTGIPPAPRGIPQIEVTFEIDANGILQVSAEDKGTGNREKIVITNDQNRLTPDDIERMIKDAEKFADDDKKLKERVEARNDL 565
F. chinensis Grp78         445 KKSQIFSTADNHTVTIQVFEGERPMTKDNILLGKFDLTGIPPAPRGVPQIEVTFEIDANGILSVSAEGKGTGNSPKITITNDQNRLTPEDIERMIKDAEVFADEDKKLKERSRNEL 564
A. californica Grp78       455 KKSQIFSTAADNQPTVTIQVFEGERPSMTKDNHLLGKFDLTGIPPAPRGVPQIEVTFEIDVNGILRKTDKGTGSKNQIVIQNDQNRLSPEDIERMINDAEKVADEDKKVKEKVDARNEL 574
C. elegans Hsp-3           451 KKSQVFSTAADNQPTVTIQVFEGERPMTKDNHQLGKFDLTGLPPAPRGVPQIEVTFEIDVNGILHVTAEDKGTGNKNKITITNDQNRLSPEDIERMINDAEKFAEDKKVKDKAEARNEL 570
C. owczarzaki Hsp70-Like   447 KKSQIFSTAADNQPTVTIQVFEGERSLTKDCHTLGKFDLTGIPPAARGVPQIEVTFEIDANGILRVSAEDKGTGKSEKITITNDQNRLTPEEIERMVNDAERFADQDKKVKERVDAKNEL 566
A. thaliana BiP-2          454 KKSQVFTTYQDQQTTVSIQVFEGERSLTKDCRLLGKFDLTGVPPAPRGTPQIEVTFEVDANGILNVKAEDKASGKSEKITITNEKGRLSQEEIDRMVKEAEEFAEEDKKVBEKIDARNAL 573
C. cohnii BiP              443 KKSQIFSTYQDNQPAVNIQVFEGERPMTKDNHLLGKFELGGIPPAPRGQPQIEVTFEIDSNGILNVCAEDKATGKGEKITITNDKGRLTEEQIEKMIKEAEQFADEDKKVKERVDAKNSF 562
S. cerevisiae Kar2p        466 KKSQIFSTAVDNQPTVMIKVYEGERAMSKDNNLLGKFELTGIPPAPRGVPQIEVTFALDANGILKVSATDKGTGKSESITITNDKGRLTQEEIDRMVEEAEKFASEDASIKAKVEARNKL 585

                               ....|....|....|....|....|....|....|....|....|....|....|....|....|....|....|....|....|....|
Consensus                  541 esYayslknqi----~dkekLggKlseedKetie~aveekieWle-snqe-a~aEdfkekkkele~ivqPI-sklYgga-gggpp----p~~~e~~e------ekDEl 621
Human Grp78                566 ESYAYSLKNQI---GDKEKLGGKLSSEDKETMEKAVEEKIEWLE-SHQD-ADIEDFKAKKKELE~IVQPIISKLYSA---GPP----PTGEEDTA------EKDEL 654
R. norvegicus Grp78        566 ESYAYSLKNQI---GDKEKLGGKLSSEDKETMEKAVEEKIEWLE-SHQD-ADIEDFKAKKKELE~IVQPIISKLYSG---GPP----PTGEEDTS------EKDEL 654
X. laevis Hspa5            567 ESYAYSLKNQI---GDKEKLGGKLSSEDKETIEKAVEEKIEWLE-SHQD-ADIEDFKAKKKELE~IVQPIVGKLYGA-G-APP----PEGAE-ET------EKDEL 655
O. latipes Bip             564 ESYAYSLKNQI---GDKEKLGGKLSDDDKETIEKAVEEKIEWME-SHQD-AETEDFQAKKKELE~EVQPITKLYSA-G-GPP----PEGAESEA------EKDEL 653
A. echinatior Hsc70-3      566 ESYAYSLKNQL----ADKEKLGSKVSEADKATMEEAIDEKIWLE-DNQD-TEPEEYKKQKKELTDIVQPIIAKLYGA-GGVP----PTGGEDD------MRDEL 656
F. chinensis Grp78         565 ESYAYSLKNQI---NDKEKLGSKLSDEDKKMDEVIEEKIWLE-DNPE-ADEEDYKTQKKELE~VQPIITKLYQQS-GEAPP----PTEDEENY------EKDEL 655
A. californica Grp78       575 ESYAYSLKNQI---GDKEKLGKAKLSDEDKKITEAVEDEKIEWLE-SNAE-AESEAPENEKKTELEG~IVQPIMTKLYEQS-GCAPP----PSGEEESE------EAEKDEI 667
C. elegans Hsp-3           571 ESYAYNLKNQI----EDKEKLGGKLDEDDKKTIEEAVEEAISLG-SNAE-ASAELKEQKKDLESKVQPIVSKLYDA-GACGE----EAPEGSD------DKDEI 661
C. owczarzaki Hsp70-Like   567 ESYAYNLRNQI----KDEDKLGGKLDEDDKETITAVDETIAWLDQSGAE-ATTEELKEKKAAFEKIVQPIVSKLYQGG-AGGAP--------SGDDD------GHDEF 655
A. thaliana BiP-2          574 ETYVYNMKNQV---SDKDKTADKIEDEKEKIEAATKEALEWID-ENQN-SEKEEYDEKLKEVEAVCNPIITAVYQRS-GGAPGAGGGESSTEEEDE------SHDEL 668
C. cohnii BiP              563 DGYIHSMRSATEGSGDNKGISEKMDEKEKILEALKDGQSWLD-SNPE-GDAEDIKEKHKEVEGICAPIVSKYYGVG-GGAG----AADEDEED------AHDEL 656
S. cerevisiae Kar2p        586 ENYAHSLKNQV-----NGDLGEKLEEDDKETLLDAANDVLEWLD-DNFETAIAEDFEDKFESLSKVAYPITSKLYGADGSGAA----DYDDEDDDGDYFEHDEL 682
```

Supplementary Data S3E

```
                          10        20        30        40        50        60        70        80        90       100       110       120
                 ....|....|....|....|....|....|....|....|....|....|....|....|....|....|....|....|....|....|....|....|....|....|....|....|
Consensus        1 m~~~k~~~l~~~l~---------------------------~~~lsv~~~~p~av~~a~~~~~~~vk~~VIGIDLGTTYSCVG~YrnGRV~IIpNdqGNRITPSYVsFtddeRki 57
T. gondii BiP    1 MTAAKKLSLFSLAA---------------------------------LFCLLSVATLRPVAASDAE---EGKVKDVVIGIDLGTTYSCVGVYRHGRVDIIPNDQGNRITPSYVAFTDDDRKI 86
N. caninum ER-Hsp70 1 MAVAHK--ILSLLA---------------------------------LFCLVGVPTLRPVAADEAE---EGQVKDVVIGIDLGTTYSCVGVYRHGRVDIIPNDQGNRITPSYVAFTDDDRKI 84
E. tenella BiP   1 LDLNKLWYLFGLSASKHTQPGDPSARPHVLYTMGVGFHGAAALPFCFFLSLFSHYPHAVRGDDTEGDGKVKDVVIGIDLGTTYSCVGVYRQGRVDIIPNDQGNRITPSYVSFAEDERKI 120
C. parvum ER-Hsp70 1 MKYLKAIIFSVF-----------------------------------LSLFLAFP--VRSSD---EKKYDGPVIGIDLGTTYSCVGIYKNGRVEIIPNEQGNRITPSYVSFTDDERLI 78
B. rodhaini Grp78 1 --------------------------------------------------------------VEGPVIGIDLGTTYSCVGIYRNGRVEIITNDLGNRITPSYVSFIGGERKV 50
B. microti Grp78 1 -----------------------------------------------VFLFLGFLVEGKT---TPSVEGPVIGIDLGTTYSCVGIYRNGRVEIITNDVGNRITPSYVSFIGGERKV 66

Consensus       58 GeaAKneAtinptnTlFDvKRLIGRrFn~~eVQ~DK~LLPYEIinKDgkPYIrvmvkG~~kelAPEEvSAmVL~KMKe~AE~fLGKeVknAVvTVPAYFNDAQRQATKDAGaIAGLNViR 168
T. gondii BiP    87 GEAAKNEATINPTNTLFDVKRLIGRRFNEKEVQKDKLLPYEIINKDGKPYIRVMVKGEPKVLAPEEVSAMVLTKMKETAEQFLGKEVNAVVTVPAYFNDAQRQATKDAGAIAGLNVIR 206
N. caninum ER-Hsp70 85 GEAAKNEATINPTNTLFDVKRLIGRRFNEKEVQKDKCLLPYEIINKDGKPYIRVMVKGQPKVLAPEEVSAMVLTKMKETAEQFLGKEVNAVVTVPAYFNDAQRQATKDAGAIAGLNVIR 204
E. tenella BiP  121 GEAAKNEAAINPTNTIFDVKRLIGRRFNEKEVQRDKELLPYEIINKDGKPYIRVMVKGQPKELAPEEVSAMVLGKMKEVAESYLGKEVNAVVTVPAYFNDAQRQATKDAGAIAGLNVIR 240
C. parvum ER-Hsp70 79 GESAKNQATINPVQTLFDVKRLIGKRRFKDDSVQKDKTLLPYEIINKDSKPYIQVSVKGEKQCLAPEEVSAMVLVKMKEIAEAYLGKEVKHAVITVPAYFNDAQRQATKDAGAIAGLNVIR 198
B. rodhaini Grp78 51 GDVAKSIATTHSKNTVFDAKRLIGRKFQDAEVQRDKKLLPYEIFDKDGKPYIRIEDEGKHEFAPEEISALVLSKMKTMAENFLGQPVTNAVVTVPAYFNDAQRQATKDAGTIAGLNVLR 170
B. microti Grp78 67 GDVAKSIATSNSKNTVFDAKRLIGKRFRDEEVQRDKKLLPYEIFDKDGRPYIRINDKGNKSFFAPEEISALVLSKMKSMAENFLGQTVKNAVVTVPAYFNDAQRQATKDAGQIAGLNVLR 186

Consensus      169 IiNEPTAAAIAyGLDkK~~EKtILVYDLGGGTFDVS~L~IDNGVFEV~ATSGDTHLGGEDFDqrVMdHFiki~k~kygkdl~sDk~alQklrreve~aKralss~~hq~kvEienLm~~vd 274
T. gondii BiP   207 IINEPTAAAIAYGLDKK-NEKTILVYDLGGGTFDVSLVIDNGVFEVLATSGDTHLGGEDFDQRVMDHFIKLVKKKYDKDLRTDKRGLQKLRREVERAKRALSSQHQAKVENLMEGVD 325
N. caninum ER-Hsp70 205 IINEPTAAAIAYGLDKK-NEKTILVYDLGGGTFDVSLVIDNGVFEVIATAGDTHLGGEDFDQRVMDHFIKVVKKKYDKDLIRTDKRALQKLRREVERAKRALSSQHQAKVEIENLMDGVD 323
E. tenella BiP  241 IINEPTAAAIAYGLDKK-DEKTILVYDLGGGTFDVSLVIDNGVFEVHATSGDTHLGGEDFDQRVMDHFLKIIQKKYNKDIRKDKSALQRLRREVERAKRALSSSHQVTVEVEGLVDGED 359
C. parvum ER-Hsp70 199 IINEPTAAAIAGLDKK-AEKSILVYDLGGGTFDVSLLTIDNGVFEVSATSGDTHLGGEDFDQRVMDHFIKIIKSLTGKDVKSDKRALQKLRREVEKSKRALSSAPQVKVEIEGLMEDVD 317
B. rodhaini Grp78 171 ILNEPTAAAIAYGLDNKDEEKNILVYDLGGGTFDVSLLTIDNGVFEVATSGDTHLGGEDFDRNVMMHFINITKSQHGIDISSSDVHAVQKLRKEVEQCKRKLSSAHSCKLEIDNLIGDFD 290
B. microti Grp78 187 ILNEPTAAAIAYGLDNKDDEKNILVYDLGGGTFDVSLLTIDNGVFEVATSGDTHLGGEDFDRNVMDHFINLTKSQHGVNIASDVQAVQSSVGRLTNVK-KVIICHSCKIEIENLIGDFN 305

Consensus      275 fseTLTRAKFeeLNaDLFqkTlkPVKqVLedadl~KSq~deIVLVGGSTRIPKIQqLIKdFFnGkEPNrGINPDEAVAYGAaVQaGIL~GEg~qdmvLLDVTPLtLGIET~GGVM~KiIn 388
T. gondii BiP   326 FSETLTRAKFEELNSDLFQKTLKPVKQVLEDADLVKSQVDEIVLVGGSTRIPKIQQLIKDFFNGKEPNRGINPDEAVAYGAAVQAGILSGEGAQDMVLLDVTPLTLGIETAGGVMAKIIN 445
N. caninum ER-Hsp70 324 FSETLTRAKFEELNADLFQNTLKPVKQVLEDADVQKSQVDEIVLVGGSTRIPKIQQLIKDFFNGKEPNRGINPDEAVAYGAAVQAGILSGEGAQDMVLLDVTPLTLGIETAGGVMAKIIN 443
E. tenella BiP  360 FSETLTRAKFEELNADLFQKTLKPVKQVLEDADLKKSQCIDEIVLVGGSTRIPKIQQLIKEFFNGKEPNRGINPDEAVAYGAAVQAGILSGEGTQDMVLLDVTPLTLGIETAGGVMAKVIN 479
C. parvum ER-Hsp70 318 ISETLTRAKFEDLNADLFRKTIEPVKVKQLEDADGIKKSEVDEIVLVGGSTRIPKIQALIKEFDQGKEPNRGINPDEAVAYGAAVQAGEGGSDLLLDVTPLTLGIETVGGVMTKLIG 437
B. rodhaini Grp78 291 FSQTLTRAKFENLNEDLFLSTIEPVKDVLTQSGLSKSDIHDIVLVGGSTRIPKIQQLIKDFFNGKEPNKGINPDEAVAYGASVQGGIIAGESHDEVVLLDVTPLSLGIETVGGVMTKIIP 410
B. microti Grp78 306 ENQTLTRAKFENLNEDLFKSTIDPVKDVLNQAKLEKSDIHDIVLVGGSTRIPKIQQLIKDFFNGKEPNKGINPDEAVAYGAAVQGGIIAGESHDELVLLDVTPLSLGIETVGGVMTKIIP 425

Consensus      388 ~NtViPTKK~~QtFSTYsDNQsaVlIQVyeGERpMTk~NhlLGkFeLtGIpPAPRgvPQIeVTF~~D~nGIlsVSAvDKGtGkSeKITITndkGRL~peeIErMIsEakEfAeEDkk~kEr 500
T. gondii BiP   446 KNTVIPTKKTQTFSTYSDNQSAVLIQVYEGERPMTKENHLLGKFELTGIPPAPRGVPQIEVTFDVDRNGILSVSAVDKGTGKSEKITITNDKGRLTPEEIERMISEAEKFAEEDKKVKER 565
N. caninum ER-Hsp70 444 KNTVIPTKKTQTFSTYSDNQSAVLIQVYEGERPMTKNNHLLGKFELTGIPPAPRGVPQIEVTFDVDRNGILSVSAVDKGTGKSEKITITNDKGRLTPEEIERMISEAEKFAEEDKKVKER 563
E. tenella BiP  480 KNTVIPTKKTQTFSTYSDNQSAVLIQVYEGERPMTKENHLLGKFELTGIPPAPRGVPQIDVTFDVDRNGIINVSAVDKGTGKSEKITITNDKGRLTPDEIERMIQEAERFADEDRKTKER 599
C. parvum ER-Hsp70 438 RNTVVPTKKSQVFSTYCDNQPAVLIQVYEGERPMTKDNNLLGKFELSGIPPAPRGVPQIEVTFEIDTDGIILQVSAKDKGTGKSEKITITNDKGRLSQEDIERMIKEAEQFAEEDKLVREK 557
B. rodhaini Grp78 411 RNSVIPTKKSQMFSTYMDNQTTVSIQVEQGERSMTKDNTPLGKEDLTGIAPAPRNTPQINVTFEIDTNGIVSVSAEDKGSKSQKITITAEQGRLSKEEIERMILESERFAQEDKEMMER 530
B. microti Grp78 426 RNTVVPTKKSQMFSTYMDNQTTVTIQVYQGERSMTQHNTLGRFBDLTGIAPAPRNTPQINVTFEIDTNGIVSVSAEDKGSKSQKITITAEQGRLSKAEIEKMIESEKYAQEDKELMER 545

Consensus      501 VdA~NaL~gYlhSM~~tvEDKDKlAdKieedDKktIldkl~~AeewL~~NP~adaee~~dKLKdvE~vCnPIIskvYgq~~~~aggAag~a~dDdYgghDEL 584
T. gondii BiP   566 VDARNALEGYLHSMKTTVEDKDKLADKIEEDDKKTILDKVTEAQEWLNTNPDADAEEETRDKLKDVEAVCNPIISKVYQSGGPGAGGAAGGADDDDYGHDEL 668
N. caninum ER-Hsp70 564 VDARNALEGYLHSMKTTVEDKDKLADKIEEDDKKTILDKVTEAQEWLNTNPDADAEETRDKLKDVEAVCNPIISKVYQTGGPGAGGAAGGADDDDYGHDEL 666
E. tenella BiP  600 VDARNALEGYLHSMRSTVEDKDKLADKIEDDDKKTIMDKITEANEWLVANPEADGEBLRDKLKDVESVCNPIISKVYQTGAPSDSGAT-STSDDDYSSHDEL 701
C. parvum ER-Hsp70 558 VDAKNALDSYVHSMRMSIEDKDKLAQKLEEEDKEKIKEALKDAEDFLSSNPDADAOEKDFVEGICNPIIAAVYGQ-----AGGAAGHAGGGDDYSGHDEL 655
B. rodhaini Grp78 531 VEAKNTDNYIASMRRTVEDKDKMADKIAQDDKTVILDNLKSAETWLIQNPEGSAEDYKSKLKSLEEVCSPIIQKLY--------ESGAAGEASNDSYA--DEL 624
B. microti Grp78 546 VEAKNTDSYIASMKRSVEDKDKLADKIEEDDKTILDALKNAETWLFQNPEADVEYKSKLKSLEEICNPIIQKLY--------QGNAAGEANYDSYG--DEL 639
```

Supplementary Data S3F

```
                              10        20        30        40        50        60        70        80        90       100       110       120
                     ....|....|....|....|....|....|....|....|....|....|....|....|....|....|....|....|....|....|....|....|....|....|....|....|
Consensus              1 M~Nt~~F~~~La~SLL~~i~a~Ds------------------------------------------------------iEGP~IGIDLGTTYSCVGVFKNGRVEILNN~LGNRIT  48
P. vivax Grp78         1 MRNTRPFIFLLALSLLNCIRGIDSN-----------------------------------------------------IEGPVIGIDLGTTYSCVGVFKNGRVEILNNELGNRIT  62
P. knowlesi ER-Hsp70   1 MRNTRPFIFLLALSLLNCIRAIDSN-----------------------------------------------------IEGPVIGIDLGTTYSCVGVFKNGRVEILNNELGNRIT  62
P. yoelii yoelii ER-Hsp70 1 MGNTKAFVLVLLVSLLKFVSAVDSAKYTYIYIWCGFGLNISSSIYMCKYVFLFSIFSTYFSFSYFLYFFIFLFSIFSHFPIFLVEGPIIGIDLGTTYSCVGVFKNGRVEILNNDLGNRIT 120
P. berghei ER-Hsp70    1 MGNSKAIVLVLFVSLLKFISAVDPA-----------------------------------------------------IEGPIIGIDLGTTYSCVGVFKNGRVEILNNDLGNRIT  62

                     ....|....|....|....|....|....|....|....|....|....|....|....|....|....|....|....|....|....|....|....|....|....|....|
Consensus             49 PSYVSFVDGERKVGEAAKLEATLHPTQTVFDVKRLIGRKFDD~EV~KDRTLLPYEIVN~EGKPNIrVQIKDK~TTFAPEQISAMVLEKMKEIAQSFLGKPVKNAVVTVPAYFNDAQRQAT 164
P. vivax Grp78        63 PSYVSFVDGERKVGEAAKLEATLHPTQTVFDVKRLIGRKFDDQEVVRDRTLLPYEIVNQEGKPNIRVQIKDKKTTFAPEQISAMVLEKMKEIAQSFLGKPVKNAVVTVPAYFNDAQRQAT 182
P. knowlesi ER-Hsp70  63 PSYVSFVDGERKVGEAAKLEATLHPTQTVFDVKRLIGRKFDDQEVVRDRTLLPYEIVNQEGKPNIRVQIKDKKTTFAPEQISAMVLEKMKEIAQSFLGKPVKNAVVTVPAYFNDAQRQAT 182
P. yoelii yoelii ER-Hsp70 121 PSYVSFVDGERKVGEAAKLEATLHPTQTVFDVKRLIGRKFDDKEVAKDRTLLPYEIVNNEGKPNIKVQIKDKPTTFAPEQISAMVLEKMKEIAQSFLGKPVKNAVVTVPAYFNDAQRQAT 240
P. berghei ER-Hsp70   63 PSYVSFVDGERKVGEAAKLEATLHPTQTVFDVKRLIGRKFDDKEVAKDRTLLPYEIVNNEGKPNIRVQIKDKPTTFAPEQISAMVLEKMKEIAQSFLGKPVKNAVVTVPAYFNDAQRQAT 182

                     ....|....|....|....|....|....|....|....|....|....|....|....|....|....|....|....|....|....|....|....|....|....|....|
Consensus            165 KDAG~IAGLNIVRIINEPTAAALAYGLDKKEETSILVYDLGGGTFDVSILVIDNGVFEVYATAGNTHLGGEDFDQRVMDYFIK~FKKK~NIDLRSDKRAIQKLRKEVEIAKRNLSVVHsT 281
P. vivax Grp78       183 KDAGTIAGLNIVRIINEPTAAALAYGLDKKEETSILVYDLGGGTFDVSILVIDNGVFEVYATAGNTHLGGEDFDQRVMDYFIKIFKKKTNIDLRSDKRAIQKLRKEVEIAKRNLSVVHST 302
P. knowlesi ER-Hsp70 183 KDAGTIAGLNIVRIINEPTAAALAYGLDKKEETSILVYDLGGGTFDVSILVIDNGVFEVYATAGNTHLGGEDFDQRVMDYFIKIFKKKTNIDLRSDKRAIQKLRKEVEIAKRNLSVVHST 302
P. yoelii yoelii ER-Hsp70 241 KDAGAIAGLNIVRIINEPTAAALAYGLDKKEETSILVYDLGGGTFDVSILVIDNGVFEVYATAGNTHLGGEDFDQRVMDYFIKMFKKKNNIDLRSDKRAIQKLRKEVEIAKRNLSVVHST 360
P. berghei ER-Hsp70  183 KDAGAIAGLNIVRIINEPTAAALAYGLDKKEETSILVYDLGGGTFDVSILVIDNGVFEVYATAGNTHLGGEDFDQRVMDYFIKMFKKKNNIDLRSDKRAIQKLRKEVEIAKRNLSVVH-T 301

                     ....|....|....|....|....|....|....|....|....|....|....|....|....|....|....|....|....|....|....|....|....|....|....|
Consensus            282 QIEIEDI~EGH~FSETLTRAKFEELNDDLFRETLEPVKKVLdDAKYEKSKIDEIVLVGGSTRIPKIQQIIK~FFNGKEPNRGINPDEAVAYGAAIQAGIILGEELQDVVLLDVTPLTLGI 398
P. vivax Grp78       303 QIEIEDIVEGHSFSETLTRAKFEELNDDLFRETLEPVKKVLDDAKYEKSKIDEIVLVGGSTRIPKIQQIIKEFFNGKEPNRGINPDEAVAYGAAIQAGIILGEELQDVVLLDVTPLTLGI 422
P. knowlesi ER-Hsp70 303 QIEIEDIVEGHSFSETLTRAKFEELNDDLFRETLEPVKKVLEDAKYEKSKIDEIVLVGGSTRIPKIQQIIKEFFNGKEPNRGINPDEAVAYGAAIQAGIILGEELQDVVLLDVTPLTLGI 422
P. yoelii yoelii ER-Hsp70 361 QIEIEDIIEGHNFSETLTRAKFEELNDDLFRETLEPVKKVLDDAKYEKSKIDEIVLVGGSTRIPKIQQIIKDFFNGKEPNRGINPDEAVAYGAAIQAGIILGEELQDVVLLDVTPLTLGI 480
P. berghei ER-Hsp70  302 QIEIEDIIEGHNFSETLTRAKFEELNDDLFRETLEPVKKVLDDAKYEKSKIDEIVLVGGSTRIPKIQQIIKDFFNGKEPNRGINPDEAVAYGAAIQAGIILGEELQDVVLLDVTPLTLGI 421

                     ....|....|....|....|....|....|....|....|....|....|....|....|....|....|....|....|....|....|....|....|....|....|....|
Consensus            399 ETVGGIMTQLIKRNTVIPTKKSQTFSTYQDNQPAVLIQVFEGERALTKDNHLLGKFEL~GIPPAQRGVPKIEVTFTVDKNGILHVEAEDKGTGKSKGITITNDKGRLSKEQIEKMINDAE 517
P. vivax Grp78       423 ETVGGIMTQLIKRNTVIPTKKSQTFSTYQDNQPAVLIQVFEGERALTKDNHLLGKFELTGIPPAQRGVPKIEVTFTVDKNGILHVEAEDKGTGKSKGITITNDKGRLSKEQIEKMINDAE 542
P. knowlesi ER-Hsp70 423 ETVGGIMTQLIKRNTVIPTKKSQTFSTYQDNQPAVLIQVFEGERALTKDNHLLGKFELTGIPPAQRGVPKIEVTFTVDKNGILHVEAEDKGTGKSKGITITNDKGRLSKEQIEKMINDAE 542
P. yoelii yoelii ER-Hsp70 481 ETVGGIMTQLIKRNTVIPTKKSQTFSTYQDNQPAVLIQVFEGERALTKDNHLLGKFELSGIPPAQRGVPKIEVTFTVDKNGILHVEAEDKGTGKSKGITITNDKGRLSKEQIEKMINDAE 600
P. berghei ER-Hsp70  422 ETVGGIMTQLIKRNTVIPTKKSQTFSTYQDNQPAVLIQVFEGERALTKDNHLLGKFELSGIPPAQRGVPKIEVTFTVDKNGILHVEAEDKGTGKSKGITITNDKGRLSKEQIEKMINDAE 541

                     ....|....|....|....|....|....|....|....|....|....|....|....|....|....|....|....|....|....|....|....|....|....|....|
Consensus            518 KFADEDKNLREKVE~KNNLDNY~Q~MKATVEDKDKLADKIEK~DKc~ILNa~K~AE~WL~NNSNAD~EaLKQKLKDVEA~CQPIIVKLYGQPGANsPPP~gDEDV~SDEL 614
P. vivax Grp78       543 KFADEDKNLREKVEAKNNLDNYLQNMKATVEDKDKLADKIEKDDKNAILNAVKDENWLSNNSNADAEALKQKLKDVEAVCQPIIVKLYGQPGANSPPPSADEDVESDEL 652
P. knowlesi ER-Hsp70 543 KFADEDKNLREKVEAKNNLDNYLQNMKATVEDKDKLADKIEKDDKDAILNAVKDENWLSNNSNADAESLKQKLKDVEAVCQPIIVKLYGQPGANSPPPSGDEDVESDEL 652
P. yoelii yoelii ER-Hsp70 601 KFADEDKNLREKVESKNNLDNYIQSMKATVEDKDKLADKIEKEDKDTILNAIKEAEDWLNNNSADSEALKQKLKDVEAICQPIIVKLYGQPGAASPPP-GDEDVDSDEL 709
P. berghei ER-Hsp70  542 KFADEDKNLREKVESKNNLDNYIQSMKATVEDKDKLADKIEKEDKDTILNTIKEAEDWLNNNSADSEALKQKLKPVEAICQPIIVKLYGQPGANTPPP-GDEDVDSDEL 650
```

Supplementary Data S3G

```
                        10        20        30        40        50        60        70        80        90       100       110       120
                        ....|....|....|....|....|....|....|....|....|....|....|....|....|....|....|....|....|....|....|....|....|....|....|....|
Consensus             1 ml?a-r-?a-r---------------------?-----------???r-yaS?a?kgavigidlgttnscvavmegkqpkv?enaEGaRTTPSvvAft?dgerLvGmpAkRQavtN  80
Human Mt-Hsp70        1 MISASRAAAARLVGAAASRGPTAA---RHQDSWNGLSHEAFR----LVSRRD-YASEAIKGAVVGIDLGTTNSCVAVMEGKQPKVLENAEGARTTPSVVAFTADGERLVGMPAKRQAVTN 112
R. norvegicus Mt-Hsp70 1 MISASRAAAARLVGTTASRSPAAA---RHQDGWNGLSHEVFR----FVSRRD-YASEAIKGAVVGIDLGTTNSCVAVMEGKQAKVLENSEGARTTPSVVAFTPDGERLVGMPAKRQAVTN 112
G. gallus Mt-Hsp70    1 MISASRAAA-RLPLLLPRGGPVPAVPGLAQTFWNGLSQNVLR----AASSRK-YASEAIKGAVVGIDLGTTNSCVAVMEGKQAKVLENSEGARTTPSVVAFTPDGERLVGMPAKRQAVTN 114
S. aurata Grp75       1 -----------------------------------------------ASEAIKGAVVGIDLGTTNSCVAVMEGKQAKVLENAEGARTTPSVIAFTAEGERLVGMPAKRQSVTN  66
X. laevis Hsp70-9     1 MLCAGGSLASPLSHVSRI---------LRKGCFNGLSQDVLQ----SVFKRD-YASEVKGAVVGIDLGTTNSCVAVMEGKQAKVLENSEGARTTPSVVAFSSEGERLVGMPAKRQAVTN 106
A. echinatior Hsc70-5 1 ----------------------------------------------------------MEGSRTTPSYVAFSKDDERLVGMPAKRQAVTN  32
M. mongolica Hsp70    1 MLQAARLCTRQVRECHGLISDPNRVRNWSTIGKNLIA------SSTWGVQCR-FKSDGVKGPVIGVDLGTTNSCVAIMEGKTPKVIENAEGVRTTPSVVAFTKGERLAGMPKRQAVTN 113
M. arenaria mortalin-2 1 MFTSARHTSRSLKKTFGT----HQARMISNICKKILAGEIKSKSGLYLLPSR-LRSSDKVKGHVIGIDLGTTNSCVAIMEGKTPKVIENAEGARTTPSVVAFTKDGERLVGMPAKRQAVTN 115
C. elegans Hsp-6      1 MLSARSFLSSAR-------------------TIARSSLMSAR-SLSSDKPKGHVIGIDLGTTNSCVSIMEGKTPKVIENAEGVRTTPSVAFTADGERLVGAPAKRQAVTN  90
C. owczarzaki Hsp70-2 1 MLKAMSVARLA----------------SSGMTMRQSAAKPIAAALSR-SYSSAVKGQVIGIDLGTTNSCVAVMEGKTPKVIENAEGARTTPSVVAFTEEGERLVGTPARRQAVTN  98
P. infestans Mt-Hsp70 1 -----------------------------MFSAAAGSEVIGIDLGTTNSCVAVMEGKTARVIENSEGARTTPSVVAILDNDERLVGMPAKRQAVTN  67
A. thaliana HSO70-2   1 MATAALLRSIRRREVVSS--PFSAYRCLSSSGKASLNSS--YLGQNFRSFSRAFSSKPAGNDVIGIDLGTTNSCVAVMEGKNPKVIENAEGARTTPSVVAFNTKGELLVGPAKRQAVTN 116
S. cerevisiae Ssc1p   1 MLAAKNILNRS------------------------SLSSSFRIATR-LQSTKVQGSVIGIDLGTTNSAVAIMEGKVPKIIENAEGSRTTPSVVAFTKEGERLVGIPAKRQAVVN  89
E. tenella Mt-Hsp70   1 MRGAVALSAARALWAAAV--P-PQPRGPPKEQRVFSAVRTAAVGTLSSLAGR-RGFSGVRGDVVGIDLGTTNSCVAVMEGSQPKVLENSEGMRTTPSVVAFTKDGQRLVGVAKRQAITN 116
D. baltica Hsp70-type 1 ------------------------------------------------------------MAKVGIDLGTTNSVVAAIEGGQPSVIINAEGLRTTPSIVAYIRQELLVGQAKRQAVIN  61

                        ....|....|....|....|....|....|....|....|....|....|....|....|....|....|....|....|....|....|....|....|....|....|....|....|
Consensus            81 p?NTfyatKRlIGrrfddpevqkd?k?vpykivrasNg-dawve--ahgk?ysPsqigafvLmKmketAenyLg??vknAViTVPAYFNDsQRQATkDAGqIsGlnVlRviNEPTAAaLa 197
Human Mt-Hsp70      113 PNNTFYATKRLIGRRYDDPEVQKDIKNVPFKIVRASNG-DAWVE--AHGKLYSPSQIGAFVLMKMKETAENYLGHTAKNAVITVPAYFNDSQRQATKDAGQISGLNVLRVINEPTAAALA 229
R. norvegicus Mt-Hsp70 113 PNNTFYATKRLIGRRYDDPEVQKDTKNVPFKIVRASNG-DAWVE--AHGKLYSPSQIGAFVLMKMKETAENYLGHTAKNAVITVPAYFNDSQRQATKDAGQISGLNVLRVINEPTAAALA 229
G. gallus Mt-Hsp70  115 PHNTFYATKRLIGRRFDDSEVKKDIKNVPFKIVRASNG-DAWVE--AHGKLYSPSQIGAFVLMKMKETAENYLGHPAKNAVTVPAYFNDSQRQATKDAGQISGLNVLRVINEPTAAALA 231
S. aurata Grp75      67 PQNTLYATKRLIGRRFDDPEVQKMKNVPYKIVRASNG-DAWVE--AHGKMYSPSQAGAFVLMKMKETAENYLTKVKNAVVTVPAYFNDSQRQATKDAGQISGLNVLRVINEPTAAALA 183
X. laevis Hsp70-9   107 PNNTFYATKRLIGRRFDDAEVQKDTKNVPFKIVKASNG-DAWVE--SHGKLYSPSQIGAFVLIKMKETAENYLGHSAKNAVITVPAYFNDSQRQATKDAGQISGLNVLRVINEPTAAALA 223
A. echinatior Hsc70-5 33 SVNTFYATKRLIGRRFEDPEVKKDMKSVTYKIVKASNG-DAWVQ-GADGKMYSPSQIGAFVLMKMKETAAYLNTSVKNAVITVPAYFNDSQRQATKDAGQIAGLNVLRVINEPTAAALA 150
M. mongolica Hsp70  114 AQNTLYATKRLIGRRFDDPEVKKDMKTVSYKIVKASNG-DAWVE--AQGKMYPSQIGAFVLVKMKETAEAYLGQPVKNAVITVPAYFNDSQRQATKDAGQIAGLNVLRIINEPTAAALA 230
M. arenaria mortalin-2 116 AANTLHATKRLIGRRFEDKEVKKDMETVPYKIVRANNG-DAWVE--AHGKTYSPSQIGAFVLMKMKETAEDNYLGQPVKNAVVTVPAYFNDSQRQATKDAGQISGLNVLRVINEPTAAALA 232
C. elegans Hsp-6     91 SANTLFATKRLIGRRYEDPEVQKDLKVVPYKIVKASNG-DAWVE--AQGKVYSPSQVGAFVLMKMKETAESYLGTTVNAVITVPAYFNDSQRQATKDAGQISGLNVLRVINEPTAAALA 207
C. owczarzaki Hsp70-2 99 PHNTFTATKRLIGROFDDSEVQRERKLVAYEIVKHTNG-DAWVK--SRDKTYSPSQIGAFVLTKMKETAESYLNTKVHNAVTVPAYFNDSQRQATKDAGQISGLNVLRVINEPTAAALA 215
P. infestans Mt-Hsp70 68 PENTFYAVKRLIGRKFEDKETQEVSKVVSYKIVKGNNGKDAWVE--AKGQKYSPSQIGSMVLTKMKETADGFLGKPITQAVVTVPAYFNDSQRQATKDAGKIAGLDVLRIINEPTAAALA 185
A. thaliana HSO70-2 117 PTNTVSGTKRLIGRKFDDPQTQKEMMKVPYKIVRAPNG-DAWVE--ANGQQYSPSQIGAFILTKMKETAEAYLGKSVTKAVVTVPAYFNDAQRQATKDAGRIACGLDVFRIINEPTAAALS 233
S. cerevisiae Ssc1p  90 PENTLFATKRLIGRRFEDAEVQRDIKQVPYKIVKHSNG-DAWVE--ARGQTYSPAQIGGFVLNKMKETAEAYLGKPVKNAVVTVPAYFNDSQRQATKDAGQIVGLNVLRVNEPTAAALA 206
E. tenella Mt-Hsp70 117 PENTEFSTKRLIGRSFDEEAIAKERKILPYKVIRADNG-DAWVE--GWGKKYSPSQIGAFVLMKMKETAESYLGRDVNQAVITVPAYFNDSQRQATKDAGKIAGLDVLRIINEPTAAALA 233
D. baltica Hsp70-type 62 PENTEFSVKRFIGSK--ESEISAEAKQLPYKVTKDSND-NIKIKCPALAKEFSPEETSAQVLRKLINDAKTYLSQDVTQAVTVPAYFNDSQRQATMDAGKIACIEVLRIINEPTAASLA 178
```

Supplementary Data S3G

```
                            ....|....|....|....|....|....|....|....|....|....|....|....|....|....|....|....|
Consensus                198 YGldk-SedkiiaVyDLGGGTFDiSiLeiqkGvFEVksTnGdTfLGGeDFDqall?hlvkeFk?etgiDltkD?mAlQRlrEAaEkAKiELsss?qTdinLPy?TmDasGpKHlnmkltR 316
Human Mt-Hsp70           230 YGLDK-SEDKVIAVYDLGGGTFDISILEIQKGVFEVKSTNGDTFLGGEDFDQALLRHIVKEFKRETGVDLTKDNMALQRVREAAEKAKCELSSSVQTDINLPYLTMDSSGPKHLNMKLTR 348
R. norvegicus Mt-Hsp70   230 YGLDK-SEDKVIAVYDLGGGTFDISILEIQKGVFEVKSTNGDTFLGGEDFDQALLRHIVKEFKRETGVDLTKDNMALQRVREAAEKAKCELSSSVQTDINLPYLTMDSSGPKHLNMKLTR 348
G. gallus Mt-Hsp70       232 YGLDK-SEDKIIAVYDLGGGTFDISILEIQKGVFEVKSTNGDTFLGGEDFDQALLQYIVKEFKRETSVDLTKDNMALQRVREASEKAKCELSSSVQTDINLPYLTMDASGPKHLNMKISR 350
S. aurata Grp75          184 YGLDK-TQDKIIAVYDLGGGTFDISVLEIQKGVFEVKSTNGDTFLGGEDFDQHLLRHIVKEFKRESGVDLTKDSMALQRVREAAEKAKCELSSSLQTDINLPYCTMDASGPKHLNMKLSR 302
X. laevis Hsp70-9        224 YGLDK-SDDKVIAVYDLGGGTFDISILEIQKGVFEVKSTNGDTFLGGEDFDQALLRHIVKQFKRESGVDLTKDNMALQRVREAAEKAKCELSSSLQTDINLPYLTMDASGPKHLNMKLSR 342
A. echinatior Hsc70-5    151 YGMDK-QEDKIIAVYDLGGGTFDISILEIQKGVFEVKSTNGDTFLGGEDFDNALVNYLVTEKKEQGIDVTKDAMQRLKEASEKAKCELSSSLQTDINLPYLTMDSSGPKHLNLKLSR 269
M. mongolica Hsp70       231 YGMDK-SEDKIIAVYDLGGGTFDISILEIQKGVFEVKSTNGDTFLGGEDFDNALVNFLVAERRDQGLDVTKDPMAMQRVKEAEKAKIELSSSMQTDINLPYLTMDASGPKHMNLKISR 349
M. arenaria mortalin-2   233 YGMDK-TGDKIIAVYDLGGGTFDISILEIQKGVFEVKSTNGDTFLGGEDFDNVLVSYLAKEFQKDQGIDVTKDNMAMQRLREAAEKAKIELSSSMLQTDINLPYLTMDAGGPKHLNMKISR 351
C. elegans Hsp-6         208 YGLDKDAGDKIIAVYDLGGGTFDVSILEIQKGVFEVKSTNGDTFLGGEDFDHALVHHLVGMEFKKEQQVDLTKDPQAMQRLREAAEKAKCELSSTTQTDINLPYITMDQSGPKHLNLKLTR 327
C. owczarzaki Hsp70-2    216 YGMDR-SDDKIIAVYDLGGGTFDVSILEIQKGVFEVKATNGDTFLGGEDFDNHLVPFLEEFKKQHGMDISKDTVALQRLREAAEKAKIELSSTNQTEVNLPYITADAKGPKHFVHKLTR 334
P. infestans Mt-Hsp70    186 YGMDK-ADGKVIAVFDLGGGTFDVSILEISGGVFEVKSTNGDTLLGGEDFDEELLRYLVNEFKKETSIDLSGDNLAMQRLREAAEKAKRELDGLAQTDISLPFITADATGPKHLNMKITR 304
A. thaliana HSO70-2      234 YGMTN-KEG-LIAVFDLGGGTFDVSVLEISNGVFEVKATNGDTFLGGEDFDNALLDFLVNEFKTTEGIDLAKDRLALQRLREAAEKAKIELSSTSQTEINLPFITADASGAKHFNITLTR 351
S. cerevisiae Ssc1p      207 YGLEK-SDSKVVAVFDLGGGTFDISILDIDNGVFEVKSTNGDTHLGGEDFDIYLLREIVSRFKTETGIDLENDRMAIQRLREAAEKAKIELSSTVSTEINLPFITADASGPKHINMKFSR 325
E. tenella Mt-Hsp70      234 YGMEK-EDGRTIAVYDLGGGTFDVSILEILGGVFEVKATNGNTSLGGEDFDQKVLQFIVNEFKKEGIDLSKDRLALQRLREAAETAKIELSSKLSTEINLPFITADQSGPKHLQVSISR 352
D. baltica Hsp70-type    179 YGLDK-KQNETTLVFDLGGGTFDVSILEVGDGIFEVLATAGDTNLGGDDFDKVLVTWLMNDFKQKEGIDLSTDIQALQRITEAAEKAMELSTVDKTNISLPFITADQTGPKHIDKELTR 297

                            ....|....|....|....|....|....|....|....|....|....|....|....|....|....|....|....|....|....|....|....|....|....|
Consensus                317 aqfEglv gdLikrtvaPcqkal?Daevsksdige VlLVGGmtRmPkvqqtVqelfgraPskaVNPDEaVAiGAAiQggvLaGdvtdvLLLDVtPLSLGiETlGGvftklI?RNTTIPtKK 436
Human Mt-Hsp70           349 AQFEGIVTDLIRRTIAPCQKAMQDAEVSKSDIGEVILVGGMTRMPKVQQTVQDLFGRAPSKAVNPDEAVAIGAAIQGGVLAGDVTDVLLLDVTPLSLGIETLGGVFTKLINRNTTIPTKK 468
R. norvegicus Mt-Hsp70   349 AQFEGIVTDLIRRTIAPCQKAMQDREVSKSDIGEVILVGGMTRMPKVQQTVQDLFGRAPSKAVNPDEAVAIGAAIQGGVLAGDVTDVLLLDVTPLSLGIETLGGVFTKLINRNTTIPTKK 468
G. gallus Mt-Hsp70       351 SQFEGIVADLIKRTVAPCQKAMQDAEVSKSDIGEVILVGGMTRMPKVQQTVQDLFGRAPSKAVNPDEAVAIGAAIQGGVLAGDVTDVLLLDVTPLSLGIETLGGVFTKLINRNTTIPTKK 470
S. aurata Grp75          303 AQFEGIVADLIRRTVAPCQKAMQDAEVSKGDIGEVLLVGGMSRMPKVQQTVQDLFGRAPSKSVNPDEAVAIGAAIQGGVLAGDVTDVLLLDVTPLSLGIETLGGVFTKLINRNTTIPTKK 422
X. laevis Hsp70-9        343 SQFEGIVGDLIKRTVAPSQKAMQDAEVGKSDIGEVLLVGGMTRMPKVQQTVQDLFGRAPSKAVNPDEAVAIGAAIQGGVLAGDVTDVLLLDVTPLSLGIETLGGVFTKLIGRNTTIPTKK 462
A. echinatior Hsc70-5    270 SKFESLVNDLIKRTVQPCQKALSDAEVTKSDIGEVLLVGGMTRVPKVQQTVQEIFGRQPSKAVNPDEAVAVGAAVQGGVLAGDVTDVLLLDVTPLSLGIETLGGVFTRLISRNTTIPTKK 389
M. mongolica Hsp70       350 AKFENLVGDLIKRTVAPCQKALKDAEVGKNEIGDVLLVGGMTRMPKVQDTVKEIFGRVPSKAVNPDEAVAVGAAIQGGVLAGGVTDILLLDVTPLSLGIETLGGVFTKLIQRNTTIPTKK 469
M. arenaria mortalin-2   352 AKFESLVDDLIKRTVGPCNKALQDAEIKKSDIGDVLLVGGMTRMPKVQQVVQEVFGRAPGKSVNPDEAVAIGAAIQGGVLAGDVTDVLLLDVTPLSLGIETLGGVFSRLITRNTTIPTKK 471
C. elegans Hsp-6         328 AKFEQIVGDLIKRTIEPCRKALHDAEVKSSQIADVLLVGGMSRMPKVQATVQEIFGKVPSKAVNPDEAVAMGAAIQGAVLAGDVTDVLLLDVTPLSLGIETLGGIMTKLIRNTTIPTKK 447
C. owczarzaki Hsp70-2    335 AKFESIVGSLVQKTIDPCRKCLKDAGLEKSQIGEVLLVGGMTRMPKVVDTVRELFGREPSKGVNPDEAVAIGAAIQGGVLAGDVTDVLLLDVTPLSLGIETFGGVFSRLINRNTTIPTKK 454
P. infestans Mt-Hsp70    305 ATFEKLVGKLIERTMGPCKKCVKDAGLDKSEINEVILVGGMSRMPKVQTTVEEFFGKKPSKVNPDEVAMGAAIQGVIRGDVKDILLLDVTPLSLGIETLGGVFTKLIPRNTTIPTKK 424
A. thaliana HSO70-2      352 SRFETLVNHLIERTRDPCKNCLKDAGISAKEVDEVLLVGGMTRVPKVQSIVAEIFGKSPSKVNPDEAVAMGAALQGGILRGDVKELLLLDVTPLSLGIETLGGVFTRLITRNTTIPTKK 471
S. cerevisiae Ssc1p      326 AQFETLTAPLVKRTVDPVKKALKDAGLSTSDISEVLLVGGMSRMPKVVETVKSLFGKDPSKVNPDEEAVAIGAAVQGAVISGEVTDVLLLDVTPLSLGIETLGGVFTRLIPRNTTIPTKK 445
E. tenella Mt-Hsp70      353 AHLEELVGALLQQSIEPCEKCIRDAGVQKADLSDVILVGGMTRMPKVAEVVKNIFHKEPSKGVNPDEAVAAGAAIQGAVLKGEIKDLLLLDVCPLSLGIETLGGVFTRLINRNTTIPTKK 472
D. baltica Hsp70-type    298 ETFEKLCEKLIDRCRIPVEKALNDARLDKSDINEVVLVGGSTRIPAIQQLVESITGKKPNQSVNPDEVVAIGAAIQAGILAGEIKDILLLDVTPLSLGVETLGGVMTKIIARNTTIPVKK 417
```

Supplementary Data S3G

```
                           ....|....|....|....|....|....|....|....|....|....|....|....|....|....|....|....|....|....|....|....|....|....|....|....|
Consensus              437 SqvfSTAadgQTqVeIkv?QGERema?dNkllGqFtLvGiPpaprGVPQiEVtFDiDanGIvhvsA?dkgtgkeqqiviqssggLskdeienMvknAEkya_eDrrrke?veavN?Aegi 556
Human Mt-Hsp70         469 SQVFSTAADGQTQVEIKVCQGEREMAGDNKLLGQFTLIGIPPAPRGVPQIEVTFDIDANGIVHVSARDKGTGREQQIVIQSSGGLSKDDIENMVKNAEKYAEDRRKKERVEAVNMAEGI 588
R. norvegicus Mt-Hsp70 469 SQVFSTAADGQTQVEIKVCQGEREMAGDNKLLGQFTLIGIPPAPRGVPQIEVTFDIDANGIVHVSARDKGTGREQQIVIQSSGGLSKDDIENMVKNAEKYAEDRRKKERVEAVNMAEGI 588
G. gallus Mt-Hsp70     471 SQVFSTAADGQTQVEIKVCQGEREMAGDNKLLGQFTLVGIPPAPRGVPQIEVTFDIDANGIVHVSARDKGTGREQQIVIQSSGGLSKDEIENMVKNAEKYAEDRRRKERVEAVNLAEGI 590
S. aurata Grp75        423 SQVFSTAADGQTQVEIKVCQGEREMAGDNKVLGQFTLVGIPPAPRGVPQIEVTFDIDANGIVHVSARDKGTGREQQIVIQSSGGLSKDEIENMVKNAEKYAEDRRRKDRVEAVNMAEGI 542
X. laevis Hsp70-9      463 SQVFSTAADGQTQVEIKVHQGEREMAGDNKLLGQFTLMPPAPRGVPQIEVTFDIDANGIVHVSARDKGTGREQQIVIQSSGGLSKDDIENMVKNAEKYAEDRRRKEQVNNAEGI 582
A. echinatior Hsc70-5  390 SQVFSTAADGQTQVEIKVHQGEREMAGDNKLLGQFSLVGIPPAPRGVPQIEVTFDIDANGIVHVSARDKGTGKEQQIVIQSSGGLSKDEIENMVKNAEQYAKADIKIKKERVEAVNQAEGI 509
M. mongolica Hsp70     470 SQVFSTAADGQTQVEIKVFQGEREMAGDNKLLGQFSLIGIPPAPRGVPQIEVTFDIDANGIVHVSARDKGTGKEQQIVIQSSGGLSKDEIENMVRNAEQFAKEDRRRKERVEAVNQAEGI 589
M. arenaria mortalin-2 472 SQVYSTAADGQTSVEIKVFQGEREMAKDNKLLGQMLSGVPPMPRGVPQIEVVFDIDANGIVNVSARDQGTGKEQQIVIQSSGGLSKDEIENMVRNAEKYAEDAQRKDMIEAVNAETM 591
C. elegans Hsp-6       448 SQVFSTAADGQTQVQIKVFQGEREMATSNKLLGQFTLAGIPPAPRGVPQVEVTFDIDANGINVSARDRGTGKEQQIVIQSSGGLSKDQIENMIKEAEKNAAEDRKRKELVEVINQAEGI 567
C. owczarzaki Hsp70-2  455 SSTFSTAADMQTSVEIKVLQGERDMAADNQALGSERLVGIPPAPRGVPQVEVTFDIDANGINVSARDKSGKEQQIVIQSSGGLTKDQIEKMIRDAEVNAAEDRRRREDIEARNQAETF 574
P. infestans Mt-Hsp70  425 SQVFSTAADSQTQVIKVLQGEREMAADNKLLGNEDLVGIPPAPRGVPQVEVSFDIDANGINVGARDKATGKEQNIVIQSSGGLSEAEIEKMADAEANAADQQRKELIEARNDADST 544
A. thaliana HSO70-2    472 SQVFSTAADNQTQVIRVLQGEREMATDNKLLGEEDLVGIPPSPRGVPQIEVTFDIDANGIVTVSARDKTTGKVQQITIRSSGGLSEDDIQKMVREAELHAQKDKERKELIDTKNTADTT 591
S. cerevisiae Ssc1p    446 SQIFSTAAAGQTSVEIRVFQGERELVRDNKLIGNETLAGIPPAPKGVPQIEVTFDIDADGIINVSARDKATNKDSSITVAGSSGLSENEIEQMVNDAEKFKSQDEARKQAIETANKADQL 565
E. tenella Mt-Hsp70    473 SQIFSTAADNQTQVIKVYQGEREMASANKLLGVPPAPRGVPQIEVTFDVDANGIMNISAVDKSTAKRQQITIQSSGGLSEAQIKQMVEDAERFKDEDQRKDLVAAKNEAETL 592
D. baltica Hsp70-type  418 SEMFSTAVDNQTNVEIHILQGERELVAGNKSLGNERLDGIPAANRGVPQIEVTFDIDVDGILSVKAKEKETGVEQSVTIQGASTLNEKEVSMLADAERYAADKEKRENIDLKNQAETL 537
```

```
                                                                                610       620       630       640       650       660       670       680       690       700
                           ....|....|....|....|....|....|....|....|....|....|....|....|....|....|....|....|....|....|....|....|....|....|....|..
Consensus              557 ?hdtE?km?efkdqlpa?ec?klk??i?k?lre?la?---kdsetgeeik?aan?lqqaslklfemaykkmaser??s-----gss-----dssd?---??e?kc?kq 647
Human Mt-Hsp70         589 IHDTETKMEEFKDQLPADECNKLKEEISKMRELLAR---KDSETGENIRQAASLQQASLKLFEMAYKKMASEREGS-----GSS-----GTGEQ---KEDQKEEKQ 679
R. norvegicus Mt-Hsp70 589 VHDTETKMEEFKDQLPADECNKLKEEISKMRELLAR---KDSETGENIRQAASLQQASLKLFEMAYKKMASEREGS-----GSS-----STGEQ---KEDQKEEKQ 679
G. gallus Mt-Hsp70     591 IHDTESKMEEFKDQLPADECNKLKEEIAKMRELLAR---KDTETGENIRQAATSLQQASLKLFEMAYKKMASERESS--------------GS------SGDQKEEKQ 675
S. aurata Grp75        543 VHDTESKMEEFKDQLPADECTKLKEEISKVRDLLAN---KDSETGENIKQAATTLQQASLKLFEMAYKKMAAEREGSSGGSGGSS-----SSGSS---EGEKKEGQQ 638
X. laevis Hsp70-9      583 IHDTESKMEEFKDQLPADECNKLKEEISKVKELLAR---KDEETGESIRNASSTLQQASLKLFEMAYKKMASERSGS-------------ESGPQ---KDDQKEEKQ 670
A. echinatior Hsc70-5  510 IHDTESKLEEFKAQLPQEECDKLKELVAKLREILAK---KDDVDPEEIKKQTNELQQASLKLFEMAYKKMAAERESQ-----------SQSQQEP-EGEKKEEKN 599
M. mongolica Hsp70     590 VHDTESKMEEFKDQLPAEECSKIREQITSVRDVLAN---KDSKTPEEIKKATNDLQQASLKLFEMAYKKMASERESS--------------NSSSS----SGESGEKKE 677
M. arenaria mortalin-2 592 IHDAEAKMNDFKEQLDQDQSDQMRKDIAAMREVIAN---KDSETPESIRTKLNELQQSTLKLFEMTYKKKMAQDQQNQQQQGGSSDSSSSDTTDE---QKEKEEKN 692
C. elegans Hsp-6       568 IHDTEAKMTEEADQLPKDEGEALRTKIADTKKILDN---KDNETPEAIKEACNTLQQQSLKLFEAAYKNMAAKNSGG-------------DAQEAKT-AEEPKKEQN 657
C. owczarzaki Hsp70-2  575 LHDTEKSIAEYGEQIGKEDMEKAKVDLNKLRDGIEK-----KAPADEIKKLLGAAQQTSLKIFEAVYQKKPNASSSS-------------DSKDKTV-DAEFK-KDN 661
P. infestans Mt-Hsp70  545 LYSTEKTLKEHEDKIDAETLEQVKTALTDLRT---K---SEGDNTEEIKAALEEVQKLSMKIGEAVYKSQQGESAASD------------DNKDENVHDAEFKDKRD 633
A. thaliana HSO70-2    592 IYSIEKSLGEYREKIPSEIAKEIEDAVADLRSASSG------DDLNEIKAKIEAANKAVSKIGEHMSGGSGGGSAPGGGSE--------GGSDQAP-EAEYBEVVK 682
S. cerevisiae Ssc1p    566 ANDTENSLKEFEGKVDKAEAQKVRDQITSLKELVARVQGGEEVNAEELKTKTEELQTSSMKLFEQLYKNDSNNNNNN-------------NGN-----NAESGETKQ 654
E. tenella Mt-Hsp70    593 VYSVEKQISDLKDKISAEDKTDLESRIQELRSAIVE------GELETIRSRVKALQELSWKVSQQAYSQSNNTSADG-------------DSSST---SSGDSSSKP 677
D. baltica Hsp70-type  538 CFEAEKELSLVKDTISTQKQESISKLISTIRQ---N---SQNDQIESLKS---SLEELKVAMKEMVASKLDSESNSD-------------PMSNL---------NDL 613
```

Supplementary Data S3H

```
                               10        20        30        40        50        60        70        80        90       100       110       120
                        ....|....|....|....|....|....|....|....|....|....|....|....|....|....|....|....|....|....|....|....|....|....|....|....|
Consensus            1  M~s~-------------------------------------------------------------------s~f~---s~~~ss~~s~~s~~~~~~sgi----------f~r~st---s   18
T. gondii Mt-Hsp70   1  MASMAAGASLAYRASYCSGAKARSISSFTSLGRAC-VETQMTPAAA---ASLRPSRFLFSSLAQNAK--SSSPF---SAGKLRSGMNT-TYERTAVLSS-----------FSRYLST---S   97
N. caninum Mt-Hsp70  1  MASLASGASLAYRASYCSGAKTRSISN-SLLGRATFANKQMTSVAAPTASLRPSGFFSDLLTQGGKAVSSSPFRNLNGGLYRNACSALNSERAGVLSS-----------FSRYLST---S  105
E. tenella Mt-Hsp70  1  MRGA------------------------------------------VALSAARALWAAAVPPQPR---GPPKEQRVFSAVRTAAVGTISSL-------------------A          47
T. parva Mt-Hsp70    1  MKSL-----------------------------------------------------------RRL---YSIFNSSRCSLELSNKLSTISGITTMLDSLRIGSRRGIFT---S         48
T. annulata Mt-Hsp70 1  MKSL-----------------------------------------------------------NRL---YSIFHSSRCSLELSNKLSTISGITTMLDSLRIGSRRGIFT---T         48
B. bovis Mt-Hsp70    1  MTLL--------------------------------------------------------SRF---FSL-GSPACRL----------GL-----------VQRSLTT---S            27
C. muris Mt-Hsp70    1  MFCT--------------------------------------------------------RYK---NNSNFGNALKSISTFLIKTSSGI----------LHRHIGT---S            38
C. parvum Mt-Hsp70   1  MSMI-------------------------------------------------------INSSF---NGVVNSSGIAA--RILKRSLPLV----------FSRYMSSKCEG          41

                               130       140       150       160       170       180       190       200       210       220       230
                        ....|....|....|....|....|....|....|....|....|....|....|....|....|....|....|....|....|....|....|....|....|....|....|
Consensus           19  a~r~~~k~~gDvvGIDLGTTNSCvA~mEGsqPKViENsEGmRTTPS~VAFtcDGQRLvGvVAKRQAvTNpENTvfaTKRlIGRrfde~a~kKEq~iLPYKivrasNgDAWvEaqgkqYSP  127
T. gondii Mt-Hsp70  98  APRMNGKARGDVVGIDLGTTNSCVAVMEGSQPKVIENSEGMRTTPSIVAFTSDGQRLVGIVAKRQAVTNPENTVFATKRLIGRRYDEDAIKKEKEILPYKIVRASNGDAWVEAQGKGYSP  217
N. caninum Mt-Hsp70 106 APRMN-KARGDVVGIDLGTTNSCVAVMEGSQPKVIENSEGMRTTPSIVAFTSDGQRLVGIVAKRQAVTNPENTVFATKRLIGRRYDEEAIKKEQILPYKIVRASNGDAWVEAQGKSYSP  224
E. tenella Mt-Hsp70  48 GRRGFSGVRGDVVGIDLGTTNSCVAVMEGSQPKVLENSEGMRTTPSVVAFTKDGQRLVGVVAKRQAITNPENTFFSTKRLIGRSFDEEAIAKERKILPYKVIRADNGDAWVEGWGKKYSP  167
T. parva Mt-Hsp70    49 TGRF-AKVCGDVVGIDLGTTNSCVAIMEGSTPKVIENAEGARTTPSVVAFTDDGQRLVGVVAKRQAVTNPENTVFATKRFIGRKDDPETKKEQQTLPYKIVRSSNNDAWIEAQGKQYSP  167
T. annulata Mt-Hsp70 49 SGRF-AKVCGDVVGIDLGTTNSCVAIMEGSTPKVIENAEGARTTPSVVAFTDDGQRLVGVVAKRQAVTNPENTVFATKRFIGRKDDPETKKEQSTLPYKIVRSSNNDAWIEAQNKQYSP  167
B. bovis Mt-Hsp70    28 RNLR-SKVCGDVVGIDLGTTNSCVAVMEGSVPKVIENSEGMRTTPSVVAFTKRFIGRRFDDDVTKKEQKTLPYKIVRASNGDAWIEAQGKQYSP                            146
C. muris Mt-Hsp70    39 ASCYSDKIGRDVIGIDLGTTNSCVAIMEGSSPKVLENSEGMRTTPSVVAFTEDGQRLIGIVAKRQATNAENTVFATKRLIGRRFDEDATKKEQNILPYKIVRAPNGDAWESHGKQYSP  158
C. parvum Mt-Hsp70   42 KKSSNRRITGDIIGIDLGTTNSCTAILEGTQPKVLENSEGMRTTPSVVAFSEDGQRLVGEVAKRQAITNPENTVYATKRLIGRRYEEFAIKKEQGILPYKIVRADNGDAWVEARGERYSP  161

                               250       260       270       280       290       300       310       320       330       340
                        ....|....|....|....|....|....|....|....|....|....|....|....|....|....|....|....|....|....|....|....|....|....|
Consensus          128 SQIgAfiL~KMkeTAEaylGRt~VkqAVITVPAYFNDSQRQATKDAGkiAgleVlRIINEPTAAALAfGmdKnDGkTiAVYDLGGGTFDiSILEILGGVFEVKATNGNTSLGGEDFDQriL  246
T. gondii Mt-Hsp70 218 SQISAFIILTKMKETAEAYIGRPVKQAVITVPAYFNDSQRQATKDAGKIAGLEVLRIINEPTAAALAFGMDKDDGKTIAVYDLGGGTFDISILEILGGVFEVKATNGNTSLGGEDFDQKIL  337
N. caninum Mt-Hsp70 225 SQISAFIILTKMKETAEAYIGRPVKQAVITVPAYFNDSQRQATKDAGKIAGLEVLRIINEPTAAALAFGMDKDDGKTIAVYDLGGGTFDISILEILGGVFEVKATNGNTSLGGEDFDQKIL  344
E. tenella Mt-Hsp70 168 SQIGAFVLLKMKETAESYLGRDVNQAVITVPAYFNDSQRQATKDAGKIAGLDVLRIINEPTAAALAYGMEKEDGRTIAVYDLGGGTFDVSILEILGGVFEVKATNGNTSLGGEDFDQKVL  287
T. parva Mt-Hsp70   168 SQIGAYILAKMKETAESYLGRTVSKAVITVPAYFNDSQRQATKDAGKIAGLEVLRIINEPTAAALAFGMDKNDGKTIAVYDLGGGTFDVSILEILGGVFEVKATNGNTSLGGEDFDQRIL  287
T. annulata Mt-Hsp70 168 SQIGAYILAKMKETAESYLGRTVSKAVITVPAYFNDSQRQATKDAGKMARVEVLRIINEPTAAALAFGMDKNDGKTIAVYDLGGGTFDVSILEILGGVFEVKATNGNTSLGGEDFDQRIL  287
B. bovis Mt-Hsp70   147 SQIGACILSKMRETAEAHLGRKVTKAVITVPAYFNDSQRQATKDAGKIAGLDVLRIINEPTAAALAFGLEKNDGKTIAVYDLGGGTFDISILEILGGVFEVKATNGNTSLGGEDFDQRIL  266
C. muris Mt-Hsp70   159 SQIGAFVLKMKQTAEAYLGRVKQAVITVPAYFNDSQRQATKDAGKIAGLDVLRIINEPTAAALAFGMDKVDGKTVAVYDLGGGTFDISILEILGGVFEVKATNGNTSLGGEDFDQRVL  278
C. parvum Mt-Hsp70  162 SQIGAFILKMKMETAETYLGRGVKHAVITVPAYFNDSQRQATKDAGSIAGLNVTRIINEPTAAALAYGMEKADGKTIAVYDLGGGTFDISILEILGGVFEVKATNGNTSLGGEDFDQRIL  281

                               360       370       380       390       400       410       420       430       440       450
                        ....|....|....|....|....|....|....|....|....|....|....|....|....|....|....|....|....|....|....|....|....|....|
Consensus          246 ~~L~dEFKK~qGIDL~kDkLALQRLREaaEtAKiELSsKtqtEiNLPFItADqsGPKHlqvkl~RaklEelv~dLLq~tvePcekCikDaGvsksdl~DvILVGGMTRMPkVte~VknIF  356
T. gondii Mt-Hsp70 338 QHLIDEFKKAQGIDLTKDKLALQRLREAAETAKIELSSKVQTEVNLPFITADQTGPKHLQVKLTRAKLEELVGGLLQQSVEPCEKCIKDAGVSKSDLSDVILVGGMTRMPKVTELVKQIF  457
N. caninum Mt-Hsp70 345 QHLIDEFKKAQGIDLTKDKLALQRLREAAETAKIELSSKVQTEVNLPFITADQSGPKHLQVKLTRAKLEELVGGLLQQSVEPCEKCIKDAGVSKSDLSDVILVGGMTRMPKVTELVKQIF  464
E. tenella Mt-Hsp70 288 QFLVNEFKKKEGIDLSKDRLALQRLREAAETAKIELSSKLSTEINLPFITADQSGPKHLQVSLSRAHLEELVGALLQQSIEPCEKCIRDAGVQKADLSDVILVGGMTRMPKVAEVVKNIF  407
T. parva Mt-Hsp70   288 NFLVDEFKKTNGIDLKKDKLALQRLRESSESAKIELSTKTQTEINLPFITADQSGPKHLLIKLSRSKLEQLTSELLEGTVDPCKKCLKDAGVNASEINDVILVGGMTRMPKVTEVVKNIF  407
T. annulata Mt-Hsp70 288 NYLVEEFKKSNGIDLKKDKLALQRLRESSESAKIELSTKTQTEINLPFITADQSGPKHLLIKLSRSKLEQLTSELLEGTVDPCKKCLKDAGVNASEINDVILVGGMTRMPKVTEVVKNIF  407
B. bovis Mt-Hsp70   267 KYLISEFKKQQGIDLTNDKLALQRLREAASAKIELSSKTQTEINLPFITADMSGPKHMQFKLTRAKLEEICDDLKGTIEPCEKCLKDAGVSKDINDIILVGGMTRMPRVGDIVQRIF   386
C. muris Mt-Hsp70   279 TFLVQEFKKNQGIDLSKDRLALQRLREAAETAKIELSSKSQTEVNLPFISADSAGPKHLHVKITRAKLEDLVHDLLSKTVVPSEQCIRDSGVSKKDISDVILVGGMTRMPKVTEIVRSIF  398
C. parvum Mt-Hsp70  282 NYLIQEFKKTQGIDLSRDKLALQRLREASETAKKELSSKTQVEINLPFITADARGPKHLQIKLSRAKYEELVDDLLKKTISPSEKCIRDSGIPKEKINDVILVGGMTRMPKVSETVKKIF  401
```

Supplementary Data S3H

```
                          ....|....|....|....|....|....|....|....|....|....|....|....|
Consensus             357 gkEpsKgvNPDEAVAmGAAIQAgVLKGEIKDLLLLDVcPLSLgIETLGGVFTrLiNRNTTIPTKKSQ~FSTAADNQTQVGIkVyQGERemAadNklLGQFdlvGIPPAPRgVPQIEVTFD 475
T. gondii Mt-Hsp70    458 GKEPSKGVNPDEAVAMGAAIQAGVLKGEIKDLLLLDVCPLSLGIETLGGVFTRLINRNTTIPTKKSQVFSTAADNQTQVGIKVYQGEREIAAANKMLGQFDLVGIPPAPRGVPQIEVTFD 577
N. caninum Mt-Hsp70   465 GKEPSKGVNPDEAVAMGAAIQAGVLKGEIKDLLLLDVCPLSLGIETLGGVFTRLINRNTTIPTKKSQVFSTAADNQTQVGIKVYQGEREMAAANKMLGQFDLVGIPPAPRGVPQIEVTFD 584
E. tenella Mt-Hsp70   408 HKEPSKGVNPDEAVAAGAAIQAGVLKGEIKDLLLLDVCPLSLGIETLGGVFTRLINRNTTIPTKKSQIFSTAADNQTQVGIKVYQGEREMASANKLLGQFDLVGIPPAPRGVPQIEVTFD 527
T. parva Mt-Hsp70     408 GKEPSKAVNPDEAVAMGAAIQAGVLKGEIKDLLLLDVCPLSLGIETLGGVFTRLINRNTTIPTKKSQIFSTAADNQTQVGIKVYQGERCMAADNQLLGQFDLVGIPPAPRGVPQIEVTFD 527
T. annulata Mt-Hsp70  408 GKEPIKAVNPDEAVAMGAAIQAGVLKGEIKDLLLLDVCPLSLRIETLGGVFTTLTNXNTTIPTKKSQIFSTAADNQTQVGIKVYQGERCMAADNQLLGQFDLVGIPPAPRSVPQIEVTFD 527
B. bovis Mt-Hsp70     387 GKEASKAVNPDEAVAMGAAIQAGVLKGEIKDLLLLDVCPLSLGIETLGGVFTRLINRNTTIPTKKSQIFSTAADNQTQVGIKVYQGERCMAADNQLLGQFELVGIPPAPRGVPQIEVTFD 506
C. muris Mt-Hsp70     399 GREPSKGINPDEAVAMGAAIQAAVLKGEIKDLLLLDVTPLSLGIETLGGVFTRLINRNTTIPTKKSQVFSTAADNQTQVGINVFQGEREIAMDNKLLGQFDLIGIPPAPRGVPQIEVTFD 518
C. parvum Mt-Hsp70    402 GREPSKGVNPDEAVAMGAAIQAGVLKGEIKDLLLLDVTPLSLGIETLGGVFTRLINRNTTIPTKKSQVFSTAADNQTQVGIKVFQGEREFAADNKLLGQFEMMGIPPAPRGVPQIEVTFD 521

                               610       620       630       640       650       660       670       680       690       700       710       720
                          ....|....|....|....|....|....|....|....|....|....|....|....|....|....|....|....|....|....|....|....|....|....|....|....|
Consensus             476 vDANGIMNisAvDKSTgkrqeITIQSSGGLS~~qiekMvk~Ae~yk~gDe~kKelvda~NeaEtliYSVekQ~tdlkDkis~adk~dL~~~i~~lrs~lsseq--------p~~i~~~~k 565
T. gondii Mt-Hsp70    578 VDANGIMNISAVDKSTGKRQEITIQSSGGLSDQIEQMVKDAEMYKEQDEKKKDAVQAKNEAETLIYSVEKQMADLKDKMTDADRTDLQEKITQLRSTLGQED--------PEPIREALK 689
N. caninum Mt-Hsp70   585 VDANGIMNISAVDKSTGKRQEITIQSSGGLSDQIEQMVKDAEMYKEQDEKKKDAVQAKNEAETLIYSVEKQMTDLKDKMTDADRTDLQEKITHLRSTLGQED--------PDPIREALK 696
E. tenella Mt-Hsp70   528 VDANGIMNISAVDKSTAKRQQITIQSSGGLSEAQIKQMVEDAERFKDEDQROKDLVAAKNEAETLVYSVEKQISDLKDKISAEDKTDLESRIQELRSALVEGE--------LETIRSRVK 639
T. parva Mt-Hsp70     528 VDANGIMNISAVDKSTGKRQEITIQSSGGLSEEEVEKMVKEASNYKEQDDRRKELVDRNESESLLYSVEKQLTDFKDKVSQAELDQLKTLSTSIKEVLTTDD--------VSAIKDKHK 639
T. annulata Mt-Hsp70  528 VDANGIMNISAVDKSTGKRQEITIQSSGGLSEEEVEKMVKEASNYKEQDERRKELVDRNESESILYSVEKQLTDLKDKVSSSELDQLRTLSTSIKEVLSSDD--------VSAIKDKHK 639
B. bovis Mt-Hsp70     507 VDANGIMNISAVDKSTGRKQEITIQSSGGLSDEQVERMVKDAEAFKQSDEQRKLLVDARNEAETLCYSVEKQLSDFKDKISEEDKKGLEEQLANLREQMSSED--------IDSLKECHK 618
C. muris Mt-Hsp70     519 IDANGIMNVSAVDKSTGKKHEITIQSSGGLSPQDIEKMIREAEEYRSSDKTKKEIIDARNDAETLIYSVEKQITDLSDKLSPSDKEELQCKIEKVKSTFSSLE--------PEEIRSETR 630
C. parvum Mt-Hsp70    522 IDANGIMNVGAIDKSTGKKHEITIQSSGGLSGAEIEKMIREAEEYRANDQAKKELIDLKNDAEFAFIYSVQNQISSLADQINTQEKDSLESKISKLQSILQESTQSSDYESAIQSTKSQLE 641

                                730       740       750       760
                          ....|....|....|....|....|....|....|....|
Consensus             566 qLqelSWk~sQqaYsq~nst~~~-sasens~~~se~~d~~s~ 593
T. gondii Mt-Hsp70    690 TLQEASWKISQQAYNQAGSTDS--SAGS-EGTGSESGDKKSS 728
N. caninum Mt-Hsp70   697 TLQEASWKISQQAYNQAGSTDS--SSTGSESGSPSGDKKSS 736
E. tenella Mt-Hsp70   640 ALQELSWKVSQQAYSQSNNT----SADGDSSTSSGDSSSKP 677
T. parva Mt-Hsp70     640 QLQELSWKVSQAAYSKSNTGSTSANASENANTSNEENDTHNK 681
T. annulata Mt-Hsp70  640 QLQELSWKVSQAAYSKSNTGATSANTSENTNTFNEENDTPNK 681
B. bovis Mt-Hsp70     619 RLQELSWKVSQQMYQGNQ------QASEKSGSDSDPESEEKK 654
C. muris Mt-Hsp70     631 QLQEVSWKISQKAYGSAYSNNT---------TDSPSGSSEGSI 664
C. parvum Mt-Hsp70    642 ELKQASWAITQKAYKPGNSDNQSSENYANHEDNSCESQDSDS 683
```

Supplementary Data S3I

```
                                         10        20        30        40        50        60        70        80        90       100       110       120
                                ....|....|....|....|....|....|....|....|....|....|....|....|....|....|....|....|....|....|....|....|....|....|....|....|
Consensus                    1  MAslNKk~ivkIlE~CvKns~LKdksRqlCTSktNlNRASGDIIGIDLGTTNSCVAIMEGKQGKVIENAEGFRTTPSVVAFTNDNQRLVGIVAKRQAITNPENTVYATKRFIGRKFDEDA 117
P. vivax Mt-Hsp70            1  MASLNKKNIVKILERCVKNSVLKDKSRQLCTSKTNLNRASGDIIGIDLGTTNSCVAIMEGKQGKVIENAEGFRTTPSVVAFTNDNQRLVGIVAKRQAITNPENTVYATKRFIGRKFDEDA 120
P. knowlesi Mt-Hsp70         1  MASLNKKSIVKILESCVKNSILKDNSRQLCTSRTNLNRASGDIIGIDLGTTNSCVAIMEGKQGKVIENAEGFRTTPSVVAFTNDNQRLVGIVAKRQAITNPENTVYATKRFIGRKFDEDA 120
P. yoelii yoelii Mt-Hsp70    1  MAGFNKNGLSRIVETCLKRNPLKNKIREICTSKVNHNRASGDIIGIDLGTTNSCVAIMEGKQGKVIENAEGFRTTPSVVAFTNDNQRLVGIVAKRQAITNPENTVYATKRFIGRKFDEDA 120

                                ....|....|....|....|....|....|....|....|....|....|....|....|....|....|....|....|....|....|....|....|....|....|....|....|
Consensus                  118  TKKEQKNLPYKIVRAPNGDAWIEAQGKKYSPSQIGACVLEKMKETAENYLGRKVHQAVITVPAYFNDSQRQATKDAGKIAGLDVLRIINEPTAAALAFGLEKSDGKVIAVYDLGGGTFDV 237
P. vivax Mt-Hsp70          121  TKKEQKNLPYKIVRAPNGDAWIEAQGKKYSPSQIGACVLEKMKETAENYLGRKVHQAVITVPAYFNDSQRQATKDAGKIAGLDVLRIINEPTAAALAFGLEKSDGKVIAVYDLGGGTFDV 240
P. knowlesi Mt-Hsp70       121  TKKEQKNLPYKIVRAPNGDAWIEAQGKKYSPSQIGACVLEKMKETAENYLGRKVHQAVITVPAYFNDSQRQATKDAGKIAGLDVLRIINEPTAAALAFGLEKSDGKVIAVYDLGGGTFDV 240
P. yoelii yoelii Mt-Hsp70  121  TKKEQKNLPYKIVRAPNGDAWIEAQGKKYSPSQIGACVLEKMKETAENYLGRKVHQAVITVPAYFNDSQRQATKDAGKIAGLDVLRIINEPTAAALAFGLEKSDGKVIAVYDLGGGTFDV 240

                                ....|....|....|....|....|....|....|....|....|....|....|....|....|....|....|....|....|....|....|....|....|....|....|....|
Consensus                  238  SILEILGGVFEVKATNGNTSLGGEDFDQRILEYFIaEFKKKENIDLKnDKLALQRLREAAETAKIELSSKTQTEiNLPFITANQTGPKHLQIKLTRAKLEELCHDLLKGTIePCEKCIKD 357
P. vivax Mt-Hsp70          241  SILEILGGVFEVKATNGNTSLGGEDFDQRILEYFIAEFKKKENIDLKNDKLALQRLREAAETAKIELSSKTQTEINLPFITANQTGPKHLQIKLTRAKLEELCHDLLKGTIDPCEKCIKD 360
P. knowlesi Mt-Hsp70       241  SILEILGGVFEVKATNGNTSLGGEDFDQRILEYFIAEFKKKENIDLKDDKLALQRLREAAETAKIELSSKTQTEINLPFITANQTGPKHLQIKLTRAKLEELCHDLLKGTIEPCEKCIKD 360
P. yoelii yoelii Mt-Hsp70  241  SILEILGGVFEVKATNGNTSLGGEDFDQRILEYFINEFKKKENIDLKNDKLALQRLREAAETAKIELSSKTQTEVNLPFITANQTGPKHLQIKLTRAKLEELCHDLLKGTIEPCEKCIKD 360

                                ....|....|....|....|....|....|....|....|....|....|....|....|....|....|....|....|....|....|....|....|....|....|....|....|
Consensus                  358  ANvkKDEINEIILVGGMTRMPKVSeTVKqIFQNnPSKSVNPDEAVALGAAIQGGVLKGEIKDLLLLDVIPLSLGIETLGGVFTKLINRNTTIPTKKSQiFSTAADNQTQVSIKVFQGERE 477
P. vivax Mt-Hsp70          361  ANVRKDEINEIILVGGMTRMPKVSETVKQIFQNNPSKSVNPDEAVALGAAIQGGVLKGEIKDLLLLDVIPLSLGIETLGGVFTKLINRNTTIPTKKSQIFSTAADNQTQVSIKVFQGERE 480
P. knowlesi Mt-Hsp70       361  ANVKKDEINEIILVGGMTRMPKVSETVKQIFQNNPSKSVNPDEAVALGAAIQGGVLKGEIKDLLLLDVIPLSLGIETLGGVFTKLINRNTTIPTKKSQIFSTAADNQTQVSIKVFQGERE 480
P. yoelii yoelii Mt-Hsp70  361  ANIKKDEINEIILVGGMTRMPKVSDTVKEIFQNSPSKSVNPDEAVALGAAIQGGVLKGEIKDLLLLDVIPLSLGIETLGGVFTKLINRNTTIPTKKSQVFSTAADNQTQVSIKVFQGERE 480

                                ....|....|....|....|....|....|....|....|....|....|....|....|....|....|....|....|....|....|....|....|....|....|....|....|
Consensus                  478  MACDNKMLGSFDLVGIPPAPRGVPQIEVTFDVDANAIINISAIDKMTNKKQQITIQSSGGLSKEEIEKMVQEAELNREKDQqKKNLTDSKNEAETLIYSSEKQLDDFKDKISDsDKdELr 597
P. vivax Mt-Hsp70          481  MACDNKMLGSFDLVGIPPAPRGVPQIEVTFDVDANAIINISAIDKMTNKKQQITIQSSGGLSKEEIEKMVQEAELNREKDQQKKNLTDSKNEAETLIYSSEKQLDDFKDKISDADKDELR 600
P. knowlesi Mt-Hsp70       481  MACDNKMLGSFDLVGIPPAPRGVPQIEVTFDVDANAIINISAIDKMTNKKQQITIQSSGGLSKEEIEKMVQEAELNREKDQQKKNLTDSKNEAETLIYSSEKQLDDFKDKISDSDKDELR 600
P. yoelii yoelii Mt-Hsp70  481  MACDNKMLGSFDLVGIPPAPRGVPQIEVTFDVDANAIINISAIDKMTNKKQQITIQSSGGLSKEEIEKMVQEAELNREKDQHKKNLTDSKNEAETLIYSSEKQLDDFKDKISDSDKEELK 600

                                ....|....|....|....|....|....|....|....|....|....|....|....|...
Consensus                  598  QkIttLREKLsseDLd~IKDATKQLQEKSWAISQEMYKNNAsQgaQQEQpk~E~KtEEnKDNA  657
P. vivax Mt-Hsp70          601  QRITTLREKLSSEDLEAIKDATKQLQEKSWAISQEMYKNNASGAQQEQPKGEGKTEENKDNA  663
P. knowlesi Mt-Hsp70       601  QKITTLREKLSTEDLDGIKDATKQLQEKSWAISQEMYKNNASQNAQQEQPKSESKTEESKDNA 663
P. yoelii yoelii Mt-Hsp70  601  QKISALREKLTSDDLDSIKDATKQLQEKSWAISQEMYKNNAQQGSQQEQNTTENKAEENKDNA 663
```

Supplementary Data S3J

```
                           10        20        30        40        50        60        70        80        90       100       110       120
                  ....|....|....|....|....|....|....|....|....|....|....|....|....|....|....|....|....|....|....|....|....|....|....|....|
Consensus           1  MaseKviGIDLGTTnScvavmegg~pk~vipNaEGaRtTPSvVaftd~gErLvGq~AKRQaitNPenTifsiKRliGR~fddp~v~kdkk~lPyki~vkad~ngda~vei~gk~ysPqeiSAm  110
R. bacterium DnaK   1  MA--KVIGIDLGTTNSCVAIMDGSQPKIVENAEGARTTPSLVAFKN-DERLVGQAAKRQAVTNPENTLFAVKRLIGRSINDPLIKKEMKHLPFNVVDGGNGAAWVQVEGEKYSPSQISAF  117
B. henselae DnaK    1  MA--KVIGIDLGTTNSCVAVMDGKNAKVIENSEGARTTPSVVAFTDGGERLVGQPAKRQAVTNPEGTIFAVKRLIGRRFDDPMVEKDKALVPYKIVKGDNGDAWVEEAGKKYSPSQISAM  118
O. anthropi DnaK    1  MA--KVIGIDLGTTNSCVSVMDGKNAKVIENAEGARTTPSIVAFTDSDERLIGQPAKRQAVTNPEGTIFAVKRLIGRRFDDPMVTKDKDLVPYQIVKGDNGDAWVEHGKKYSPSQVSAM  118
Chelativorans DnaK  1  MA--KVIGIDLGTTNSCVAVMDGKDAKVIENAEGARTTPSIVAFTDSDERLVGQPAKRQAVTNPENTFFAIKRLIGRRFEDPMVEKDKKLVPYKILKADNGDAWVESHGTKYSPSQISAM  118
M. alhagi DnaK      1  MA--KVIGIDLGTTNSCVAVMEGGDARVIANAEGNRTTPSIVAFTDGDERLVGQPAKRQAVTNPENTIFAVKRLIGRRYDDPVTEKDKKLVPYKILVGDNGDAWVEAGAKKQSPSQISAM  118
D. indicum DnaK     1  MG--KVIGIDLGTTNSCVAVMEGGDAKVIANAEGNRTTPSIVAFTDKDERLVGLTAKRQAVTNPEGTIFAIKRYIGRKYDSDIVQRDVKSLPFKLSGASNGDVRIGAQGKDYSPQEISAM  118
Leptospirillum DnaK 1  MG--KVIGIDLGTTNSCVSIMEGGEPVVIPNQEGARITPSVVAFTDKGEVLVGQVAKRQAITNPENTVYSVKRLIGRKFDSEEVAHAMKRLPYKVVKAPNGDAHVEIRGKVYSPAEISAK  118
T. yellowstonii DnaK 1 MG--KAIGIDLGTTNSVVAVVGGEPVVIPNQEGGRTTPSVVAFTDKGERLVGQAAKRQAITNPENTIFSIKRLMGRKYNSQEVQEAKKRLPYKIVEAPNGDAHVEIMGKRYSPPEISAM  118
N. defluvii DnaK    1  MG--KVIGIDLGTTNSCVAIMSGGDPVVIANAEGSRTTPSVVGITDKNERLVGQIAKRQAITNPENTIFSVKRLMGRKFRSKEVQEAMKRLPYKVVEADNGDAHVELRGKRYSPPEVSAM  118
M. tarda DnaK       1  MAKEKILGIDLGTTFSCMAIMEAGKPIIIPNEGARTTPSVVAFTGERLVGQORTTIQSIKRKMG----------------------TSEKVIDDKSYTPQEISAM  96
M. marisnigri DnaK  1  MVSEKVLGIDLGTTNSCMAIMEGGRATVIANAEGRTTPSVVAFSKEGERLVGNVAKRQAITNPNRTVQSIKRRMC----------------------TNEKVTIGDKTYTPQEISAM  96
M. petrolearius DnaK 1 MASNKVLGIDLGTTNSCMSIMEGGKPVVIPNAEGARTTPSVVAFSKDGERLIGSVAKRQAVTNPDRTIISIKRDMG----------------------TDRKIKDDKVFTPQEISAM  96
M. mahii DnaK       1  MG--KILGIDLGTTNSCMAVIEGGEPTVLPNAEGSRTTPSVVGESKKGEKLVGQVAKRQMVANPNNTVSSIKRHIG----------------------EGDYQVTLNDKEYTPQEISAM  95
M. zhilinae DnaK    1  MG--KILGIDLGTTNSCMAVIEGGKPTVIPNAEGGRTTPSVVGFSKKGEKLVGQVAKRQLIANPENTVYSIKRHIG----------------------KADYKVTLRDKDYTPQEISAM  95
M. mazei DnaK       1  MA--KILGIDLGTTNSCVAVMEGGEAVVIPNAEGSRTTPSVVGESKKGEKLVGQVAKRQAISNPDNTVYSIKRHMC----------------------EANYKVTLNGKDYTPQEISAM  95

Consensus          111  iLqKmketAESYlGEkvt~aViTVPAYFnDaQRqaTKdAGkIAGleVlRIINEPTAaaLAYGldk~~gd~tiaVyDLGGGTFDvSileigdGvFEVksTnGdthLGGdDFD~rivdyl~d  224
R. bacterium DnaK  118  ILQKMKETAECYLGETVSEAVITVPAYFNDAQRQATKDAGKIAGLNVLRIINEPTAAALAYGLDKEG-GKTIAVYDLGGGTFDVSIMEIDDGLFEVKSTNGDTALGGEDFDMRLVEYLAE  236
B. henselae DnaK   119  ILQKMKETAESYLGEKVEQAVITVPAYFNDAQRQATKDAGKIAGLEVLRIINEPTAAALAYGLDKKD-GKTIAVYDLGGGTFDISVLEIGDGVFEVKSTNGDTFLGGEDFDMRLVGYFAD  237
O. anthropi DnaK   119  ILQKMKETAESYLGEKVTQAVITVPAYFNDAQRQATKDAGKIAGLEVLRIINEPTAAALAYGLDKNE-GKTIAVYDLGGGTFDVSILEIGDGVFEVKSTNGDTFLGGEDFDMRLVEYIVA  237
Chelativorans DnaK 119  ILQKMKETAESYLGEKVTQAVITVPAYFNDAQRQATKDAGKIAGLEVLRIINEPTAAALAYGLDKKE-GKTIAVYDLGGGTFDISILEIGDGVFEVKSTNGDTFLGGEDFDMRLVEYIAA  237
M. alhagi DnaK     119  ILQKMKETAEAYLGEKVTQAVITVPAYFNDAQRQATKDAGKIAGLEVLRIINEPTAAALAYGLDKKE-GKTIAVYDLGGGTFDISVLEIGDGVFEVKSTNGDTFLGGEDFDMRLVEYLAA  237
D. indicum DnaK    119  VLQKMRQTAEDYLGETVTEAVITVPAYFNDSQRQATKDAGKIAGLEVLRIINEPTAAALAYGMDKKG-DEVVAVFDLGGGTFDVSILEISEGVFEVKSTNGDTHLGGEDFDQRILNYVAD  237
Leptospirillum DnaK 119 ILLKLPKQAAEDYLGEKVTEAVITVPAYFNDAQRQDTKNAGAIAGLNVLRIINEPTAASLAYGLDSKK-EEKIAVYDLGGGTFDISILEIGDGVFEVKSTNGDTYLGGDDFDLKIIDYLVD  237
T. yellowstonii DnaK 119 ILLKLQAAEDYLGEKVTEAVITVPAYFDDSQRQATKDAGRIAGLNVLRIINEPTAAALAYGLDKKK-EEKIAVYDLGGGTFDISILEIGEGVIEVKATNGDTYLGGDDFDIRVMDWLIE  237
N. defluvii DnaK   119  ILQKMRQTAEDYLGEKVTEAVVTVPAYFDSSQRQATKDAGRIAGLNVLRIINEPTAAALAYGLDKKK-DERIAVYDLGGGTFDVSILEIGDGVFEVKSTNGDTYLGGDDFDERVMDWLVE  237
M. tarda DnaK       97  ILQKLKADAEAYLGEKITKAVITVPAYFNDAQRQATKDAGRIAGLEVMRIINEPTASALAYGIDKEN-DATVLVYDLGGGTFDVSILTLGDGVFEVKATAGNNHLGGDDFDNRIIDYLVE  215
M. marisnigri DnaK  97  ILQMKLDAEAYLGEKIAKAVITVPAYFNDAQRQATKDAGTIAGLEVLRIINEPTAASALAYGIDREG-DSIVLVYDLGGGTFDVSILQLGDGVFEVKSTACNNRLGGDDFDTRVVDYLAD  215
M. petrolearius DnaK 97 ILQKLKADAEAYLGEKKAVITVPAYFNDAQRNATKDAGKIAGLEVLRIINEPTAASALAYGIDKA-DAIVLVEDLGGGTFDVSILTLGDGVFEVQATAGDNHLGGDDFDHLVKYLVD  215
M. mahii DnaK       96  ILRKLKEDAESYIGETITDTVITVPAYFNDSQRQATKDAGKIAGLEVLRIINEPTAASLAYGLDKETGDHKILVYDLGGGTFDVSILESGDTHLGGDDFDDRIVNHIIS  215
M. zhilinae DnaK    96  ILRKMKEDAESYLGEKVTIDEAVITVPAYFDDSQRQATKDAGKIAGFEVKRIINEPTAASLAYGLDSEGDQKILVYDLGGGTFDVSVLELGDGVFEVKSTGDTKLGGDDFDQKIVDYVVS  215
M. mazei DnaK       96  ILQKLKADAEAYLGETIKQAVITVPAYFNDSQRQATKDAGAIAGLEVLRIINEPTAASLAYGLDKGDIDQKILVYDLGGGTFDVSILEGGGVFEVKSTSGDTHLGGDDFDQRVIDYILA  215

Consensus          225  EFkKeeGidL~~~Dkm~alQRlkeAAEkAKiELSss~qteiNlPfitad~sGGqPkhl~iklRakfeqlv~DLigrtlepck~aLkDAgl~agdidevvLvGGmTRmPkvqe~~Vk~ffgKe  335
R. bacterium DnaK  237  EFKKENGVDLTKDKMALQRLKEAAEKAKIELSSSAQTEINQPFISMDPNGGQPLHLMKLTRAKLESLVADLIKRSLKPCAALKDAGLSKSDIDEIVLVGGMTRMPKVLEAVTEFFGKE  356
B. henselae DnaK   238  EFKKEEGIDLKNDKLALQRLKEAAEKAKIELSSSQQTEINLPFITADQPGG-PKHLTMKLTRAKFBSLVADLIKVRTVEPCKAALKDAGLKAGEIDEVVLVGGMTRMPKIQEVVQSFFGKD  355
O. anthropi DnaK   238  EFKKESGIDLKNDKLALQRLKEAAEKAKIELSSAQQTEVNLPFITADQTG--PKHIAIKLSRAKFESLVDDLIVQRTIEPCKAALKDAGLKAGEIDEVVILGGMTRMPKIQEVVKAFFGKE  355
Chelativorans DnaK 238  EFKKEQGIDLKNDKLALQRLKEAAEKAKIELSSQQTEINLPFITADQTG--PKHLAIKLTRAKFESLVEDIVTRTIEPCRAALKDAGLSAGEIDEVVLVGGMTRMPKVQETVKNFFGKE  355
M. alhagi DnaK     238  EFKKEQGIDLKNDKLALQRLKEAAEKAKIELSSSQTEINLPFITADQSC--PKHLVILKTRAKFESLVDDLVDLVLVPVERVLTIEPCRAALKDAGLKAGEIDEVVLVGGMTRMPKVQKQFFGKE  355
D. indicum DnaK    238  EFKKESGIDLRNDKMALQRLKEAAEKAKHELSGSMETEINLPFITMDASG--PKHIQIKLTRAKLEQLVEDLIERTIEPCRKALKDAGLSKASDIDEVVLVGGMTRMPKVQQKVEFFGKE  355
Leptospirillum DnaK 238 EFKKENGIDLKKDKMALQRLKEAAEKAKIELSTALETEINLPFYITADQTG--PKHLVLKLSRSRLEQLVGDLIQHSLEPVRKALDAGMNTGMIDEVVLVGGQTRMPKVQEAVKSFFGKE  355
T. yellowstonii DnaK 238 EFKKQEGIDLRKDRMALQRLKEAAERAKIELSSSAQMETEINLPFITADASG--PKHIMLKTRAKLEQLVDDLIQKSLEPCKKALSDAGLISQSQIDEVILVGGQRPKVQKVVQDFFGKE  355
N. defluvii DnaK   238  EFKKDQGIDLRKDRMALQRLKEAAERAKIELSSSQEKTEINLPFITADASC-----PKHLMVTKLTRAKLEQLVEDLIQKRTIEPCKRALDAGVSAKDIQEVVLVGGMTRMPKIQVVKEFFGKE  355
M. tarda DnaK       216 EFQKKEGINIRSDPIAMQRLRDAAENAKIELSQKMKTNINLPYITTGPDG--PKFETDIDLTRAKFEQLIGDVESTVIPVKQALSDAKLTPDQIDFVLLVGGSTRVPLVQETVRKILGKE  333
M. marisnigri DnaK  216 EFRKKEGIDLRNDPVAMQRIRDAAENAKIELSTVQKTNINLPYITTTESC--PKFLDVDLTRAKFEQLIGDVDSTLGPVKQALSDAKLGADDIDHILLVGGSTRVPLVQETVKKVLKKE  333
M. petrolearius DnaK 216 EFKKQEGIDLSSDKMAMQRLRDAAENAKKELSTVQKTNINLPYITTDSSG--PKFETDIDLTRAKFEQLIGDIVDRTIGPVKQALSDAKTAKDIDFVLLVGGSTRVPLVQERVKALLAKE  333
M. mahii DnaK      216  EFKKEEGIDLSGDKAAMQRFKDAAEKAKIELSGVGTTNINLPFITADSNG--QPKHVDIDIDITRAQFEKMTEDLIEKTFEKMTEDLIEKLHSMRQSLSDAKLTPDQIDEVLILGGSTRMPAVVRTVKEIGKD  334
M. zhilinae DnaK   216  EFKKEEGIDLSKDRSALQRITLKDAAEKAKIELSGVTTNINLPFVTADSNG--QPKHIDIDITRTQFEKMTEDIVKLTLHSMRQSLSDAKLTPDIDKVLLVGGSTRMPAVYETVKKEIGKD  334
M. mazei DnaK      216  EFKKSEGIDLSKDKAVLQRLKDAAEKAKIELSGVANTNINLPFLTVGTDG--EPKHMIDIDLTRAQFQKMTEDLLEKTLVSMRRALSDAKLTPNDLDKVILVGGSTRMPAVVPLVENRTGKK  334
```

Supplementary Data S3J

```
                              ....|....|....|....|....|....|....|....|....|....|....|....|....|....|....|....|....|....|....|....|....|....|....|....|
Consensus              336    PhkgvNPDEvVA~GAaiQagvl~Gevkdvll LLDVtPLsLgIETlGgvfTklIeRNTTIPtkksQvFsTAaDnQtsVtihVfQGERemAadNk~LGqF~L~gIPPAPRG~PQiEVtFDIDa 449
R. bacterium DnaK      357    PHKGVNPDEVVALGAAIQAGVLQGDVKDVLLLDVTPLSLGIETLGGVFTRLIDRNTTIPTKKAQVFSTAEDNQNAVTIRVFQGEREMAADNKVLGQFNLEDIPPAPRGLPQIEVAFDIDA 476
B. henselae DnaK       356    PHKGVNPDEVVAMGAAIQGGVLQGDVKDVLLLDVTPLSLGIETLGGVFTRLIERNTTIPTKKSQVFSTADDNQNAVTIRVFQGEREMAANDKLLGQFDLVGIPPAPRGVPQIEVTFDIDA 475
O. anthropi DnaK       356    PHKGVNPDEVVAMGAAIQAGVLQGDVKDVLLLDVTPLSLGIETLGGVFTRLIERNTTIPTKKSQTFSTADDNQSAVTIRVFQGEREMAADNKMLGQFDLVGIPPAPRGVPQIEVTFDIDA 475
Chelativorans DnaK     356    PHKGVNPDEVVAMGAAIQAGVLQGDVKDVLLLDVTPLSLGIETLGGVFTRLIDRNTTIPTKKSQVFSTAEDNQNAVTIRVFQGEREMAADNKVLGQFDLVGIPPAPRGMPQIEVTFDIDA 475
M. alhagi DnaK         356    PHKGVNPDEVVALGAAIQAGVLQGDVKDVLLLDVTPLSLGIETLGGVFTRLIERNTTIPTKKSQVFSTAEDSQSAVTIRVFQGEREMAADNKALGQFDLVGIPPAPRGVPQIEVTFDIDA 475
D. indicum DnaK        356    PHKGVNPDEVVAIGAAIQAGVLKGDVKDVLLLDVTPLSLGIETLGGVMTKVIERNTTIPTKRSQVFSTAADNQTSVSVHVLQGEREMAADNKSLGRFDLVEIPPAPRGMPQIEVTFDIDA 475
Leptospirillum DnaK    356    PHKGVNPDEVVAMGAAIQAGVLQGDVKDVLLLDVTPLSLGIETLGGVFTRLIERNTTIPAKKKSQVFTTAADNQSAVTIRVFQGEREMAADNKLLGQFDLEGIPPAPRGVPQIEVTFDIDS 475
T. yellowstonii DnaK   356    PHKGVNPDEVVAVGAAIQAAILKGEVKEVLLLDVTPLSLGIETLGGVFTKIIERNTTIPTKKSQIFTTAADNQTAVTIKVYQGEREMAADNKLVQFELVGIPPAPRGIPQIEVTFDIDA 475
N. defluvii DnaK       356    PHRGVNPDEVVAGAVQGGVLKGEVKDVLLLDVTPLSLGIETLGGVFTKLIERNTTIPTKKSQVFSTADDNQTAVTIRVFQGEREMANDNKLLGQFDLVGIPPAPRGMPQVEVSFDIDA 475
M. tarda DnaK          334    PDKGINPDECVALGAAIQAGAVLSGETKDIVLLDVTPLTGIETLGGIATKLIERNTTIPTRKSQIFSTAADGQTSVEIHVVQGERALAKDNFTLGRFTLTGIPPAPRGVPQIEVTFDIDA 453
M. marisnigri DnaK     334    PDKGINPDECVALGAAIQAGVLTGETKDVLLLDVTPLTGIETLGGIATKLIERNTTIPTRKSQIFSTAADGQTSVEIHVVQGERALAKDNFTLGRFQLTGIPPAPRGIPQIEVTFDIDA 453
M. petrolearius DnaK   334    PDKGINPDECVAGASIQGGVLTGEAKDVLLLDVTPLTLSIETLGGIATKLIERNTTIPTKKSQIFTTAADNTSVEIHVVQGERAMASDNFTLGRFQLTGIPPAPRGIPQIEVTFDIDA 453
M. mahii DnaK          335    PYKNINPDEAVGAAIQAGVMGGEVEDVLLLDVTPLTLGIETMGGVATPLIERNTTIPTRKSQVFSTAADNQSSVEIHVLQGERGIASANKTLGRFVLDGIPPAPRGMPQIEVTFDIDA 454
M. zhilinae DnaK       335    PYKNINPDEAVAMGAAIQAGVLAGEKDVLLLDVIPLTLGIETLGDVATPLIERNTTIPTKKSQIFSTAADNQTSVEIHVLQGERGIASANKTLGRFTLEGIPPAPRGVPQVEVTFDIDS 454
M. mazei DnaK          335    PYKNINPDEAVAIGAAIQAGVLGGEVKDVLLLDVTPLTLGIETLGGIATPLIQRNTTIPTKKSQIFSTAADNQPSVEIHVLQGERGIASENKTLGRFILDGIPPAPRGIPQIEVTFDIDA 454

                              ....|....|....|....|....|....|....|....|....|....|....|....|....|....|....|....|....|....|....|....|....|....|....|....|
Consensus              450    NGIvhVsAKD~gTgkeqsI~IqPasgglSde~eiekmvkdAe~haeeDkkrre~vEarNqae~lihstEks~lke~gdki~s~dk~~iesaia~lkkale~DTg~da~aikak~e~Laeasmk 557
R. bacterium DnaK      477    NGIVSVSAKDKGTNKEQKITIQ-ASGGLSDDIEQMVKDABENAESDKKRELVEAKNSABSLLHSTEKSVEEHGDKVDPTTIEVIEMAMNLTESLE--SDDASKINSRAQDLTEASMK 593
B. henselae DnaK       476    NGIVHVSAKDKGTGKEHQIRIQ-ASGGLSDADIEKMVKDAEAHAEEDKKRRDAVEABALIHSTEKSLTEYGDKVSTEEKEQIETAISDLKSVLD--STDTEEVKAKMQKLAEVSMK 592
O. anthropi DnaK       476    NGIVNVTAKDKGTGKEHQIRIQ-ASGGLSDADIEKMVKDAEANAEDKKRDAVEAKNQGPSLVHSTEKSLSEYGDKVSADDKKALEDALASLKTSLE--GDDAEDIKAKTQTLABASMK 592
Chelativorans DnaK     476    NGIVNVSAKDKGTGKEHQIRIQ-ASGGLSDAEIEKMVKDAEANAEADKKRRETVEVKNQQBALIHSTEKSLKDYGDKVSEDDRKAIENAIABLKTATE--GEDADAIRAKTTALEASMK 592
M. alhagi DnaK         476    NGIVHVSAKDKGTGKEHQIRIQ-ASGGLSDDEIEKMVKDAEANADKKRRALVEANKQSVTHVSEKSKEYGDKVSEADRTAISDAIAALKTATE--GDDVDDINAKSQSLAEVSMK 592
D. indicum DnaK        476    NGIVSVSAKDKGTGKEQQSIVIR-DSSGGLSDEEIEKMVKDAELHADEDKRKKERIETLNQDTLVYSTEKAMGEHGDKLQPEDKGRIEEKLAALKNLIQDTGASKEQLEAATKELSEASMK 594
Leptospirillum DnaK    476    NGIVHVGAKDKGTGKEQNIHIT-AQGGLSKPEEIDRLIREABSHAAEDKARRERIELRNNLETLVYSTEKSINEIGDKRISDVERGSIREATESAKGKLT--SEDKDVLQQAFKDLETQSHK 592
T. yellowstonii DnaK   476    NGILHVSAKDLATGKEQSIRIT-ASSGLSEEEIKKMIREABAHAEEDRRKQIAEARNEADNMIYTVEKTIRDMGDRISEDERKRIEEAIEKCRRIKD-TSNDVNEIKAAVEELAKASHK 593
N. defluvii DnaK       476    NGIVHVAAKDLATQKESIKIT-ASSLKSEVDKIVKDAQSHTEDKKRRVAEARNQADSLLYSTEKNLSEHGDKIGEDDKTKITEAIAGIRKAME--CDDPAAIETATQTLTTASHK 592
M. tarda DnaK          454    NGILHVSAKDIGTGNEQAITTK-GDKKLSEEIKRMVEDARKFEEDDRKKREEIEIRNSADNAIFTAELLKDNAATIEPADREKVESGIADLRKAIG-GDQVDDIKAKMESLTESVYN 570
M. marisnigri DnaK     454    NGIVHVSAKDIGTGNEQSITIKPQDSRPSEABIQRMDEAKKFEADQKNREEIEIRNTADTAIFTAERAMKDAGDSIEAADKKIESAIADLKKALE--CDDLEAIKQKMDALTEAVYA 571
M. petrolearius DnaK   454    NGIINVSAKDIGTGNQQAITTS-GDKRLSKDEIDEMVNKAFDFEDADKKREEIEIKNSADNSVFAAEKLLKDSADKIEADKGRIESALEKVKNTME--GEDIEALKKDVEELQEAVFA 570
M. mahii DnaK          455    NGILHVSAKDRGTGKEQSITIE-KPGGLSDEEIDQMVKDAEHAEEDKAKKEEVEARNNAETLVSSAEKALKDAGDIASEEQKTKVESAISDINTALE--GSDIEAIKSKTETLQAMYE 571
M. zhilinae DnaK       455    NGILHVSAKDIGTGKQQSISID-KPGGLSDEEIEQMVRKDAEHAEEDRQRKEAVETRNTAESLINSAERTIKEAEDIATDDQKSRVESEIEELRKALE--CDDIEEIKKKTESLQNAVYE 571
M. mazei DnaK          455    NGILHVSAKDIGTGKQQSISIQ-KPGGLSDDEIERMVKDAEMHAEEDRRKKEEVEIRNNABALINAAEKTIKEAGDIATEDQKSKVNAAIEDLKKALE--CKDAEDIKAKTEALQESVYP 571

                              ....|....|....|....|....|....|....|....|....|....|....|
Consensus              558    l~~amYgkaqaaa~a~~~a~gg~g~ag~teaa~~dedvvdAdfe~evddDDDKkkkk 601
R. bacterium DnaK      594    LGEAIYKAQSEAPEAPEG----MGDDDAGEPRGVDEDIVDADFEDLGK-----DDQ 640
B. henselae DnaK       593    LGQAMYEASQAA---------------TPNTETDTKSDDVVDADFEEIND-----KKK 630
O. anthropi DnaK       593    LGQAMYEAAQAAETGS---------AGGSEEAASNDDVVDADYEEIDDD---KKKS 636
Chelativorans DnaK     593    LGQAVYEASQAENAA---------GGTEETGGAKDDVVDADFEEIDEDD----KKSA 636
M. alhagi DnaK         593    LGQAMYEASQKEAAEA---------DAAADAAKDGADVVDADFEEINEDDDKKKSA 639
D. indicum DnaK        595    LGEIVYQQAQQEQQA GENQADASGDAGSSEGNKADDDVVDAEFEDMDD----KEKK 646
Leptospirillum DnaK    593    LAELVYKASQSTPGGEP---GAGTAGGSSAQGEDPTVVDAEYEDVNQ-----KKN 640
T. yellowstonii DnaK   594    VAEELYMKKAGASQQG---------AGSTTQSKKEEDVIEAEVED-------KDNK 632
N. defluvii DnaK       593    LAEEMYKKASAAAGAGAGADAAASSGDGGAQAKTDEKVVDAEFEEVDK-----DKK 643
M. tarda DnaK          571    ITTKIYQKAQAEROQAGASGNPGGPGTKAEPKDDTVVDADFKVKD--------E 618
M. marisnigri DnaK     572    LTTKMYQQAQAAAAQQQQTEG---------AAKKDDTVVDADYEV---------KE 609
M. petrolearius DnaK   571    ATTKIYQKAEAEKAKAGEA---------GSAGEDETVVDADYEVKDD-----EKKE 614
M. mahii DnaK          572    ISALLYQKAQEEAEASGQAAGGEGASASDSEGGSDEDVVDADFEEVDD-----NDKK 623
M. zhilinae DnaK       572    VSSAMYQKAQEEAAAQQQQEG---GAETAGSESQPDEDVVDADYEVVDE-----DKK 620
M. mazei DnaK          572    ISTAMYQKAQQAQQA----AGGEGGAAGTDARGPDETVVDADYEVVDD----EKRK 619
```

Supplementary Data S4

| Model | Dope Z-score | GDT_TS | ProSA | Verify3D |
|---|---|---|---|---|
| PfHsp70-1_1YUW.pdb | -0.90 | 79.64 | -10.15 | 236.31 |
| PfHsp70-1_2KHO.pdb | -0.73 | 63.12 | -10.67 | 240.87 |
| PfHsp70-1_3D2F.pdb | -1.06 | 76.68 | -11.59 | 255.80 |
| PfHsp70-x_PfA(Hybrid) | -1.40 | 80.4 | -11.6 | 230.2 |
| PfHsp70-x_PfB(Hybrid) | -1.17 | 75.7 | -10.7 | 219.6 |
| PfHsp70-x_PfE(Hybrid) | -1.17 | 66.1 | -10.5 | 216.2 |
| PfHsp70-x_JA1(Hybrid) | -1.36 | 77.7 | -11.0 | 228.6 |
| PfHsp70-x_JA2(Hybrid) | -1.43 | 80.7 | -12.0 | 217.9 |
| PfHsp70-x_JB1(Hybrid) | -1.32 | 79.26 | -11.56 | - |
| PfHsp70-x_JB4(Hybrid) | -1.24 | 78.62 | -11.51 | - |
| PfHsp70-x_JC5(Hybrid) | -1.20 | 74.57 | -11.36 | - |
| PfHsp70-x_JC13(Hybrid) | -1.08 | 74.20 | -11.18 | - |
| PfHsp70-x_PfA(Hybrid-EM) | -1.47 | 80.9 | -11.8 | 221.5 |
| PfHsp70-x_PfB(Hybrid-EM) | -1.29 | 74.0 | -11.1 | 215.3 |
| PfHsp70-x_PfE(Hybrid-EM) | -1.31 | 66.3 | -10.8 | 215.7 |
| PfHsp70-x_JA1(Hybrid-EM) | -1.44 | 75.3 | -11.1 | 223.3 |
| PfHsp70-x_JA2(Hybrid-EM) | -1.51 | 82.0 | -12.0 | 224.5 |
| PfHsp70-x_JB1(Hybrid-EM) | -1.31 | 79.32 | -11.72 | - |
| PfHsp70-x_JB4(Hybrid-EM) | -1.27 | 81.10 | -11.67 | - |
| PfHsp70-x_JC5(Hybrid-EM) | -1.28 | 76.40 | -11.55 | - |
| PfHsp70-x_JC13(Hybrid-EM) | -1.25 | 75.85 | -11.59 | - |

Supplementary Data S4

| | | | | |
|---|---|---|---|---|
| PfHsp70-x_PfA(Haddock) | -1.35 | 78.0 | -11.1 | 218.4 |
| PfHsp70-x_PfB(Haddock) | -1.33 | 79.4 | -10.8 | 218.6 |
| PfHsp70-x_PfE(Haddock) | -1.31 | 77.7 | -10.8 | 215.6 |
| PfHsp70-x_JA1(Haddock) | -1.48 | 76.7 | -10.8 | 225.2 |
| PfHsp70-x_JA2(Haddock) | -1.49 | 85.8 | -11.6 | 212.1 |
| PfHsp70-x_JB1(Haddock) | -1.32 | 71.98 | -10.83 | - |
| PfHsp70-x_JB4(Haddock) | -1.31 | 77.48 | -10.75 | - |
| PfHsp70-x_JC5(Haddock) | -1.29 | 68.62 | -10.92 | - |
| PfHsp70-x_JC13(Haddock) | -1.20 | 65.78 | -10.93 | - |

**Supplementary Data S5:**

```
CrHsp70-4  1  MVGLG------------LIATLVAASALASIPQAKAASPTTDKLGTVIGIDL  40
CrHsp70-5  1  MAQWK------------AAVLLL---ALACASYGFGVWAEEEKLGTVIGIDL  37
OlHsp70-3  1  MPIKRASYRNAVVCACASLFLF---AVCALPIN---AENPTEITGTVIGIDL  46
HsGrp78    1  MKLSL------------VAAMLL---LLSAARAE--EEDKKEDVGTVVGIDL  35
PfHsp70-2  1  MKQIR-----------PYILLL---IVSLLKFI---SAVDSNIEGPVIGIDL  35
PfHsp70-x  1  MKTKICSYIH------YIVLFL---IATTTVHT--ASNNAEESEVAIGIDL  40
```

Supplementary Data S6:

Supplementary Data S7A

| PfHsp70-x | PfA | PfB | PfE | HmDNAJA1 | HmDNAJA2 | DnaJB4 | DnaJB1 | DnaJC5 | DnaJC13 |
|---|---|---|---|---|---|---|---|---|---|
| **Hydrophobic Interactions** | | | | | | | | | |
| **I246** | V121 | A123 | | | | | | P51 | |
| **F247** | | | | | | | | P48 | P1337 |
| **I410** | P111, F127 | | | | P37 | | P33 | | |
| **L411** | P111 | P117 | P109 | P35 | P37 | P33 | P33 | P44, F57 | P1333 |
| **Hydrogen Bonds** | | | | | | | | | |
| **L200** | D112 | D118 | D110 | | D38 | D34 | D34 | H43 | H1332 |
| **R201** | D112 | P117 | P109 | D36 | D38 | D34 | D34 | | |
| **L411** | H110 | | | | H36 | H32 | H32 | | |
| **A418** | K128 | | | K46 | K48 | K46 | K46 | | |
| **V419** | E124, K128 | | | K46 | K48 | | K46 | | E1344 |
| **K420** | E124 | D130 | | | | E42 | | | |
| **Ionic Interactions** | | | | | | | | | |
| **K189** | | D118 | D110 | D36 | | D34 | | D45 | |
| **R201** | D112 | D118 | D110, D115, E122 | D36 | D38 | D34 | D34 | D45 | D1334 |
| **K420** | | D130 | D115, E116, E122 | E43 | D45 | E42 | E42 | | E1344 |

Supplementary Data S7B

| PfHsp70-x | J-Conserved | PfA | PfB | PfE | HmDNAJA1 | HmDNAJA2 | DnaJB4 | DnaJB1 | DnaJC5 | DnaJC13 |
|---|---|---|---|---|---|---|---|---|---|---|
| **Hydrophobic Interactions** | | | | | | | | | | |
| **I246** | - | V22 | A18 | | | | | | P22 | |
| **F247** | - | | | | | | | | P16 | P16 |
| **I410** | - | P12, F28 | | | | P12 | | P12 | | |
| **L411** | **P12** | P12 | P12 | P12 | P12 | P12 | P12 | P12 | P12, F28 | P12 |
| **Hydrogen Bonds** | | | | | | | | | | |
| **L200** | **D13** | D13 | D13 | D13 | | D13 | D13 | D13 | H11 | H11 |
| **R201** | **D13** | D13 | P12 | P12 | D13 | D13 | D13 | D13 | | |
| **L411** | **H11** | H11 | | | | H11 | H11 | H11 | | |
| **A418** | **K29** | K29 | | | K29 | K29 | K29 | K29 | | |
| **V419** | **K29** | E25, K29 | | | K29 | K29 | | K29 | | E29 |
| **K420** | **E25** | E25 | D25 | | | | | E25 | | |
| **Ionic Interactions** | | | | | | | | | | |
| **K189** | **D13** | | D13 | D13 | D13 | | D13 | | D13 | |
| **R201** | **D13** | D13 | D13 | D13, D18, E25 | D13 | D13 | D13 | D13 | D13 | D13 |
| **K420** | **E25** | | D25 | D18, E19, E25 | E26 | D26 | E25 | E25 | | E29 |

Supplementary Data S7C

| PfHsp70-x | PfA | PfB | PfE | DnaJA1 | DnaJA2 | DnaJB4 | DnaJB1 | DnaJC5 | DnaJC13 |
|---|---|---|---|---|---|---|---|---|---|
| **Hydrophobic Interactions** | | | | | | | | | |
| **I246** | P111, F127 | A112, M113, F133 | M105 | Y26, A30, F45 | P41 | A28, L29, F45 | | A39, L40 | A1328 |
| **L411** | M107 | | L113 | | | | | P48 | P1337 |
| **V419** | Y134 | | | | | L29 | P39 | L40 | |
| **Hydrogen Bonds** | | | | | | | | | |
| **R201** | D112, K113 | P117, D118, H120 | H108, P109, H112, D115 | P35, D36, N40 | P37, D38 | P33, D34, E42, | H32, D34 | P44, D45, N47 | P1333, D1334 |
| **N204** | D112 | D118 | D110 | D36 | D38 | D34 | D34 | D45 | D1334 |
| **T207** | D112 | D118 | D110 | D36 | D38 | D34 | D34 | D45 | D1334 |
| **D216** | | | | | | K35 | K35 | K46 | K1335 |
| **E243** | | K109 | K111 | | K39 | | K35 | 36 | |
| **D244** | | | | R27 | | R25 | | 36 | Q1329 |
| **F247** | | H116 | | | D38, K39 | H32 | D34, K35 | H43 | H1332 |
| **K420** | | | | H34 | | K35 | E42 | K46 | K1330, K1335 |
| **D421** | H110 | | | K37 | | K35 | K37 | K46 | K1335 |
| **Ionic Interactions** | | | | | | | | | |
| **R185** | D112 | D118 | D115, E122 | D36 | | D34 | | | |
| **R201** | D112 | D118 | D115 | D36 | D38 | D34, E42 | D34 | D45 | D1334 |
| **D216** | | K119 | | | | K35 | K37 | K46 | K1335 |
| **E243** | | K109 | R101, K111 | | K39 | | K35 | 36 | |
| **D244** | | | | R27 | | R25 | | 36 | |
| **E248** | | E138 | R101, D110, K111 | | K33, K39 | | K35 | | |
| **D421** | | | | | | K35 | K37 | K46 | K1335 |

Supplementary Data S7D

| PfHsp70-x | J-Conserved | PfA | PfB | PfE | DnaJA1 | DnaJA2 | DnaJB4 | DnaJB1 | DnaJC5 | DnaJC13 |
|---|---|---|---|---|---|---|---|---|---|---|
| **Hydrophobic Interactions** | | | | | | | | | | |
| I246 | A7, M8, F28 | P12, F28 | A7, M8, F28 | M8 | Y26, A30, F28 | P22 | A7, L8, F28 |  | A7, L8 | A7 |
| L411 | - | M8 |  | L16 |  |  |  |  | P16 | P16 |
| V419 | - | Y35 |  |  |  |  | L8 | P22 | L8 |  |
| **Hydrogen Bonds** | | | | | | | | | | |
| R201 | P12, D13 | D13, K14 | P12, D13, H15 | H11, P12, H15, D18 | P12, D13, N23 | P12, D13 | P12, D13, E25 | H11, D13 | P12, D13, N15 | P12, D13 |
| N204 | D13 | D13 | D13 | D13 | D13 | D13 | D13 | D13 | D13 | D13 |
| T207 | D13 | D13 | D13 | D13 | D13 | D13 | D13 | D13 | D13 | D13 |
| D216 | K14 |  |  |  |  |  | K14 | K14 | K14 | K14 |
| E243 | R4, K14 |  | K4 | K14 |  | K14 |  | K14 | R4 |  |
| D244 |  |  |  |  | R4 |  | R4 |  | R4 | Q8 |
| F247 | H11 |  | H11 |  |  | D13, K14 | H11 | D13, K14 | H11 | H11 |
| K420 | K14 |  |  |  | H11 |  | K14 | E25 | K14 | K9, K14 |
| D421 |  | H11 |  |  | K14 |  | K14 | K16 | K14 | K14 |
| **Ionic Interactions** | | | | | | | | | | |
| R185 | D13 | D13 | D13 | D18, E25 | D13 |  | D13 |  |  |  |
| R201 | D13 | D13 | D13 | D18 | D13 | D13 | D13, E25 | D13 | D13 | D13 |
| D216 | K14 |  | K14 |  |  |  | K14 | K16 | K14 | K14 |
| E243 | R4, K14 |  | K4 | R4, K14 |  | K14 |  | K14 | R4 |  |
| D244 |  |  |  |  | R4 |  | R4 |  | R4 |  |
| E248 | K14 |  | E33 | R4, D13, K14 |  | K8, K14 |  | K14 |  |  |
| D421 | K14 |  |  |  |  |  | K14 | k14 | K14 | K14 |

**Supplementary Data S8:**

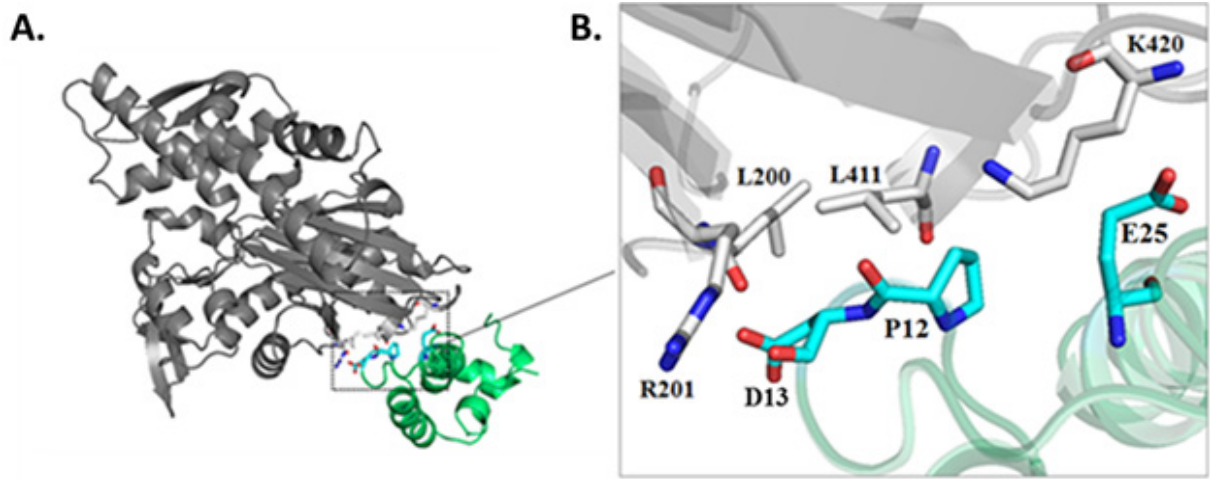

**Supplementary Data S9:**

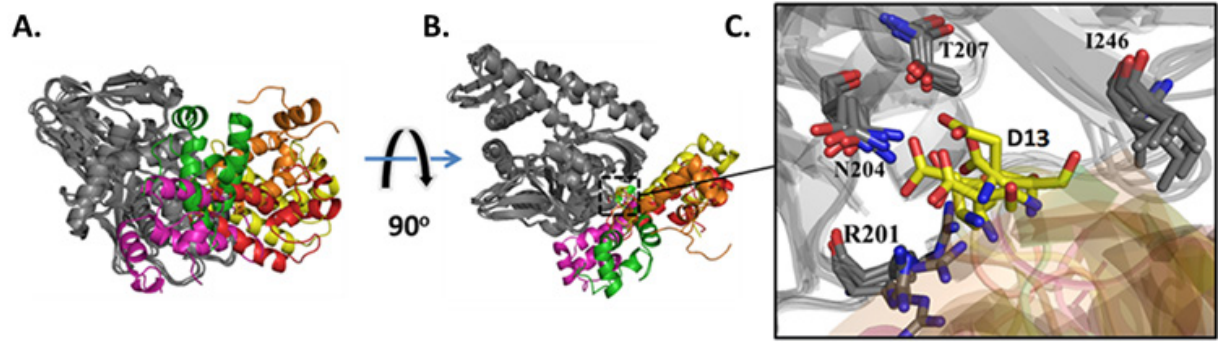

Supplementary Data S10A

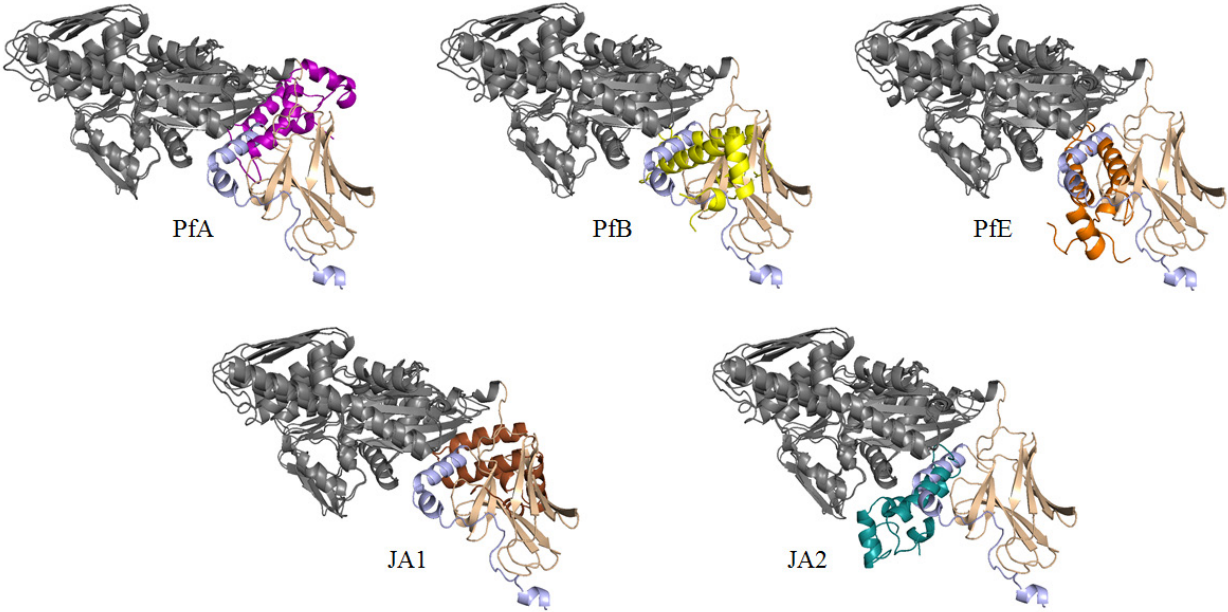

Supplementary Data S10B

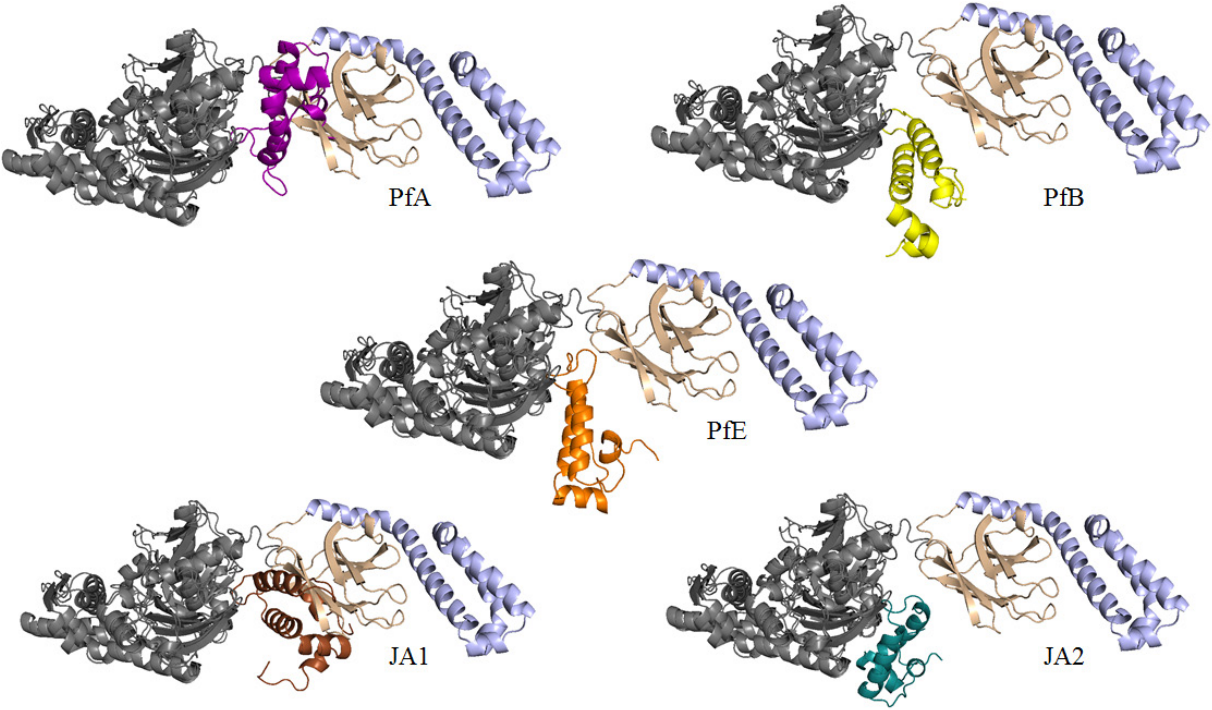